\documentclass[longauth]{aa} % for the abstract without structuration 
\usepackage{cases}
\usepackage[normalem]{ulem}
\usepackage{txfonts}
\usepackage{multirow}
\usepackage{xcolor}
\usepackage{lscape}
\usepackage[breaklinks]{hyperref}
\bibpunct{(}{)}{;}{a}{}{,}
\begin{document}
   \title{Physical properties of the ambient medium and of dense cores in the Perseus star-forming region derived from \textit{Herschel}\thanks{{\it Herschel} is an ESA space observatory with science instruments provided by European-led Principal Investigator consortia and with important participation from NASA.} Gould Belt Survey observations}

   \author{S. Pezzuto\inst{1}\and M. Benedettini\inst{1} \and J. Di Francesco\inst{2,3} \and P. Palmeirim\inst{4} \and S. Sadavoy\inst{5} \and E. Schisano\inst{1} \and G. Li Causi\inst{1} \and Ph. Andr\'e\inst{6} \and D. Arzoumanian\inst{4} \and J.-Ph. Bernard\inst{7} \and S. Bontemps\inst{8} \and D. Elia\inst{1} \and E. Fiorellino\inst{9,10,11,1} \and J.M. Kirk\inst{12} \and V. K\"onyves\inst{12} \and B. Ladjelate\inst{6} \and A. Men’shchikov\inst{6} \and F. Motte\inst{13,6} \and L. Piccotti\inst{14,15,1} \and N. Schneider\inst{16} \and L. Spinoglio\inst{1} \and D. Ward-Thompson\inst{12} \and C. D. Wilson\inst{17}
          }

   \institute{INAF - IAPS, via Fosso del Cavaliere, 100, I-00133 Roma, Italy\\
              \email{stefano.pezzuto@inaf.it}
    \and Department of Physics and Astronomy, University of Victoria, Victoria, BC, V8P 5C2, Canada
    \and NRC Herzberg Astronomy and Astrophysics, 5071 West Saanich Road, Victoria, BC, V9E 2E7, Canada
    \and Instituto de Astrof\'{\i}sica e Ci\^encias do Espa\c{c}o, Universidade do Porto, CAUP, Rua das Estrelas, PT4150-762 Porto, Portugal
    \and Department of Physics, Engineering Physics, and Astronomy, Queen's University, Kingston, ON K7L 3N6, Canada
    \and Laboratoire AIM, CEA/DSM-CNRS-Universit\'e Paris Diderot, IRFU/Service d'Astrophysique, CEA Saclay, 91191 Gif-sur-Yvette, France
    \and Universit\'e de Toulouse, UPS-OMP, IRAP, 31400 Toulouse, France
    \and Laboratoire d’Astrophysique de Bordeaux, Univ. Bordeaux, CNRS, B18N, all\'ee G. Saint-Hilaire, 33615 Pessac, France
    \and INAF - Osservatorio Astronomico di Roma, via di Frascati 33, 00078, Monte Porzio Catone, Italy
    \and Dipartimento di Fisica, Universit\`a di Roma Tor Vergata, via della Ricerca Scientifica 1, 00133, Roma, Italy
    \and ESO/European Southern Observatory, Karl-Schwarzschild-Str. 2, D-85748 Garching bei Munchen, Germany
    \and Jeremiah Horrocks Institute, University of Central Lancashire, Preston, Lancashire PR1 2HE, UK
    \and Universit\'e Grenoble Alpes, CNRS, Institut de Plan\'etologie et d’Astrophysique de Grenoble, 38000 Grenoble, France
    \and Observatorio Astron\'omico Ram\'on Mar\'{\i}a Aller, Universidade de Santiago de Compostela, Santiago de Compostela, E-15782, Galiza, Spain
    \and Instituto de Matem\'aticas and Departamento de Matem\'atica Aplicada, Universidade de Santiago de Compostela, Santiago de Compostela, E-15782, Galiza, Spain
    \and I. Physik. Institut, University of Cologne, Germany
    \and Department of Physics and Astronomy, McMaster University, Hamilton Ontario L8S 4H7 Canada}

   \date{}

  \abstract{The complex of star-forming regions in Perseus is one of the most studied due to its proximity (about 300 pc). In addition, its regions show different activity of star formation and ages, with low-mass and intermediate-mass stars forming. In this paper, we present analyses of images taken with the \textit{Herschel} ESA satellite from 70~$\mu$m to 500~$\mu$m. From these images, we first constructed column density and dust temperature maps. Next, we identified compact cores in the maps at each wavelength, and characterise the cores using modified blackbody fits to their spectral energy distributions (SEDs): we identified 684 starless cores, of which 199 are bound and potential prestellar cores, and 132 protostars. We also matched the \textit{Herschel}-identified young stars with \textit{GAIA} sources to model distance variations across the Perseus cloud. We measure a linear gradient function with right ascension and declination for the entire cloud. This function is the first quantitative attempt to derive in an analytical form the gradient in distances going from West to East Perseus. From the SED fits, mass and temperature of cores were derived. The core mass function can be modelled with a log-normal distribution that peaks at 0.82~$M_\sun$ suggesting a star formation efficiency of 0.30 for a peak in the system initial mass function of stars at 0.25~$M_\sun$. The high-mass tail can be modelled with a power law of slope $\sim-2.32$, close to the Salpeter's value. We also identify the filamentary structure of Perseus and discuss the relation between filaments and star formation, confirming that stars form preferentially in filaments. We find that the majority of filaments where star formation is ongoing are transcritical against their own internal gravity because their linear masses are below the critical limit of 16~$M_\sun\,\mathrm{pc}^{-1}$ above which we expect filaments to collapse. We find a possible explanation for this result, showing that a filament with a linear mass as low as 8~$M_\sun\,\mathrm{pc}^{-1}$ can be already unstable. We confirm a linear relation between star formation efficiency and slope of dust probability density function and a similar relation is also seen with the core formation efficiency. We derive a lifetime for the prestellar core phase of $1.69\pm0.52$~Myr for the whole Perseus but different regions have a wide range in prestellar core fractions, hint that star-formation has started only recently in some clumps. We also derive a free-fall time for prestellar cores of 0.16~Myr.}

\titlerunning{The Perseus population of dense cores}

   \maketitle

\section{Introduction}
The aim of this paper is to derive a catalogue of cold compact cores, dust and gas condensations of $\la0.1$~pc in size, in the Perseus molecular complex and to present and discuss their physical properties. This work uses \textit{Herschel} observations and is part of the ``Herschel Gould Belt Survey'' \citep[HGBS, ][]{2010A&A...518L.102A}, which aims at probing the origin of the stellar initial mass function (IMF) by finding the physical mechanisms influencing star formation at its earliest stages. The HGBS team has already published core catalogues for many clouds: Aquila \citep[][the reference paper for HGBS]{Aquila}; L1495 in Taurus \citep{2016MNRAS.459..342M}; Lupus \citep{milena2}; Corona~Australis \citep{rcra}; Orion~B \citep{orionB}, Ophiuchus \citep[][in press]{bilOph}; Cepheus (Di Francesco et al., ApJ, in press).

The analysis of these data have led to the idea that gravitational fragmentation of the cloud filaments is the dominant physical mechanism generating prestellar cores within interstellar filaments \citep[see ][and references therein]{2014prpl.conf...27A}. It was already suggested in the past that the filamentary structure of the interstellar medium is fragmenting to form cores \citep[see, e.g.,][]{2009ApJ...704..183R}, but only with \textit{Herschel} it became clear that filament fragmentation is the dominant mode of (solar-type) star formation, because the physical properties and the spatial distribution of both the diffuse dust and of the compact cold cores in the clouds can be derived simultaneously, with sufficiently high spatial resolution, 36\arcsec. Moreover, the large areas observed combined with the high sensitivity of \textit{Herschel} instruments, allow to construct catalogues containing a factor 2--9 more cores than in previously ground-based surveys \citep[e.g.,][]{doug,stanke,enoch,nutter,2007A&A...462L..17A}.

The first prestellar core mass functions (CMFs) in star-forming regions, \citep[e.g., ][]{motte,1998ApJ...508L..91T} revealed a similarity between the shape of the CMF and the initial mass function (IMF) of the stars. The \textit{system} IMF shows \citep{IMF2005} a log-normal distribution with a peak at $0.25M_\sun$, and a high-mass tail for $M\ge1M_\sun$ modelled with a power-law function d$N/$d$\log(M)\propto M^{-1.35}$, also known as the Salpeter slope \citep{salpeter}.

A quantitative assessment of the similarity between CMF and IMF was derived for the Pipe nebula by \citet{2007A&A...462L..17A} who found that the peak of the CMF was about 3 times higher than the peak of the IMF. Assuming a one-to-one correspondence between cores and stars, the shift in the peaks of CMF and IMF was interpreted by \citet{2007A&A...462L..17A} as an indication of a star formation efficiency (SFE) per single core of about 0.3, meaning that 30\% of the core mass eventually ends in the star mass. Later studies have confirmed that values around 0.3--0.4 are typical in many other star-forming regions even if different values are also reported (see Section~\ref{secCMF}).% \citep[][found later 0.22 for Ophiuchus]{rath}.}

The similarity between CMF and IMF led to the hypothesis that the IMF is a consequence of the physical mechanisms that produce the prestellar core population. Testing this hypothesis requires to derive the CMF in many star-forming regions in the most possible uniform way with respect to how observations are executed as well as to how cores are detected and their intensities are measured. Only in this way different CMFs can be compared consistently \citep[see, e.g.,][]{swiftBeau}.

Another issue on a proper CMF derivation is that it is very important to assign each core its temperature rather than using a constant temperature for all the cores in a cloud, assumption that brings to ill-determined core masses. Unless we determine both mass and temperature at same time, the stability of cores cannot be investigated on a purely photometric basis, and the derived CMF can contain collapsing cores and transient structures that are not forming stars, as pointed out already 20 years ago by \citet{doug}.

Deriving the CMF for the most important nearby star-forming regions in a robust and uniform way is the main aim of the HGBS.

The Gould Belt is a complex of neutral gas and star-forming molecular clouds covering an area of $\sim2.6\times10^{5}$~pc$^2$ \citep[e.g.,][]{GB}; the molecular clouds are located within distances $d\la500$~pc from the Sun, close enough to observe individual cores. Among them, the system of clouds in Perseus  (Fig.~\ref{gianluca}) are at distances in the range $\sim230$~pc -- 320~pc \citep{hirota8,hirota1,strom} with many sites of active low-mass as well as intermediate-mass star formation. The review by \citet{bally} gives a complete presentation of the region, including its surrounding, with a long list of references; here we limit the literature references to papers that are pertinent to our work or when their results are used in this study.

Perseus was observed as part of the \textit{Spitzer} ``Cores to Disks'' (c2d) program \citep{evans} with the InfraRed Array Camera \citep[IRAC,][]{jorge} and the Multiband Imaging Photometer \citep[MIPS,][]{rebull}. The analysis of both datasets was later presented by \citet{c2d2015}. Perseus was also mapped in the sub-mm with SCUBA \citep{2005A&A...440..151H} and later with SCUBA-2 \citep{SCUBA2}, and at 1.1~mm with BOLOCAM \citep{enoch}.

Previous publications related to HGBS data are the study of B1-E \citep{sarahB1} and of the relation between \textit{Spitzer} Class~0 and the distribution of dense gas \citep[][see also Sect.~\ref{PDFFit} of this work]{sarah2}; \citet{FHSC} discussed the properties of two sources, B1-bN and B1-bS, that were proposed to be first hydrostatic cores candidates (see Sect.~\ref{stability}).

The paper is organised as follows. In Sect.~2, we give an overview of the observations of Perseus and present our data reduction and analysis. In Sect.~3, we derive a first quantitative estimate of the distance gradient across the cloud using using \textit{Gaia} Data Release~2 (DR2) results. Then, we derive and discuss the column density and temperature maps, and the filamentary structure of the region. In Sect.~4, we present the source extraction and the physical properties of the cores derived via SED fitting. We also discuss the core stability against their internal gravity. In Sect.~5, we present the CMF and link the cores to the filaments. We also analyse the core and star formation efficiencies and estimate the lifetimes of the different phases of star formation. In Sect.~6, we summarise our conclusions.

We include several appendices with the full source catalogue, completeness testing, region definitions, and the catalogue of sources found in our maps that are not related to star formation like, e.g., galaxies.

\subsection{Nomenclature}
In this paper, we present a catalogue of protostars and starless cores. We define a core as a compact over density of gas and dust that is round in shape and exceeds the density of its local diffuse background emission appearing as a local maximum in an intensity image or in a column density map. Cores are most easily identified using mid-infrared to mm wavelengths \citep{2007prplJ,2007prpl.conf...33W,2014prpl.conf...27A}.

A core is defined starless if there is no internal source of energy (e.g., a protostar). Starless cores are instead warmed by the interstellar radiation field and their spectral energy distribution (SED) can be modelled as a single modified blackbody $I_\nu\propto\nu^\beta\mathrm{B}_\nu(T)$ at temperature $T$. We use our source extraction technique and the modified blackbody fits to derive the radius, $R$, mass, $M$, and temperature $T$ of the cores, and we use these physical properties to determine the dynamical state of the cores. If the core self-gravity exceeds its pressure support, we consider the source to be bound. We define bound, starless cores as prestellar cores. These objects can collapse, or are collapsing, and will likely form one or more stars. If the core's self-gravity is insufficient to balance its internal pressure, the core is classified as unbound and may be a transient structure that will dissipate in the future unless it is confined by another mechanism (e.g., pressure confined).

Once a star forms, it warms up the surrounding envelope whose emission, in the first phases, is still resembling a modified blackbody for $\lambda\ga100$~$\mu$m while at shorter wavelengths the SED is no longer compatible with a such a model. For this reason, a core with compact 70~$\mu$m emission is considered protostellar because a central source in the centre must be present to warm the dust. Note, however, that in principle a starless core can be detected at 70~$\mu$m if it is sufficiently warm ($T\ga20$~K), so the shape of the SED at short wavelengths determines if the object is already a protostar. On the other hand, an object undetected at 70~$\mu$m is always considered a starless core even if the lack of detection in the PACS band(s) may be just a matter of sensitivity.

The focus of this paper is on starless cores. Protostars are touched upon in the text when necessary, for instance to derive star formation efficency in Sect.~5.3, but a fully discussion on the protostars in Perseus is postponed to a forthcoming paper.

\section{Observations and data reduction}
Perseus was observed with \textit{Herschel} \citep{2010A&A...518L...1P} as part of the HGBS \citep{2010A&A...518L.102A} in two overlapping mosaics: the Western field (mainly NGC1333, B1, L1448, L1455) and the Eastern field (L1468, IC348, B5). Results from these observations were initially presented in \citet{sarahB1,sarah2}, \citet{FHSC}, and \citet{zari}.

Both fields were observed with PACS \citep{PACS} at 70~$\mu$m (blue) and 160~$\mu$m (red), and with SPIRE \citep{SPIRE}, at 250~$\mu$m (PSW), 350~$\mu$m (PMW) and 500~$\mu$m (PLW), in parallel mode with the telescope scanning at a speed of 60$\arcsec$/s. The total area common to both instruments is about 13 square degrees. Table~\ref{log} gives the observation log.

\begin{table}
\caption[]{The log of the observations. OBSID: identifier in the Herschel Science Archive; Date: start of the observation; Centre: the J2000 central coordinates of each field; Size: requested size.\label{log}}
\begin{tabular}{lcc}
\hline
&West Perseus&East Perseus\\\hline
OBSID&1342190326&1342214504\\
&1342190327&1342214505\\
Date&09/02/2010&19/02/2011\\
Centre&3$^\mathrm{h}$29$^\mathrm{m}$34$^\mathrm{s}$ +30$^\mathrm{d}$54\arcmin07\arcsec&3$^\mathrm{h}$42$^\mathrm{m}$51$^\mathrm{s}$ +32$^\mathrm{d}$00\arcmin49\arcsec\\
Size&135\arcmin$\times$150\arcmin&90\arcmin$\times$184\arcmin\\\hline
\end{tabular}
\end{table}

\begin{figure*}
\centering
\includegraphics[scale=.7]{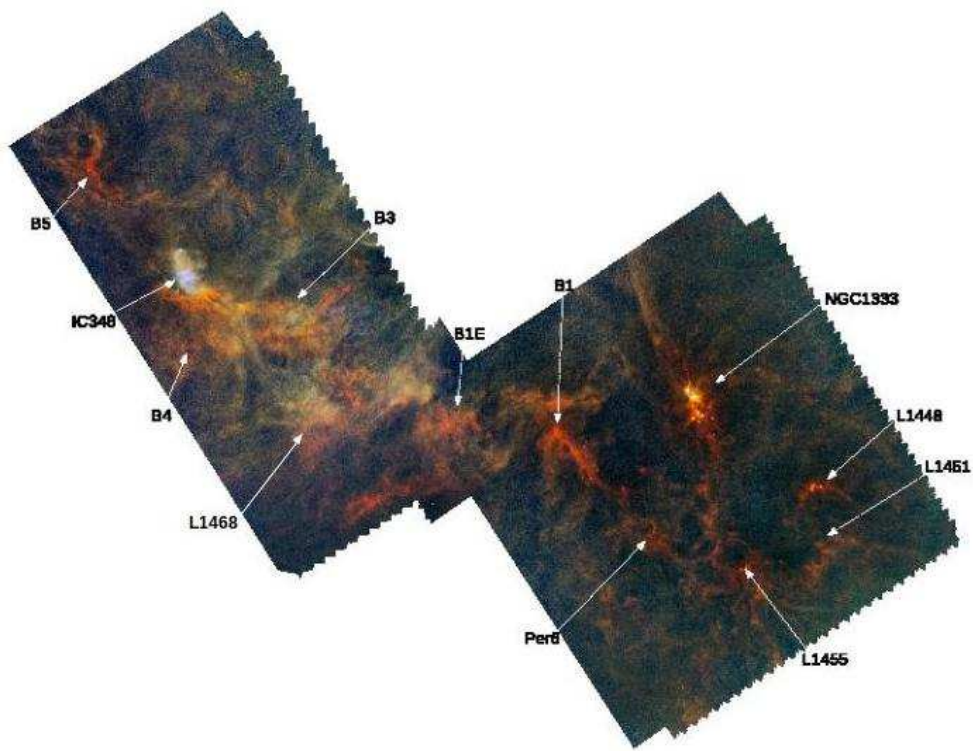}
\caption{An RGB composite of the star-forming region in Perseus. Blue is PACS 70~$\mu$m; green is PACS 160~$\mu$m; red is SPIRE 250~$\mu$m; North is up, East is to the left. The latter two maps have been processed to simulate having the highest resolution of the 70~$\mu$m. The resulting image is then not meant to be used for science analysis. Labels identify on the map the approximate position of main sub-regions. Monochromatic \textit{Herschel} intensity maps are reproduced in Appendix~D with coordinate grids.\label{gianluca}}
\end{figure*}

A composite RGB image using PACS bands and SPIRE PSW band is shown in Fig.~\ref{gianluca}. All the maps have been reported to the same spatial resolution of the 70~$\mu$m map by using the method presented in \citet{gianlu}.

Each field was observed twice along two almost orthogonal directions to remove better instrumental $1/f$ noise. The \textit{Herschel} raw data were reduced with \textsl{HIPE} \citep{HIPE} version~10. For PACS, we reduced the data to Level~1 with \textsl{HIPE}, using version~45 of the set of calibration files. Level~1 data consist of the calibrated timelines of the detectors to which the celestial coordinates were added. Afterwards, the data were exported outside \textsl{HIPE} and maps were generated with \textsl{Unimap} \citep{unimap} version 6.5.3. The adopted spatial grid is 3\arcsec\ and 4\arcsec\ for the blue and red bands, respectively.

For SPIRE, we used the calibration files version 10.1 and maps were generated with the \textsl{destriper} module in \textsl{HIPE}. We adopted the default values for the spatial grids: 6\arcsec, 10\arcsec\ and 14\arcsec\ for PSW, PMW and PLW bands, respectively. The SPIRE maps are given in Jy/beam that we converted into MJy/sr using the nominal beamsizes of 426~arcsec$^2$, 771~arcsec$^2$ and 1626~arcsec$^2$, respectively. Additional details on the PACS and SPIRE data processing can be found in \citet{Aquila}.

The alignment of each image was checked by comparing the \textit{Herschel} position at 70~$\mu$m of 22 bright and isolated sources, with their corresponding positions reported in the Spitzer catalogue at 24~$\mu$m \citep{evans}, 11 each for the Western and Eastern part. For the Western field we found a shift in RA of $\Delta\alpha$=4\farcs82$\pm$0\farcs56 that was added to our coordinates. We also applied a similar shift to the red map. SPIRE maps required no correction. The final residuals in the five maps run from $-$1/4 of a pixel in the PSW map to +1/5 of a pixel in the red map for right ascension, and from 1/10 of a pixel in the PSW map to 1/3 of a pixel in the blue map for declination.

The PACS and SPIRE maps were made using a common coordinates system, at each band, for both Perseus East and West. A zero-level offset was applied to each map, obtained by correlating the \textit{Herschel} data with \textit{Planck} and \textit{IRAS} data \citep{JPB}. %, and added. Afterwards, the final maps were obtained by summing pixel by pixel the West and the East portion at each wavelength; the zero-level offsets are written as keywords in the header of the FITS file, two for each map.
All five maps, shown in the Appendix~\ref{immagini}, are available on the HGBS archive\footnote{\url{http://gouldbelt-herschel.cea.fr/archives}}.

\section{Data analysis}
\subsection{The distance to Perseus}\label{distPer}
Given that physical parameters, like mass and radius, depend on the distance to the sources, we start this section with a discussion on the distance to Perseus.

For IC~348, in the Eastern part of the complex, \citet{herbig} discussed a variety of distance measurements in literature and concluded that the value $316\pm22$~pc derived by \citet{strom} is the most reliable. This distance was later confirmed indirectly by \citet{ripepi} who detected $\delta$-Scuti oscillations in one star of IC~348, and concluded that the derived star properties were not compatible with distances $\la$260~pc, but were compatible with 316~pc. On the other hand, the Western half of Perseus regions is known to be closer, which led to the conclusions that the molecular clouds in Perseus form a chain with increasing distance from West to East \citep[see][]{cernis}. This conjecture was definitively confirmed with the measure of maser parallaxes for a couple of sources in NGC1333 and L1448 \citep{hirota8,hirota1} giving a distance of $\sim240$~pc.

To gain some insight on this gradient in distance, we exploited the \textit{GAIA} archive second data release (DR2) of astrometric data \citep{gaia}. We looked for sources in our protostar catalogue (see Appendix~\ref{appCat}) spatially close to a source in the \textit{GAIA} archive with an angular separation of less than 10\arcsec. Objects with negative parallax were excluded. All sources had spatial correspondences of $<8$\arcsec, with only two having separations $>6$\arcsec\ (6.5\arcsec\ and 7.4\arcsec).  Conversely, 18 are spatially separated by $<2$\arcsec and 7 are separated by $<1$\arcsec. Parallaxes were converted into distances following \citet{luri}. In particular, we used the \verb+DistanceEstimatorApplication.py+ tool\footnote{\url{https://repos.cosmos.esa.int/socci/projects/GAIA/} \url{repos/astrometry-inference-tutorials/browse/single} \url{-source/tutorial}} with the median of the \textit{Exponentially Decreasing Space Density Prior} distribution option. The tool returns also the 5\% and 95\% quantiles that are not symmetrical around the median (this reflects the obvious prior that the distance must be positive). Since the difference between (median-5\% quantile) and (95\% quantile-median) is negligible, we adopted the larger (95\%-median), for the distance uncertainty. No attempt was made to exclude projection effects. Hence, some associations between our sources and \textit{GAIA} stars may not be physical. On the other hand, it is unlikely that a proximity of few arcseconds between a \textit{GAIA} source at the distance expected for Perseus and a protostar is only due to projection.

In Table~\ref{distanze} we show distances and uncertainties for the 28 sources possibly associated with our protostars. We also report the distances for the maser sources in L1448 \citep[][h1 in the table]{hirota1} and NGC1333 \citep[][h2]{hirota8} derived using the python tool reported above to convert the parallax to distance.

\begin{table*}
\centering
\caption{Sources from our protostar catalogue with \textit{GAIA} counterpart within 10\arcsec; we listed also h1 (L1448 C) and h2 (SVS13 in NGC1333) which host masers from which an astrometric distance was obtained \citep{hirota8,hirota1}. For all the sources, the distance in parsec was derived from parallax according to \citet{luri}.\label{distanze}}
\begin{tabular}{rccc@{$\pm$}c|cccl@{$\pm$}c|cccc@{$\pm$}c}
\hline
ID&RA&Dec&\multicolumn{2}{c|}{Distance}&ID&RA&Dec&\multicolumn{2}{c|}{Distance}&ID&RA&Dec&\multicolumn{2}{c}{Distance}\\\hline
 1)&51.27800&31.11428&299  &10&10)&52.23893&31.23733&291  &10 &19)&55.41336&31.60310&300  &11\\
 2)&51.28956&30.77220&285  &65&h2)&52.26552&31.26772&240  &36 &20)&55.49376&31.81556&312  &25\\
h1)&51.41199&30.73479&237  &26&11)&52.37169&31.30924&315  &24 &21)&55.72903&31.72932&330  &13\\
 3)&51.45769&31.17290&292  &11&12)&52.47598&31.34729&296  &17 &22)&55.73202&31.97893&313  &10\\
 4)&51.61836&31.20151&292  &14&13)&52.71848&30.90509&327  &66 &23)&56.11088&32.07526&330  &19\\
 5)&51.94876&30.20121&250  &29&14)&52.82618&30.82677&302  &12 &24)&56.14806&32.15777&333  &89\\
 6)&52.15382&31.29370&374  &60&15)&52.87012&30.51453&295.1&8.6&25)&56.31799&32.10562&308  &23\\
 7)&52.18035&31.29248&205  &36&16)&53.14148&31.01549&307  &19 &26)&56.33484&32.10952&310  &20\\
 8)&52.21291&31.30397&296  &31&17)&55.19518&32.53161&308  &13 &27)&56.45099&32.40305&311  &13\\
 9)&52.23529&31.30954&345  &40&18)&55.28775&31.74389&326  &28 &28)&56.94595&33.06746&343  &76\\\hline
\end{tabular}
\end{table*}

\begin{figure}
\centering
\includegraphics[scale=0.4]{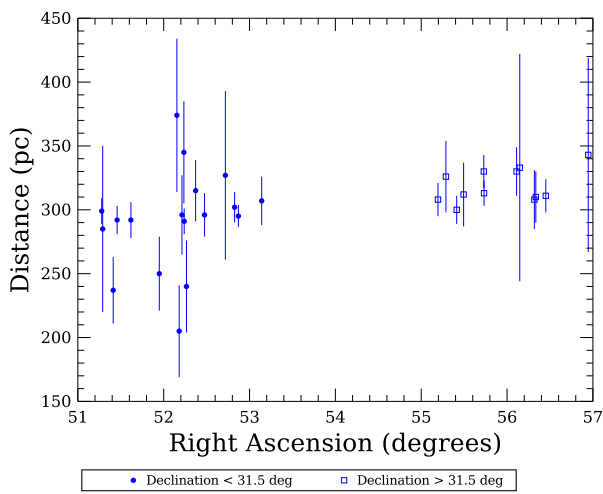}
\caption{Relation between distance and Right Ascension for the 30 sources in Table~\ref{distanze}.\label{dRA}}
\end{figure}

In Fig.~\ref{dRA} the 30 distances are plotted as a function of their right ascension. Starting from the observational finding that a distance gradient exist going from West to East, and noting that a gradient also exists going from South to North, see the different symbols of the points in the figure, we made a multivariate linear fit to the data and we found the following relation
\begin{equation}
d(\mathrm{pc})=281.8\pm3.4+(5.13\pm0.65)\times(\alpha-\alpha_0)+(5.5\pm1.9)\times(\delta-\delta_0)\label{fitDistanza}
\end{equation}
$\alpha_0=51\fdg27800$ and $\delta_0=30\fdg20121$ are the smallest right ascension and declination of sources in Table~\ref{distanze}, source~1 and source~5, respectively. Such a relation $d\equiv d(\alpha,\delta)$ is the first quantitative estimate of the distance gradient in Perseus even if the large scatter in distances could be due to false associations between our protostars and \textit{GAIA} sources. In particular, if all the associations in NGC1333 are physical, an exceptional depth of more than 100~pc, or an overlap along the line of sight of at least two different clouds, is implied (see also Appendix~\ref{GaiaCom}). Further, there are no data points (no associations) in the range $53\fdg14\le\alpha\le55\fdg20$, so the middle part of the fit is not constrained. A refinement of Equation~(\ref{fitDistanza}) will be attempted after our protostar catalogue will separate Class~II pre-main sequence stars, that are more easily detected by \textit{GAIA}, from younger sources, Class~0 objects, that are more obscured in the visible light and, for this reason, more difficult to be present in the \textit{GAIA} archive. An association between a Class~0 source and a \textit{GAIA} object may be due to a projection effect.

Recent distance measurements to the main Perseus clouds have been reported by \citet{zucker} and \citet{ortiz}. Table~\ref{confDist} compares these distances: Col.~P gives the distance based on our Equation~(\ref{fitDistanza}) using right ascension and declination from \citet{zucker}, Col.~Z gives the \citet{zucker} distances, and Col.~O gives the \citet{ortiz} values. The distances from \citet{zucker} are derived by combining data from PAN-STARRS1, 2MASS, \textit{Gaia} and the \element[][12]{C}\element[][]{O} data from the COMPLETE survey \citep{COMPLETE} (we use the \element[][13]{C}\element[][]{O} data in Sect.~\ref{secMasse} and Appendix~\ref{GaiaCom}). \citet{ortiz} measure distances using VLBA and \textit{Gaia} data. Uncertainties in Col.~P, derived through error propagation from Equation~(\ref{fitDistanza}), can be misleading. They are small because they give the (estimated) distance of each point in the sky assuming a 2D geometry for Perseus. These uncertainties do not catch the dispersion in distance caused by the depth of each cloud, and they do not take into account that in the region with right ascension between $\sim53^\circ$ and $\sim55^\circ$, there are no protostars, and then we do not have indications on the distance. Uncertainties in Col.~Z are the sum in quadrature of the random and sistematic errors given by \citet{zucker}.

\begin{table}
\caption{Comparison between distances, in pc, derived by different author: RA and Dec are in degrees as quoted by \citet{zucker}; P distance based on our Equation~(\ref{fitDistanza}); Z distance from \citet{zucker}; O distance from \citet{ortiz}.\label{confDist}}
\begin{tabular}{lccccc}
\hline
Name&RA&Dec&P&Z&O\\\hline
L1451&51.0&30.5&282.3$\pm$3.5&279$\pm$14&\\
L1448&51.2&30.9&285.5$\pm$3.7&288$\pm$14&\\
NGC1333&52.2&31.2&292.3$\pm$4.0&299$\pm$14&293$^{+24}_{-21}$\\
B1&53.4&31.1&297.9$\pm$4.1&301$\pm$15&\\
IC348&55.8&31.8&314.0$\pm$5.5&295$\pm$15&320$^{+28}_{-24}$\\
B5&56.9&32.9&325.7$\pm$7.2&302$\pm$16&\\\hline
\end{tabular}
\end{table}

We find excellent agreement (< 3\% differences) between our measurements and those of \citet{zucker} and \citet{ortiz}. Only IC 348 and B5 show more significant differences but still < 10\% and, in any case, within the uncertainties.

\citet{zucker2} derived a distance map of Perseus and other molecular clouds through per-star distance–extinction estimates improved with \textit{GAIA} parallaxes. The result for a small set of lines of sight is reported in Table~\ref{confZuc} where we compare their distances with ours: the agreement is very good, better than 5\% in all cases.

\begin{table}
\caption{Comparison between the distance, in pc, derived in this paper (column P) and that derived by \citet{zucker2} (column Z) for the four lines of sight reported in their Table~3 for the Perseus cloud.\label{confZuc}}
\begin{tabular}{cccc}
\hline
RA&Dec&P&Z\\\hline
51.5455&30.6747&285.7$\pm$3.5&289$\pm$15\\
52.0097&30.1235&285.1$\pm$3.4&279$\pm$14\\
52.0315&30.6110&287.9$\pm$3.5&289$^{+16}_{-15}$\\
52.4212&31.3616&294.0$\pm$4.1&281$^{+15}_{-14}$\\\hline
\end{tabular}
\end{table}

\begin{figure*}
\centering
\sidecaption
\includegraphics[scale=0.8]{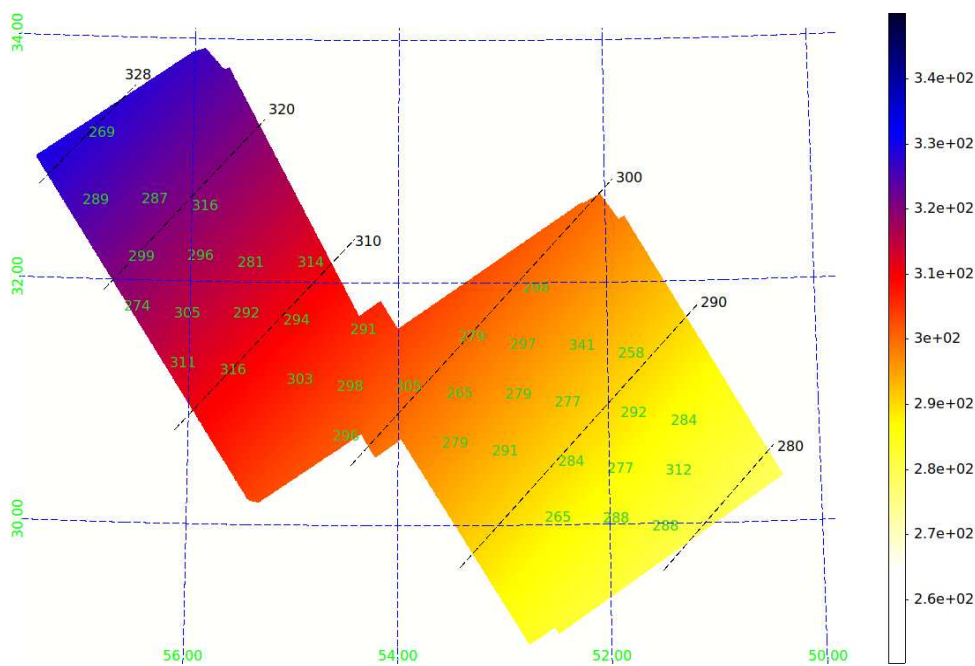}
\caption{Color-coded 2D visualisation of Equation~(\ref{fitDistanza}). Color bar is in parsec, coordinates grid in degrees. Black dashed lines connect points at the same distance, shown in the labels. Green labels are the distances derived by \citet{zucker2} in boxes of 0.84$^{\circ^2}$, the labels are at the centre of the boxes.\label{distFits}}
\end{figure*}

The full set of distances derived by \citet{zucker2} is shown in Fig.~\ref{distFits} where we give a graphical representation of Equation~(\ref{fitDistanza}): our distance was computed for each point of the column density map (see Sect.~\ref{cMap}) and coded in colour according to the bar on the right of the figure. Dashed black lines connect points at the same distance, reported at the top of each line. The coordinates grid, in blue, is in step of 2$^\circ$. \citet{zucker2} give the distance averaged over boxes of 0.84~deg$^2$: they are shown in green at the position of the centre of each box.

While, on one hand, \citet{zucker2} computed accurate distances to Perseus that are more precise than those given by our Equation~(\ref{fitDistanza}), it is not easy, on the other hand, to extract a distance for a given point in space from their map; for instance, B5 is put at 302~pc in \citet{zucker}, distance that can be hardly derived from the grid of distances shown in Fig.~\ref{distFits}. Conversely, our Equation~(\ref{fitDistanza}) has been derived in a less rigorous way, but it gives immediately a reference distance at any point $(\alpha,\delta)$ in space, distance that captures the distance gradient in the complex, and matches very well with the distances derived by other authors.

For the rest of the paper, we adopted the mean of the 30 sources, 300~pc, as a \textbf{representative} value of the distance for the whole Perseus complex \citep[294~pc in ][]{zucker2}. We will, however, show how the physical description of the whole region changes when distances are varied according to our formula.

\subsection{Dust temperature and column density maps}\label{cMap}
Dust emission in the far-infrared can be modelled as a modified blackbody $I_\nu=\kappa_\nu\Sigma B_\nu(T_\mathrm{d})$, where $\kappa_\nu$ is the dust opacity per unit mass expressed as a power law normalised to the value $\kappa_0$ at a reference frequency $\nu_0$ so $\kappa_\nu=\kappa_0(\nu/\nu_0)^\beta$~cm$^2$g$^{-1}$. $\Sigma$ is the gas surface-density distribution $\mu_{\mathrm{H}_2}m_{\mathrm H}N(\mathrm{H}_2)$ with $\mu_{\mathrm{H}_2}$ the molecular weight, $m_{\mathrm H}$ the mass of the hydrogen atom and $N(\mathrm{H}_2)$ the column density assuming the gas is fully molecular hydrogen. Finally, %\footnote{In the text we use cm$^{-2}$ as units of $N(\mathrm{H}_2)$ instead of molecules~cm$^{-2}$.}
$B_\nu(T_\mathrm{d})$ is the Planck function at the dust temperature $T_\mathrm{d}$. Following the GBS conventions \citep{Aquila}, we adopted an opacity of 0.1~cm$^2$g$^{-1}$ at 300~$\mu$m (gas-to-dust ratio of 100), a dust opacity index $\beta = 2$, and $\mu_{\mathrm{H}_2}=2.8$ representing the mean weight per hydrogen molecule \citep{mmw}.

We used the four intensity maps from 160~$\mu$m to 500~$\mu$m to derive $N(\mathrm{H}_2)$ and $T_\mathrm{d}$ maps. The 70~$\mu$m map was not included because emission at this wavelength may include contributions from very small grains (VSGs), which can not be modelled with a simple single-temperature modified blackbody. Indeed, \citet{schnee} found that in Perseus contribution from VSGs may elevate the 70~$\mu$m emission by 70\% at 17~K and by 90\% below 14~K. Contribution from VSGs is not expected for $\lambda>100$~$\mu$m \citep{lidraine}.

The three intensity maps shortwards of 500~$\mu$m were degraded to the spatial resolution of the PLW band (36\farcs1), then the four maps were reported to a common coordinate system using 14\arcsec\ pixels at all wavelengths.

The SED fitting procedure was executed pixel by pixel with a code that takes as input the four images in FITS format. The code creates a grid of models, as in \citet{FHSC}, by varying only the temperature, in the range $5\le T_\mathrm{d}(\mathrm{K})\le 50$, in step of 0.01~K. For each temperature $T_j$, the code computes the intensity at far-infrared wavelengths for a fixed column density of 1~cm$^{-2}$.

Since $I_\nu$ is linear with $N(\mathrm{H}_2)$, we can compute the column density at each pixel using a straightforward application of the least-squares technique
\begin{equation}
N(\mathrm{H}_2)_j=\frac{\sum_i f_iq_i(T_j)/\sigma_i^2}{\sum_i q^2_i(T_j)/\sigma_i^2}\label{minQua}
\end{equation}
where $f_i$ is the observed SED $I_\nu$ at each frequency $i$ and $q_i(T_j)$ is the synthetic SED model at frequency $i$ and temperature $T_j$. The uncertainty $\sigma_i$ is 10\% and 20\% for SPIRE and PACS respectively \citep{Aquila}. Index $i$ runs from 1 to 4, index $j$ runs from 1 to 4501, the number of models in the grid.

At each pixel of the intensities maps, the model $j$ with the smallest residuals is kept as best-fit model. To be consistent with the other HGBS papers, we run the code without applying colour corrections but in Sect.~3.3 we show how much the results depend on this assumption and on the choice of $\beta=2$.

The code, available on request, outputs the column density map, the temperature map, and the map of the uncertainty of $N(\mathrm{H}_2)$.

Another column density map at higher spatial resolution (i.e., 18\farcs2 corresponding to the SPIRE 250~$\mu$m band) was also obtained using the method described in \citet{pedro}. The procedure is based on a multi-scale decomposition technique. The high-resolution column density map, used to optimise the source extraction described in Sect.~\ref{estrazione}, is available for downloading on the HGBS website. This map, shown in Appendix~D, has an higher spatial resolution than those previously obtained with the same \textit{Herschel} data at 36\farcs1 \citep{sarah2,zari}.

\begin{figure*}
\sidecaption
\centering
\begin{tabular}{c}
\includegraphics[scale=.75]{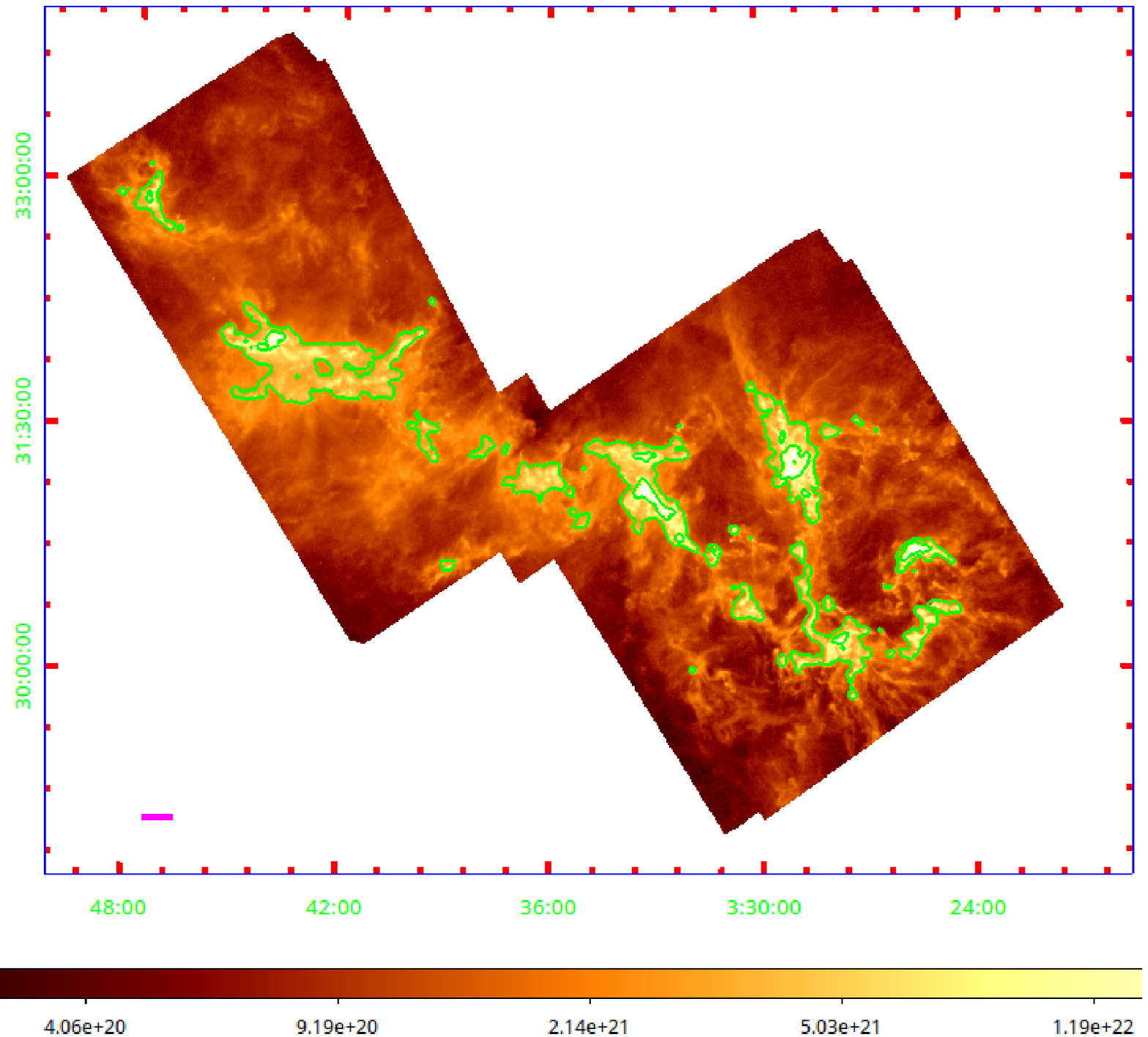}\\
\includegraphics[scale=.75]{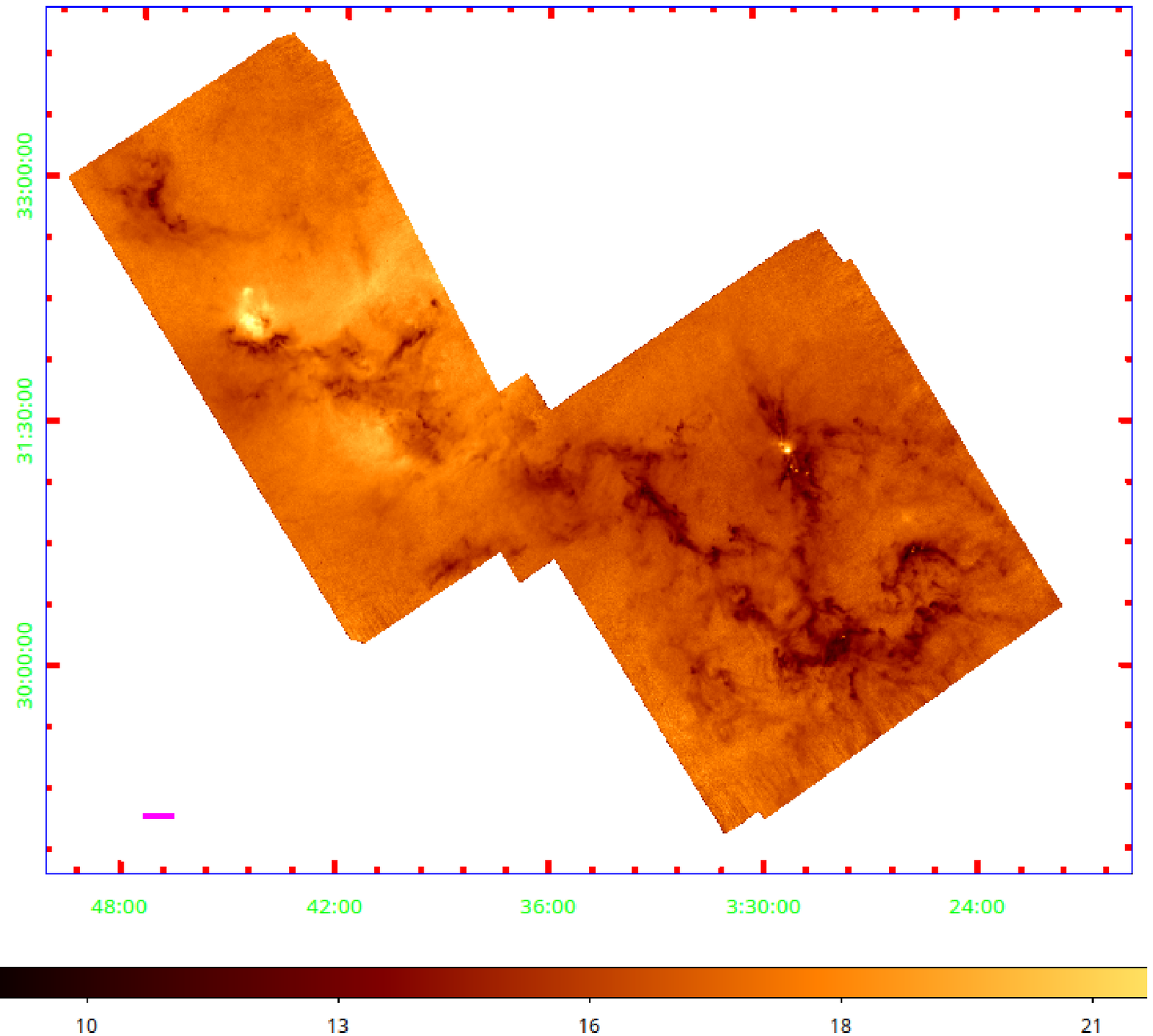}\\
\end{tabular}
\caption{Top panel: column density map with contours at $3\times 10^{21}$~cm$^{-2}$ and $10^{22}$~cm$^{-2}$. Bottom panel: dust temperature map.  In both panels the magenta line in the bottom left corner shows the angular scale corresponding to 1~pc at 300~pc; J2000.0 coordinates grid is shown. Both maps have a spatial resolution of 36\farcs1. The anticorrelation between $N(\mathrm{H}_2)$ and $T_{\mathrm{d}}$ is evident: regions at high/low column density are cold/warm, with few exceptions in IC348 and NGC1333.\label{NHT}}
\end{figure*}

The column density and temperature maps of Perseus at the PLW spatial resolution are shown in the two panels of Fig.~\ref{NHT}. The top panel shows the column density map and the bottom panel shows the temperature map. Contour levels in the column density map are $3\times 10^{21}$~cm$^{-2}$ and $10^{22}$~cm$^{-2}$. The first level follows quite well the border of the known main regions, even if B3 and B4 form a single complex with IC348, while the other marks the densest part of the clouds.

\begin{figure}
\centering
\includegraphics[scale=0.4]{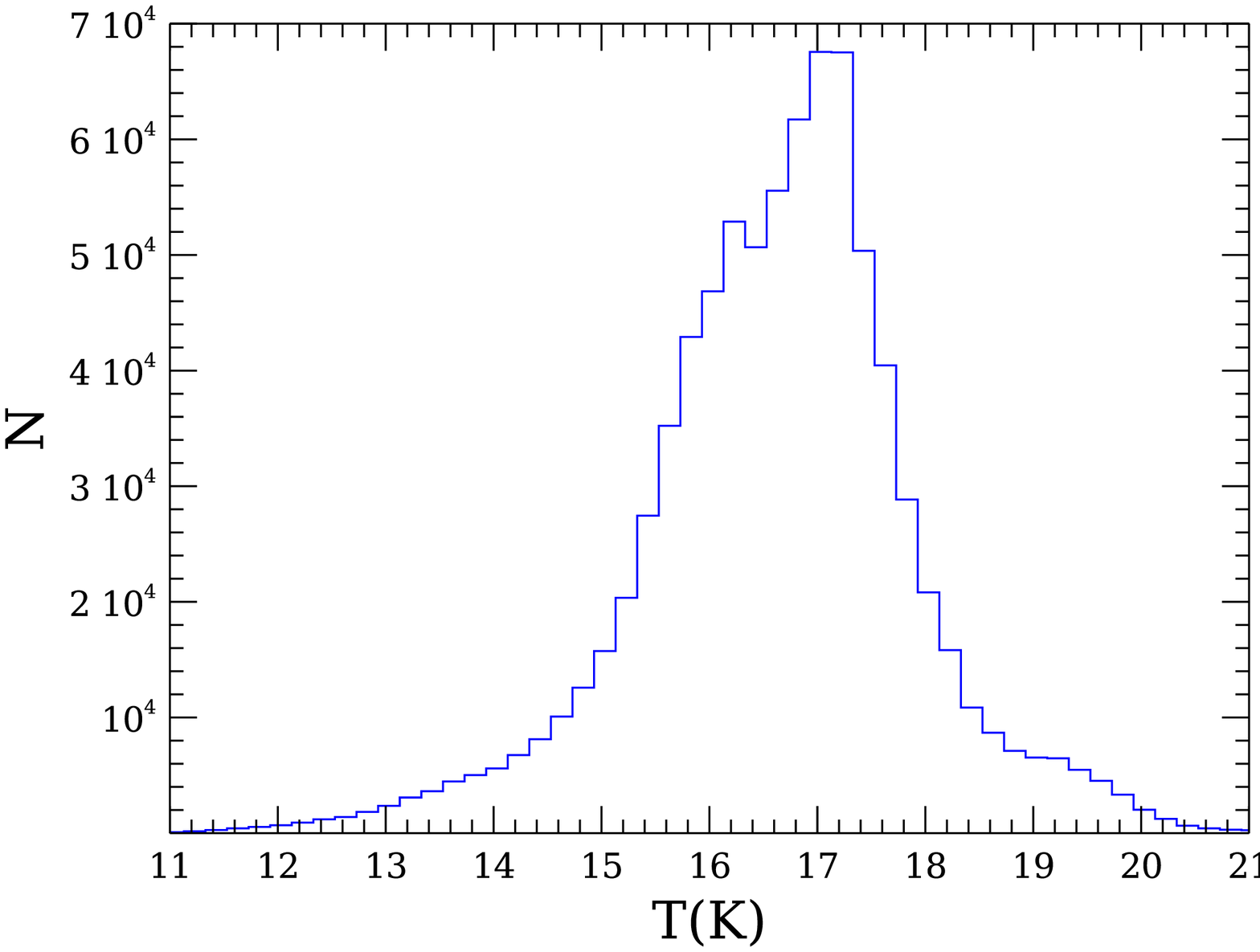}
\caption{Histogram of dust temperature in the range 11 -- 21~K (minimum is 10.1~K, maximum is 28.3~K). \label{istoDust}}
\end{figure}

The histogram of $T_\mathrm{d}$ is shown in Fig.~\ref{istoDust}. There are two peaks seen at 16.4~K and 17.1~K, and a third, very broad one, a plateau indeed, at 19.2~K. The first two peaks correspond to the temperatures of the diffuse medium in West and East Perseus, respectively. In other words, the dust temperature is slightly lower in the Western half of Perseus than in the East. The peak at 19.2~K reflects the inner parts of NGC1333 and IC348, and a few other regions that we discuss below.

To check our results, we compared the \textit{Herschel} $N(\mathrm{H}_2)$ map with the all-sky \textit{Planck} map of the optical depth $\tau_\mathrm{P}$ at 850~$\mu$m \citep{planck}\footnote{\texttt{HFI\_CompMap\_ThermalDustModel\_2048\_R1.20.fits} available at \url{http://pla.esac.esa.int/pla}}. First we convolved our map to the \textit{Planck} resolution of 5\arcmin\ and projected the result onto the \textit{Planck} grid. We then computed the ratio $r=\tau_\mathrm{H}/\tau_\mathrm{P}$ where $\tau_\mathrm{H}=0.1(3/8.5)^2\mu m_\mathrm{H}N(\mathrm{H}_2)$. Because of the convolution, the \textit{Herschel} column density and, as a consequence, the optical depth, have very low values close to map borders, with $\tau_\mathrm{H}$ as low as $6\times10^{-21}$. To exclude these points, we made the comparison in the region where $\tau_\mathrm{H}\ge1.38\times10^{-5}=\mathrm{min}(\tau_\mathrm{P})$.

\begin{figure}
\centering
\includegraphics[scale=0.4]{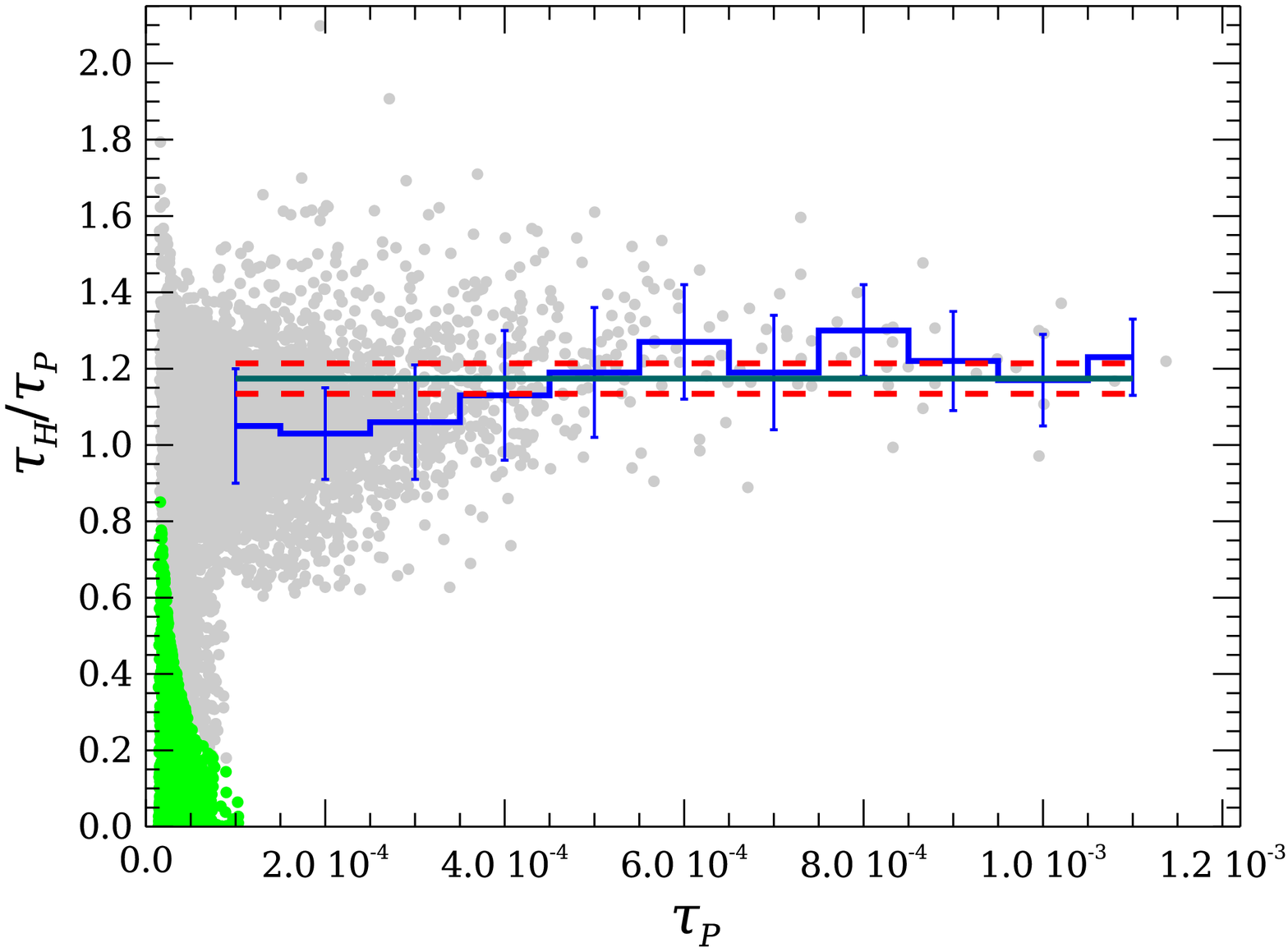}
\caption{With the grey points we show here the ratio of $\tau_\mathrm{H}$, the optical depth derived from our \textit{Herschel} observations, over $\tau_\mathrm{P}$, the optical depth derived from \textit{Planck} observations \citep{planck}, both $\tau$'s computed at 850~$\mu$m. The ratio is against $\tau_\mathrm{P}$. Green points correspond to low values of $\tau_\mathrm{H}$ due to the convolution applied to the \textit{Herschel} column density map (see text). Mean values $\pm$ one standard deviation of the grey points, excluding the green ones, are shown with the blue histogram. The dark green and the two red lines show the weighed mean of the ratio: $1.174\pm0.040$.\label{rappT}}
\end{figure}

In Figure~\ref{rappT} we show in grey the ratios $r$ vs. $\tau_\mathrm{P}$; with green points we highlight the values corresponding to $\tau_\mathrm{H}<1.38\times10^{-5}$. The blue histogram shows the mean ratios, averaged in bins of $10^{-4}$, excluding the green points. The weighed mean of the histogram values is $\bar{r}=1.174\pm0.040$ (shown as dark green line in Figure~\ref{rappT}, with the two red dashed lines giving $r\pm1\sigma$) and all the blue points fall in the red-lines region. This implies that the histogram is compatible with a constant ratio for $r$. On the other hand, an increasing trend of $r$ with $\tau_\mathrm{P}$ seems present in the figure, with $r\la1.05$ for $\tau_\mathrm{P}<3\times10^{-4}$, and increasing up to 1.30 when $\tau_\mathrm{P}\sim8\times10^{-4}$. In particular, 98\% of the points of the optical depth map have $1.38\times10^{-5}\le\tau_\mathrm{P}\le3\times10^{-4}$, and for them the ratio $r$ is $1.05\pm0.15$. A change in $r$ might witness a change in opacity at high column density.

\subsubsection{On the possible contribution of VSG}
\begin{figure*}
\centering
\includegraphics[scale=1.1]{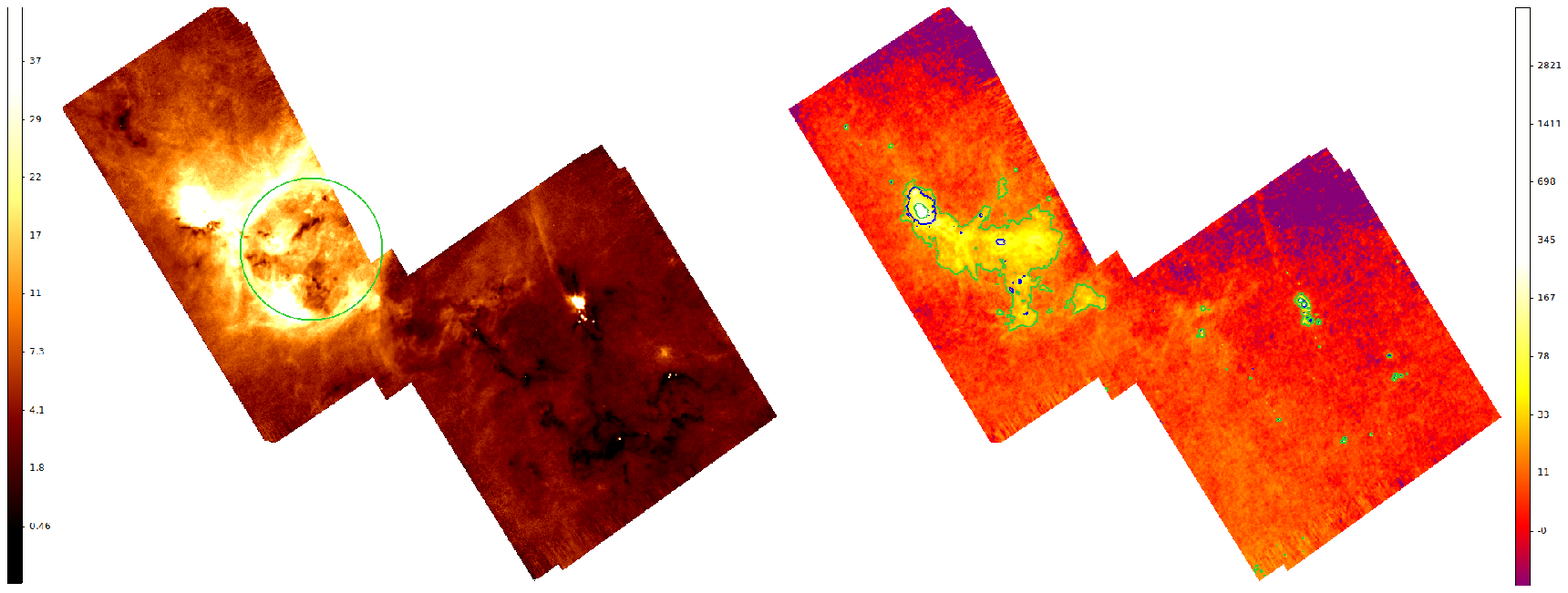}
\caption{Left panel: the inferred 70~$\mu$m intensity map extrapolated from the column density and temperature maps obtained for $\lambda\ge160$~$\mu$m. The green circle shows the bubble found by \citet{shell} \citep[CP5 in the list by][]{bubbles}. Right panel: difference between the observed 70~$\mu$m map and the extrapolated map. Green contours are 20~MJy~sr$^{-1}$ and 200~MJy~sr$^{-1}$; blue contours show the region where $T_\mathrm{PACS only}\ge20$~K (see text). Colourbars are in MJy~sr$^{-1}$.\label{diff70}}
\end{figure*}

As written, the 70~$\mu$m map was not used to derive the column density map, because of possible VSG emission at this wavelength. We can, however, estimate this contribution by computing the intensity map at 70~$\mu$m from the column density and temperature maps obtained for $\lambda\ge160$~$\mu$m. Fig.~\ref{diff70} compares the extrapolated 70~$\mu$m map with the difference map between the extrapolated and observed data. All maps have been convolved to 36\farcs1 resolution. The most striking feature of Fig.~\ref{diff70} is that the dense subregions in Perseus are expected to be dark against the brighter diffuse background emission if given sufficient sensitivity. Such 70~$\mu$m dark clouds are typically classified as infrared dark clouds \citep[IRDCs,][]{IRDC}. The observed data (see Appendix~\ref{immagini}), however, do not show the subregions as silhouettes.

The right panel of Fig.~\ref{diff70} shows the difference between the observed 70~$\mu$m map, degraded in spatial resolution and projected onto the 500~$\mu$m spatial grid, and the computed 70~$\mu$m map. If the far-infrared dust emission can be extrapolated by a modified blackbody distribution, then the difference map should approach zero everywhere. This is indeed true for most of Perseus with the exception of NGC1333 and the complex IC348/B3/B4/L1468, with differences as high as $10^3$~MJy~sr$^{-1}$ in NGC1333.

As first hypothesis we can suppose that there is not a population of VSG and that the differences in observed and computed 70~$\mu$m flux, are due to a poor estimation of $T_\mathrm{d}$. In fact, for a modified blackbody, $T$, $\beta$ and the wavelength where the SED peaks $\lambda_\mathrm{peak}$ are related through the relation \citep{grey}
\begin{equation}
\lambda_\mathrm{peak}=\frac{1.493}{T(K)(3+\beta)}\,\,\,\mathrm{cm} \label{maxGB}
\end{equation}
If $T=18.7$~K, the SED peaks at $\lambda_\mathrm{peak}=160$~$\mu$m if $\beta=2$. At higher temperatures, $\lambda_\mathrm{peak}$ moves shortwards of 160~$\mu$m so that the peak wavelength, and then $T$, is poorly determined from our dataset built for $\lambda\ge160$~$\mu$m. To quantify this effect, we derived the column density and temperature maps using only PACS data. To this aim we used an additional PACS intensity map at 100~$\mu$m, also observed as part of the HGBS and that will be the subject of a future paper focused on Class~0 objects.

The 70~$\mu$m and 100~$\mu$m maps were degraded to the red band spatial resolution, projected onto the spatial grid of the latter image, and then fitted following the same pixel-by-pixel technique described previously. Since the 100~$\mu$m observation covered a smaller area than the PACS/SPIRE parallel-mode observations, and because the PACS intensities have larger uncertainties in the zero-level of the diffuse emission, the resulting $N(\mathrm{H}_2)$ and $T$ maps have small coverage and higher noise when compared to the ``nominal'' maps obtained for $\lambda\ge160$~$\mu$m.

The new temperature map, $T_\mathrm{PACS only}$, was then degraded to the 500~$\mu$m spatial resolution and projected onto the spatial grid of the nominal $T$ map, $T_\mathrm{160 + SPIRE}$.

\begin{figure}
\centering
\includegraphics[scale=.4]{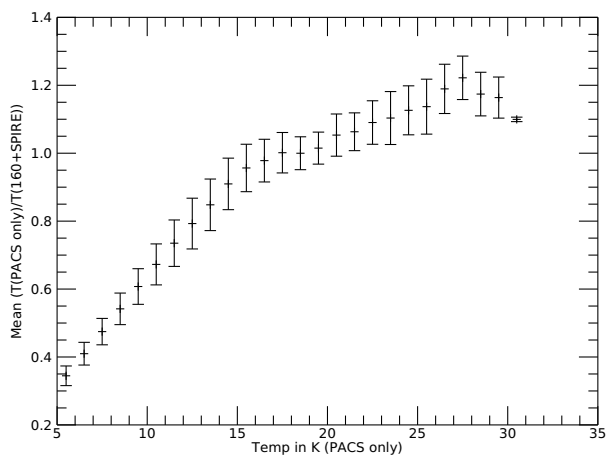}
\caption{Change in dust temperature when PACS data only are used (70~$\mu$m, 100~$\mu$m and 160~$\mu$m) instead of PACS 160~$\mu$m plus SPIRE bands. Error bars are one standard deviation. The point at $T\sim30$~K is not reliable because mean of few values.\label{TPTS}}
\end{figure}

In Fig.~\ref{TPTS}, we show, as a function of $T_\mathrm{PACS only}$, the mean of the ratio $T_\mathrm{PACS only}/T_\mathrm{160 + SPIRE}$ in bins of 1$^\circ$~K. At low temperatures $T_\mathrm{PACS only}$ is a poor estimate of $T$, as expected. More interesting, however, is the trend for $T_\mathrm{PACS only}>20$~K. In this regime, corresponding to $\lambda_\mathrm{peak}\la150$~$\mu$m, the peak of the SED falls outside the range of wavelengths used to derive $T_\mathrm{160 + SPIRE}$. On the other hand, the PACS bands span the SED peak, making $T_\mathrm{PACS only}$ a more reliable measurement of the dust temperature, at least as long as $T\la43$~K (above this temperature the SED peak moves at $\lambda<70\,\mu$m). From the figure, we see that $T_\mathrm{160 + SPIRE}$ starts being colder than $T_\mathrm{PACS only}$ for $T>20$~K (for $T_\mathrm{PACS only}>28$~K, the value of the mean is likely affected by the small number of points).

The blue contours in the right panel of Fig.~\ref{diff70} show the regions where $T_\mathrm{PACS only}\ge 20$~K. These regions are confined to the inner parts of IC348 and NGC 1333. Clearly, underestimation of $T_\mathrm{d}$ cannot explain the large difference where the observed emission at 70~$\mu$m is higher than that computed from the 160 + SPIRE column density and temperature maps. The VSG population in Perseus may be the cause of this excess. If so, however, it is clear that VSGs are detected primarily in the Eastern field. We also note that the bubble CP5 \citep{bubbles} for which \citet{shell} derived $T\sim29$~K with IRAS 60~$\mu$m and 100~$\mu$m data, is not visible in the right panel of Fig.~\ref{diff70}, suggesting that its 70~$\mu$m emission seen by \textit{Herschel} is actually compatible with a modified blackbody at lower temperatures, i.e., less than $\sim$20~K. HD~278942, thought to be the driving source of the bubble, has a distance of $\sim$520~pc \citep{gaia} excluding that this star can be indeed related to the shell.

The fact that $T$ can be underestimated when $T>20$~K could have, in principle, an impact on the core temperature derived from SED fits when a source is not visible in the PACS blue band. For the starless cores, however, almost all of their temperatures are $<$20~K with only a few exceptions. Sources with temperature $>20$~K are detected at 70~$\mu$m so that the observed SED covers the region of the peak intensity.

\subsubsection{Masses and temperatures of the Perseus subclouds}\label{secMasse}
From the intensity maps at the different wavelengths or the column density map, it is not clear how to define the borders of individual subregions of Perseus. The contour at $3\times10^{21}$~cm$^{-2}$ shown in Fig.~\ref{NHT} provides a good first order estimate of the dense material, but it does not separate the individual subregions well. Further, our choice of $3\times10^{21}$~cm$^{-2}$ does not have a physical meaning and one could adopt another level of $N(\mathrm{H}_2)$ and measure a different mass.

To find a solution for the border definition, we started drawing a polygon enclosing each $3\times10^{21}$~cm$^{-2}$ contour. Indeed, a polygon is
much easier to handle in a computer code given the fact that it is defined with much fewer vertices. Hence, we used such polygons to get a first guess for the borders. Then we visualized the \element[][13]{CO} 1--0 \citep{COMPLETE} spectrum inside each area whose perimeter was varied by eye to find locations with one velocity component. This approach, however, was possible only in a few cases because most of regions have multiple CO components. So, in the end, we defined the different subclouds trying to have one bright line with few \textit{contaminants}. The areas so defined are shown in Fig.~\ref{zoneNH2} while in Figs.~\ref{mappe13COa} to \ref{mappe13COb} of Appendix~\ref{GaiaCom} they are overplotted onto the \element[][13]{CO} intensity maps at different velocity components to make it clear why a certain region was defined with that border. We introduced additional zones not associated with already known subregions, naming them HPZ\#, for Herschel Perseus Zone number \#. In Appendix~\ref{GaiaCom} we also give the coordinates of the corners for all regions in DS9 format.

\begin{figure*}
\centering
\includegraphics[scale=0.8]{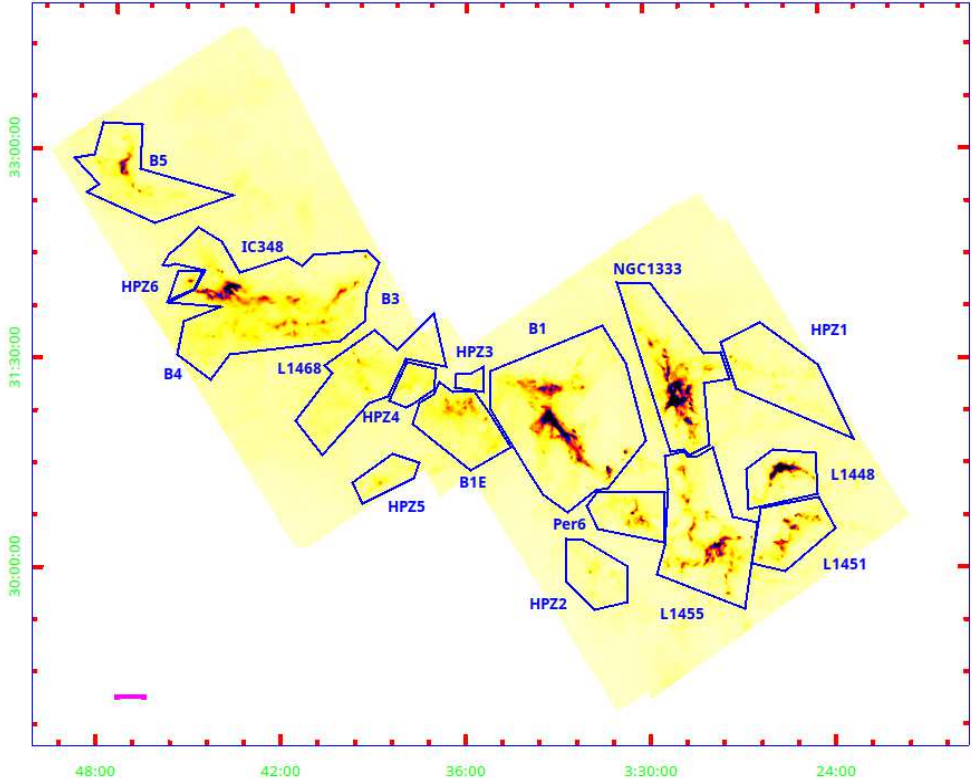}
\caption{The regions identified in Perseus overplotted on the column density map. The magenta line in the bottom left corner shows the angular scale corresponding to 1~pc at 300~pc; J2000.0 coordinates grid is shown. Coordinates of the regions are given in Appendix~\ref{GaiaCom}.\label{zoneNH2}}
\end{figure*}

In Table~\ref{masse} we report the main physical properties of each identified sub-region. Cols. labelled RA and Dec give, in degrees, the position of the peak in column density reported in fourth Col. (for HPZ1, the peak is located on the far-east side of the subregion, so we computed the geometrical centroid of the zone). In the next Cols. we give mass, area and median temperature for the whole region and within the denser part where $N(\mathrm{H}_2)>3\times10^{21}$~cm$^{-2}$. The last two Cols. show the effect of assigning to each point of the column density map a different distance, following Equation~(\ref{fitDistanza}). Specifically, $M_{d(\mathrm{\alpha,\delta})}$ shows how the total mass of the cloud is affected, and Col. $\bar{d}$ gives the distance at which $M_{d(\mathrm{\alpha,\delta})}=M_{\bar{d}}$, where $\bar{d}=\sqrt{\sum N_i(\mathrm{H}_2)d^2(\alpha_i,\delta_i)/\sum N_i(\mathrm{H}_2)}$, with $i$ running over all the pixels of each subregion. The total mass, area, median temperature, and the mean distance for the entire cloud within specific column density ranges are given in Table~\ref{masseTot}. From the values of $\bar{d}$ in this table we conclude that our choice of 300~pc as representative distance is reasonable.

\begin{table*}
\caption{The properties of the sub-regions of the Perseus molecular cloud. Coordinates give the position of the peak in column density reported under Peak. The next six Cols. give the mass, area, and median temperature, with two values for each parameter. The first value refers to the entire subregion within the boundaries shown in Fig.~\ref{zoneNH2}. The second value applies only for the data with $N(\mathrm{H}_2)>3\times10^{21}$~cm$^{-2}$. Column labeled $M_{d(\mathrm{\alpha,\delta})}$ gives the mass of each subregion using a modified distance according to Equation~(\ref{fitDistanza}), with $\bar{d}$ being the mean distance for the entire region. The last three Cols. are the number of cores detected: Pro(tostars), Pre(stellar cores) and U(n)B(ound) cores.\label{masse}}
\begin{tabular}{lcc@{\hspace{2em}}c@{\hspace{2em}}rrrrccrcrrr}
\hline
\rule{0pt}{1.\normalbaselineskip} Name&RA&Dec&Peak&\multicolumn{2}{c}{$M$}&\multicolumn{2}{c}{Area}&\multicolumn{2}{c}{Med($T$)}&\multicolumn{1}{c}{$M_{d(\mathrm{\alpha,\delta})}$}&$\bar{d}$&Pro&Pre&UB\\
&\multicolumn{2}{c}{J2000}&($10^{21}$~cm$^{-2}$)&\multicolumn{2}{c}{($M_\odot$)}&\multicolumn{2}{c}{(arcmin$^2$)}&\multicolumn{2}{c}{(K)}&($M_\odot$)&(pc)\\\hline
\rule{0pt}{1.\normalbaselineskip}L1451&51.32&30.32&\phantom{0}16.9\phantom{0}&276&125&746&140&15.1&13.3&246&283&0&19&8\\
L1448&51.40&30.76&106\phantom{.00}&313&194&610&122&15.2&13.2&285&285&4&18&7\\
L1455&51.92&30.20&\phantom{0}37.0\phantom{0}&740&375&1897&420&15.0&13.3&677&286&8&36&17\\
NGC1333&52.29&31.23&138\phantom{.00}&1060&727&1661&538&15.2&13.8&1021&294&44&23&23\\
Perseus6&52.64&30.44&\phantom{0}19.1\phantom{0}&203&74&579&93&15.5&13.6&191&290&5&4&2\\
B1&53.34&31.13&112\phantom{.00}&1443&780&3444&657&15.8&14.2&1423&297&15&31&38\\
B1E&53.98&31.24&\phantom{00}9.65&389&157&887&215&15.6&14.7&395&301&0&8&9\\
L1468&54.91&31.53&\phantom{00}7.87&455&49&1391&79&17.4&15.7&479&307&0&4&18\\
IC348\tablefootmark{a}&55.98&32.01&\phantom{0}57.6\phantom{0}&1568&881&3258&1082&17.2&16.5&1730&314&26&41&51\\
B5&56.62&32.49&\phantom{0}23.9\phantom{0}&410&106&1225&122&16.4&14.6&484&325&3&6&6\\
HPZ1\tablefootmark{b}&51.37&31.40&\phantom{00}7.98&275&8&1282&13&15.9&14.3&257&289&0&1&9\\
HPZ2&52.98&30.01&\phantom{00}6.73&116&3&575&5&16.1&13.9&108&289&0&1&5\\
HPZ3&53.89&31.40&\phantom{00}3.88&24&1&92&2&16.3&14.4&24&302&0&0&1\\
HPZ4&54.54&31.34&\phantom{00}4.70&93&26&243&44&17.5&17.1&96&305&1&0&1\\
HPZ5&54.72&30.66&\phantom{00}6.33&90&12&315&18&15.9&14.3&92&302&0&1&0\\
HPZ6&56.32&32.08&\phantom{0}11.4\phantom{0}&40&17&101&23&16.8&15.9&46&318&2&2&2\\\hline
\end{tabular}
\tablefoot{
\tablefoottext{a}{Includes regions B3 and B4;}
\tablefoottext{b}{coordinates of the geometrical centre.}}
\end{table*}

\begin{table}
\caption{The properties of the Perseus molecular cloud as a whole in three different regimes of column density; mass $M$, area and median temperature Med($T$); column labeled $M_{d(\mathrm{\alpha,\delta})}$ gives the mass as in Table~\ref{masse} using Equation~(\ref{fitDistanza}) for the distance; $\bar{d}$ is the mean distance.\label{masseTot}}
\begin{tabular}{crccrc}
\hline
\rule{0pt}{1.\normalbaselineskip} $N($H$_2)$&$M$&Area&Med($T$)&\multicolumn{1}{c}{$M_{d(\mathrm{\alpha,\delta})}$}&$\bar{d}$\\
($10^{21}$~cm$^{-2}$)&($M_\odot$)&(degrees$^2$)&(K)&($M_\odot$)&(pc)\\\hline
$\le3$\tablefootmark{a}&8116&11.5&16.8&8300&303\\
$>3$&3546&1.0&14.5&3560&300\\
$>10$&954&303\tablefootmark{b}&12.7&937&296\\\hline\end{tabular}
\tablefoot{
\tablefoottext{a}{Minimum value in our map is $2.51\times10^{20}$~cm$^{-2}$;}
\tablefoottext{b}{area is in arcmin$^2$.}}
\end{table}

The mass for $N($H$_2)>1\times10^{22}$~cm$^{-2}$ is 954~$M_\sun$ at 300~pc. With the same set of data \citet{sarah2} found 1171~$M_\sun$ when adopting 235~pc, that translates into 1908~$M_\sun$ at our distance of 300~pc, a factor 2 higher than our value. The main difference between the two analyses is that we adopted a newer version of the calibration files with slightly different SPIRE beamsizes. Since the mass enclosed within a certain contour scales with the area defined by that contour, small variations in column density can cause large variations in mass. In particular, scaling by 30\% the column density derived by \citet{sarah2}, the area decreases from 600~arcmin$^2$ to 303~arcmin$^2$ and the mass reduces to 944~$M_\sun$, in agreement with the values derived by us.

We also compare the \textit{Herschel}-derived column density maps with the near-infrared extinction maps based on star counts \citep[see, e.g.,][]{laurent,nicolaAv}. Such a comparison, however, is not immediate. \citet{milena} found large discrepancies between the masses derived with these two methods in Lupus, due to the uncertainties on the opacity law used to compute the \textit{Herschel} column density map. The situation is more difficult for Perseus because of the known variability of $R\equiv A_\mathrm{V}/\mathrm{E(B-V)}$. \citet{foster} found a strong correlation between $R$ and $A_\mathrm{V}$ in Perseus, with the former increasing from $\sim$3 to $\sim$5 when the latter goes from 2~mag to 10~mag. Without a-priori knowledge of the relation $R=R(A_\mathrm{V})$, it is not possible to translate $N($H$_2)$ to $A_\mathrm{V}$. Moreover, $A_\mathrm{V}$ maps are also subject to a zero-level uncertainty, in the sense that stars must be counted relative to a fiducial zone where $A_\mathrm{V}=0$~mag. Since Perseus is a large cloud, it is difficult to find such a clean region.

To investigate further, we derived the mass enclosed within a certain contour of $A_\mathrm{V}$ using both our column density map and a 2MASS-based extinction map (Cambr\'esy 2015, private communication), taking into account the different spatial resolution and pixel size. We use the nominal conversion $N(\mathrm{H}_2)=9.4\times10^{20}A_\mathrm{V}$~mag$^{-1}$~cm$^{-2}$ from \citet{bsd}, which assumes fully ionised hydrogen and $R=3.1$. The mass found for $A_\mathrm{V}>3$~mag is 3620~$M_\sun$ in our map and 9980~$M_\sun$ in the extinction map, a factor 2.76 higher. For $A_\mathrm{V}>5$~mag, however, the ratio decreases to 2.35 (1830~$M_\sun$ and 4290~$M_\sun$). The comparison improves if we derive the mass enclosed within the same area instead of the same $A_\mathrm{V}$. The area is found projecting the $A_\mathrm{V}$ contour derived on the extinction map onto our column density map. In this way we measure in our map 5523~$M_\sun$ and 2548~$M_\sun$ for $A_\mathrm{V}>3$~mag and 5~mag, respectively, with a ratio mass(extinction map)/mass(column density map) of 1.81 and 1.68 for the two contours. A qualitative similar trend is found in Orion~B \citep{orionB}. All in all, the masses derived from our \textit{Herschel-based} column density map are accurate to within a factor of $\sim1.5-2$ for $A_\mathrm{V}>3$~mag. At low column densities, $A_\mathrm{V}<3$~mag, we may overestimate the dust opacity and, consequently, underestimate the masses by a factor of $\sim2-3$ \citep{milena}.

To test the hypothesis that $R$ can play a role in causing column density discrepancies, we varied $R$ until a good agreement was found between the extinction map and thermal dust map. We chose a fiducial extinction level of $A_\mathrm{V}=7$~mag to make this comparison, so that any uncertainties in both maps from the zero-point corrections are negligible relative to the measured column of dust. We then searched iteratively for the value of $R$ that equalises the masses found from the two maps. We find a best match for $R=3.95$ which implies $N(\mathrm{H}_2)=7.34\times10^{20}A_\mathrm{V}$~mag$^{-1}$~cm$^{-2}$. With this conversion, the cloud mass for $A_\mathrm{V}>7$~mag is roughly 1600~$M_\sun$ from both maps. Assuming the nominal value $R=3.1$, the thermal dust map gives a mass of 1213~$M_\sun$ and the extinction map gives 2046~$M_\sun$.

Another study of the extinction in the Perseus molecular cloud has been done by \citet{zari} using the same set of Herschel images. They give the mass enclosed within a set of $A_\mathrm{K}$ contours based on the relation $A_\mathrm{K}=1.67\times10^{22}[2N(\mathrm{H}_2)+N(\mathrm{H})]$~mag~cm$^{-2}$ that becomes $A_\mathrm{K}=8.35\times10^{21}\,N(\mathrm{H}_2)$~mag~cm$^{-2}$ assuming fully molecular hydrogen. To take into account the different mean molecular weight ($1.37\times2$ instead of our value 2.8) and the different adopted distance (240~pc instead of 300~pc), we increased their masses by 1.53.

Finally, we had to consider the different dust opacity, which was more tricky. By combining their relation between $A_\mathrm{K}$ and $N(\mathrm{H}_2)$ with their Equation~(4) that relates $A_\mathrm{K}$ with the optical depth at 850~$\mu$m, we derived
\begin{equation}
\kappa_{850}=\frac{A_\mathrm{K}-\delta}{8.35\times10^{21}\mu_{\mathrm{H}_2}m_{\mathrm H}A_\mathrm{K}}\label{kappaAK}
\end{equation}
The fact that their zero-point $\delta$ is not zero, makes $\kappa_{850}$ a function of $A_\mathrm{K}$; on the other hand, if $A_\mathrm{K}\gg\|\delta\|=0.05$~mag, then $\kappa_{850}\rightarrow0.0065$~cm$^2$~g$^{-1}$. Since we have assumed a dust opacity index $\beta=2$, $\kappa_{300}=0.052$~cm$^2$~g$^{-1}$; and because $I_\nu\propto\kappa_\nu\Sigma$, lowering $\kappa_\nu$ by a factor $0.052/.1=0.52$, implies increasing our values of $N(\mathrm{H}_2)$ by the factor $1/0.52=1.92$.

We downloaded \citet{zari}'s $\tau_{850}$ map from CDS and converted it in $A_\mathrm{K}$ following their prescription. Our $N(\mathrm{H}_2)$ was projected onto theirs, to take into account the different pixel size, and then we computed the mass enclosed within the $A_\mathrm{K}$ contours used by \citet{zari}. For $A_\mathrm{K}>0.2$~mag the ratio between the mass $M_\mathrm{Z}$, derived by \citet{zari}, and $M_{\mathrm{P}}$, the mass estimated from our column density map, is 1.72 instead of the expected value of 1.92. This small difference, 10\% in mass, can be due to the different way to derive the zero-level of the intensity maps, combined with the fact that for 0.6~mag, the zero-point $\delta$ in Equation~(\ref{kappaAK}) is not yet negligible. The ratio of the masses increases with increasing $A_\mathrm{K}$ until, for $A_\mathrm{K}>6.4$~mag, it reaches 1.93, as expected.

We finish this section by showing how the mass derived from the intensity maps depends on the spatial resolution of the column density map, on the dust opacity index $\beta$ fixed to 2 and on neglecting colour corrections. In Table~\ref{RMass} we report the masses enclosed within different $N(\mathrm{H}_2)$ levels and for $A_\mathrm{V}>7$~mag, converted in column density value using $R=3.1$. The first set of Cols. show the masses from our data for $\beta=2$ and the second set show our data for $\beta=1.7$, which is the the peak of the distribution of $\beta$ in the Perseus region \citep{planck}. For each $\beta$ case, we measure masses first without applying a colour correction (No CC) and then with applying the colour correction (see below for details).  Finally, in the middle Col. for the $\beta=2$ set, we also give the mass when the column densities are convolved to the Planck resolution (5\arcmin) with 90\arcsec\ pixel size.

\begin{table}
\caption{Mass cloud derived with two dust emissivity indices $\beta$, without and with colour corrections applied, and with a different spatial sampling of the column density map. The value $\beta=2$ is the HGBS standard, $\beta=1.7$ is the peak of the distribution of $\beta$ in the Perseus region \citep{planck}. Masses in Cols.~(2), (3) and (5) were derived without applying colour corrections, while in Cols.~(4) and (6) the reported masses were computed with colour corrections. In all cases but Col.~(3), the column density map was created on a grid of size 14\arcsec\ and spatial resolution of 36\arcsec, default values for HGBS. In Col.~(3) the column density map was projected onto the grid used for \textit{Planck} data: 90\arcsec\ pixels and 5\arcmin\ resolution. The 7~mag contour has been drawn assuming $R=3.1$.\label{RMass}}
\begin{tabular}{lccc|cc}
\hline
\rule{0pt}{1.\normalbaselineskip}&\multicolumn{3}{c|}{$\beta=2$}&\multicolumn{2}{c}{$\beta=1.7$}\\\cline{2-6}
\rule{0pt}{1.\normalbaselineskip}&\multicolumn{2}{c}{No CC}&CC&\multicolumn{1}{c}{No CC}&CC\\
&\multicolumn{5}{c}{(Masses in $10^3M_\sun$)}\\
\multicolumn{1}{c}{(1)}&(2)&(3)&(4)&(5)&(6)\\\hline
\rule{0pt}{1.\normalbaselineskip}$N(\mathrm{H}_2)\le3\times10^{21}$&8.12&8.38&7.27&6.41&5.66\\
$N(\mathrm{H}_2)>3\times10^{21}$&3.55&3.28&2.59&1.73&1.29\\
$N(\mathrm{H}_2)>10^{22}$&0.95&0.59&0.66&0.72&0.37\\
$A_\mathrm{V}>7$~mag&1.54&1.14&1.14&0.74&0.56\\\hline
\end{tabular}
\end{table}

In general, smoothing the map decreases the column density in the densest parts and increases the column density in the more diffuse regions surrounding the denser material.  For example, at $N(\mathrm{H}_2)>10^{22}$~cm$^{-2}$, we recover only 62\% of the mass measured in the smoothed Herschel map at 5\arcmin\ resolution compared to the original map at 36\arcsec\ resolution, whereas for $N(\mathrm{H}_2)<3\times10^{21}$~cm$^{-2}$, we find a slightly higher mass in the 5\arcmin\ resolution map.

With $\beta=1.7$, the column density and cloud mass decreases substantially. We find values that are roughly 75--80\% what was obtained with $\beta=2$.

Table~\ref{RMass} shows that the colour correcions (CC) decrease the estimated masses by 70--90\%. Such correcions are in general necessary because the flux calibration of PACS and SPIRE was performed assuming that the SED of a source displays a flat $\nu F_\nu$ spectrum\footnote{This is a common assumption when calibrating a photometer.}. For all other kinds of SEDs, the derived fluxes must be colour-corrected according to the intrinsic source spectrum. To compute the column density with color corrections applied, our fitting code integrates the synthetic SEDs over the PACS and SPIRE response filters during the generation of the grid. In this way, each model has its own CC built-in.

In the rest of this paper we used the maps at the spatial resolution of 500~$\mu$m derived as in the other HGBS works: no CC applied and $\beta$ fixed to 2.

\subsubsection{The filamentary structure of Perseus}\label{filDescr}
One of the main results of \textit{Herschel} in the field of star formation is the discovery of the deep link between the filamentary structure of molecular clouds and the sites where stars form. In particular, star-forming cores are found preferentially in denser filaments \citep{2010A&A...518L.102A,HIGALSDP,rayner}.

Based on \textit{Herschel} data, \citet{danae} derived two distinct core mass functions in L1641, part of the Orion~A complex, for sources inside and outside of filaments. The mass distribution of the sources on the filaments was found to peak at 4~$M_\sun$ with a CMF at higher masses modelled with a power law d$N/$d$\log M\propto M^{-1.4}$. The mass distribution of the sources off the filaments has the peak at 0.8~$M_\sun$ and a flat CMF at masses lower than $\sim4\,M_\sun$.

\citet{arabindo} found a possible link between the 1D power spectrum of filaments and the origin of the high-mass tail in the core mass function (very close to Salpeter's law d$N/$d$M\propto M^{-2.5}$).

A complete study of the filaments in Perseus is not in the scope of this paper, nevertheless, given the aforementioned results, it is important to give here at least a short summary of the filaments main properties that will be later discussed in Sect.~\ref{discussione} where the relation with the core population will be addressed.

We used the filament detection algorithm\footnote{The code is freely available at the following URL \url{http://vialactea.iaps.inaf.it/vialactea/eng/tools.php}} of \citet{eugenio2020,eugenio} to identify filamentary structures and their properties across our column density map. Here we summarise how the algorithm works adopting the nomenclature described in those papers.

By thresholding the minimum eigenvalue of the Hessian of the column density map, the code is able to find and encompass regions where there are maximum variations of the contrast, i.e., the bright features on the map. Filamentary structures are picked among these features through selection criteria on the elongation and coverage of the region of interests.% \citep[][Schisano et al., submitted]{eugenio}.

Once the regions of interest are identified, the \textsl{IDL}\footnote{\textsl{IDL} is registered trademark of Harris Corporation.} morphological operator \textsl{THIN} is applied to them. The result of the operator is the region ``skeleton''. Because of the nested morphology of filamentary features, the skeleton in each region is composed by one or more ``branches''. The spine of the filament is defined as the group of consecutive branches, connected each other, tracing the longest possible path over the filamentary region.

\citet{eugenio} found that the area of the filament is underestimated by a mere thresholding of the eigenvalue, that only traces where the emission is concave downwards. They also found that by enlarging the border by three pixels on both direction perpendicular to the spine gives a better estimate of the position where filaments merge with the background. We verified that such approach is valid also for our case and we applied it to our filament sample.

The code was run on the 36\arcsec-resolution column density map. Among the identified features, we selected as filaments the elongated regions having a spine longer than 12 pixels, corresponding to 0.24~pc at 300~pc, or about 4 times the spatial resolution at 500~$\mu$m.

\begin{figure*}
\centering
\includegraphics[scale=1.5]{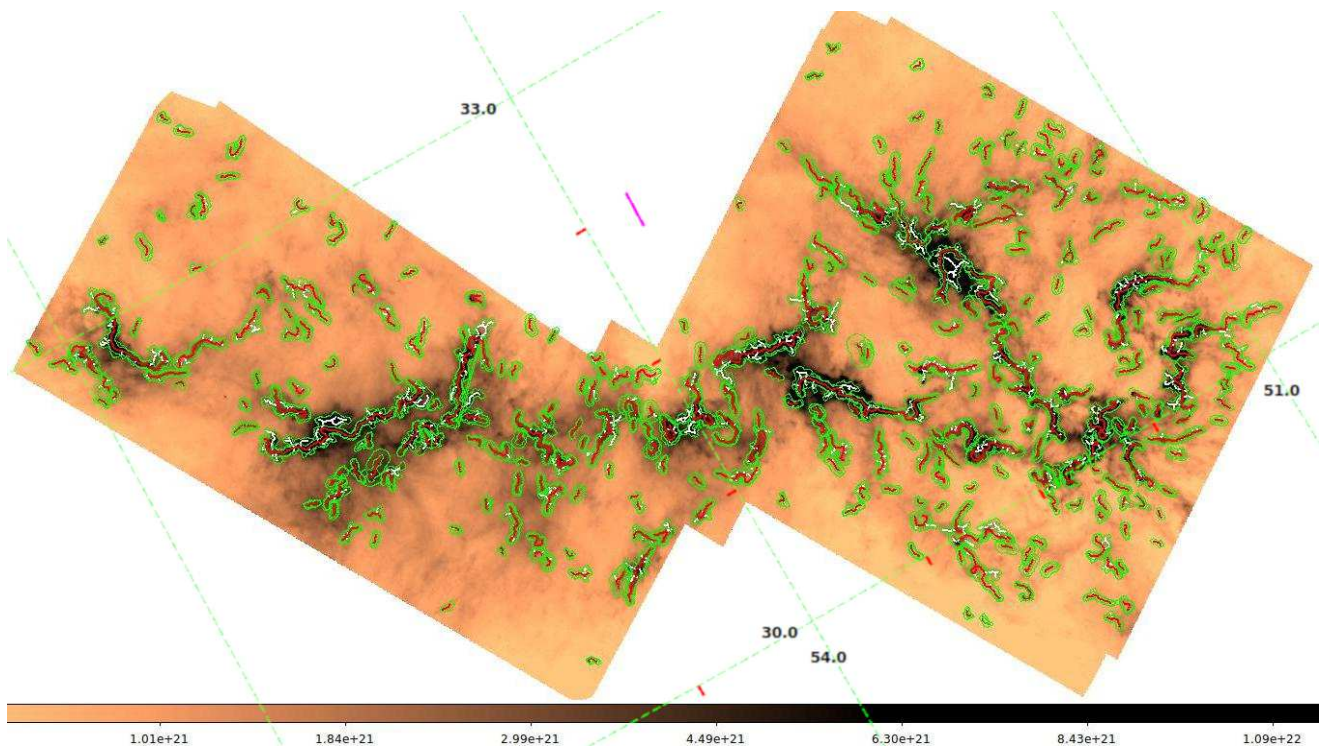}%\\
\caption{Network of filaments overplotted on the column density map. Red lines show the spine of the filaments, white lines show the branches, and the green contours show the width of the filaments and branches. Map has been rotated by 28$^\circ$. The magenta line in the centre shows the angular scale corresponding to 1~pc at 300~pc; J2000.0 coordinates grid is shown.\label{filaments}}
\end{figure*}

The result of the extraction is shown in Fig.~\ref{filaments} where the filaments are overlapped on the column density map. Green lines show the filament borders, red and white lines are the spines and the branches, respectively.

\begin{figure*}
\centering
\begin{tabular}{cc}
\includegraphics[scale=.5]{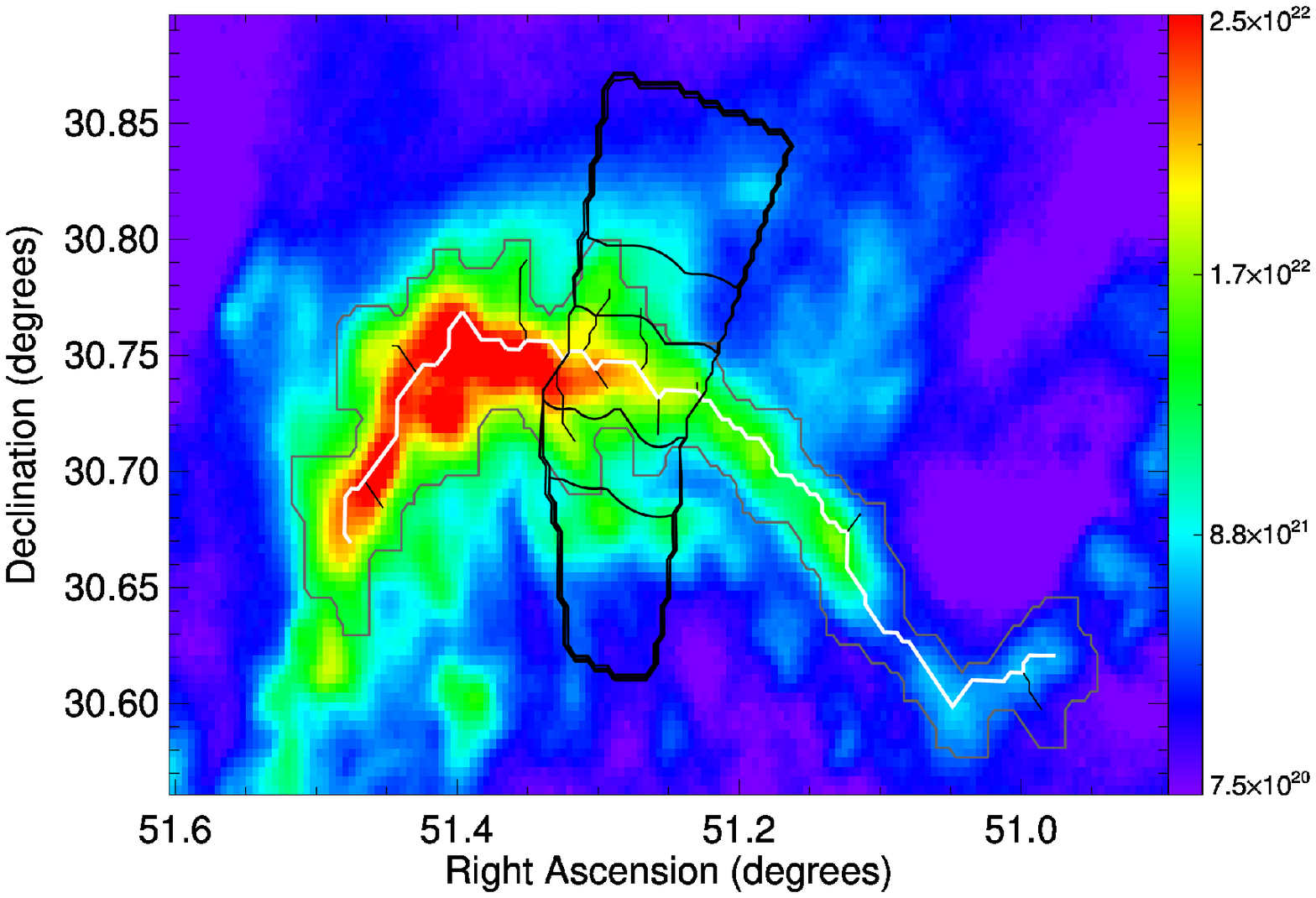}&\includegraphics[scale=.5]{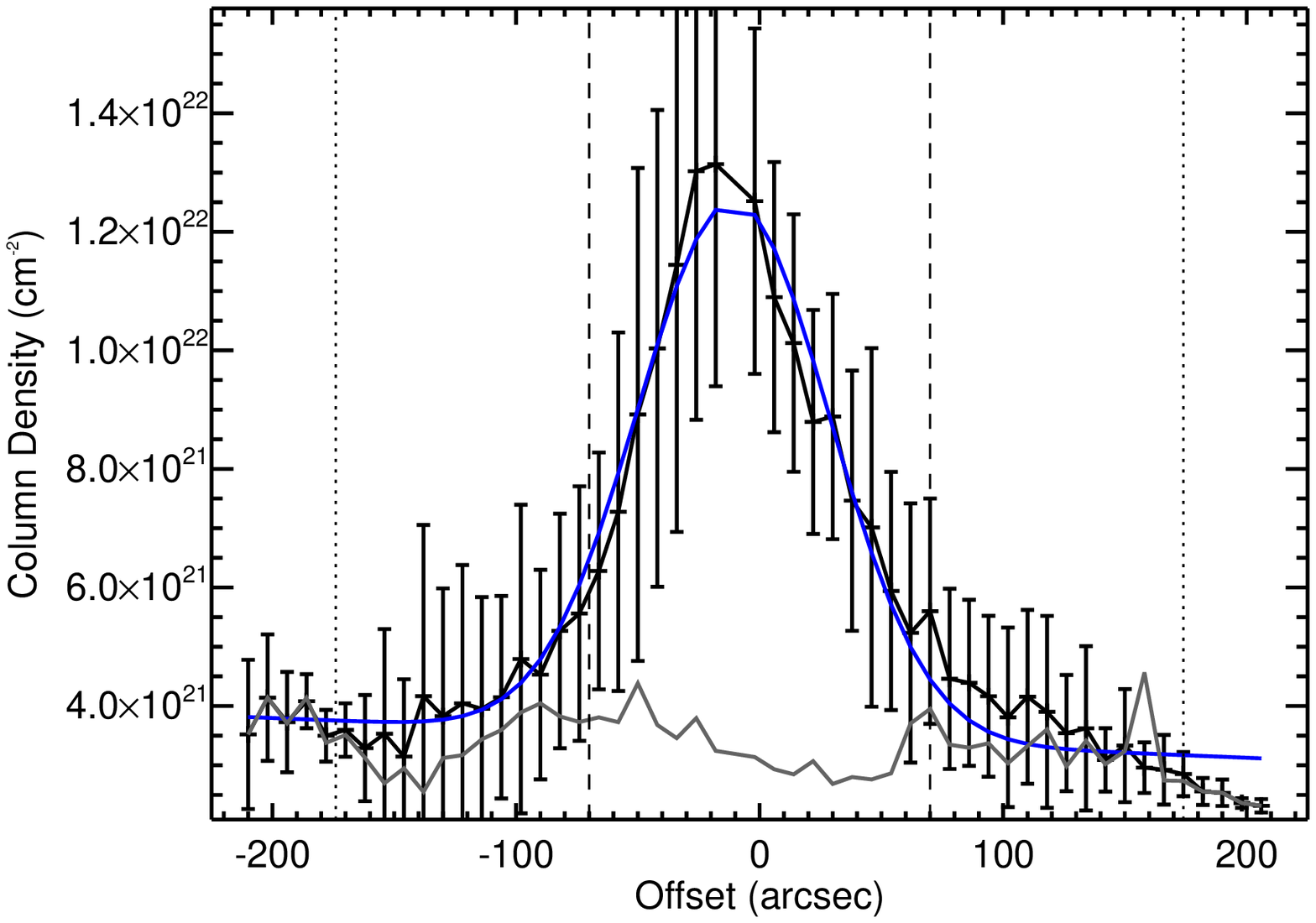}
\end{tabular}
\caption{Left panel: zoom-in of the filament in the region of L1448. The white line is the spine, while the grey line is the border of the filament. The short black lines are branches. To explain how the radial profile of the filament is derived, we defined, with the thick black line, an arbitrary region. Right panel: the radial profile of the filament averaged longitudinally within the black region shown in the left panel. The error bars are the standard deviation of the column density averaged within 10 pixels. The blue line is the Gaussian fit, and the grey line the estimated background. The dashed and dotted vertical lines are explained in the text.\label{esempioFilamento}}
\end{figure*}

In the left panel of Fig.~\ref{esempioFilamento} we show in details all the features previously defined: the filament border (thin grey line), the main spine (white line) and the system of branches (short black lines), for the case of the L1448 cloud. The identified filament runs along almost all the cloud. The thick black line defines an arbitrary region $R$ used in the right panel of Fig.~\ref{esempioFilamento} to show the radial profile of the filament, averaged along the filament itself, and the to assess, at least in one case, the validity of the 3-pixels border expansion.

As written, in fact, the filament-detection code expands radially the filament border by three pixels to identify the position where the filament merges with the surroundings. The average radial profile inside $R$ is shown in the right panel of Fig.~\ref{esempioFilamento}. The error bars are the standard deviation of the column density for each bin of distance from the spine. Due to the irregular shape of expansion, that reflects the irregular filament profile, the border does not have a constant offset from the spine. The minimum and the maximum radial distance from the spine reached by the filament border over $R$, $r_{1}$ and $r_{2}$, respectively, are shown with thin black lines, running parallel to the spine, in the left panel of Fig.~\ref{esempioFilamento}. For distances $|r|\le r_1$, within the two vertical dashed black lines in the right panel of Fig.~\ref{esempioFilamento}, all the pixel of the map belong to the filament area, while for distances $|r|\ge r_2$, marked with the two dotted black lines, all the pixel are external to the filament and contains only background pixel. At intermediate distances $r_1\le|r|\le r_2$, to be or to be not part of the filament becomes a local property, depending on the position along the spine.

The background is estimated from linear interpolation of the pixels that are outside the filament area, where $|r|\ge r_2$, or, locally, where $r_1\le|r|\le r_2$. The background so estimated is shown with a thick grey line in the right panel of Fig.~\ref{esempioFilamento}. The blue line in the same figure is the Gaussian fit to the radial profile. The filament width is estimated as the FWHM of the fit. The Gaussian fit gives another estimate of the background, but we prefer to use the linear interpolation because the Gaussian fit is less suited to catch possible asymmetries, as is visible in our example. These asymmetries are stronger in the wings of the profile rather than in the inner part, so that it is reasonable to use the Gaussian FWHM to estimate the width.

\begin{figure}
\centering
\includegraphics[scale=0.4]{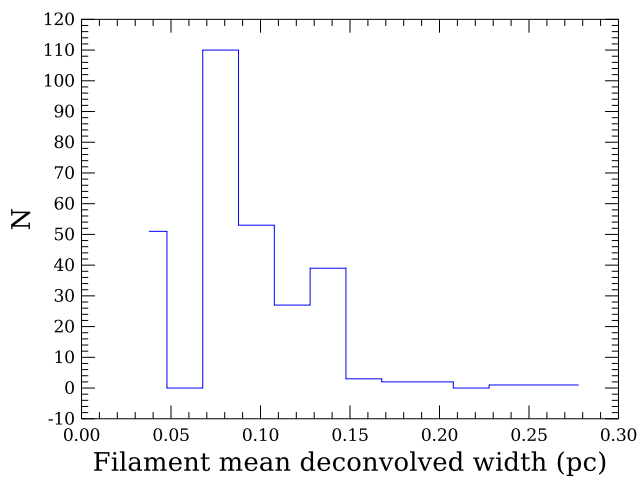}%&\includegraphics[scale=0.4]{meanLength.eps}
\caption{Histogram of the filament widths, averaged over the spine (bin size 1 pixel or 0.02~pc at $d=300$~pc).\label{meanWidth}}
\end{figure}

In %the left panel of 
Fig.~\ref{meanWidth} we show the filament widths averaged over the spine and deconvolved by the FWHM at 500~$\mu$m, 36\arcsec. The width is given in pc assuming a distance of 300~pc. The peak of the distribution is $\sim0.08$~pc, consistent with the finding of \citet{doris2019} of a characteristic width of $0.10\pm0.03$~pc for the filaments in 8 HGBS regions.

An important physical parameter for filaments is the mass per unit length. Theoretical models of isothermal infinite cylindrical filaments confined by the external pressure of the ambient medium predict the existence of a maximum equilibrium value to the mass per unit length \citep{ostriker}:
\begin{equation}
M_\mathrm{line,max}=\frac{2kT}{\mu m_\mathrm{H}G}=16.4\left(\frac{T}{10\,\mathrm{K}}\right)\,M_\sun\,\mathrm{pc}^{-1}\label{mCrit}
\end{equation}
Above this value the filament is unstable and can fragment to form cores.

In our case, $M_\mathrm{line}=M/L$ where the mass $M$ of the filament is $\mu m_\mathrm{H}\sum_i n_i(\mathrm{H}_2)A$ with $A$ the area of one pixel in cm$^2$, and the sum is over all the pixel in the filament area, while $L$ is the spine length. Note that $A$ depends on the distance as $d^2$ while $L$ on $d$, so $M/L$ increases with $d$ and, for a constant $d=300$~pc, is likely overestimated in the Western half of Perseus, and underestimated in the Eastern half.

\begin{figure}
\includegraphics[scale=0.4]{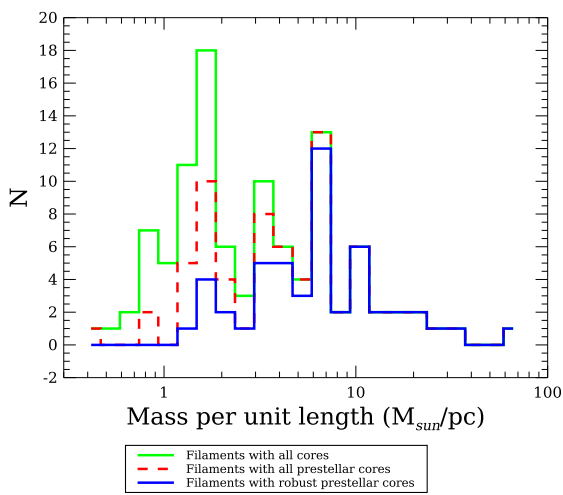}
\caption{Histograms of mass per unit length for filaments with cores (green line) and prestellar cores (blue and red lines, see text).\label{mSuL}}
\end{figure}

In Fig.~\ref{mSuL} we show with the green line, the distribution of the mass per unit length for all the filaments we find in Perseus, that contain at least one core of any type (for core definition and extraction, see Sect.~\ref{estrazione} below). The mass per unit length distribution peaks at $\sim1.7\,M_\sun$~pc$^{-1}$, well below the typical value of 16~$M_\sun$~pc$^{-1}$ of Equation~(\ref{mCrit}).If we consider that 10~K is not representative of the dust temperature and that we should adopt $T\ga12$~K \citep[see Fig.~\ref{istoDust}; see also][where a range between 11~K and 30~K is found for the dust temperature of filaments in many HGBS regions]{doris2019} the peak in the distribution is more than a factor 10 smaller than the maximum linear mass.

An important issue here is the background subtraction because, if the background is overestimated, the linear mass of the filament is underestimated. To address this problem we made use of the results found by \citet{hacar}. They derive 14 kinematically-coherent structures in NGC1333 through observations of the \element[][]N$_2$\element[+][]H (1–0) line. These structures, named fibres, are similar to what we call branches. The comparison between fibres and branches, however, is not easy because of the different tracers used, molecular gas vs. dust emission, and because of the different algorithm used to derive the structures. Only in one case we found a fibre and a branch that have a reasonable spatial overlap with a similar length. This happens for fibre number 9: \citet{hacar} derived a length of 0.47~pc and $M_\mathrm{lin}=58.5\,M_\sun$pc$^{-1}$, while we found a length of 0.50~pc and $M_\mathrm{lin}=62.5\,M_\sun$pc$^{-1}$, so, at least for this case, the background estimate looks reliable.

Since only prestellar cores are likely star-forming in nature (see Sect.~\ref{stability}), it is perhaps more reasonable to consider the filaments that contain prestellar cores instead of the filaments that contain only unbound cores. In this case (blue histogram in Fig.~\ref{mSuL}), the peak of the distribution increases to $\sim7\,M_\sun$~pc$^{-1}$. This value is still less than the value $M_\mathrm{lin}/M_\mathrm{lin,max}$.
Adding the \textit{candidate} prestellar cores (see again Sect.~\ref{stability}) to enlarge the samples of bound cores, changes the shape of the histogram (dashed red line), but only in the region of smaller masses. In any case, the majority of filaments are below $M_\mathrm{lin,max}$. Note, however, that \citet{FM} found that fragmentation and core formation can occur in filaments when $M_\mathrm{lin}/M_\mathrm{lin,max}>0.5$, close to our value.

Similarly, \citet{milena2} found that the majority of filaments with bound cores in Lupus have $M_\mathrm{lin}<M_\mathrm{lin,max}$. The authors suggest that in a filament the mass per unit length should be considered locally instead of giving one value for a whole filament. In fact, if we compute $M_\mathrm{lin}$ for the single branches instead of considering the entire filament, we derive a much broader distribution toward high values, $\ga130\,M_\sun$~pc$^{-1}$, of $M_\mathrm{lin}$. Nonetheless, the median of the distribution is $\sim1.5\,M_\sun$~pc$^{-1}$ with a peak at less than $1\,M_\sun$~pc$^{-1}$.

Note, however, that the formula given in Equation~(\ref{mCrit}) gives the maximum line-mass for isothermal equilibrium when the non-dimensional radius $\xi$ of an infinite length filament goes to infinity. In the more general case of finite $\xi$, however, Equation~(\ref{mCrit}) reads \citep{ostriker}
\begin{equation}
M(\xi)=\frac{2kT}{\mu m_\mathrm{H}G}\frac1{1+8/\xi^2}\le16.4\left(\frac{T}{10\,\mathrm{K}}\right)\,M_\sun\,\mathrm{pc}^{-1}\label{mCrit2}
\end{equation}
where the equality holds only for $\xi\rightarrow\infty$. So, in the more general case of finite $\xi$, {the mass $M(\xi)$ of a self-gravitating cylinder in thermal equilibrium is smaller than the asymptotic $M_\mathrm{lin,max}$ for the same temperature.

The non-dimensional radius $\xi$ can be transformed back to a physical radius once we know $T$ and $\rho_0$, i.e., the temperature and central density of the filament, respectively. Alternatively, one can estimate $\rho_0$ from the observed values. For example, if we assume $T=16$~K and $M_\mathrm{lin,max}=8\,M_\sun$~pc$^{-1}$, Equation~(\ref{mCrit2}) solved for $\xi$ gives $\xi\sim2.15$. Then, if we assume 0.08~pc to be the typical filament radius, solving Equation~(45) in \citet{ostriker} for the central density yields $\rho_0\sim4.3\times10^{-20}$~g~cm$^{-3}$ or $n_0\sim9\times10^3$~cm$^{-3}$ for $\mu=2.8$. Strictly speaking, 0.08~pc is the width of the filament, defined as the FWHM of the Gaussian fit. If we define, instead, the radius as 1.29$\times$FWHM \citep{FHSC}, then the typical filament radius becomes $\sim0.10$~pc and $n_0\sim5.5\times10^3$~cm$^{-3}$. In any case, we conclude that as long as $n\ga n_0$, a filament with radius 0.08~pc and $T=16$~K has $M_\mathrm{lin,max}=8\,M_\sun$~pc$^{-1}$, similar to the peak value we find in the histogram of Fig.~\ref{mSuL}.

\subsubsection{The probability density function of the column density map}\label{PDFFit}
The probability density function (PDF) derived from the column density map has been used to probe the physics governing the diffuse medium \citep[see][for a review on PDF]{nicola}. Briefly, the PDF shows a log-normal behaviour at low densities due to the turbulence in the cloud, while a power-law tail develops at higher densities as a consequence of self-gravity and star-formation activity.

In Fig.~\ref{PDFlow}, we present the PDF for the whole molecular cloud. We model the observed distribution with a log-normal and a power-law functions, fitting each curve independently.

\begin{figure}
\centering
\includegraphics[scale=0.4]{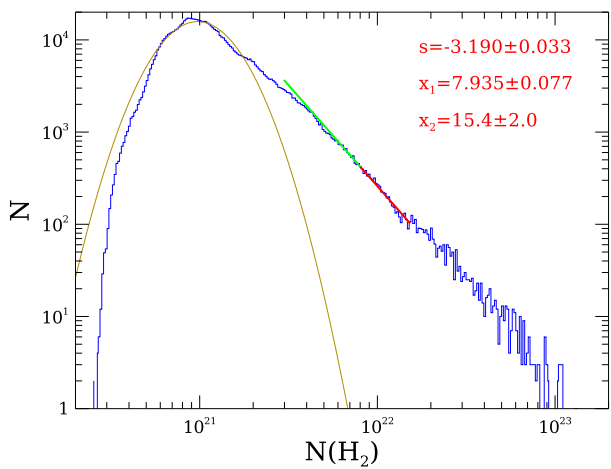}
\caption{The column density PDF for the whole Perseus molecular cloud: $s$ is the slope of the power-law fit, shown in red, over the interval $x_1$--$x_2$, both in $10^{21}$~cm$^{-2}$. As shown with the green line, the fit can be extended down to $\sim4\times10^{21}$~cm$^{-2}$. The brown line is a log-normal fit to the low-density PDF whose parameters are given in the text. Note that $N(\mathrm{H}_2)\sim1.6\times10^{21}$~cm$^{-2}$ is the smallest column density having a closed contour in our map.\label{PDFlow}}
\end{figure}       

The brown line shows the best log-normal fit to the low-column density part of the distribution. The peak is at $N(\mathrm{H}_2)=(9.738\pm0.061)\times 10^{20}$~cm$^{-2}$. The significance of this fit is, however, limited by the fact that the smallest value of column density having a closed contour in our map is $N(\mathrm{H}_2)\sim1.6\times10^{21}$~cm$^{-2}$. As a consequence, the shape of the PDF in the region where the log-normal fit is derived, can be distorted by data incompleteness \citep[][argued that the overall log-normal shape is actually created rather than distorted by incompleteness]{alves}.

Moving now to the high column density tail in the PDF, \citet{2013ApJ...763...51F} showed that the following relation exists between $s$, the power-law slope in the PDF, and $q$, the power-law slope $\rho(r)\propto r^{-q}$ of the volume density for cores:
\begin{equation}
q=1-\frac2s \label{sKappa}
\end{equation}
so that if $s$ is derived, hints on how the star-formation process in a cloud is proceeding can be obtained.

To find the exponent of a power-law, the easiest and often used solution is to linearise the problem in log-log space, getting an equation of the kind $\log y=\log c-\gamma\log x$. The unknowns $c$ and $\gamma$ are immediately found by fitting a straight-line to the dataset. This procedure, however, should be avoided for many reasons \citep{bauke,clauset}, for instance it violates the assumption that uncertainties on the dependent variable follow a Gaussian distribution. This assumption is the base for least-square fit.

Our strategy instead was to fit directly a function $y=cx^\gamma$ using non-linear fitting routine. We used the linearisation scheme described above to obtain only a first estimate of the parameters. Since we do not know the extent of the interval over which the fit should be done, we treated the interval extremes, $N(\mathrm{H}_2)_\mathrm{min}$ and $N(\mathrm{H}_2)_\mathrm{max}$, as free parameters.

In this way, it is not possible to compare directly the $\chi^2$ corresponding to each model, because the number of data points is not constant. Furthermore, the use of the reduced $\chi^2$ with non-linear models is questionable because the degrees of freedom in this case are generally not known\footnote{The usual assumption of $N-n$ degrees of freedom if $N$ is the number of data and $n$ is the number of parameters may be valid, but in general is not.}. To derive the best model we adopted the cross-validation method \citep{andrae2}. Namely, for a given $N(\mathrm{H}_2)_\mathrm{min}$ and $N(\mathrm{H}_2)_\mathrm{max}$, we computed the best $c,\gamma$ excluding one point of the dataset and from the derived best-fit model we computed the difference between the expected value and the model-derived value. This procedure is repeated for all the points in the dataset and the product of all the differences gives an estimate of the likelihood for that model. Then, we looked for the maximum likelihood value among all the models ($N(\mathrm{H}_2)_\mathrm{min}$,$N(\mathrm{H}_2)_\mathrm{max},c,\gamma$).

The validity of this strategy is limited by the fact that often the PDF does not show a clear power-law trend. As a consequence, the fit procedure tends to minimize the interval $N(\mathrm{H}_2)_\mathrm{min}$--$N(\mathrm{H}_2)_\mathrm{max}$. To impose physical constraints to the fit, we varied $N(\mathrm{H}_2)_\mathrm{min}$ around the value reported in Col.~$N(\mathrm{H}_2)$ of Table~\ref{SFE}, discussed later on in the paper. This value  in column density is the smallest background column density\footnote{The background column density is computed by \textsl{getsources} at the position of each core in the high-resolution column density map, and reported in the catalogue, see Table~\ref{esempioCat}.} found for prestellar cores in each sub-cloud. Since the power-law tail should trace the region where gravity is strong enough for cores to form, it looks reasonable to impose that $N(\mathrm{H}_2)_\mathrm{min}$ is not much different from the minimum background column density.

To derive an uncertainty for $N(\mathrm{H}_2)_\mathrm{min}$ and $N(\mathrm{H}_2)_\mathrm{max}$, dubbed $x_1$ and $x_2$ in Fig.~\ref{PDFlow}, we created the PDF histogram with five different binsizes from 0.8 to 1.2, in steps of 0.1, in units of $10^{20}$~cm$^{-2}$ (the one shown in the figure corresponds to the choice $1\times10^{20}$~cm$^{-2}$). The power-law slopes and limits were found for the five histograms and similar solutions were sought. The red line in Fig.~\ref{PDFlow} shows the power-law whose slope is the weighed average of the five slopes and is measured over an interval of $7.935\times10^{21}$~cm$^{-2}$ and $1.54\times10^{22}$~cm$^{-2}$. These intervals are the mean of the five starting and end points over which the fit extends.

For the whole Perseus cloud, the minimum background column density for prestellar cores is $9.4\times10^{20}$~cm$^{-2}$ (see Table~\ref{SFE}). We could not find any good power-law fit starting from such small values. Indeed, the best fit we found (slope --3.190) is limited to $7.9\times10^{21}$~cm$^{-2}$ but it is nonetheless a good solution down to $\sim4\times10^{21}$~cm$^{-2}$, as it can be seen from the green line that extends the best fit to smaller column densities.

Our slope $-3.190\pm0.033$ is in good agreement with the value --3 found by \citet{zari}, and translates into $q=1.6269\pm0.0065$, similar to the 1.5 slope expected for spherically symmetric free-fall collapse. According to \citet{nicola}, this slope should be quite insensitive to the spatial resolution and to the histogram bin size.

\citet{sarah2} found a correlation between the slope of the PDFs and the star formation efficiency. We will discuss this topic later on in the paper. Here, we show the results for the individual subregions defined in Fig.~\ref{zoneNH2}. Fig.~\ref{PDFtutte} shows the PDFs for each of the subregions that hosts at least one protostar. In many cases, the low-density portion cannot be fit with a log-normal function, while the high density tail shows a wide range of values (labels $s$ in Fig.~\ref{PDFtutte}). The slopes $s$ for the single clouds are different from the one found for the whole Perseus region, as predicted by \citet{2018ApJ...859..162C}. Through Equation~(\ref{sKappa}), the interval in slopes translates into an interval for $q$ limited by 1.5, in IC348, and 2.3, in HPZ6 (for this region, however, the result of the fit is not meaningful).

\begin{figure*}
\sidecaption
\centering
\begin{tabular}{cc}
\includegraphics[scale=.29]{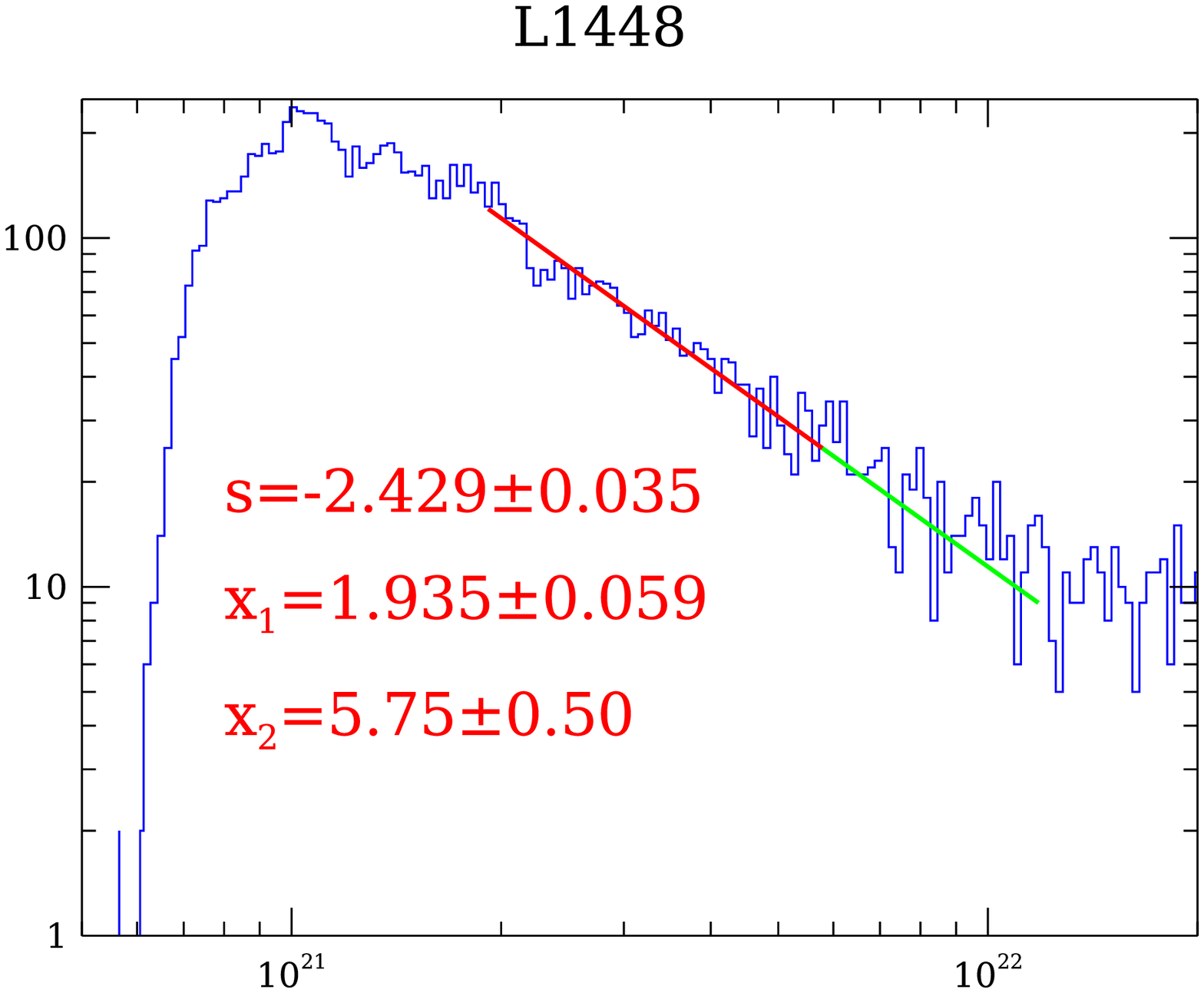}&
\includegraphics[scale=.29]{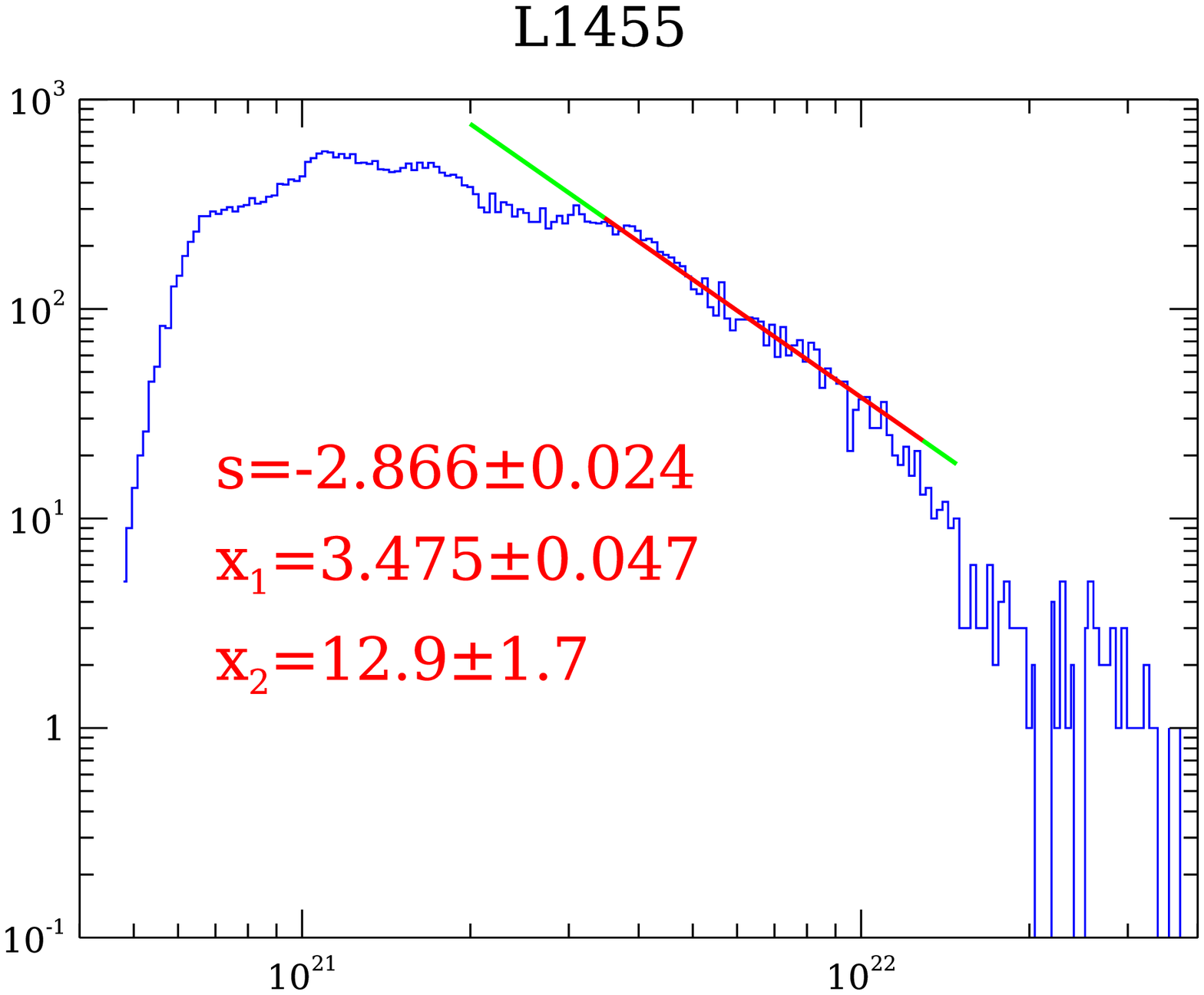}\\
\includegraphics[scale=.29]{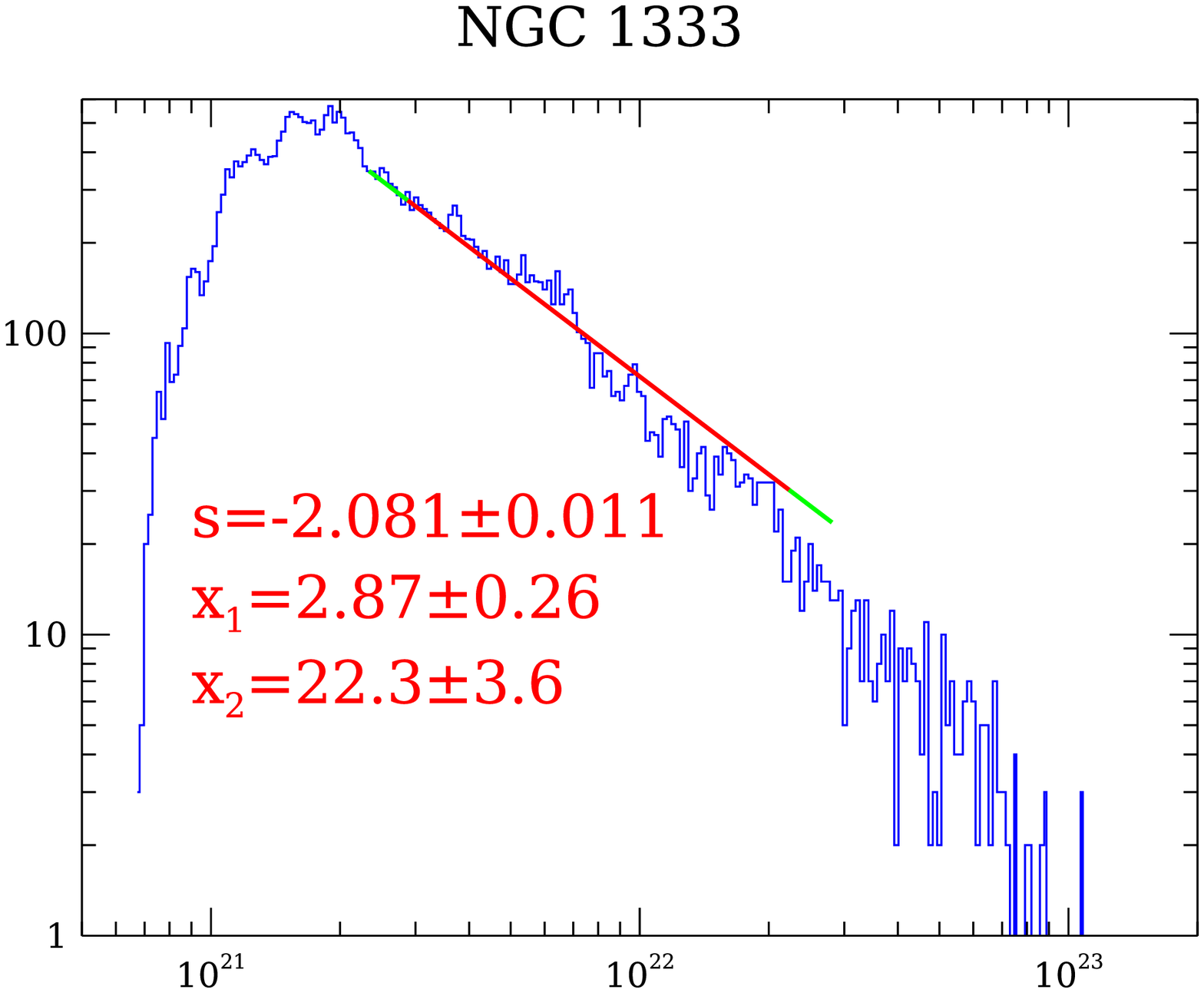}&
\includegraphics[scale=.29]{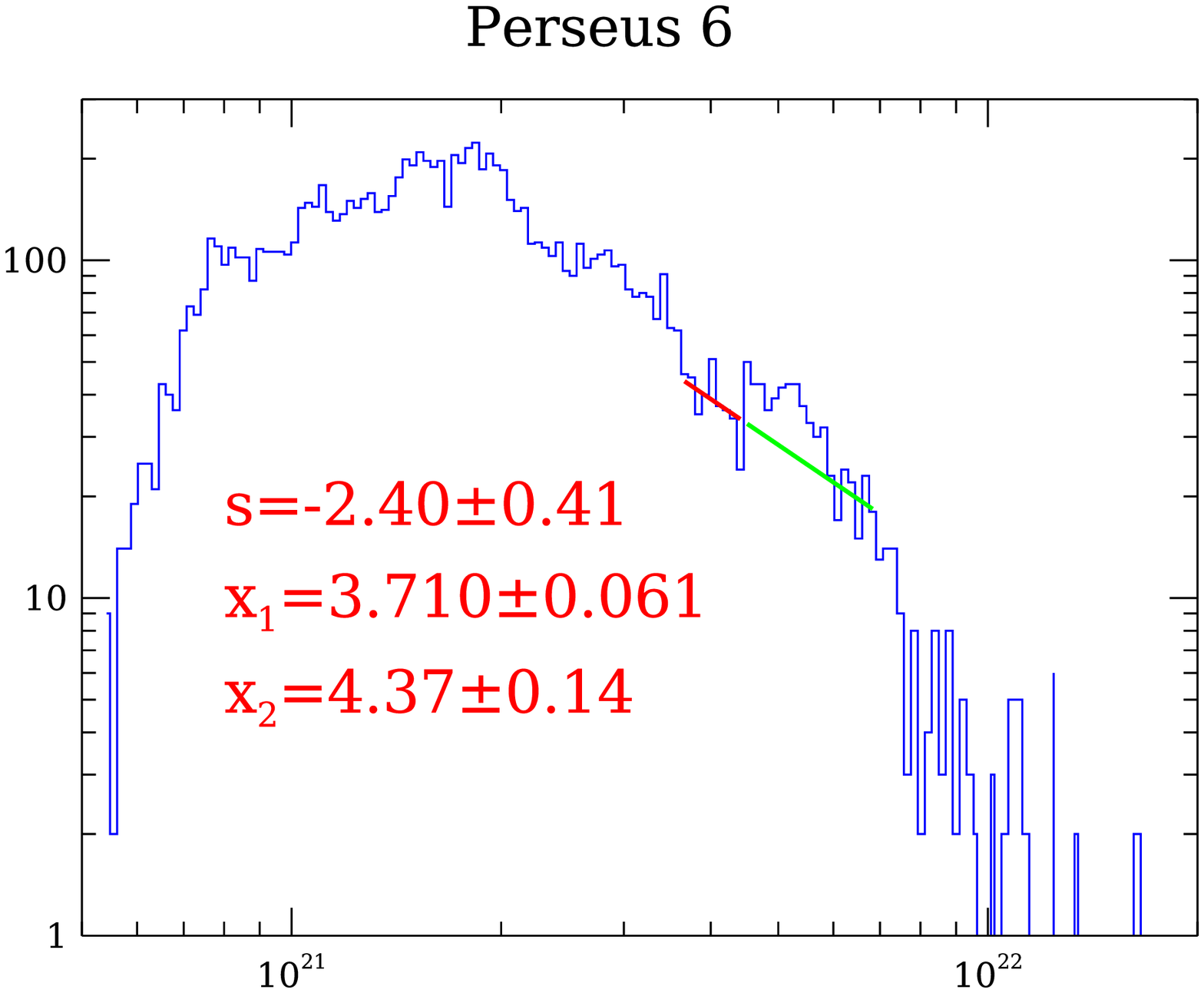}\\
\includegraphics[scale=.29]{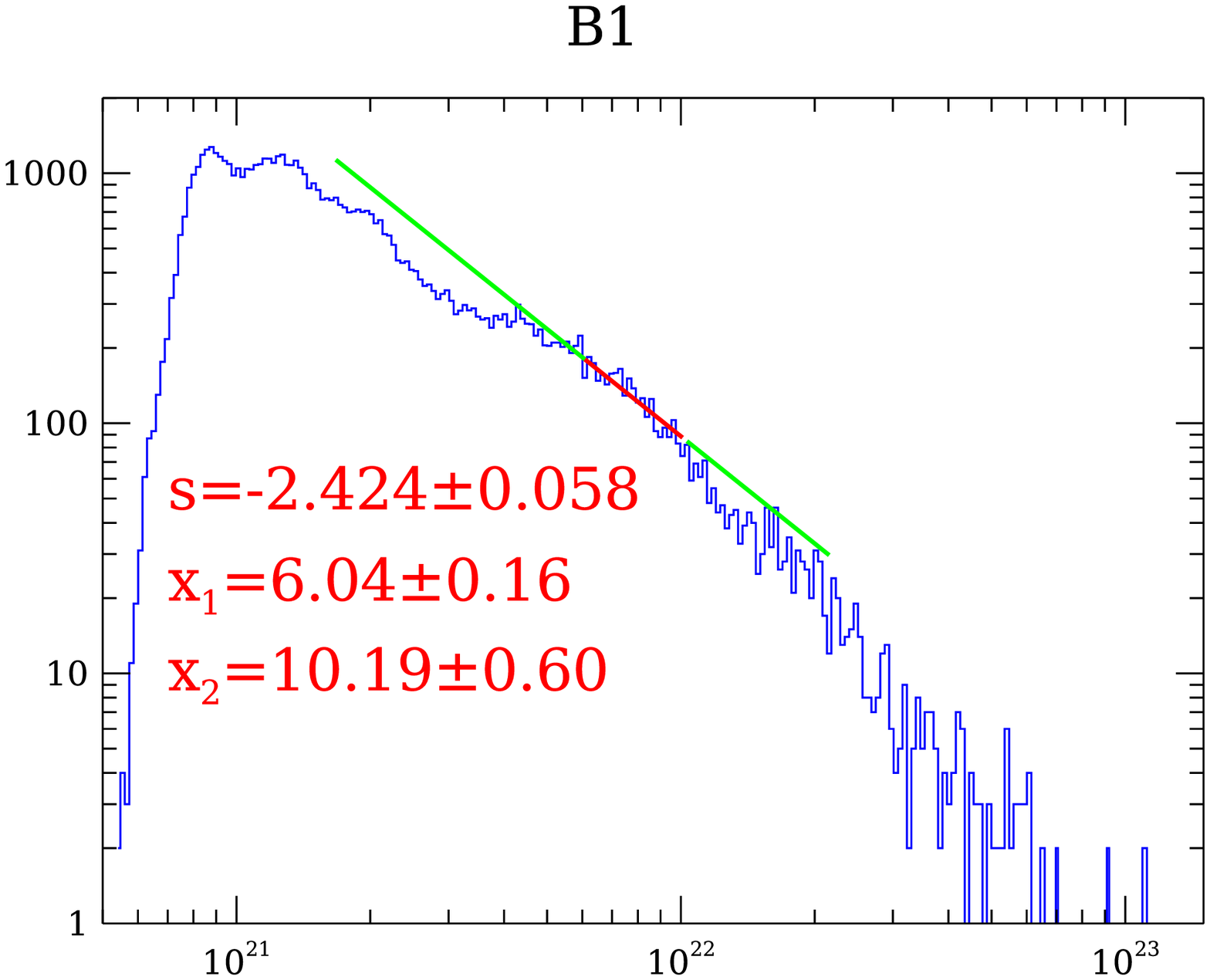}&
\includegraphics[scale=.29]{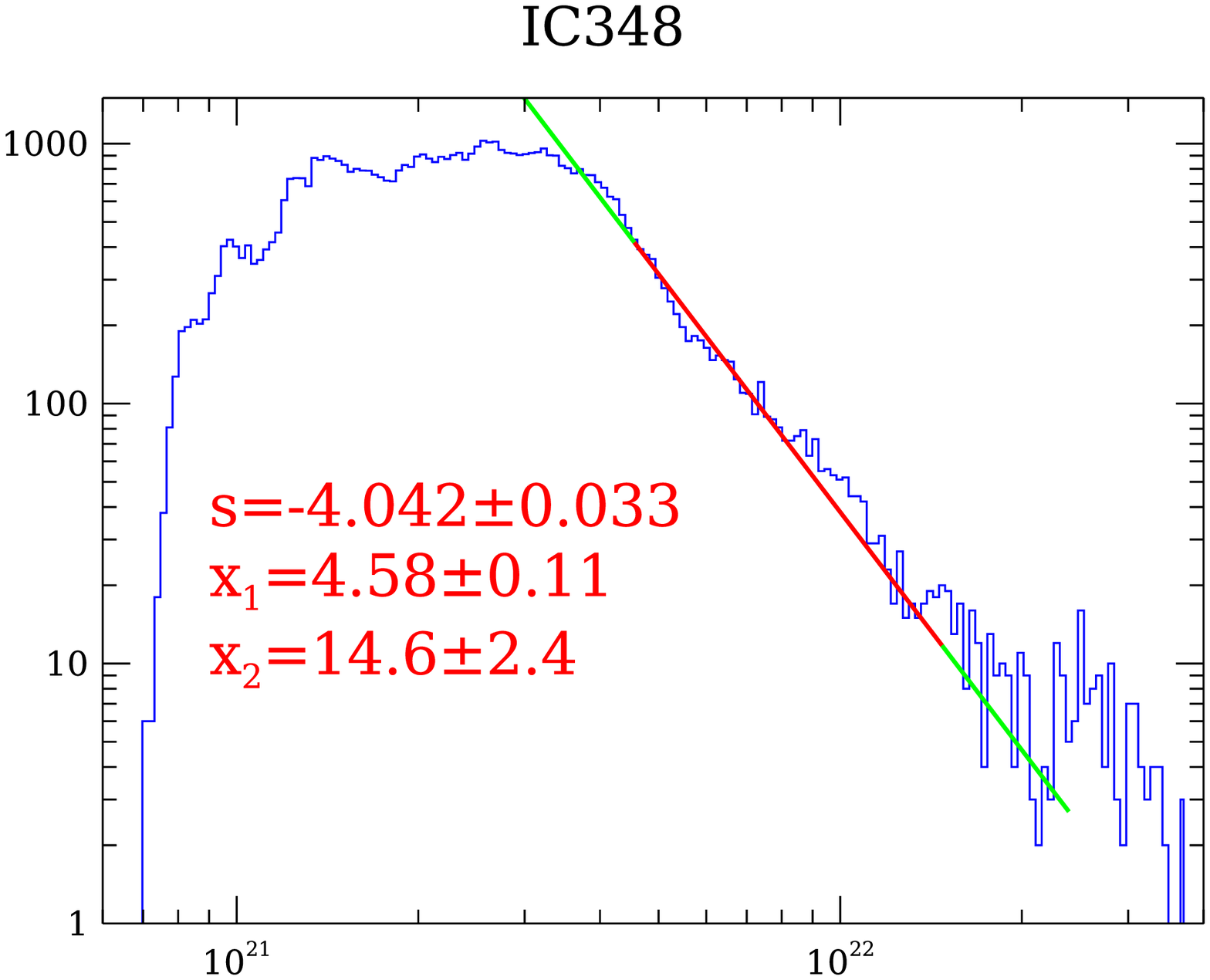}\\
\includegraphics[scale=.29]{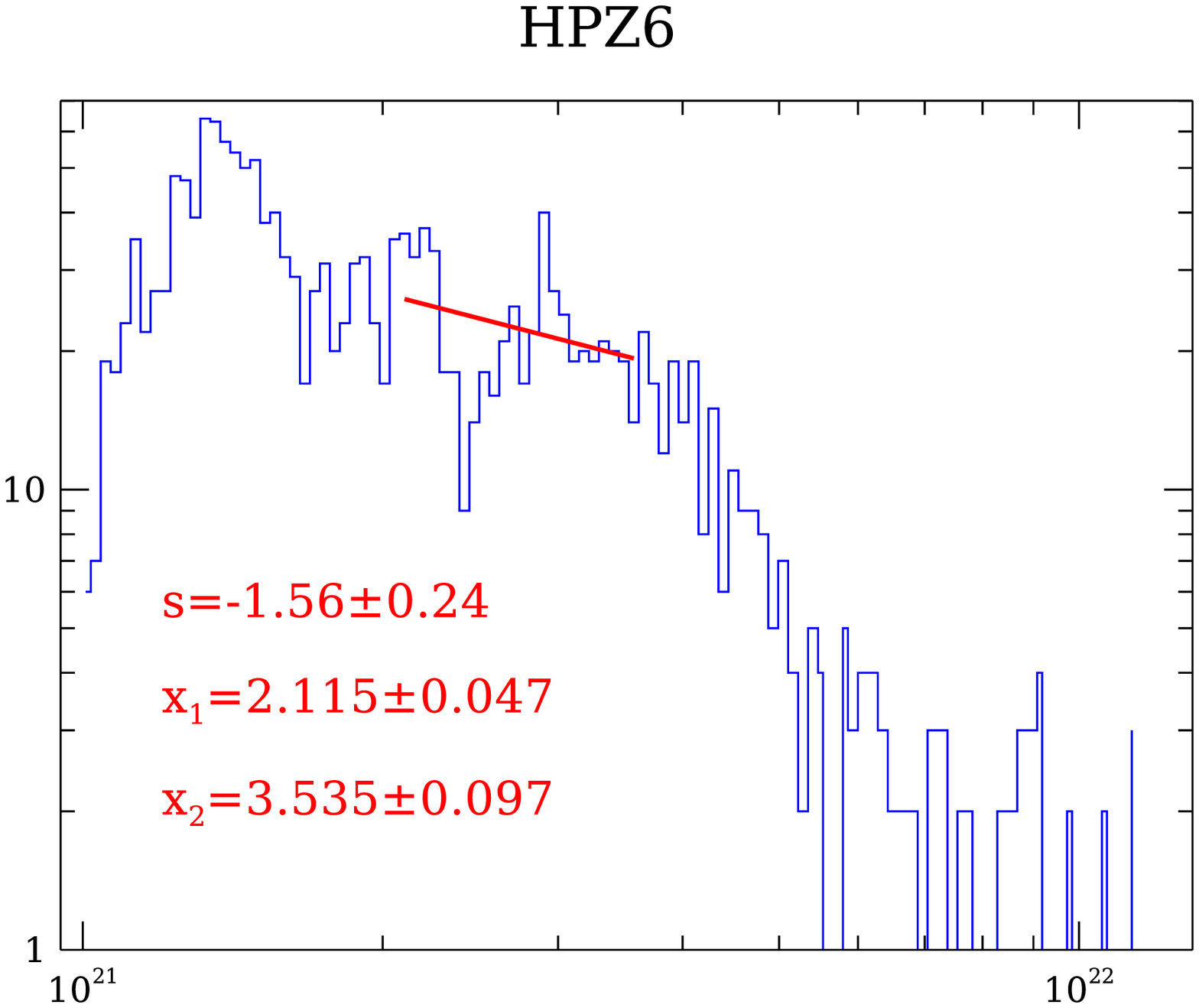}&
\includegraphics[scale=.29]{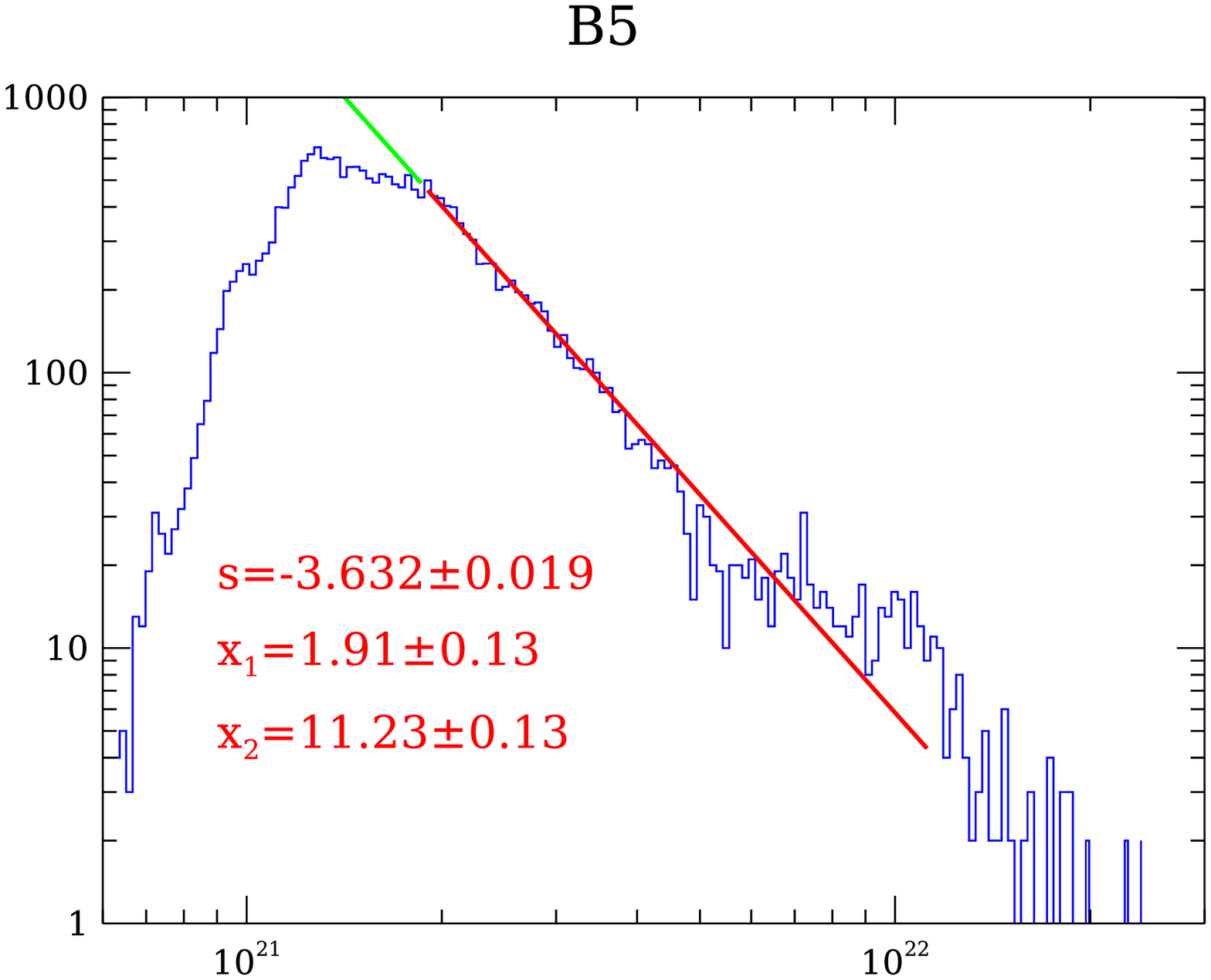}\\
\end{tabular}
\caption{The PDF's for the subregions identified in Perseus where protostars were detected. In each panel $s$ gives the slope of the power-law fit, shown in red, in the interval delimited by $x_1$ and $x_2$, both in $10^{21}$~cm$^{-2}$; the fitting procedure is detailed in the text. The \textit{x}-axis is $N(\mathrm{H}_2)$ in cm$^{-2}$. Green lines extend the best-fit over the interval of column densities where protostars were found (see Col.~$N($H$_2)$ in Table~\ref{SFEProto}).\label{PDFtutte}}
\end{figure*}

It is not possible to compare directly our slopes with those found by \citet{sarah2}, given the different strategy adopted. To make the comparison, we derived the slope of the power law following their method: first, we fixed $N(\mathrm{H}_2)_\mathrm{min}$ to $7\times10^{21}$~cm$^{-2}$, 30\%  (see Sect.~\ref{secMasse}) of $1\times10^{22}$~cm$^{-2}$, the value adopted in Sadavoy et al. Second, we made a linear fit in the log-log plane. The slopes are compared in Table~\ref{confronti}.

\begin{table}
\caption{Comparison of the PDF power-law slopes found by \citet{sarah2} (Col.~S) and in this paper (Col.~P). Column P$_\mathrm{S}$ reports the slope found with our data following Sadavoy et al.'s procedure.\label{confronti}}
\begin{tabular}{lccc}
\hline
\rule{0pt}{1.\normalbaselineskip}Region&S&P$_\mathrm{S}$&P\\\hline
\rule{0pt}{1.\normalbaselineskip}L1448&$-1.4\pm0.3$&$-0.982\pm0.053$&$-2.429\pm0.035$\\
L1455&$-2.9\pm0.2$&$-2.67\pm0.11$&$-2.866\pm0.024$\\
NGC1333&$-1.8\pm0.2$&$-1.643\pm0.041$&$-2.081\pm0.011$\\
B1&$-2.2\pm0.2$&$-1.983\pm0.052$&$-2.424\pm0.058$\\
B1E&$-9.6\pm1.0$&$-7.8\pm1.2$&$-4.26\pm0.30$\\
IC348&$-2.5\pm0.2$&$-2.308\pm0.082$&$-4.042\pm0.033$\\
B5&$-2.8\pm0.7$&$-2.34\pm0.14$&$-3.632\pm0.019$\\\hline
\end{tabular}
\end{table}

In this table we report both the slopes found with our data according to the procedure written above (Col.~P) and the the slopes found still with our data but following the same procedure as in Sadavoy et al. (Col.~P$_\mathrm{S}$). Naming $s\pm\sigma_s$ and $p\pm\sigma_p$ the values reported in Col.~S and Col.~P$_\mathrm{S}$, respectively, we see that $s>p$ always. In three cases $s-\sigma_s<p<s$ and in other three case the intervals $s-\sigma_s$ and $p+\sigma_p$ overlap. Only for L1448 we find a difference slightly larger than $1\sigma_s$: $p=s-1.4\sigma_s$. We conclude that the slopes reported in Fig.~\ref{PDFtutte} differ from those reported in \citet{sarah2} because of the different strategy adopted to fit the power-law tail in the PDF. When used in the rest of this paper, the slopes of the PDFs are those reported in Fig.~\ref{PDFtutte}.

\section{Physical properties of the cores}
\subsection{Source extraction}\label{estrazione}
Sources were extracted from the intensity images using release 1.140127 of \textsl{getsources} \citep{sasha,getfilaments}, a multi-scale, multi-wavelength source extraction algorithm. Its application to the fields of the HGBS is described in \citet{Aquila}. We followed their strategy to detect starless cores and protostars by running twice \textsl{getsources} with two separate set of parameters to optimize the extraction. In the following we summarize the extraction procedure.

For starless cores, we extracted sources on the SPIRE intensity maps and the high spatial-resolution column density map \citep[][see also Sect.~\ref{cMap}]{pedro}. We also use the 160~$\mu$m data via a temperature-corrected map, which is obtained by convolving the PACS red image to the resolution of the SPIRE 250~$\mu$m band and then combining them to derive a two-colour column density map \citep{Aquila}. The purpose of this approach is to avoid strong gradients of intensity that can be found in proximity to photon-dominated regions (PDRs) or hot sources. For protostars, only the 70~$\mu$m band was used for detection.

After source detection is completed, flux intensity, or simply flux, is measured for each source by \textsl{getsources} at all \textit{Herschel} wavelengths. In particular, at 160~$\mu$m we use the intensity map, and not the colour-corrected map, used only for detection. To both catalogues, we applied the selection criteria used for Aquila \citep{Aquila}. For the starless cores:
\begin{itemize}
\item Column density detection significance greater than 5;
\item Global detection significance over all wavelengths greater than 10;
\item Global ``goodness'' $\ge1$;
\item Column density measurement with signal-to-noise ratio (SNR) greater than 1 in the high-resolution column density map;
\item Monochromatic detection significance greater than 5 in at least two bands between 160~$\mu$m and 500~$\mu$m;
\item Flux measurement with SNR $>$1 in at least one band between 160~$\mu$m and 500~$\mu$m for which the monochromatic detection significance is simultaneously greater than 5.
\end{itemize}

For prostostars, the criteria are:
\begin{itemize}
\item Monochromatic detection significance greater than 5 in the 70~$\mu$m band;
\item Positive peak and integrated flux densities at 70~$\mu$m;
\item Global ``goodness'' $\ge1$;
\item Flux measurement with SNR $>$ 1.5 in the 70~$\mu$m band;
\item FWHM source size at 70~$\mu$m smaller than 1.5 times the 70~$\mu$m beam size (i.e., $<$12\farcs6 )
\item Estimated source elongation $<$1.30 at 70~$\mu$m.
\end{itemize}

The two catalogues were cross-matched to look for associations within 6\arcsec. We found 60 sources present in both catalogues. These sources were removed from the cores analysis being considered as candidate protostars. In Appendix~\ref{appCat} we explain in more detail how the SEDs were built for the objects detected in both catalogues. Sources with no counterpart at 70~$\mu$m were classified as tentative starless cores. We then visually inspected the intensity maps and column density map toward each candidate to be sure that the source is not an artefact. This step removes non-existent sources due to local fluctuations in the maps and also finds candidate protostars present only in the starless cores catalogue (e.g., protostars that are not well detected at 70~$\mu$m). The reason for this depends on the slightly different criteria used to build the catalogues (see Appendix~\ref{appCat} for details).

To remove known galaxies we queried the NED\footnote{\url{http://ned.ipac.caltech.edu/}} and SIMBAD\footnote{\url{http://simbad.u-strasbg.fr/simbad/}} archives.

Starless cores not visible at 70~$\mu$m but with a SIMBAD counterpart at shorter wavelengths were removed from the catalogue because starless cores are not expected to be detected $\la160$~$\mu$m. We use a distance criterion of 6\arcsec. Since the association with a SIMBAD source may be only spatial and not physical, when a starless core with no detection at 70~$\mu$m has also a (sub)mm counterpart within 6\arcsec\ beside a counterpart at short wavelengths, the source was left in our catalogue.

We also scanned the \textit{WISE}\footnote{\url{https://irsa.ipac.caltech.edu/cgi-bin/Gator/nph-scan?} \url{mission=irsa&submit=Select&projshort=WISE}} \citep{wise} archive within a radius of 6\arcsec\ from our objects, imposing to have SNR greater than 0 in bands~3 and 4. This approach is quite conservative because the threshold of SNR is very low. Moreover, there remains the possibility that two different sources appear close each other in projection. In any case, sources with a WISE counterpart were removed.

All the starless cores removed from our catalogue are reported in Appendix~\ref{addCat}.

In the end, we were left with 816 candidates: 684 starless cores and 132 protostars. A search for counterparts within 6\arcsec\ in the \textit{c2d} archive \citep{evans}\footnote{\url{http://vizier.u-strasbg.fr/viz-bin/VizieR?-source=J/}
\url{ApJS/181/321}} shows no associations for the starless cores, while about 70\% of our prostostars has a possible counterpart. The catalogue, presented in Appendix~\ref{appCat} and available as on-line material, reports data for all 816 sources, both starless cores and protostars. In the following we discuss the physical properties of the starless cores while the protostars will be the subject of a forthcoming paper.

As an additional check on the reliability of the detected sources, we made a second extraction with a completely different source identification algorithm, namely \textsl{CuTEx} \citep{cutex}. This method identifies sources via the second-order differentiation of the signal image in four different directions, looking for minima of the curvature. Since \textsl{CuTEx} makes detection and measurements on single band images, it is not straightforward to make comparisons with extractions by \textsl{getsources}, which builds its catalogue by combining the results of monochromatic detections. Given that the peak of the SED of cold cores is expected around 250~$\mu$m, we looked for agreement from both \textsl{getsources} and PSW \textsl{CuTEx} catalogues. Indeed, we find that 467 (68\%) of \textsl{getsources} cores have a \textsl{CuTEx}-detected object located within the 250~$\mu$m elliptical extent measured by \textsl{getsources}.

We find excellent agreement (98\%) between the protostar catalogues derived with \textsl{getsources} and \textsl{CuTEx}. Since to extract the candidate protostars both algorithms use only the 70~$\mu$m data, where diffuse emission is less intense, it is not surprising that the two methods found similar objects.

\subsection{SED fitting}\label{SEDfit}
Once the source catalogue has been produced, we fitted the measured SEDs using a grid of theoretical models. This approach, already used in \citet{milena2} and equivalent to that used in Sect.~\ref{cMap}, consists of generating a number of modified blackbodies
\begin{equation}
F_\nu=\frac{\kappa_\nu M\mathrm{B}_\nu(T)}{d^2};\,\,\,\,\,\,\kappa_\nu=\kappa_0\left(\frac{\lambda_0}\lambda\right)^\beta\label{theoSED}
\end{equation}
where $M$ is the mass of the isothermal emitting dusty envelope at temperature $T$ and distance $d$. For the dust opacity, we used the same parametrisation as for the column density map: $\kappa_0=0.1$~cm$^2$g$^{-1}$ at $\lambda_0=300$~$\mu$m with $\beta=2$.

The grid of models was constructed fixing $M=1\,M_\odot$ and $d=1$~pc, while $T$ was varied in the range 5--50~K in steps of 0.01~K (4501 models). The distance is scaled to the required value, 300~pc for Perseus, during the fitting procedure.

Using a grid instead of a non-linear fitting procedure has two advantages. First, it is not necessary to give initial guesses of the parameters, which are more or less arbitrary. Second, as long as the grid is large and dense, we can be sure to find the global minimum and not a local one.

Since Equation~(\ref{theoSED}) is linear in $M$, the mass can be found with a straightforward application of the least squares method when comparing a given SED to the grid of models, as shown in Equation~(\ref{minQua}). Among the 4501 models, the best fit corresponds to the pair ($T,M$) with the minimum $\chi^2$. Uncertainty in $T$ is derived from the range in temperatures whose models have $\chi^2\le\chi^2_\mathrm{min}+1$ \citep{andrae}. In practice, if $T_1$ and $T_2$ are minimum and maximum temperatures of these models then we give $T_{-(T-T_1)}^{+(T_2-T)}$.

Uncertainty in $M$ was derived both according to the least squares method, and from the variation in mass corresponding to $T_1$ and $T_2$. Clearly, if $T_1<T_2$ then $M_2<M_1$. The larger of the two uncertainties was then adopted.

For SEDs with less than three significant flux measurements (see Appendix~\ref{appCat}), the fitting procedure was not applied. We also did not look for the best fit in cases where $F_\nu(350\,\mu\mathrm{m})<F_\nu(500\,\mu\mathrm{m})$. For all these cases, we adopted a fiducial temperature equal to $10.4\pm1.0$~K, the median temperature of all the reliable SEDs, excluding the protostars. The uncertainty of this median is arbitrary large (the median $\Delta T/T$ for reliable sources is $\sim3$\%). For these sources, the mass was computed from the measured integrated flux density at the longest significant wavelength, assuming $T=10.4$~K. The number of sources having reliable SEDs is 519 cores and 85 protostars. For the latter, however, Equation~(\ref{theoSED}) is not an appropriate model, even if it can give an estimate of the envelope properties. As noted earlier, discussion of the protostars catalogue is beyond the scope of this paper.

\subsection{Core classification and stability}\label{stability}
As explained in Sect.~\ref{estrazione}, we classify a core as protostellar if it is in emission at 70~$\mu$m. This approach is physically plausible because in general cold cores are not visible at this wavelength, often not even at 160~$\mu$m. On the contrary, once the source is visible at 70~$\mu$m, a star has already formed, warming up its surrounding envelope. This rule, however, is not always valid. In principle, any core emits in the PACS bands, and its detection at 70~$\mu$m depends on its physical properties combined with the sensitivity of the instrument. Indeed, we find seven objects (ID \#48, 692, 745, 754, 756, 761, 769 in the catalogue) with 70~$\mu$m emission that is either uncertain or at a level compatible with, or below, the best-fit SED found with Equation~(\ref{theoSED}). These sources were classified post facto as starless cores.

First Hydrostatic Cores (FHSC) are expected to have excess 70~$\mu$m with respect to a modified blackbody, even if they can not be considered genuine Class~0 objects. They have masses comparable to a giant planet and radii are only few astronomical units \citep{larson}. Source \#500 is indeed such a candidate \citep[B1-bS,][]{FHSC}. In our catalogue, however, we classified this source as a protostar because its properties are not compatible with those of the starless cores.

Finally, source \#777 was first classified as starless because its 70~$\mu$m emission was an upper limit. From visual inspect, however, this source is visible at 70~$\mu$m but since the uncertainty on the flux is higher than the flux itself, the code used to make the SED fitting classified it as a starless, putting an upper limit in the blue band. We re-defined this source as protostar.

Among the starless sources, we want to identify cores that are not stable against their own gravity. As a result, they are likely to be collapsing or on the verge of collapse, forming in the near future one or more stars. As done in other HGBS papers \citep[e.g.,][]{Aquila}, we adopted the Bonnor-Ebert (BE) critical mass as an indicator of the maximum mass a core can have to be stable against gravitational collapse. For an isothermal sphere consisting of particles of mass $m$, the BE mass is \citep{bonnor}:
\begin{equation}
M_\mathrm{BE,crit}=\frac{2.4R_\mathrm{BE}KT}{\mu m_\mathrm{H} G}\label{BEcrit}
\end{equation}
where $\mu=2.33$ is the mean molecular weight per free particle \citep{mmw}, $R_\mathrm{BE}$ is the BE radius estimated as the deconvolved geometrical mean radius of each source measured in the high-resolution column density map, $K$ is the Boltzmann constant and $G$ is the gravitational constant. We imposed a deconvolved radius of 6\farcs1, one third of the FWHM at 250~$\mu$m, to sources that are unresolved. Note that in the above equation, non-thermal effects like turbulence or magnetic fields are not taken into account. If the source mass is $M$, we define an object as bound (or prestellar) if it has $M_\mathrm{BE,crit}/M\equiv\alpha_\mathrm{BE}\le 2$ \citep[see][]{Aquila}. Note that in the catalogue tables presented in Appendix~\ref{appCat}, we used the short name \textit{prestellar} for bound cores and \textit{starless} for unbound cores, even if both are starless, to be consistent with the other HGBS catalogues.

Since for unresolved cores the deconvolved radius tends to zero causing $M_\mathrm{BE,crit}\rightarrow0$, \citet{Aquila} used a set of simulated bound cores to find that bound, unresolved cores can be identified if $\alpha_\mathrm{BE}\le 5(\mathrm{HPBW}/\mathrm{FWHM})^{0.4}$, where FWHM is the non-deconvolved source radius estimated in the high-resolution column density map and HPBW is the linear map resolution, 18\farcs2. Sources that fulfilled this criterion were named \textit{candidate} bound cores while cores having $\alpha_\mathrm{BE}\le 2$ were named \textit{robust}. In this paper we adopt the same nomenclature.

We simulated bound cores as in \citet{Aquila} to derive the completeness limit in mass of our survey. In Appendix~\ref{complSec} we present our simulations. We find that for sources with mass above $\sim0.32\,M_\sun$ we are complete at $\ga$90\% in general, but this limit depends on the contrast between source intensity and background level. We used the simulations also to verify that the above formula for candidate bound cores works in our case too.

\subsection{Mass, temperature and radius}
In Fig.~\ref{MR} we show the relation between mass and radius for the whole population of 519 cores. The bound sources are in blue, unbound sources in green, and candidate bound sources in red.

\begin{figure}
\includegraphics[scale=0.4]{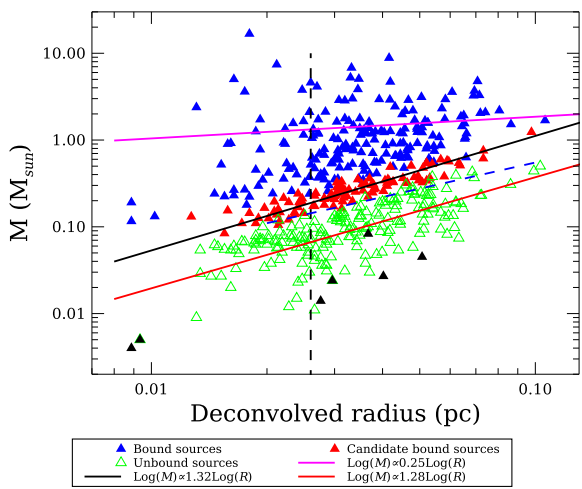}
\caption{Relation between mass and radius for all the 519 detected cores: vertical dashed line corresponds to 0.026~pc, the linear distance corresponding to the FWHM (18\farcs2) of the high-resolution column density map at 300~pc. The three power laws are fit to our data. Black triangles are hot starless cores discussed in the text; also the blue dashed line is discussed in the text.\label{MR}}
\end{figure}

At any radius, bound cores are more massive than unbound cores: this is not a consequence of using Equation~(\ref{BEcrit}) as stability criterion. A massive starless core can be unbound if warm enough and, conversely, a low-mass starless core can be bound if cold enough. The fact that bound cores are more massive implies some relation between $M$ and $T$, an anticorrelation that is in fact present in our data (see below Fig.~\ref{relTM}).

For the population of unbound cores, we find $M\propto R^{1.28}$, shown with the red line in Fig.~\ref{MR}, while for candidate bound cores, we find $M\propto R^{1.32}$, the solid black line in the same figure. Bound cores do not show a clear relation between mass and radius: indeed, the best-fit slope is $M\propto R^{0.25}$. The lack of correlation is not surprising given that: 1) Equation~(\ref{BEcrit}) is a relation between mass, temperature and radius: when, as in our case, $T$ is not assumed constant, as it was in general before \textit{Herschel}, there is not an obvious relation between $M$ and $R$; 2) if a bound core is collapsing, it is not even expected to be described by Equation~(\ref{BEcrit}). In fact, our stability criterion is not $M=M_\mathrm{BE}$, but rather $M$ greater than $M_\mathrm{BE}$ by some factor. Coming back to the original paper by Bonnor, in particular Fig.~1 in \citet{bonnor}, in the plot $p-V$ bound cores should populate the curve on the left of point A, where instability is dominating. Equation~(\ref{BEcrit}), on the contrary, describes the condition of a core that is found exactly on point A.

In Fig.~\ref{MR} the dashed blue line shows the $M-R$ relation for cores at $T=6.7$~K having $\alpha_\mathrm{BE}=2$. This temperature is the lowest we derived for prestellar cores. We consider this line to be the lower-limit for bound cores in our sample. Note, however, that for $T=6.7$~K and $\beta=2$, the modified black body distribution peaks at $\lambda>430\,\mu$m, which means that we may lack colder, bound cores in our data because we imposed that $F_\nu(350\,\mu\mathrm{m})>F_\nu(500\,\mu\mathrm{m})$ in Sect.~\ref{SEDfit}.

The smallest radius in our sample, 0.0089~pc or 6\farcs1 at 300~pc, simply reflects our choice to impose a deconvolved radius of 6\farcs1, i.e., one third of the FWHM at 250~$\mu$m, to sources that are unresolved. The largest observed radius is 0.11~pc for bound cores. During the detection and measurement of the sources with \textsl{getsources}, we imposed a maximum radius of 220\arcsec\ (0.32~pc), which is 3--4 times larger than the largest core size that we detect. Thus, this size constraint does not affect our results. Instead, we find that the cores in Perseus have a maximum size of $\sim0.1$~pc. This limiting size could be physical or may reflect another, unaccounted for bias. For instance, larger cores may have smaller density contrasts between their peaks and the background, which makes them more difficult to detect.

In Fig.~\ref{NT} we show the relation between cores temperature $T$ and their mean column density $N(\mathrm{H}_2)$ (defined as $M/\pi R^2$, where $T$, $M$ and $R$ were derived from source SED fitting). We modelled the trend seen with the equation $T=s\log N(\mathrm{H}_2) + b$. Two different slopes are found for prestellar and unbound cores: $-1.27\pm0.21$ and $-4.32\pm0.41$ for bound and unbound, respectively. This result is in contrast with what was found by \citet{2016MNRAS.459..342M} in L1495. They found there that the trend from the unbound populations continues smoothly when entering the bound region, with a single slope indeed enough to describe the overall trend. Their slope itself, $-3.6$, is different from the value of $-4.32$ we find for unbound cores. Such a difference could be due to the different environments in which cores are embedded.

\begin{figure}
\includegraphics[scale=0.4]{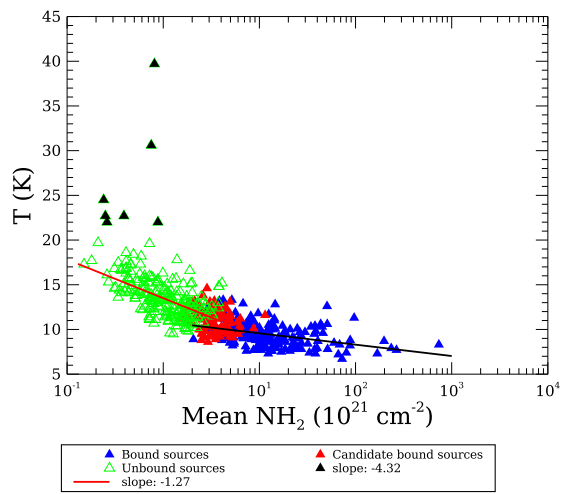}
\caption{Relation between $T$ and mean column density of the sources, where symbols are defined as in the previous figure. The two lines are linear fits of the form $T=s\log N(\mathrm{H}_2) + b$. The slopes $s$ are given in the legend and in the text where we also report the uncertainties. Black triangles are hot starless cores discussed in the text.\label{NT}}
\end{figure}

In deriving the $M-T$ slope for the unbound cores, we excluded seven objects that are clearly warmer than other cores, i.e., with $T>20$~K. Including these points the result is $s=-5.43\pm0.67$. These sources are the seven starless cores that show emission at 70~$\mu$m (see Sect.~\ref{stability}). As noted earlier, detection in the PACS blue band does not necessarily mean that a source has evolved into protostar, but it requires the object to be warm. For these sources the peak of the SED moves toward shorter $\lambda$ \citep[for $T>20$~K the peak moves at $\lambda<144$~$\mu$m for $\beta=2$,][]{grey} while at low $T$ the peak moves at longer wavelengths so that the emission in the PACS blue band becomes too faint to be detected. Moreover, the SED increases at higher $T$ as $L\propto T^{4+\beta}$ \citep[see][]{grey}. In principle, one could have bright enough starless core emission at 70~$\mu$m by also increasing the mass, but the intensity depends only linearly with $M$ so that huge masses are required to make an unbound, and cold, core visible at short $\lambda$. In fact, these seven sources are warm and have small masses (see black triangles in Fig.~\ref{NT}).

In principle, there is no reason why warm unbound cores should not exist. Their position in Fig.~\ref{NT} seems, however, to define a class of sources independent from the other unbound cores. It is possible that these sources have poorly-determined intensities so that $T$ was not estimated correctly. For example, they could be objects falling very close to strong sources, so that their envelope is unusually warm. Alternatively, these sources could have some intrinsic properties that make them different from the others. Since our goal here is a statistical description of the cold core population in Perseus, we do not investigate further this small sample.

Finally, in Fig.~\ref{relTM} we show the relation $M$ vs. $T$, which shows similar distinctions between the populations that were seen in Fig.~\ref{NT}. The anticorrelation between $M$ and $T$ is evident (see the discussion about Fig.~\ref{MR}).

\begin{figure}
\includegraphics[scale=0.4]{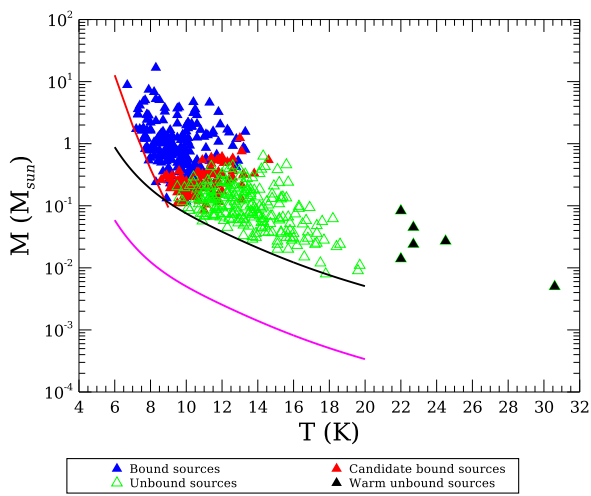}
\caption{Relation between $T$ and $M$ of the sources. The warmest unbound source at $\sim40$~K is not shown to limit the extension of the x-axis. The three lines are explained in the text.\label{relTM}}
\end{figure}

For any given temperature, there is a mass below which cores are not detected. To investigate from a quantitative point of view this relation, we studied if, and to what extent, the sensitivity of the \textit{Herschel} instruments plays any role. First we constructed a SED with the SPIRE-PACS Parallel Mode 1-$\sigma$ instrument sensitivity for a point source\footnote{\url{http://herschel.esac.esa.int/Docs/PMODE/html/}
\url{ch02s03.html}. We adopted the sensitivity when scanning in the nominal direction that were slightly worse than those in the orthogonal direction.}, divided by $\sqrt{2}$ because of the two directions. The sensitivities are then, in mJy, 14.8 and 33.2 for PACS, 8.91, 7.42 and 10.6 for SPIRE. We then made a fit to this SED with the same routine and grid of models used for the detected cores, and the same distance of 300~pc. The best fit model has $T=23.9$~K and $M=1.7\times10^{-4}\,M_\sun$. Then, we derived for each $T$, the mass that scales the model to the SED of sensitivities. For instance, at 20~K the mass is $M=3.4\times10^{-4}\,M_\sun$, at 10~K $M=5.3\times10^{-3}\,M_\sun$ and so on. To first approximation, without considering measurements uncertainties, these values give an estimate of the smallest detectable mass for a given $T$ compatible with \textit{Herschel} sensitivities; clearly, the match between models and sensitivities-based SED is worse and worse as $T$ decreases.

All these values give the magenta line in Fig.~\ref{relTM}, well described as $M\propto T^{-4}$. This line, derived from \textit{Herschel} sensitivities for a point source, can be considered as the locus of the smallest detectable mass of a compact source, as a function of temperature. The line is well below the detected masses: indeed, in star forming regions it is well known that the \textit{Herschel} detection of cores is not driven by the sensitivity but by the contrast between core and diffuse medium emission. Interestingly, a simple scaling by the arbitrary factor of 15, found by eye, gives the black line that reproduces well the smallest detected mass of starless cores for $T\ga 8$~K.

We can see that for $T\la 9$~K unbound cores are not detected: indeed, the minimum temperature in our catalogue for an unbound core is 9.5~K. This can be an observational bias because at low temperatures $M_\mathrm{BE}$ decreases while the smallest detectable mass of a source increases. So, an unbound cold core must have high mass to be detected, and then large radius not to be unbound, but low $T$ and high $R$ means a faint and diffuse object that could remain undetected.

The magenta line does not work with bound cores for $T\la8$~K. The smallest detected mass scales as $T^{-12.2}$, shown in Fig.~\ref{relTM} with a red line. This slope was found by choosing by eye two arbitrary bound cores, with $T=8.9$~K and $M=0.13\,M_\sun$, and with $T=7.2$~K, $M=1.73\,M_\sun$. A possible explanation for this behaviour is based on the finding that more massive prestellar cores are found in regions of high column density (see Fig.~\ref{distroNH2} in Section~\ref{filStars}). As a consequence, a bound core is detected if massive enough to show up against the bright diffuse medium, and the minimum mass does no longer reflect the law $M\propto T^{-4}$ that is more related to the instrument sensitivity. This conclusion is consistent with the result shown in Appendix~\ref{appComplMass} where we discuss the completeness in mass.

\section{Discussion}\label{discussione}
\subsection{Core mass function}\label{secCMF}
In Fig.~\ref{cmf} the histograms shows the core mass functions (CMF) for robust prestellar cores (blue line), and for robust plus candidate prestellar cores (red line) in Perseus in the form d$N$/d$\log M$. The histograms use logarithmic bin sizes of 0.1. The most massive candidate prestellar core has $M=1.24\, M_\sun$, so that at high mass blue and red lines overlap.

\begin{figure}
\includegraphics[scale=0.4]{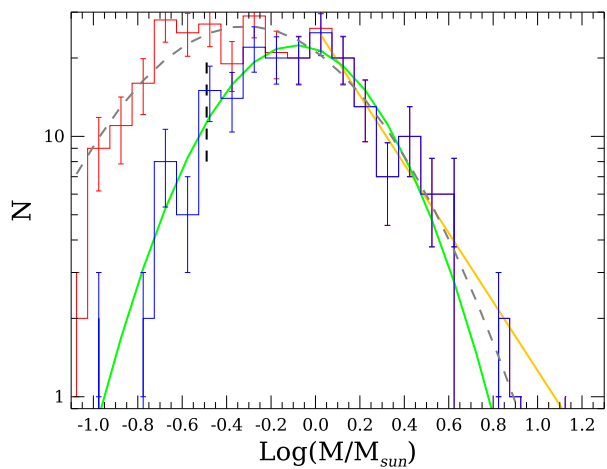}
\caption{The core mass function (d$N$/d$\log M$) for the 199 robust prestellar cores (blue histogram), and for the 299 robust and candidate prestellar cores (red histogram) in Perseus. Above $M\sim1.2$ the two distributions overlap. Histograms use a bin size of $\log M=0.1$. Green and dashed gray lines show a log normal fit, orange line a power-law fit (see text for details). The dashed black line shows the completeness limit at $\sim0.32\,M_\sun$.\label{cmf}}
\end{figure}

The results found by \citet{swiftBeau} show that with 199 sources it is difficult to discern between a power-law tail or a log-normal function to model the CMF shape. Thus, we fit to the CMF shown in Fig.~\ref{cmf} both a power-law and a log-normal separately.

The high-mass tail was fit with a power-law for $M>1\,M_\sun$. Given the limited number of sources, the size of the bin of the histogram may play a role. Hence, we followed the prescription of \citet{binSizeHisto} and used a variable bin size with a fixed number of values in each bin. In this procedure, $N$ values, here the core masses, are arranged in $m$ bins whose size is chosen such that each bin contains $n=N/m$ values. This means that each bin will have its own different size. \citet{binSizeHisto} have shown that this procedure is robust even when $n=1$, i.e., when each bin contains one single value, or the number $m$ of bins is as low as 3.

There are 96 prestellar cores with $M>1\,M_\sun$. We constructed different histograms having 4, 6, 8, 12, 16 and 24 sources per bin. For the fit we adopted the functional form d$N$/d$M\propto M^\alpha$ for the CMF and we find $\alpha=-2.321\pm0.035$, mean of the slope for the six cases, in agreement at $1\sigma$ level with Salpeter's slope of $-$2.35. In Fig.~\ref{cmf} the CMF is shown in the form d$N$/d$\log M\propto M^{\alpha+1}$ so the power-law best-fit, orange line, has a slope $-1.321$.

Another survey of cold cores in Perseus was performed by \citet{enoch} at 1.1~mm at CSO using BOLOCAM, with a spatial resolution of 31\arcsec. They fit the high-mass part of their CMF with two power laws merging at 2.5~$M_\sun$. They also tried a fit with only one slope for $M>0.8\,M_\sun$. Although they find a single power-law slope for d$N$/d$M$ of $-2.1\pm0.1$, which is close to the Salpeter value, they note that the CMF is better fit with two power laws. In this respect, it is worth mentioning that the three most massive cores in their survey, namely Bolo~48, Bolo~8 and Bolo~43 with masses of 25.6~$M_\sun$, 11.3~$M_\sun$ and 10.8~$M_\sun$ (for an assumed dust temperature of 10~K, distance of 250~pc, and a different opacity law), respectively, were detected by PACS in its blue band (ID 330, 68 and 303 in our catalogue, see Appendix~A). For this reason, we classified them as protostars, and did not include them in our CMF. It is possible that removing these sources would have made the BOLOCAM-derived CMF better described with a single slope.

\citet{sarah2010} also produced CMFs of Perseus using SCUBA observations at 850~$\mu$m from the SCUBA Legacy Survey. They found a d$N$/d$\log M$ slope of $-0.95\pm0.20$, which is within 2-$\sigma$ of Salpeter. Like \citet{enoch}, \citet{sarah2010} also assumed a fixed dust temperature (11~K based on \element[][]{NH}$_3$ observations of cores in Perseus). 

The SCUBA Legacy catalogue was incomplete, however. The sample did not include B5 and also missed several other high density regions. To account for the missing areas, \citet{sarah2010} produced predicted CMFs. The predicted CMF was still consistent with Salpeter, and they estimated the number of predicted starless cores to be 232, which isn't far off from the number of prestellar cores we found. Even if the predicted number of 232 will include some unbound cores, SCUBA 850~$\mu$m data are most sensitive to dense cores that are likely to be bound rather than fluffy unbound objects.  

Our CMF in Fig.~\ref{cmf} for robust prestellar cores (blue histogram) can be fit with a log-normal function (green line) with peak at $-0.086\pm0.028$ (or $M_\mathrm{peak}=0.820\pm0.053\,M_\sun$ and width $0.347\pm0.028$}. The system IMF for field stars in the Galaxy disc was parametrised by \citet{IMF2005} in terms of a log-normal function with a peak at 0.25~$M_\sun$ and $\sigma=0.55$. If we assume that the IMF of Perseus will resemble that of field stars, we derive $\epsilon_\mathrm{core}$, the star formation efficiency per single core, of 0.30. Adding the 199 candidate prestellar cores give the CMF shown as red histogram in Fig.~\ref{cmf}. Since these cores have $M<1.2\,M_\sun$, the effect is that of enlarging the distribution and shifting the peak to a smaller value. The log-normal function (dashed gray line) has peak at $-0.314\pm0.033$ (or $M_\mathrm{peak}=0.485\pm0.037\,M_\sun$) and width $0.471\pm0.036$; $\epsilon_\mathrm{core}$ increases to 0.52. In case not all 199 candidate prestellar cores are indeed bound, especially at low masses, we can reasonably estimate $\epsilon_\mathrm{core}$ in the range 0.3 -- 0.4. A similar value, 0.4, was found in Aquila \citep{Aquila}, but higher values are also reported in HGBS regions \citep[e.g.,][ with $\epsilon_\mathrm{core}\sim1$]{2016MNRAS.459..342M,milena2}.

Another estimate of $\epsilon_\mathrm{core}$ in Perseus was derived by \citet{mercimek} using data from literature. They found $\epsilon_\mathrm{core}\sim0.16$, about half of what we find, but using a different method: they estimated $\epsilon_\mathrm{core}$ as the ratio between the total mass of protostars (Class~0 and Class~I objects) over the total mass of starless cores. To compare these two values for $\epsilon_\mathrm{core}$ is not easy because the two methods depend on different assumptions. Here we limit to note that, on one side, \citet{mercimek} considered their value a lower limit; on the other side, our value can be lowered if the peak of the IMF is chosen differently.

In fact, instead of using the peak of the IMF for field stars, we can compare the peak of our CMF with the IMF derived in two regions in Perseus: \citet{IC348IMF} fit the IMF of IC348 with a log-normal function finding $M_\mathrm{peak}=0.21\pm0.02\,M_\sun$ and $\sigma=0.52\pm0.03$; \citet{ICNGCIMF} derived instead the median mass of the IMF for IC348 and NGC1333, finding for both clouds the value $\bar{M}=0.27\,M_\sun$. Since the IMF has a high-mass tail, clearly $\bar{M}>M_\mathrm{peak}$, but even assuming $M_\mathrm{peak}=\bar{M}$ we conclude that $0.26\la\epsilon_\mathrm{core}\la0.56$.

\citet{enoch} also fit their CMF with a log-normal distribution for $M>0.8\,M_\sun$. They could not estimate from their data the dust temperature as we did, however, so they gave two values for $M_\mathrm{peak}$: 0.9~$M_\sun$ and 0.3~$M_\sun$, assuming $T_\mathrm{d}=10$~K and 20~K, respectively. If we want to compare their results with ours, we have to take into account both the different opacities (0.0074~cm$^2$g$^{-1}$ following our opacity law instead of 0.0114~cm$^2$g$^{-1}$) and the different distance (300~pc here vs. 250~pc). Considering both differences, the two values for $M_\mathrm{peak}$ become $\sim2$~$M_\sun$ and $\sim0.7$~$M_\sun$ for $T_\mathrm{d}=10$~K and 20~K, respectively. Our $M_\mathrm{peak}$ is closer to the value corresponding to 20~K even if our range in $T_\mathrm{d}$ (see Fig.~\ref{relTM}) suggests that $T_\mathrm{d}=10$~K should be more appropriate.

\begin{figure}
\includegraphics[scale=0.4]{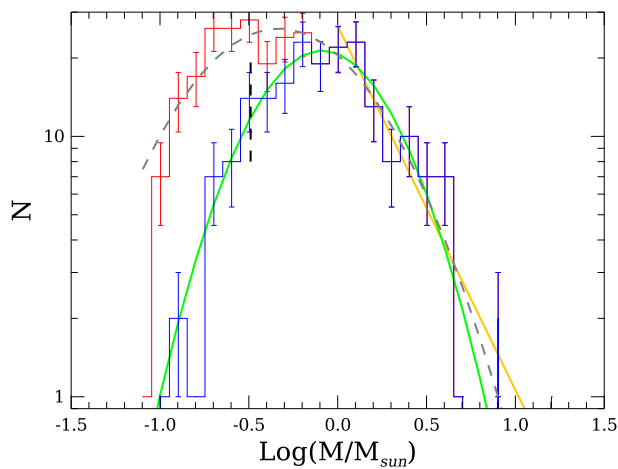}
\caption{The prestellar core mass function (d$N$/d$\log M$) for sources in Perseus when each core is assigned a different distance based on Equation~(\ref{fitDistanza}). Lines as in Fig.~\ref{cmf}.\label{cmfDist}}
\end{figure}

In Sect.~\ref{distPer}, we derived a relation between distance and source coordinates. In Fig.~\ref{cmfDist}, we show the resulting CMF for prestellar cores when each source is assigned a different distance. Note that physical radius increases with $d$ while mass increases as $d^2$, so, in general, $\alpha_\mathrm{BE}\propto d^{-1}$. As a consequence, objects that are closer will have in general lower $\alpha_\mathrm{BE}$ and objects that are further will have higher $\alpha_\mathrm{BE}$. Anyway, we find that changing the distance affects our prestellar core counts only minimally as the number of robust prestellar sources stays at 199 while candidate bound cores increases by 1, from 100 to 101.

Including a varying distance, the log-normal fits do not change significantly: the peak for the robust prestellar cores is $M_\mathrm{peak}=0.816\pm0.038\,M_\sun$ while that for all the bound cores, roubust and candidate, is $M_\mathrm{peak}=0.464\pm0.034\,M_\sun$. Within the uncertainties the two sets of values (with fixed and varying distance) are equivalent. The slope of the high-mass power law, on the contrary, changes: for $M>1.03M_\sun$ there are 92 sources and the average of the slopes found with 4 and 23 bins becomes $-2.398\pm0.026$, different from Salpeter's value at 1.8$\sigma$. Even if Equation~(\ref{fitDistanza}) is only an approximation, nonetheless it highlights the importance of assuming a non uniform distance for a complex like Perseus.

\subsection{Link between filamentary structure and star formation}\label{filStars}
One of the key result from \textit{Herschel} observations is the deep link between filamentary structures and the star formation in clouds. Numerous studies have shown that star formation preferentially occurs within filaments, with dense prestellar cores primarily in denser filaments \citep[e.g,][and references therein]{2014prpl.conf...27A}. This correlation has led to a possible paradigm, where filaments fragment into prestellar cores by gravitational instability \citep{2019A&A...629L...4A}.

We derived the number of robust prestellar cores inside a filament in two ways: in the most conservative approach, a core is considered internal to the filament if its distance from the spine is less than the filament width, actually 1.29$\times$FWHM. In the relaxed approach, we simply require that the core is inside the filament border (refer to Sect.~\ref{filDescr} and Fig.~\ref{esempioFilamento}). The first, conservative sample consists of 167 bound cores (84\% of the total) with a median mass of 1.0~$M_\sun$. The second, relaxed sample contains an additional 23 (12\%) bound cores, with a median mass of 0.96~$M_\sun$. The remaining 9 (4\%) prestellar cores have median mass of 0.47~$M_\sun$. So, bound cores are preferentially found within filaments, where the more massive cores are.

For the unbound cores, the percentages for the three samples (conservative, relaxed and out of filaments) are 47\%, 14\% and 39\%, respectively, with median mass of 0.10~$M_\sun$, 0.15~$M_\sun$, and 0.073~$M_\sun$.

\begin{figure*}
\centering
\includegraphics[scale=0.4]{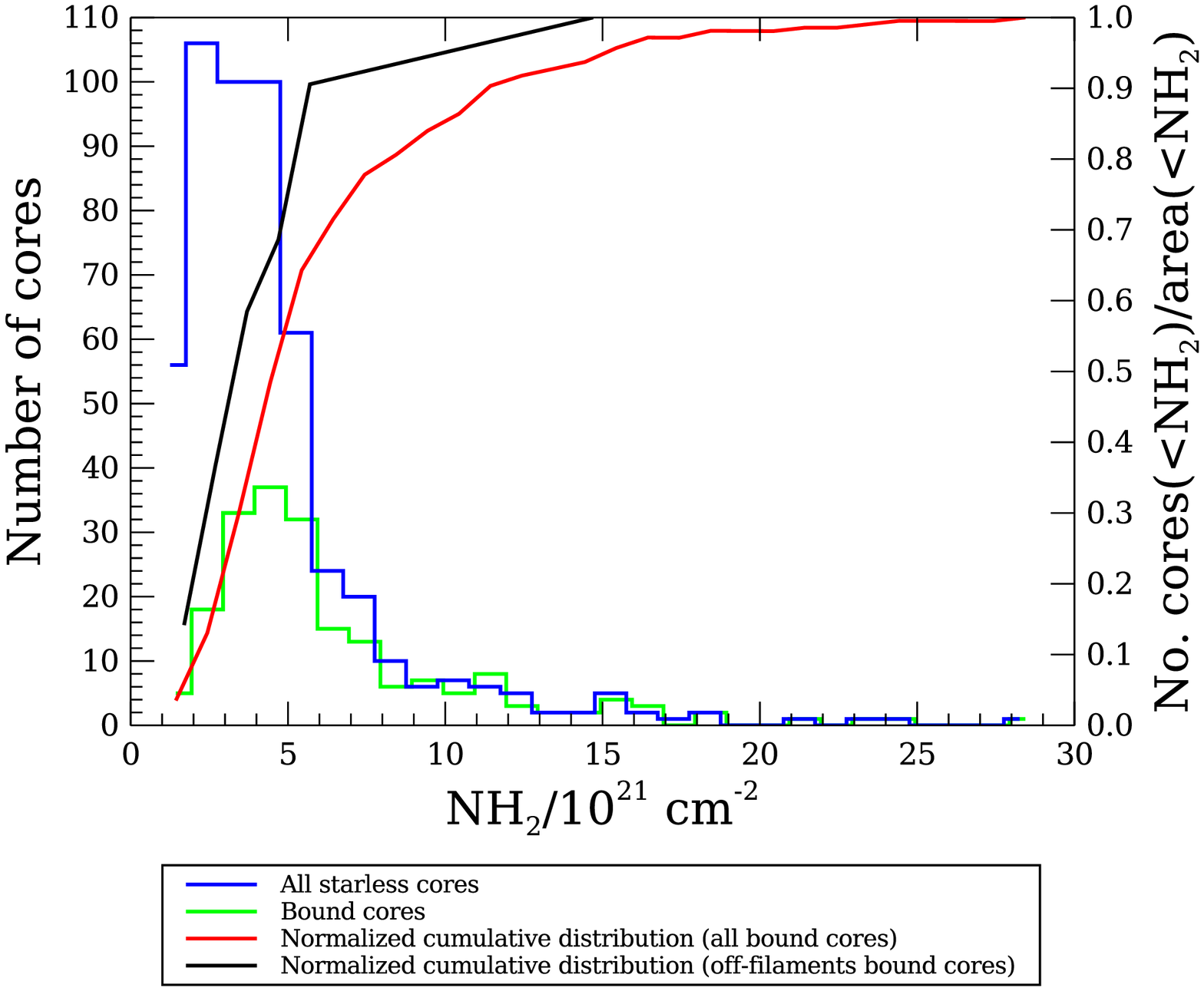}
\includegraphics[scale=0.4]{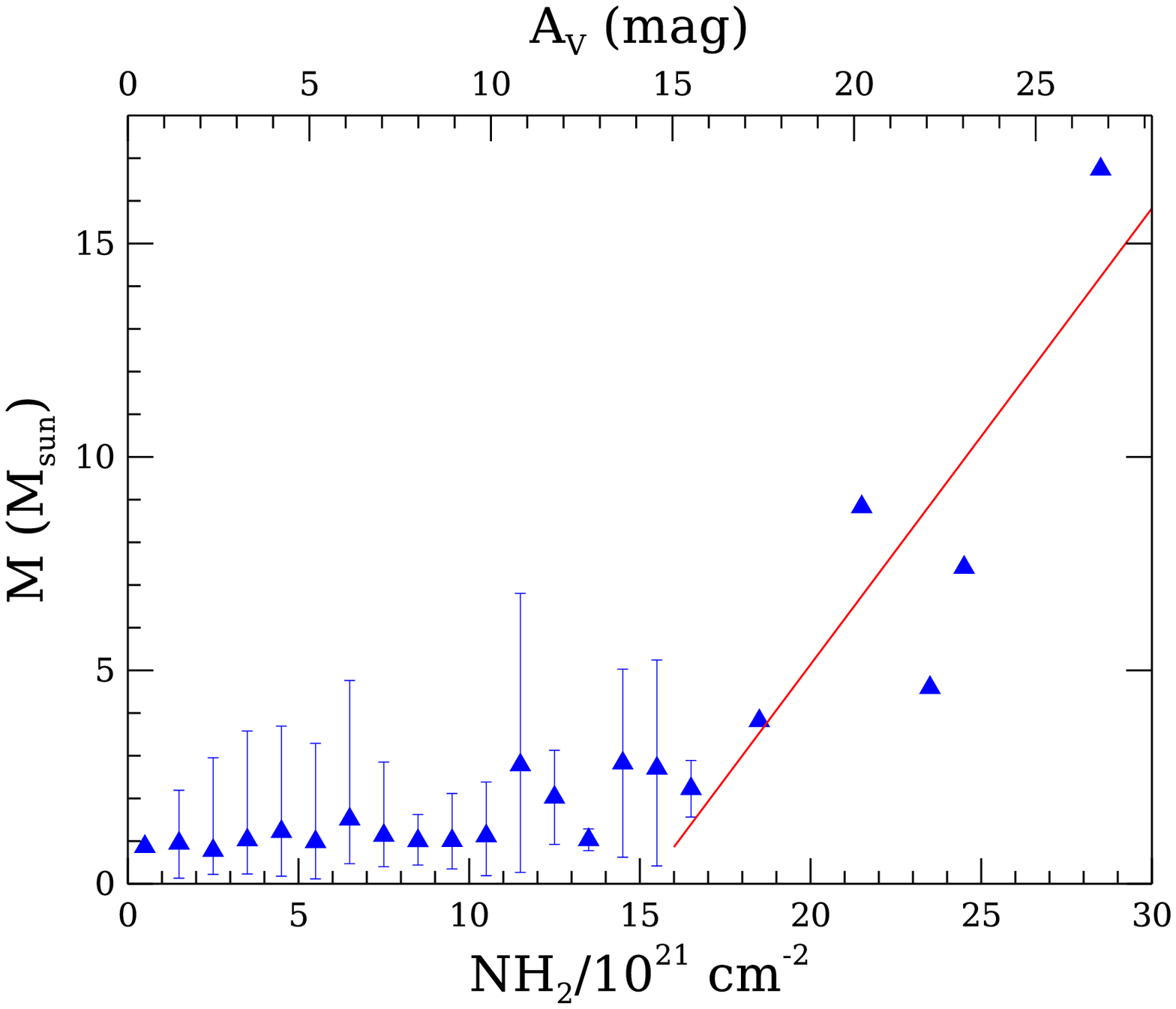}
\caption{Left panel: comparison of robust bound cores and starless cores. Histograms, left $y-$axis, show the distributions of all starless cores (blue) and all bound cores (green). The red and black lines, right $y-$axis, show cumulative surface density distributions for all bound cores and the off-filament bound cores, respectively. Right panel: prestellar core mass vs. background $N($H$_2)$. Triangles are the mean mass averaged over bins of $10^{21}$~cm$^{-2}$.\label{distroNH2}}
\end{figure*}

Figure~\ref{distroNH2} compares the background column density, measured by getsources, for the 519 starless cores (blue line) and 199 robust prestellar cores (green line, adding the 100 candidate prestellar cores shifts the distribution upwards but does not change the conclusions). The low number of cores at high density reflects to some extent the fact that regions with high $N($H$_2)$ have an intrinsically smaller area than regions at low density. To check for this geometrical bias, we also plot in the same figure the cumulative distribution of bound cores normalised to area. For those histograms, for each $N($H$_2)$ bin we computed the number of cores found at this level or below, and divided by the total area in the image where the density is below that value. The red curve shows the distribution for all 199 robust prestellar cores and the black curve shows the distribution for the 9 off-filament prestellar cores.

This cumulative distribution shows a few interesting features. First, about 90\% of prestellar cores are found above $\sim2\times10^{21}$~cm$^{-2}$, in line with the value of $\sim$2~mag found in Lupus \citep{milena2}. Higher values are instead found in Aquila \citep[7~mag,][]{Aquila} and in L1495 \citep[6~mag,][]{2016MNRAS.459..342M}. Second, we find two separate regimes in the distribution: an almost linear regime up to $\sim5\times10^{21}$~cm$^{-2}$ below which $\sim$60\% of bound sources are found, followed by a flatter distribution.

The distribution of the off-filament prestellar cores shows a different trend (clearly, the on-filament distribution overlaps almost perfectly with the cumulative distribution for all bound cores). For the off-filament cores, all but one sources are found at $N($H$_2)<5.5\times10^{21}$~cm$^{-2}$.

One might speculate that off-filament sources are less massive than on-filament sources because the former are found at lower background column density than the latter. This is not, however, the case, because if we consider the median of the mass for sources that have $N($H$_2)<5.5\times10^{21}$~cm$^{-2}$, we find for on-filament sources $\sim0.9\,M_\sun$. For off-filament sources, the median of the mass is only 0.24~$M_\sun$.

The right panel of Fig.~\ref{distroNH2} shows the mass of prestellar cores as a function of the background column density. We find a relatively flat mass distribution up to a threshold of $15\times10^{21}$~cm$^{-2}$. Above this threshold, sources are found only on-filament (actually, 8 out of 9 off-filament bound cores are below $5.5\times10^{21}$~cm$^{-2}$) and their masses grow with the background column density with a slope $1.06\pm0.23\, M_\sun\,10^{-21}$~cm$^2$, if a linear trend is assumed (red line in figure).

The distributions shown in Fig.~\ref{distroNH2} raise the question of why we do not detect prestellar cores below $0.9\times10^{21}$~cm$^{-2}$. For example, the prestellar core with the smallest level of column density has a mass of 0.87~$M_\sun$, 2.7 times above the completeness limit. The area of the map at column densities lower than this limit is 40\% of the total map. If cores were distributed randomly, we would expect 90 prestellar cores of any mass to be found at low column density values\footnote{Strictly speaking, background column density and column density are not the same thing in the sense that the former is estimated, the latter is measured; on the other hand, the column density is exactly the background column density if a source would be put at that position.}. The fact that we do not find bound cores at such low column densities implies that column densities $N($H$_2)\ga10^{21}$~cm$^{-2}$ or $A_\mathrm{V}\ga1$ is necessary for prestellar cores to form.

The possible existence of an $A_\mathrm{V}$ threshold for star formation in Perseus was already investigated in the past with contrasting results.  \citet{2005A&A...440..151H} found no column density cutoff for dense cores based on \element[][]{C}\element[][18]{O} (1-0) observations, whereas \citet{2006ApJ...646.1009K} found that submillimetre clumps were detected only at $A_\mathrm{V}>5$--7~mag. \citet{sarah2010} found that 94\% of SCUBA-identified sources were located at $N($H$_2)>3\times10^{21}$~cm$^{-2}$. Above the same level of column density, we find 63\% of our sources.

\begin{figure}
\centering
\includegraphics[scale=0.4]{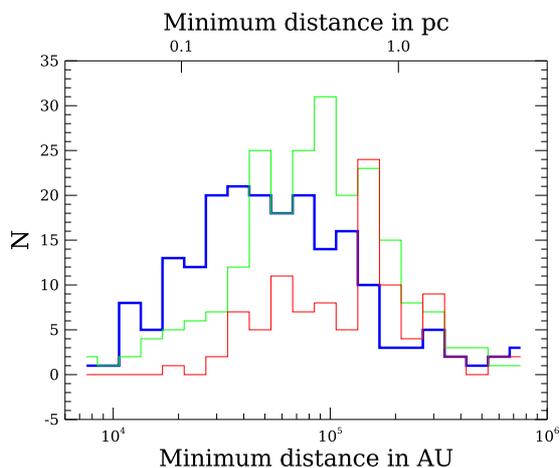}
\caption{Minimum distance between bound cores (blue), unbound cores (green), and tentative bound cores (red).\label{minDist}}
\end{figure}

In Fig.~\ref{minDist} we show the histogram of the minimum distance between cores. For robust bound cores, the distribution has a median value of 0.29~pc with a broad peak in the range 0.15--0.37~pc. For unbound cores, however, the distribution has a median of 0.47~pc with a broad peak shifted to slightly larger separations than in the previous case, between 0.23--0.73~pc. For the tentative bound cores the median is 0.74~pc.

\begin{figure*}
\centering
\includegraphics[scale=1]{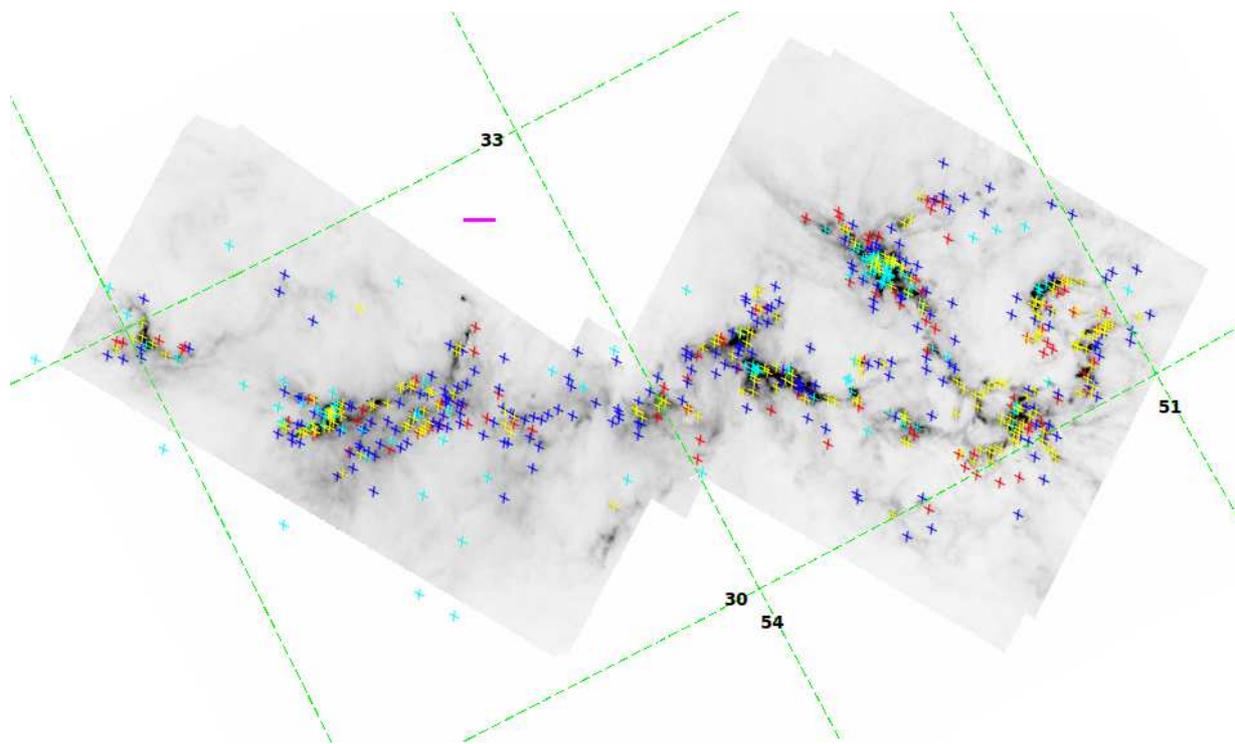}
\caption{Column density map overlaid with the positions of the 299 prestellar cores (yellow crosses and red crosses are robust and tentative cores, respectively), 220 unbound cores (blue hashes), and 132 protostars (cyan asterisks, note that these sources were detected on the 70~$\mu$m map that does not overlap completely with the SPIRE field of view, so five cores are found outside the map). Blue regions are from Fig.~\ref{zoneNH2}. The map is rotated by 28$^\circ$, with the top and bottom corners cut out. The magenta line in the centre shows the angular scale corresponding to 1~pc at 300~pc; J2000.0 coordinates grid is shown. \label{posizione}}
\end{figure*}

If cores were distributed uniformly across the observed region, one would expect a mean distance among bound cores of 0.69~pc, a factor 2.4 larger than the observed value of 0.29~pc. For starless cores, the mean distance is 0.65~pc, a factor 1.4 larger than 0.47~pc. This difference implies that prestellar cores cluster to some extent. This clustering is evident in Fig.~\ref{posizione}, which shows the position of all the prestellar cores (red crosses) and of the unbound cores (green crosses). We find that sources do not appear uniformly distributed, but with few exceptions they are confined inside the regions defined in Fig.~\ref{zoneNH2}.

\subsection{SFR, SFE and CFE}
The star formation rate (SFR) and star formation efficiency (SFE) are two fundamental properties derived from observations that any star-formation theory has to address, explain and predict. SFR is usually estimated as the total mass of young stellar objects (YSOs) divided by the mean duration for star formation \citep[e.g., 2~Myr,][]{evans}.

The total mass of YSOs is usually computed as the number $N$ of YSOs multiplied by the representative value of 0.5~$M_\sun$. Clearly, it is necessary to determine as precisely as possible $N$. Indeed, \textit{Herschel} observations aimed to achieve better completeness for starless cores rather than YSOs, so from our data alone we cannot derive a reliable SFR for Perseus.

The SFE is defined as the total mass of YSOs divided by the total mass $M_\mathrm{cloud}$ of the cloud, i.e., the mass of protostars plus gas. For each subcloud, the SFE depends on how the border is defined. As we have already discussed in Sect.~\ref{secMasse}, defining such borders is not obvious. \citet{sarah2} computed dense cloud masses for the Perseus subclouds using a column density threshold of $N($H$_2)=10^{22}$~cm$^{-2}$, chosen to select material primarily found in the high column density power-law tail. Here we follow another approach based on the local properties of each subcloud. Namely, we find for each subcloud the minimum background column density at the positions of the protostars in each region and compute $M_\mathrm{cloud}$ for $N($H$_2)>$min. In Table~\ref{SFEProto}, we report for each zone the number of protostars, the minimum background column density where a protostar is found, the mass of the subcloud within this minimum value, the exponent $s$ of the power-law fit to the associated PDF (see Sect.~\ref{PDFFit}) and the respective SFE(\%)$=\sum M_\mathrm{YSO}/(\sum M_\mathrm{YSO}+M_\mathrm{cloud})$. The number of protostars associated to the whole Perseus cloud is 126 instead of 132 because five sources fall in the region observed with PACS only, and one source is in a region masked out for 250~$\mu$m extraction because close to the border of the SPIRE map. In both cases the column density was not computed. L1451, B1E and L1468 were not included because no protostars were found.

\begin{table}
\centering
\begin{tabular}{lccccc}
\hline Region&N&$N($H$_2)$&$M$&$s$&SFE\\\hline
L1448  &4      &17.1&87   &$-2.429\pm0.035$&2.2\\
L1455  &8      &2.14&477  &$-2.866\pm0.024$&0.8\\
NGC1333&44     &4.26&613  &$-2.081\pm0.011$&3.5\\
Per6   &5      &1.68&146  &$-2.40\pm0.41$&1.4\\
B1     &15     &0.99&1336 &$-2.424\pm0.058$&0.6\\
IC348  &26     &2.06&1241 &$-4.042\pm0.033$&1.0\\
B5     &3      &2.96&108  &$-3.632\pm0.019$&1.4\\
HPZ6   &2      &2.71&20   &$-1.56\pm0.24$  &4.8\\
Perseus&126&0.64&11229&$-3.190\pm0.033$&0.6\\\hline
\end{tabular}
\caption{$N$: number of protostars (126 instead of 132 for whole Perseus because for six protostars the column density could not be computed, see text); $N($H$_2)$: minimum background column density in $10^{21}\,\mathrm{cm}^{-2}$; $M$: mass in $M_\sun$ for $N($H$_2)>$min; $s$ slope of PDF; SFE in \%.\label{SFEProto}}
\end{table}

For the whole Perseus cloud the minimum column density associated with a protostar is $0.64\times10^{21}$~cm$^{-2}$, below the limit of $1.6\times10^{21}$~cm$^{-2}$ that defines the last closed contour in the density column map. For this reason, the total mass of the cloud for $N($H$_2)>$min is likely underestimated. If we consider only protostars above the last closed column density contour, we find 113 protostars whose minimum $N($H$_2)$ is $1.65\times10^{21}$~cm$^{-2}$. The total mass is now 6198~$M_\sun$ and the SFE becomes 0.9\%.

\begin{figure}
\centering
\includegraphics[scale=0.4]{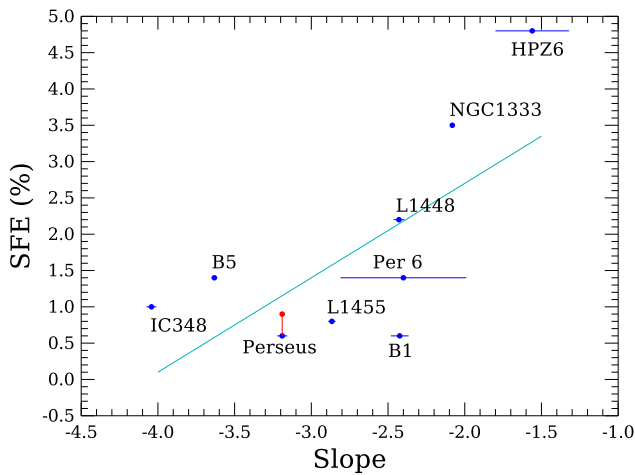}
\caption{Star formation efficiency as a function of the PDF slope for the different subclouds in Perseus, and for the whole cloud. The red point is an alternative SFE for Perseus (see text). The cyan line is a linear fit without taking into account uncertainties on the $x$-axis.\label{figuraSFE}}
\end{figure}

\citet{sarah2} found a linear correlation between SFE and slope $s$ of the PDF, SFE$\propto0.02s$, and they discussed the implications of such correlation. They identified candidate Class~0 sources through their detectability at 70~$\mu$m, excluding objects that could be more evolved, Class~I and Class~II sources. Since our column density map and slopes $s$ are different from those derived by \citet{sarah2}, as discussed in Sections~\ref{secMasse} and \ref{PDFFit}, one may wonder if their result still holds with our data. A complete answer to this question cannot be addressed in this paper because in our work we used the criterion of 70~$\mu$m visibility to pick up potential protostars in our catalogue, without separating sources in Class~0, I or II (see Sect.~\ref{estrazione}). Nonetheless, to give at least an indication that \citet{sarah2}'s linear correlation is present also in our data, we made a linear fit using fifth and sixth columns of Table~\ref{SFEProto}, finding SFE$\propto0.013s$, see Fig.~\ref{figuraSFE}. So, even with the caveat on the used sample, we confirm the correlation between SFE and $s$.

Since the \textit{Herschel}-based catalogue is more complete in starless cores than in YSOs, it is more reliable to derive from our sample the core mass efficiency (CFE) than the SFE. Similarly to the latter, the former is computed as CFE(\%)$=\sum M_\mathrm{pre}/(\sum M_\mathrm{pre}+M_\mathrm{cloud})$ where we use only prestellar cores because unbound cores are less likely to form stars. For the CFE we used the core masses computed during SED fitting instead of assuming a constant 0.5~$M_\sun$ value as done for SFE. The results are reported in Table~\ref{SFE}. Column~N gives the number of prestellar cores per subcloud. In a few cases there is also an uncertainty for N that will be explained below. N$_{\mathrm{II}}$ is the number of Class~II objects given by \citet{c2d2015}. They counted N$_{\mathrm{II}}$ also for the rest of the cloud (RC) defined as whole Perseus without IC348 and NGC1333. In Table ~\ref{SFE} we define RC as whole Perseus without the 4 subregions for which the number of Class~II objects was found. Column~$N($H$_2)$ has the same meaning as in Table~\ref{SFEProto}: the minimum background column density at the position of the prestellar cores in each region. Mass $M$, CFE and Area are then computed for $N($H$_2)>$min. The ratio $M$/Area gives the surface density reported in Col.~$\Sigma$(gas). Finally, $t_\mathrm{pre}$ is the estimated lifetime of the prestellar core phase, derived by scaling the number of bound cores to the number of Class~II objects, and assuming a constant SFR, i.e., $t_\mathrm{pre}=N_\mathrm{pre}/N_\mathrm{II}\times t_\mathrm{II}$. As before we fixed $t_\mathrm{II}$ to the usually adopted value of 2\,Myr for the duration of the Class~II.

\begin{table*}
\centering
\begin{tabular}{lccccccccccc}
\hline Region&N&N$_{\mathrm{II}}$&$N($H$_2)$&$M$&CFE&Area&$\Sigma$(gas)&$t_\mathrm{pre}$\\%&$N_-$&$N_+$
&&&$(10^{21}\,\mathrm{cm}^{-2})$&$(M_\sun)$&(\%)&(pc$^2$)&$(M_\sun\,$pc$^{-2}$)&(Myr)\\\hline
L1451  &19 &  &2.15&165 &13&1.77&93 \\
L1448  &18 &  &2.37&213 &10&1.24&172\\
L1455  &$36_{-4}^{+8}$& 4& 1.55&575 &7.9&7.44&77 &$18.0\pm$4.0\\%&5&9(5,4)
NGC1333&$23_{-1}^{+10}$&59&3.18&708 &7.4&3.82&186&$0.77\pm0.34$\\%&5&6(4,2)
Per6   &4  &  &3.59&56  &6.9&0.47&120\\
B1     &$31_{-2}^{+6}$& 8&1.67&1053&3.9&10.6&99&7.7$\pm1.5$\\%&3&6(3,3)
B1E    &8  &  &3.11&146 &4.4&1.47&99&\\
L1468  &4  &  &2.64&82  &4.3&1.11&73&\\
IC348  &$41_{-2}^{+6}$&104&3.62&660&8.5&5.25&126 &$0.78\pm0.12$\\%&2&9(4,5)
B5     &6  &  &2.93&109 &5.8&0.98&111 &\\
HPZ1   &1  &  &3.61&4.5  &11&0.04&106 &\\
HPZ2   &1  &  &2.08&13  &10&0.21&61 &\\
HPZ5   &1  &  &2.55&17  &2.0&0.23&76 &\\
HPZ6   &2  &  &3.99&9.5  &21&0.08&126 &\\
Perseus&$196_{-17}^{+60}$&231&1.67&5960&4.5&76 &77 &$1.69\pm0.52$\\%&28&56(36,20)
RC&$68_{-9}^{+30}$&56&&&&&&2.4$\pm1.1$\\\hline%&13&26(20,6)
\end{tabular}
\caption{$N$: number of prestellar cores; N$_{\mathrm{II}}$: number of Class~II objects from \citet{c2d2015}; $N($H$_2)$: minimum background column density; $M$: mass for $N($H$_2)$ larger than minimum; CFE: core formation efficiency; Area: area for $N($H$_2)$ larger than minimum; $\Sigma$(gas): surface density; $t_\mathrm{pre}$: lifetime of prestellar phase. RC (rest of cloud) refers to Perseus without the 4 regions with independent lifetime estimates \citep[note that ][define RC whole Perseus excluding only IC348 and NGC1333]{c2d2015}.\label{SFE}}
\end{table*}

The uncertainty in $t_\mathrm{pre}$ was derived based on $\alpha_\mathrm{BE}$, the criterion to define bound a core (see Sect.~\ref{stability}). We first derived the uncertainty $\Delta\alpha_\mathrm{BE}$ from the uncertainties in mass and temperature of each source. Then, amoung the bound cores we counted the number of sources having $\alpha_\mathrm{BE}+\Delta\alpha_\mathrm{BE}>2$: within the uncertainty these sources could be unbound cores. Let's call this number $N_-$. Similarly, unbound or candidate bound sources with reliable SED having $\alpha_\mathrm{BE}-\Delta\alpha_\mathrm{BE}\le2$ could be potential prestellar cores. We define $N_+$ the number of these sources. In Table~\ref{SFE} we report the number of prestellar cores as $N_{N_-}^{N_+}$ for the subclouds with N$_{\mathrm{II}}$ estimated. For the other soubclouds, which do not have an estimate of N$_{\mathrm{II}}$, $t_\mathrm{pre}$ was not computed. The uncertainty in $N$ is assumed as max($N_+,N_-$). This uncertainty is then propagated to $t_\mathrm{pre}$.

The resulting lifetime for the whole Perseus cloud is $t_\mathrm{pre}=1.69\pm0.52$~Myr that overlaps, within $1\sigma$, with the value $1.2\pm0.3$~Myr found in Aquila by \citet{Aquila}.

We see that for L1455 and B1 $t_\mathrm{pre}$ is much higher than for the other regions. This can be explained as if both clouds are very young such that star formation started very recently. In this case, in fact, the number of Class~II objects is less than expected assuming a constant SFR \citep[for a similar case reported in Lupus~I and Lupus~IV, see][]{kazy}. On the contrary, in NGC1333 and IC348 we observe a deficit of prestellar cores compared to Class~II objects. This can be interpreted as a decelerating SFR in these two regions, with IC348 having the shortest lifetime, in line with the known fact that this cloud is the oldest of the Perseus complex. We do not have the number of Class~II stars for the other regions, but the relative high $t_\mathrm{pre}$ for RC might suggest that star formation started recently also in the other subclouds.

\begin{figure}
\centering
\includegraphics[scale=0.46]{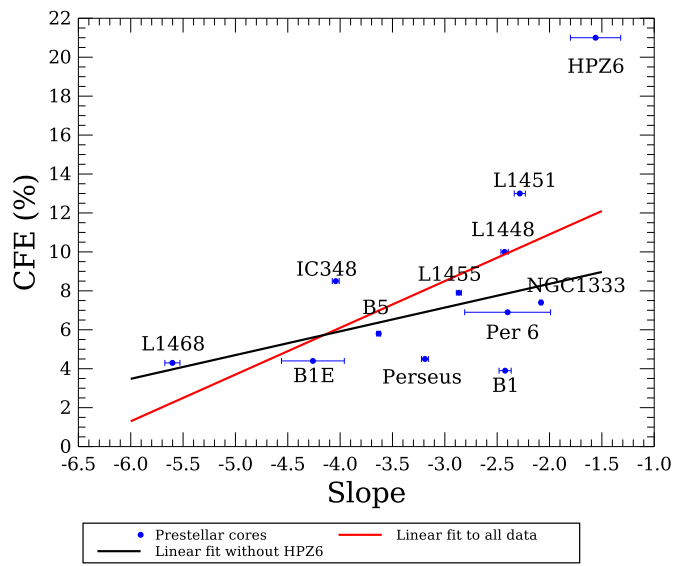}
\caption{Core formation efficiency as a function of the power-law slope of the PDFs for the different subclouds in Perseus, and for the whole cloud.\label{figuraCFE}}
\end{figure}

In Fig.~\ref{figuraCFE} we show the CFE as a function of the PDF power-law exponent. Similarly to the case of SFE, also for CFE we can consider the threshold in column density of $1.6\times10^{21}$~cm$^{-2}$. In doing so, $N$ decreases to 196 with a CFE of 4.5\%, this is the value used in figure. The red line is a linear fit to all the data with a slope of 2.4. If the value of HPZ6, looking like an outlier, is not used, the result is the black line with a slope 1.22.

The mean value of all the subclouds excluding Perseus as a whole is $8.2\pm4.8$\% or $7.2\pm2.8$\% if we exclude also the HPZ\#6 zone: the CFE of the whole cloud is within 1$\sigma$ from these values.% 

Another time scale important to compare with our estimate of the prestellar phase duration is the free-fall time, i.e., the time required for a uniform sphere of density $\rho$ and pressureless gas, i.e., gas at zero temperature, to collapse to infinite density \citep[e.g.,][]{krumholz}
\begin{equation}
t_\mathrm{ff}=\sqrt{\frac{3\pi}{32G\rho}}\label{tff}
\end{equation}
where $G$ is the gravitational constant and $\rho$ is the uniform density.

\begin{figure}
\centering
\includegraphics[scale=0.4]{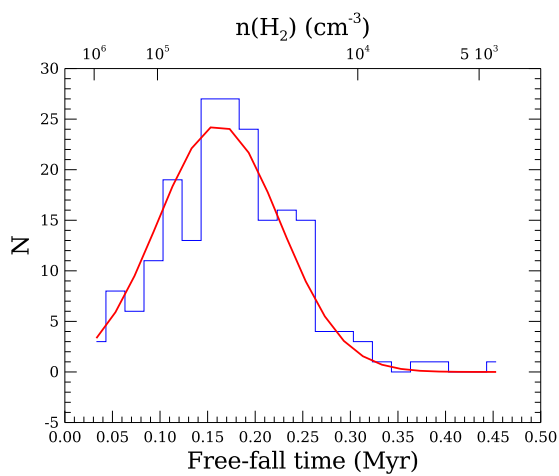}
\caption{Distribution of free-fall time for robust prestellar cores. Top-axis is the uniform density for spherical cores ($3M/4\pi R^3$). The red line is a Gaussian fit.\label{figuratFF}}
\end{figure}

In Fig.~\ref{figuratFF} we show the distribution of $t_\mathrm{ff}$ for the 199 robust bound cores. The Gaussian fit to the distribution, plotted in red, has peak at $0.16\pm 0.01$~Myr, and $\sigma=0.06\pm0.01$~Myr, where we assigned and uncertainty of 0.1~Myr, half size of the histogram bin, to both parameters. For a star of 0.25~$M_\sun$, peak of the IMF, this implies a mean mass accretion rate $\dot{M}\sim1.6\times10^{-6}\,M_\sun\,$yr$^{-1}$.

Our peak of 0.16~Myr is close to the value of 0.10~Myr derived by \citet{mercimek} using the sample of 69 cores defined by \citet{sarah2010}.

\section{Conclusions}
In this paper, we analysed photometric observations of the Perseus complex executed with PACS at 70~$\mu$m and 160~$\mu$m, and SPIRE at 250~$\mu$m, 350~$\mu$m and 500~$\mu$m.

We summarize our main conclusions here:
\begin{enumerate}
\item using distances to protostars from \textit{GAIA} DR2, we find a linear correlation between coordinates ($\alpha,\delta)$ in the cloud and distance across whole Perseus. This correlation is the first estimate of such a trend that, qualitatively, was already suspected. Distances estimated from our relation do not take into account the depth of the region at any point and are meant to give representative values; but they are not limited to a few lines of sigths and are in very good agreement with other recent estimates reported in literature found with more robust ways. We confirm that 300~pc is a good mean distance value for Perseus;
\item we produce dust temperature and column density maps. The dust temperature is in the range 10--28~K. Two main peaks in $T_\mathrm{d}$ distribution are present at 16.4~K and 17.1~K, corresponding to the average temperature of the diffuse medium in West and East Perseus, respectively. A small bump is present at 19.2~K associated with the inner part of NGC1333 and IC348. The column density range is $2.5\times10^{20}$--$1.4\times10^{23}$~cm$^{-2}$ with the smallest closed contour at $1.6\times10^{21}$~cm$^{-2}$. $T_\mathrm{d}$ and $N($H$_2)$ are anticorrelated throughout the complex, with the exception of the inner part of NGC1333 and IC348 that are both warm and dense. We also show that a lack of data at $\lambda<160\,\mu$m makes $T_\mathrm{d}$ underestimated by more than 10\% for $T_\mathrm{d}>22$~K;
\item extrapolating the column density and temperature maps to 70~$\mu$m shows that most of the cloud would appear as infrared dark clouds if observed with enough sensitivity. We also find excess emission at 70~$\mu$m above the extrapolated fluxes, which we attribute to very small dust grains;
\item we find that the optical depth derived from our column density map, and extrapolated at 850~$\mu$m, agrees with that measured with \textit{Planck} for $\tau_\mathrm{P}<3\times 10^{-4}$, where $\tau_\mathrm{P}$ is the optical depth measured by \textit{Planck}. At higher $\tau_\mathrm{P}$, the \textit{Herschel} optical depth is higher than \textit{Planck}'s. This can be due to a change in the dust opacity. Our column density map, on the contrary, disagrees with the column density maps derived from near-infrared extinction maps if masses within the same $A_{\mathrm V}$ contour are compared; a better agreement is found when masses are derived from same areas in the two maps. The comparison is, in any case, not easy, either because of the varying dust opacity, or because in star-forming regions $R\equiv A_{\mathrm V}/E(\mathrm{B}-\mathrm{V})>3.1$ and in Perseus it is also variable throughout the complex;
\item we find a typical filament width of 0.08~pc for an assumed distance of 300~pc, confirming the general result of previous HGBS studies that a characteristic filament width of 0.1~pc exists in star-forming regions;%. The distribution of filament orientations shows a peak at 145$^\circ$ in agreement with what was found in dust polarization studies but we cannot confirm the presence of a second peak previously observed at 71$^\circ$;
\item on average, star-forming filaments in Perseus have mean line masses significantly below the critical value of 16~$M_\sun\,$pc$^{-1}$, above which an isothermal filament at 10~K and infinite radius becomes unstable  to radial collapse. But we suggest that with a proper internal volume density a filament with a line mass as low as 8~$M_\sun\,$pc$^{-1}$, i.e., half the critical line mass, can fragment.
\end{enumerate}

We extracted compact cores from the intensity maps and used SED fitting to determine their masses and temperatures. We find:
\begin{enumerate}
\setcounter{enumi}{6}
\item 199 starless cores have masses larger than their corresponding Bonnor-Ebert masses, above which an isothermal sphere of gas cannot provide sufficient support against gravity. These cores are thus thought to be collapsing, or close to collapse, and forming stars. We name these sources robust bound or prestellar cores. Other 100 cores may be also collapsing and we name these candidate prestellar cores. On the other hand, 220 sources have masses below their corresponding Bonnor-Ebert threshold so they are thought to be not dominated by gravity. We name these objects unbound cores. Next, we name 132 cores protostars because they show, through emission at 70~$\mu$m, that an internal source of energy should be already present. Finally, the remaining 165 sources cannot be fit satisfactorily with a modified blackbody because their SEDs are too uncertain;
\item on the plane $M-R$ starless cores do not trace a clear path but show a large scatter with a faint hint of a correlation between $M$ and $R$. Ignoring the scatter and fitting anyway a power-law, for unbound sources we find $M\propto R^{1.28}$. A similar relation is found for candidate prestellar cores: $M\propto R^{1.32}$. For bound sources, however, we find $M\propto R^{0.25}$, a relation that is shallower than expected for Bonnor-Ebert spheres ($M\propto R$). We suggest that our slope 0.25 likely reflects the lack of a correlation for the prestellar cores as a whole, probably a hint that these sources are indeed collapsing. From $M$ and $R$, we derived a mean column density for sources that appears anticorrelated with the temperature, with different exponents for bound and unbound sources;
\item $M$ and $T$ are also anticorrelated. We have shown that the, for a given $T$, the smallest detectable mass of an unbound core is given by a combination of the \textit{Herschel} instruments sensitivity, that gives a relation $M\propto T^{-4}$, scaled by an empirical factor that is likely due to the presence of the bright diffuse medium in which cores are embedded. For bound cores, however, the relation is much steeper $M\propto T^{-12}$, probably a consequence of the fact that more massive, and colder, bound cores are found in brighter regions of the clouds;
\item the prestellar core mass function (CMF) can be modelled with a log-normal curve having a peak at $M\sim0.82\,M_\sun$, above the completeness limit in mass estimated at 0.32~$M_\sun$. Assuming a one-to-one mapping of the CMF over the IMF, this peak implies a core mass efficiency of 0.30. The high-mass tail of the CMF can be modelled with a power law of slope $-2.321\pm0.035$ in agreement, at 1$\sigma$, with Salpeter's IMF value of --2.35. The two distributions, log-normal and power-law, were modelled separately because the number of bound cores is not large enough to make a combined fit; for the same reason, we cannot conclude if one distribution alone can fit the whole dataset;
\item more than 84\% of robust bound cores are found within filaments versus 47\% for unbound cores. On-filament bound sources are on average more massive than the off-filament ones. No bound cores are found below $0.9\times10^{20}$~cm$^{-2}$. Should sources be distributed uniformly on the map, 90 prestellar cores of any mass should have been found at column densities lower than that value. We propose that a limit of $\sim1\times10^{21}$~cm$^{-2}$ in column density is required to form bound cores;
\item we find a correlation between the star formation efficiency and slope of the column density PDF for different Perseus subclouds. The correlation has a slope of 1.3\%. We also find a correlation with the core formation efficiency, with a slope between 1.2\% and 2.4\%, depending on wether we use all the subclouds or we exclude from the sample one extreme value;
\item assuming a constant star formation rate we derived a lifetime for the prestellar phase $t_\mathrm{pre}$ of 1.69~Myr for Perseus but with significant scatter for the individual subclouds;
\item we find a typical free fall time of 0.16~Myr that implies a mean mass accretion rate of $\dot{M}\sim1.6\times10^{-6}\,M_\sun\,$yr$^{-1}$ to form a star of 0.25~$M_\sun$, peak of the IMF.
\end{enumerate}

The catalogue of all 816 detected sources is given in the Appendix~A. We also list in Appendix~E sources that were excluded from our analysis, e.g, galaxies.

\begin{acknowledgements}
We thank the anonymous referee for the valuable comments that improved the readability of the paper.
SB and NS ackowledge support by the french ANR and the german DFG through the project "GENESIS" (ANR-16-CE92-0035-01/DFG1591/2-1).
PP and DA acknowledge support by FCT/MCTES through national funds (PIDDAC) by this grant UID/FIS/04434/2019.
PP also acknowledge the support from the fellowship SFRH/BPD/110176/2015 funded by FCT (Portugal) and POPH/FSE (EC).

PACS has been developed by a consortium of institutes led by MPE (Germany) and including UVIE (Austria); KU Leuven, CSL, IMEC (Belgium); CEA, LAM (France); MPIA (Germany); INAF-IFSI/OAA/OAP/OAT, LENS, SISSA (Italy); IAC (Spain). This development has been supported by the funding agencies BMVIT (Austria), ESA-PRODEX (Belgium), CEA/CNES (France), DLR (Germany), ASI/INAF (Italy), and CICYT/MCYT (Spain).

SPIRE has been developed by a consortium of institutes led by Cardiff University (UK) and including Univ. Lethbridge (Canada); NAOC (China); CEA, LAM (France); IFSI, Univ. Padua (Italy); IAC (Spain); Stockholm Observatory (Sweden); Imperial College London, RAL, UCL-MSSL, UKATC, Univ. Sussex (UK); and Caltech, JPL, NHSC, Univ. Colorado (USA). This development has been supported by national funding agencies: CSA (Canada); NAOC (China); CEA, CNES, CNRS (France); ASI (Italy); MCINN (Spain); SNSB (Sweden); STFC (UK); and NASA (USA).

HIPE are joint developments by the Herschel Science Ground Segment Consortium, consisting of ESA, the NASA Herschel Science Center, and the HIFI, PACS and SPIRE consortia.

Part of this work has received support from the European Research Council (ERC Advanced Grant Agreement no. 291294 -- ORISTARS) and from the 
French national programs of CNRS/INSU on stellar and ISM physics (PNPS and PCMI).

This research has made use of: the SIMBAD database, operated at CDS, Strasbourg, France; the NASA/IPAC Extragalactic Database (NED) and Infrared Science Archive which are operated by the Jet Propulsion Laboratory, California Institute of Technology, under contract with the National Aeronautics and Space Administration; data from the ESA mission {\it Gaia} (\url{https://www.cosmos.esa.int/gaia}), processed by the {\it Gaia}
Data Processing and Analysis Consortium (DPAC, \url{https://www.cosmos.esa.int/web/gaia/dpac/consortium}). Funding for the DPAC has been provided by national institutions, in particular the institutions participating in the {\it Gaia} Multilateral Agreement.

Figures showing the column density and temperature maps were done with \textsl{DS9} \citep{ds9}.

\end{acknowledgements}

\bibliographystyle{aa} % style aa.bst
\bibliography{perseus.bib} % your references Yourfile.bib

\begin{thebibliography}{101}
\expandafter\ifx\csname natexlab\endcsname\relax\def\natexlab#1{#1}\fi

\bibitem[{{Alves} {et~al.}(2007){Alves}, {Lombardi}, \&
  {Lada}}]{2007A&A...462L..17A}
{Alves}, J., {Lombardi}, M., \& {Lada}, C.~J. 2007, \aap, 462, L17

\bibitem[{{Alves} {et~al.}(2017){Alves}, {Lombardi}, \& {Lada}}]{alves}
{Alves}, J., {Lombardi}, M., \& {Lada}, C.~J. 2017, \aap, 606, L2

\bibitem[{{Alves de Oliveira} {et~al.}(2013){Alves de Oliveira}, {Moraux},
  {Bouvier}, {Duch{\^e}ne}, {Bouy}, {Maschberger}, \& {Hudelot}}]{IC348IMF}
{Alves de Oliveira}, C., {Moraux}, E., {Bouvier}, J., {et~al.} 2013, \aap, 549,
  A123

\bibitem[{{Andrae}(2010)}]{andrae}
{Andrae}, R. 2010, ArXiv e-prints [\eprint[arXiv]{1009.2755}]

\bibitem[{{Andrae} {et~al.}(2010){Andrae}, {Schulze-Hartung}, \&
  {Melchior}}]{andrae2}
{Andrae}, R., {Schulze-Hartung}, T., \& {Melchior}, P. 2010, ArXiv e-prints
  [\eprint[arXiv]{1012.3754}]

\bibitem[{{Andr{\'e}} {et~al.}(2019){Andr{\'e}}, {Arzoumanian}, {K{\"o}nyves},
  {Shimajiri}, \& {Palmeirim}}]{2019A&A...629L...4A}
{Andr{\'e}}, P., {Arzoumanian}, D., {K{\"o}nyves}, V., {Shimajiri}, Y., \&
  {Palmeirim}, P. 2019, \aap, 629, L4

\bibitem[{{Andr{\'e}} {et~al.}(2014){Andr{\'e}}, {Di Francesco},
  {Ward-Thompson}, {Inutsuka}, {Pudritz}, \& {Pineda}}]{2014prpl.conf...27A}
{Andr{\'e}}, P., {Di Francesco}, J., {Ward-Thompson}, D., {et~al.} 2014, in
  Protostars and Planets VI, ed. H.~{Beuther}, R.~S. {Klessen}, C.~P.
  {Dullemond}, \& T.~{Henning}, 27

\bibitem[{{Andr{\'e}} {et~al.}(2010){Andr{\'e}}, {Men'shchikov}, {Bontemps},
  {K{\"o}nyves}, {Motte}, {Schneider}, {Didelon}, {Minier}, {Saraceno},
  {Ward-Thompson}, {di Francesco}, {White}, {Molinari}, {Testi}, {Abergel},
  {Griffin}, {Henning}, {Royer}, {Mer{\'{\i}}n}, {Vavrek}, {Attard},
  {Arzoumanian}, {Wilson}, {Ade}, {Aussel}, {Baluteau}, {Benedettini},
  {Bernard}, {Blommaert}, {Cambr{\'e}sy}, {Cox}, {di Giorgio}, {Hargrave},
  {Hennemann}, {Huang}, {Kirk}, {Krause}, {Launhardt}, {Leeks}, {Le Pennec},
  {Li}, {Martin}, {Maury}, {Olofsson}, {Omont}, {Peretto}, {Pezzuto}, {Prusti},
  {Roussel}, {Russeil}, {Sauvage}, {Sibthorpe}, {Sicilia-Aguilar}, {Spinoglio},
  {Waelkens}, {Woodcraft}, \& {Zavagno}}]{2010A&A...518L.102A}
{Andr{\'e}}, P., {Men'shchikov}, A., {Bontemps}, S., {et~al.} 2010, \aap, 518,
  L102

\bibitem[{{Arce} {et~al.}(2011){Arce}, {Borkin}, {Goodman}, {Pineda}, \&
  {Beaumont}}]{bubbles}
{Arce}, H.~G., {Borkin}, M.~A., {Goodman}, A.~A., {Pineda}, J.~E., \&
  {Beaumont}, C.~N. 2011, \apj, 742, 105

\bibitem[{{Arzoumanian} {et~al.}(2019){Arzoumanian}, {Andr{\'e}},
  {K{\"o}nyves}, {Palmeirim}, {Roy}, {Schneider}, {Benedettini}, {Didelon}, {Di
  Francesco}, {Kirk}, \& {Ladjelate}}]{doris2019}
{Arzoumanian}, D., {Andr{\'e}}, P., {K{\"o}nyves}, V., {et~al.} 2019, \aap,
  621, A42

\bibitem[{{Bally} {et~al.}(2008){Bally}, {Walawender}, {Johnstone}, {Kirk}, \&
  {Goodman}}]{bally}
{Bally}, J., {Walawender}, J., {Johnstone}, D., {Kirk}, H., \& {Goodman}, A.
  2008, {The Perseus Cloud}, ed. B.~{Reipurth}, Vol.~4, 308

\bibitem[{{Bauke}(2007)}]{bauke}
{Bauke}, H. 2007, European Physical Journal B, 58, 167

\bibitem[{{Benedettini} {et~al.}(2018){Benedettini}, {Pezzuto}, {Schisano},
  {Andr{\'e}}, {K{\"o}nyves}, {Men'shchikov}, {Ladjelate}, {Di Francesco},
  {Elia}, {Arzoumanian}, {Louvet}, {Palmeirim}, {Rygl}, {Schneider},
  {Spinoglio}, \& {Ward-Thompson}}]{milena2}
{Benedettini}, M., {Pezzuto}, S., {Schisano}, E., {et~al.} 2018, \aap, 619, A52

\bibitem[{{Benedettini} {et~al.}(2015){Benedettini}, {Schisano}, {Pezzuto},
  {Elia}, {Andr{\'e}}, {K{\"o}nyves}, {Schneider}, {Tremblin}, {Arzoumanian},
  {di Giorgio}, {Di Francesco}, {Hill}, {Molinari}, {Motte}, {Nguyen-Luong},
  {Palmeirim}, {Rivera-Ingraham}, {Roy}, {Rygl}, {Spinoglio}, {Ward-Thompson},
  \& {White}}]{milena}
{Benedettini}, M., {Schisano}, E., {Pezzuto}, S., {et~al.} 2015, \mnras, 453,
  2036

\bibitem[{{Bernard} {et~al.}(2010){Bernard}, {Paradis}, {Marshall}, {Montier},
  {Lagache}, {Paladini}, {Veneziani}, {Brunt}, {Mottram}, {Martin},
  {Ristorcelli}, {Noriega-Crespo}, {Compi{\`e}gne}, {Flagey}, {Anderson},
  {Popescu}, {Tuffs}, {Reach}, {White}, {Benedettini}, {Calzoletti},
  {Digiorgio}, {Faustini}, {Juvela}, {Joblin}, {Joncas}, {Mivilles-Deschenes},
  {Olmi}, {Traficante}, {Piacentini}, {Zavagno}, \& {Molinari}}]{JPB}
{Bernard}, J.-P., {Paradis}, D., {Marshall}, D.~J., {et~al.} 2010, \aap, 518,
  L88

\bibitem[{{Bobylev}(2016)}]{GB}
{Bobylev}, V.~V. 2016, Astronomy Letters, 42, 544

\bibitem[{{Bohlin} {et~al.}(1978){Bohlin}, {Savage}, \& {Drake}}]{bsd}
{Bohlin}, R.~C., {Savage}, B.~D., \& {Drake}, J.~F. 1978, \apj, 224, 132

\bibitem[{{Bonnor}(1956)}]{bonnor}
{Bonnor}, W.~B. 1956, \mnras, 116, 351

\bibitem[{{Bresnahan} {et~al.}(2018){Bresnahan}, {Ward-Thompson}, {Kirk},
  {Pattle}, {Eyres}, {White}, {K{\"o}nyves}, {Men'shchikov}, {Andr{\'e}},
  {Schneider}, {Di Francesco}, {Arzoumanian}, {Benedettini}, {Ladjelate},
  {Palmeirim}, {Bracco}, {Molinari}, {Pezzuto}, \& {Spinoglio}}]{rcra}
{Bresnahan}, D., {Ward-Thompson}, D., {Kirk}, J.~M., {et~al.} 2018, \aap, 615,
  A125

\bibitem[{{Cambr{\'e}sy}(1999)}]{laurent}
{Cambr{\'e}sy}, L. 1999, \aap, 345, 965

\bibitem[{{Carey} {et~al.}(1998){Carey}, {Clark}, {Egan}, {Price}, {Shipman},
  \& {Kuchar}}]{IRDC}
{Carey}, S.~J., {Clark}, F.~O., {Egan}, M.~P., {et~al.} 1998, \apj, 508, 721

\bibitem[{{\v{C}ernis}(1993)}]{cernis}
{\v{C}ernis}, K. 1993, Baltic Astronomy, 2, 214

\bibitem[{{Chabrier}(2005)}]{IMF2005}
{Chabrier}, G. 2005, in Astrophysics and Space Science Library, Vol. 327, The
  Initial Mass Function 50 Years Later, ed. E.~{Corbelli}, F.~{Palla}, \&
  H.~{Zinnecker}, 41

\bibitem[{{Chen} {et~al.}(2018){Chen}, {Burkhart}, {Goodman}, \&
  {Collins}}]{2018ApJ...859..162C}
{Chen}, H. H.-H., {Burkhart}, B., {Goodman}, A., \& {Collins}, D.~C. 2018,
  \apj, 859, 162

\bibitem[{{Chen} {et~al.}(2016){Chen}, {Di Francesco}, {Johnstone}, {Sadavoy},
  {Hatchell}, {Mottram}, {Kirk}, {Buckle}, {Berry}, {Broekhoven-Fiene},
  {Currie}, {Fich}, {Jenness}, {Nutter}, {Pattle}, {Pineda}, {Quinn}, {Salji},
  {Tisi}, {Hogerheijde}, {Ward-Thompson}, {Bastien}, {Bresnahan}, {Butner},
  {Chrysostomou}, {Coude}, {Davis}, {Drabek-Maunder}, {Duarte-Cabral}, {Fiege},
  {Friberg}, {Friesen}, {Fuller}, {Graves}, {Greaves}, {Gregson}, {Holland},
  {Joncas}, {Kirk}, {Knee}, {Mairs}, {Marsh}, {Matthews}, {Moriarty-Schieven},
  {Mowat}, {Pezzuto}, {Rawlings}, {Richer}, {Robertson}, {Rosolowsky},
  {Rumble}, {Schneider-Bontemps}, {Thomas}, {Tothill}, {Viti}, {White},
  {Wouterloot}, {Yates}, \& {Zhu}}]{SCUBA2}
{Chen}, M.~C.-Y., {Di Francesco}, J., {Johnstone}, D., {et~al.} 2016, \apj,
  826, 95

\bibitem[{{Clauset} {et~al.}(2009){Clauset}, {Rohilla Shalizi}, \&
  {Newman}}]{clauset}
{Clauset}, A., {Rohilla Shalizi}, C., \& {Newman}, M.~E.~J. 2009, SIAM Review,
  51, 661

\bibitem[{{COMPLETE team}(2012)}]{10075_2012}
{COMPLETE team}. 2012, FCRAO Perseus 13CO cubes and map

\bibitem[{{di Francesco} {et~al.}(2007){di Francesco}, {Evans}, {Caselli},
  {Myers}, {Shirley}, {Aikawa}, \& {Tafalla}}]{2007prplJ}
{di Francesco}, J., {Evans}, N.~J., I., {Caselli}, P., {et~al.} 2007, in
  Protostars and Planets V, ed. B.~{Reipurth}, D.~{Jewitt}, \& K.~{Keil}, 17

\bibitem[{{Elia} \& {Pezzuto}(2016)}]{grey}
{Elia}, D. \& {Pezzuto}, S. 2016, \mnras, 461, 1328

\bibitem[{{Enoch} {et~al.}(2006){Enoch}, {Young}, {Glenn}, {Evans}, {Golwala},
  {Sargent}, {Harvey}, {Aguirre}, {Goldin}, {Haig}, {Huard}, {Lange},
  {Laurent}, {Maloney}, {Mauskopf}, {Rossinot}, \& {Sayers}}]{enoch}
{Enoch}, M.~L., {Young}, K.~E., {Glenn}, J., {et~al.} 2006, \apj, 638, 293

\bibitem[{{Evans} {et~al.}(2009){Evans}, {Dunham}, {J{\o}rgensen}, {Enoch},
  {Mer{\'{\i}}n}, {van Dishoeck}, {Alcal{\'a}}, {Myers}, {Stapelfeldt},
  {Huard}, {Allen}, {Harvey}, {van Kempen}, {Blake}, {Koerner}, {Mundy},
  {Padgett}, \& {Sargent}}]{evans}
{Evans}, II, N.~J., {Dunham}, M.~M., {J{\o}rgensen}, J.~K., {et~al.} 2009,
  \apjs, 181, 321

\bibitem[{{Federrath} \& {Klessen}(2013)}]{2013ApJ...763...51F}
{Federrath}, C. \& {Klessen}, R.~S. 2013, \apj, 763, 51

\bibitem[{{Fischera} \& {Martin}(2012)}]{FM}
{Fischera}, J. \& {Martin}, P.~G. 2012, \aap, 542, A77

\bibitem[{{Foster} {et~al.}(2013){Foster}, {Mandel}, {Pineda}, {Covey}, {Arce},
  \& {Goodman}}]{foster}
{Foster}, J.~B., {Mandel}, K.~S., {Pineda}, J.~E., {et~al.} 2013, \mnras, 428,
  1606

\bibitem[{{Gaia Collaboration} {et~al.}(2016){Gaia Collaboration}, {Prusti},
  {de Bruijne}, {Brown}, {Vallenari}, {Babusiaux}, {Bailer-Jones}, {Bastian},
  {Biermann}, {Evans}, \& et~al.}]{gaia}
{Gaia Collaboration}, {Prusti}, T., {de Bruijne}, J.~H.~J., {et~al.} 2016,
  \aap, 595, A1

\bibitem[{{Griffin} {et~al.}(2010){Griffin}, {Abergel}, {Abreu}, {Ade},
  {Andr{\'e}}, {Augueres}, {Babbedge}, {Bae}, {Baillie}, {Baluteau}, {Barlow},
  {Bendo}, {Benielli}, {Bock}, {Bonhomme}, {Brisbin}, {Brockley-Blatt},
  {Caldwell}, {Cara}, {Castro-Rodriguez}, {Cerulli}, {Chanial}, {Chen},
  {Clark}, {Clements}, {Clerc}, {Coker}, {Communal}, {Conversi}, {Cox},
  {Crumb}, {Cunningham}, {Daly}, {Davis}, {de Antoni}, {Delderfield}, {Devin},
  {di Giorgio}, {Didschuns}, {Dohlen}, {Donati}, {Dowell}, {Dowell}, {Duband},
  {Dumaye}, {Emery}, {Ferlet}, {Ferrand}, {Fontignie}, {Fox}, {Franceschini},
  {Frerking}, {Fulton}, {Garcia}, {Gastaud}, {Gear}, {Glenn}, {Goizel},
  {Griffin}, {Grundy}, {Guest}, {Guillemet}, {Hargrave}, {Harwit}, {Hastings},
  {Hatziminaoglou}, {Herman}, {Hinde}, {Hristov}, {Huang}, {Imhof}, {Isaak},
  {Israelsson}, {Ivison}, {Jennings}, {Kiernan}, {King}, {Lange}, {Latter},
  {Laurent}, {Laurent}, {Leeks}, {Lellouch}, {Levenson}, {Li}, {Li},
  {Lilienthal}, {Lim}, {Liu}, {Lu}, {Madden}, {Mainetti}, {Marliani}, {McKay},
  {Mercier}, {Molinari}, {Morris}, {Moseley}, {Mulder}, {Mur}, {Naylor},
  {Nguyen}, {O'Halloran}, {Oliver}, {Olofsson}, {Olofsson}, {Orfei}, {Page},
  {Pain}, {Panuzzo}, {Papageorgiou}, {Parks}, {Parr-Burman}, {Pearce},
  {Pearson}, {P{\'e}rez-Fournon}, {Pinsard}, {Pisano}, {Podosek}, {Pohlen},
  {Polehampton}, {Pouliquen}, {Rigopoulou}, {Rizzo}, {Roseboom}, {Roussel},
  {Rowan-Robinson}, {Rownd}, {Saraceno}, {Sauvage}, {Savage}, {Savini},
  {Sawyer}, {Scharmberg}, {Schmitt}, {Schneider}, {Schulz}, {Schwartz},
  {Shafer}, {Shupe}, {Sibthorpe}, {Sidher}, {Smith}, {Smith}, {Smith},
  {Spencer}, {Stobie}, {Sudiwala}, {Sukhatme}, {Surace}, {Stevens}, {Swinyard},
  {Trichas}, {Tourette}, {Triou}, {Tseng}, {Tucker}, {Turner}, {Vaccari},
  {Valtchanov}, {Vigroux}, {Virique}, {Voellmer}, {Walker}, {Ward}, {Waskett},
  {Weilert}, {Wesson}, {White}, {Whitehouse}, {Wilson}, {Winter}, {Woodcraft},
  {Wright}, {Xu}, {Zavagno}, {Zemcov}, {Zhang}, \& {Zonca}}]{SPIRE}
{Griffin}, M.~J., {Abergel}, A., {Abreu}, A., {et~al.} 2010, \aap, 518, L3

\bibitem[{{Hacar} {et~al.}(2017){Hacar}, {Tafalla}, \& {Alves}}]{hacar}
{Hacar}, A., {Tafalla}, M., \& {Alves}, J. 2017, \aap, 606, A123

\bibitem[{{Hatchell} {et~al.}(2005){Hatchell}, {Richer}, {Fuller},
  {Qualtrough}, {Ladd}, \& {Chandler}}]{2005A&A...440..151H}
{Hatchell}, J., {Richer}, J.~S., {Fuller}, G.~A., {et~al.} 2005, \aap, 440, 151

\bibitem[{{Herbig}(1998)}]{herbig}
{Herbig}, G.~H. 1998, \apj, 497, 736

\bibitem[{{Hirota} {et~al.}(2008){Hirota}, {Bushimata}, {Choi}, {Honma},
  {Imai}, {Iwadate}, {Jike}, {Kameya}, {Kamohara}, {Kan-Ya}, {Kawaguchi},
  {Kijima}, {Kobayashi}, {Kuji}, {Kurayama}, {Manabe}, {Miyaji}, {Nagayama},
  {Nakagawa}, {Oh}, {Omodaka}, {Oyama}, {Sakai}, {Sasao}, {Sato}, {Shibata},
  {Tamura}, \& {Yamashita}}]{hirota8}
{Hirota}, T., {Bushimata}, T., {Choi}, Y.~K., {et~al.} 2008, \pasj, 60, 37

\bibitem[{{Hirota} {et~al.}(2011){Hirota}, {Honma}, {Imai}, {Sunada}, {Ueno},
  {Kobayashi}, \& {Kawaguchi}}]{hirota1}
{Hirota}, T., {Honma}, M., {Imai}, H., {et~al.} 2011, \pasj, 63, 1

\bibitem[{{Johnstone} {et~al.}(2000){Johnstone}, {Wilson}, {Moriarty-Schieven},
  {Joncas}, {Smith}, {Gregersen}, \& {Fich}}]{doug}
{Johnstone}, D., {Wilson}, C.~D., {Moriarty-Schieven}, G., {et~al.} 2000, \apj,
  545, 327

\bibitem[{{J{\o}rgensen} {et~al.}(2006){J{\o}rgensen}, {Harvey}, {Evans},
  {Huard}, {Allen}, {Porras}, {Blake}, {Bourke}, {Chapman}, {Cieza}, {Koerner},
  {Lai}, {Mundy}, {Myers}, {Padgett}, {Rebull}, {Sargent}, {Spiesman},
  {Stapelfeldt}, {van Dishoeck}, {Wahhaj}, \& {Young}}]{jorge}
{J{\o}rgensen}, J.~K., {Harvey}, P.~M., {Evans}, II, N.~J., {et~al.} 2006,
  \apj, 645, 1246

\bibitem[{{Joye} \& {Mandel}(2003)}]{ds9}
{Joye}, W.~A. \& {Mandel}, E. 2003, in Astronomical Society of the Pacific
  Conference Series, Vol. 295, Astronomical Data Analysis Software and Systems
  XII, ed. H.~E. {Payne}, R.~I. {Jedrzejewski}, \& R.~N. {Hook}, 489

\bibitem[{{Kauffmann} {et~al.}(2008){Kauffmann}, {Bertoldi}, {Bourke}, {Evans},
  \& {Lee}}]{mmw}
{Kauffmann}, J., {Bertoldi}, F., {Bourke}, T.~L., {Evans}, N.~J., I., \& {Lee},
  C.~W. 2008, \aap, 487, 993

\bibitem[{{Kirk} {et~al.}(2006){Kirk}, {Johnstone}, \& {Di
  Francesco}}]{2006ApJ...646.1009K}
{Kirk}, H., {Johnstone}, D., \& {Di Francesco}, J. 2006, \apj, 646, 1009

\bibitem[{{K{\"o}nyves} {et~al.}(2020){K{\"o}nyves}, {Andr{\'e}},
  {Arzoumanian}, {Schneider}, {Men'shchikov}, {Bontemps}, {Ladjelate},
  {Didelon}, {Pezzuto}, {Benedettini}, {Bracco}, {Di Francesco}, {Goodwin},
  {Rygl}, {Shimajiri}, {Spinoglio}, {Ward-Thompson}, \& {White}}]{orionB}
{K{\"o}nyves}, V., {Andr{\'e}}, P., {Arzoumanian}, D., {et~al.} 2020, \aap,
  635, A34

\bibitem[{{K{\"o}nyves} {et~al.}(2015){K{\"o}nyves}, {Andr{\'e}},
  {Men'shchikov}, {Palmeirim}, {Arzoumanian}, {Schneider}, {Roy}, {Didelon},
  {Maury}, {Shimajiri}, {Di Francesco}, {Bontemps}, {Peretto}, {Benedettini},
  {Bernard}, {Elia}, {Griffin}, {Hill}, {Kirk}, {Ladjelate}, {Marsh}, {Martin},
  {Motte}, {Nguy{\^e}n Luong}, {Pezzuto}, {Roussel}, {Rygl}, {Sadavoy},
  {Schisano}, {Spinoglio}, {Ward-Thompson}, \& {White}}]{Aquila}
{K{\"o}nyves}, V., {Andr{\'e}}, P., {Men'shchikov}, A., {et~al.} 2015, \aap,
  584, A91

\bibitem[{{Krumholz}(2015)}]{krumholz}
{Krumholz}, M.~R. 2015, arXiv e-prints, arXiv:1511.03457

\bibitem[{{Ladjelate} {et~al.}(2020){Ladjelate}, {Andr{\'e}}, {K{\"o}nyves},
  {Ward-Thompson}, {Men'shchikov}, {Bracco}, {Palmeirim}, {Roy}, {Shimajiri},
  {Kirk}, {Arzoumanian}, {Benedettini}, {Di Francesco}, {Fiorellino},
  {Schneider}, \& {Pezzuto}}]{bilOph}
{Ladjelate}, B., {Andr{\'e}}, P., {K{\"o}nyves}, V., {et~al.} 2020, arXiv
  e-prints, arXiv:2001.11036

\bibitem[{{Larson}(1969)}]{larson}
{Larson}, R.~B. 1969, \mnras, 145, 271

\bibitem[{{Li} \& {Draine}(2001)}]{lidraine}
{Li}, A. \& {Draine}, B.~T. 2001, \apj, 554, 778

\bibitem[{{Li Causi} {et~al.}(2016){Li Causi}, {Schisano}, {Liu}, {Molinari},
  \& {Di Giorgio}}]{gianlu}
{Li Causi}, G., {Schisano}, E., {Liu}, S.~J., {Molinari}, S., \& {Di Giorgio},
  A. 2016, in \procspie, Vol. 9904, Space Telescopes and Instrumentation 2016:
  Optical, Infrared, and Millimeter Wave, 99045V

\bibitem[{{Luri} {et~al.}(2018){Luri}, {Brown}, {Sarro}, {Arenou},
  {Bailer-Jones}, {Castro-Ginard}, {de Bruijne}, {Prusti}, {Babusiaux}, \&
  {Delgado}}]{luri}
{Luri}, X., {Brown}, A.~G.~A., {Sarro}, L.~M., {et~al.} 2018, \aap, 616, A9

\bibitem[{{Ma{\'{\i}}z Apell{\'a}niz} \& {{\'U}beda}(2005)}]{binSizeHisto}
{Ma{\'{\i}}z Apell{\'a}niz}, J. \& {{\'U}beda}, L. 2005, \apj, 629, 873

\bibitem[{{Marsh} {et~al.}(2016){Marsh}, {Kirk}, {Andr{\'e}}, {Griffin},
  {K{\"o}nyves}, {Palmeirim}, {Men'shchikov}, {Ward-Thompson}, {Benedettini},
  {Bresnahan}, {di Francesco}, {Elia}, {Motte}, {Peretto}, {Pezzuto}, {Roy},
  {Sadavoy}, {Schneider}, {Spinoglio}, \& {White}}]{2016MNRAS.459..342M}
{Marsh}, K.~A., {Kirk}, J.~M., {Andr{\'e}}, P., {et~al.} 2016, \mnras, 459, 342

\bibitem[{{Men'shchikov}(2013)}]{getfilaments}
{Men'shchikov}, A. 2013, \aap, 560, A63

\bibitem[{{Men'shchikov} {et~al.}(2012){Men'shchikov}, {Andr{\'e}}, {Didelon},
  {Motte}, {Hennemann}, \& {Schneider}}]{sasha}
{Men'shchikov}, A., {Andr{\'e}}, P., {Didelon}, P., {et~al.} 2012, \aap, 542,
  A81

\bibitem[{{Mercimek} {et~al.}(2017){Mercimek}, {Myers}, {Lee}, \&
  {Sadavoy}}]{mercimek}
{Mercimek}, S., {Myers}, P.~C., {Lee}, K.~I., \& {Sadavoy}, S.~I. 2017, \aj,
  153, 214

\bibitem[{{Molinari} {et~al.}(2011){Molinari}, {Schisano}, {Faustini},
  {Pestalozzi}, {di Giorgio}, \& {Liu}}]{cutex}
{Molinari}, S., {Schisano}, E., {Faustini}, F., {et~al.} 2011, \aap, 530, A133

\bibitem[{{Molinari} {et~al.}(2010){Molinari}, {Swinyard}, {Bally}, {Barlow},
  {Bernard}, {Martin}, {Moore}, {Noriega-Crespo}, {Plume}, {Testi}, {Zavagno},
  {Abergel}, {Ali}, {Anderson}, {Andr{\'e}}, {Baluteau}, {Battersby},
  {Beltr{\'a}n}, {Benedettini}, {Billot}, {Blommaert}, {Bontemps}, {Boulanger},
  {Brand}, {Brunt}, {Burton}, {Calzoletti}, {Carey}, {Caselli}, {Cesaroni},
  {Cernicharo}, {Chakrabarti}, {Chrysostomou}, {Cohen}, {Compiegne}, {de
  Bernardis}, {de Gasperis}, {di Giorgio}, {Elia}, {Faustini}, {Flagey},
  {Fukui}, {Fuller}, {Ganga}, {Garcia-Lario}, {Glenn}, {Goldsmith}, {Griffin},
  {Hoare}, {Huang}, {Ikhenaode}, {Joblin}, {Joncas}, {Juvela}, {Kirk},
  {Lagache}, {Li}, {Lim}, {Lord}, {Marengo}, {Marshall}, {Masi}, {Massi},
  {Matsuura}, {Minier}, {Miville-Desch{\^e}nes}, {Montier}, {Morgan}, {Motte},
  {Mottram}, {M{\"u}ller}, {Natoli}, {Neves}, {Olmi}, {Paladini}, {Paradis},
  {Parsons}, {Peretto}, {Pestalozzi}, {Pezzuto}, {Piacentini}, {Piazzo},
  {Polychroni}, {Pomar{\`e}s}, {Popescu}, {Reach}, {Ristorcelli}, {Robitaille},
  {Robitaille}, {Rod{\'o}n}, {Roy}, {Royer}, {Russeil}, {Saraceno}, {Sauvage},
  {Schilke}, {Schisano}, {Schneider}, {Schuller}, {Schulz}, {Sibthorpe},
  {Smith}, {Smith}, {Spinoglio}, {Stamatellos}, {Strafella}, {Stringfellow},
  {Sturm}, {Taylor}, {Thompson}, {Traficante}, {Tuffs}, {Umana}, {Valenziano},
  {Vavrek}, {Veneziani}, {Viti}, {Waelkens}, {Ward-Thompson}, {White},
  {Wilcock}, {Wyrowski}, {Yorke}, \& {Zhang}}]{HIGALSDP}
{Molinari}, S., {Swinyard}, B., {Bally}, J., {et~al.} 2010, \aap, 518, L100

\bibitem[{{Motte} {et~al.}(1998){Motte}, {Andre}, \& {Neri}}]{motte}
{Motte}, F., {Andre}, P., \& {Neri}, R. 1998, \aap, 336, 150

\bibitem[{{Nutter} \& {Ward-Thompson}(2007)}]{nutter}
{Nutter}, D. \& {Ward-Thompson}, D. 2007, \mnras, 374, 1413

\bibitem[{{Ortiz-Le{\'o}n} {et~al.}(2018){Ortiz-Le{\'o}n}, {Loinard}, {Dzib},
  {Galli}, {Kounkel}, {Mioduszewski}, {Rodr{\'{\i}}guez}, {Torres}, {Hartmann},
  {Boden}, {Evans}, {Brice{\~n}o}, \& {Tobin}}]{ortiz}
{Ortiz-Le{\'o}n}, G.~N., {Loinard}, L., {Dzib}, S.~A., {et~al.} 2018, \apj,
  865, 73

\bibitem[{{Ostriker}(1964)}]{ostriker}
{Ostriker}, J. 1964, \apj, 140, 1056

\bibitem[{{Ott}(2010)}]{HIPE}
{Ott}, S. 2010, in Astronomical Society of the Pacific Conference Series, Vol.
  434, Astronomical Data Analysis Software and Systems XIX, ed. Y.~{Mizumoto},
  K.-I. {Morita}, \& M.~{Ohishi}, 139

\bibitem[{{Palmeirim} {et~al.}(2013){Palmeirim}, {Andr{\'e}}, {Kirk},
  {Ward-Thompson}, {Arzoumanian}, {K{\"o}nyves}, {Didelon}, {Schneider},
  {Benedettini}, {Bontemps}, {Di Francesco}, {Elia}, {Griffin}, {Hennemann},
  {Hill}, {Martin}, {Men'shchikov}, {Molinari}, {Motte}, {Nguyen Luong},
  {Nutter}, {Peretto}, {Pezzuto}, {Roy}, {Rygl}, {Spinoglio}, \&
  {White}}]{pedro}
{Palmeirim}, P., {Andr{\'e}}, P., {Kirk}, J., {et~al.} 2013, \aap, 550, A38

\bibitem[{{Pezzuto} {et~al.}(2012){Pezzuto}, {Elia}, {Schisano}, {Strafella},
  {Di Francesco}, {Sadavoy}, {Andr{\'e}}, {Benedettini}, {Bernard}, {di
  Giorgio}, {Facchini}, {Hennemann}, {Hill}, {K{\"o}nyves}, {Molinari},
  {Motte}, {Nguyen-Luong}, {Peretto}, {Pestalozzi}, {Polychroni}, {Rygl},
  {Saraceno}, {Schneider}, {Spinoglio}, {Testi}, {Ward-Thompson}, \&
  {White}}]{FHSC}
{Pezzuto}, S., {Elia}, D., {Schisano}, E., {et~al.} 2012, \aap, 547, A54

\bibitem[{{Piazzo} {et~al.}(2015){Piazzo}, {Calzoletti}, {Faustini},
  {Pestalozzi}, {Pezzuto}, {Elia}, {di Giorgio}, \& {Molinari}}]{unimap}
{Piazzo}, L., {Calzoletti}, L., {Faustini}, F., {et~al.} 2015, \mnras, 447,
  1471

\bibitem[{{Pilbratt} {et~al.}(2010){Pilbratt}, {Riedinger}, {Passvogel},
  {Crone}, {Doyle}, {Gageur}, {Heras}, {Jewell}, {Metcalfe}, {Ott}, \&
  {Schmidt}}]{2010A&A...518L...1P}
{Pilbratt}, G.~L., {Riedinger}, J.~R., {Passvogel}, T., {et~al.} 2010, \aap,
  518, L1

\bibitem[{{Planck Collaboration} {et~al.}(2014){Planck Collaboration},
  {Abergel}, {Ade}, {Aghanim}, {Alves}, {Aniano}, {Armitage-Caplan}, {Arnaud},
  {Ashdown}, {Atrio-Barandela}, \& et~al.}]{planck}
{Planck Collaboration}, {Abergel}, A., {Ade}, P.~A.~R., {et~al.} 2014, \aap,
  571, A11

\bibitem[{{Poglitsch} {et~al.}(2010){Poglitsch}, {Waelkens}, {Geis},
  {Feuchtgruber}, {Vandenbussche}, {Rodriguez}, {Krause}, {Renotte}, {van
  Hoof}, {Saraceno}, {Cepa}, {Kerschbaum}, {Agn{\`e}se}, {Ali}, {Altieri},
  {Andreani}, {Augueres}, {Balog}, {Barl}, {Bauer}, {Belbachir}, {Benedettini},
  {Billot}, {Boulade}, {Bischof}, {Blommaert}, {Callut}, {Cara}, {Cerulli},
  {Cesarsky}, {Contursi}, {Creten}, {De Meester}, {Doublier}, {Doumayrou},
  {Duband}, {Exter}, {Genzel}, {Gillis}, {Gr{\"o}zinger}, {Henning},
  {Herreros}, {Huygen}, {Inguscio}, {Jakob}, {Jamar}, {Jean}, {de Jong},
  {Katterloher}, {Kiss}, {Klaas}, {Lemke}, {Lutz}, {Madden}, {Marquet},
  {Martignac}, {Mazy}, {Merken}, {Montfort}, {Morbidelli}, {M{\"u}ller},
  {Nielbock}, {Okumura}, {Orfei}, {Ottensamer}, {Pezzuto}, {Popesso},
  {Putzeys}, {Regibo}, {Reveret}, {Royer}, {Sauvage}, {Schreiber}, {Stegmaier},
  {Schmitt}, {Schubert}, {Sturm}, {Thiel}, {Tofani}, {Vavrek}, {Wetzstein},
  {Wieprecht}, \& {Wiezorrek}}]{PACS}
{Poglitsch}, A., {Waelkens}, C., {Geis}, N., {et~al.} 2010, \aap, 518, L2

\bibitem[{{Polychroni} {et~al.}(2013){Polychroni}, {Schisano}, {Elia}, {Roy},
  {Molinari}, {Martin}, {Andr{\'e}}, {Turrini}, {Rygl}, {Di Francesco},
  {Benedettini}, {Busquet}, {di Giorgio}, {Pestalozzi}, {Pezzuto},
  {Arzoumanian}, {Bontemps}, {Hennemann}, {Hill}, {K{\"o}nyves},
  {Men'shchikov}, {Motte}, {Nguyen-Luong}, {Peretto}, {Schneider}, \&
  {White}}]{danae}
{Polychroni}, D., {Schisano}, E., {Elia}, D., {et~al.} 2013, \apjl, 777, L33

\bibitem[{{Rayner} {et~al.}(2017){Rayner}, {Griffin}, {Schneider}, {Motte},
  {K{\"o}nyves}, {Andr{\'e}}, {Di Francesco}, {Didelon}, {Pattle},
  {Ward-Thompson}, {Anderson}, {Benedettini}, {Bernard}, {Bontemps}, {Elia},
  {Fuente}, {Hennemann}, {Hill}, {Kirk}, {Marsh}, {Men'shchikov}, {Nguyen
  Luong}, {Peretto}, {Pezzuto}, {Rivera-Ingraham}, {Roy}, {Rygl},
  {S{\'a}nchez-Monge}, {Spinoglio}, {Tig{\'e}}, {Trevi{\~n}o-Morales}, \&
  {White}}]{rayner}
{Rayner}, T.~S.~M., {Griffin}, M.~J., {Schneider}, N., {et~al.} 2017, \aap,
  607, A22

\bibitem[{{Rebull} {et~al.}(2007){Rebull}, {Stapelfeldt}, {Evans},
  {J{\o}rgensen}, {Harvey}, {Brooke}, {Bourke}, {Padgett}, {Chapman}, {Lai},
  {Spiesman}, {Noriega-Crespo}, {Mer{\'{\i}}n}, {Huard}, {Allen}, {Blake},
  {Jarrett}, {Koerner}, {Mundy}, {Myers}, {Sargent}, {van Dishoeck}, {Wahhaj},
  \& {Young}}]{rebull}
{Rebull}, L.~M., {Stapelfeldt}, K.~R., {Evans}, II, N.~J., {et~al.} 2007,
  \apjs, 171, 447

\bibitem[{{Ridge} {et~al.}(2006{\natexlab{a}}){Ridge}, {Di Francesco}, {Kirk},
  {Li}, {Goodman}, {Alves}, {Arce}, {Borkin}, {Caselli}, {Foster}, {Heyer},
  {Johnstone}, {Kosslyn}, {Lombardi}, {Pineda}, {Schnee}, \&
  {Tafalla}}]{COMPLETE}
{Ridge}, N.~A., {Di Francesco}, J., {Kirk}, H., {et~al.} 2006{\natexlab{a}},
  \aj, 131, 2921

\bibitem[{{Ridge} {et~al.}(2006{\natexlab{b}}){Ridge}, {Schnee}, {Goodman}, \&
  {Foster}}]{shell}
{Ridge}, N.~A., {Schnee}, S.~L., {Goodman}, A.~A., \& {Foster}, J.~B.
  2006{\natexlab{b}}, \apj, 643, 932

\bibitem[{{Ripepi} {et~al.}(2002){Ripepi}, {Palla}, {Marconi}, {Bernabei},
  {Arellano Ferro}, {Terranegra}, \& {Alcal{\'a}}}]{ripepi}
{Ripepi}, V., {Palla}, F., {Marconi}, M., {et~al.} 2002, \aap, 391, 587

\bibitem[{{Rom{\'a}n-Z{\'u}{\~n}iga} {et~al.}(2009){Rom{\'a}n-Z{\'u}{\~n}iga},
  {Lada}, \& {Alves}}]{2009ApJ...704..183R}
{Rom{\'a}n-Z{\'u}{\~n}iga}, C.~G., {Lada}, C.~J., \& {Alves}, J.~F. 2009, \apj,
  704, 183

\bibitem[{{Roy} {et~al.}(2015){Roy}, {Andr{\'e}}, {Arzoumanian}, {Peretto},
  {Palmeirim}, {K{\"o}nyves}, {Schneider}, {Benedettini}, {Di Francesco},
  {Elia}, {Hill}, {Ladjelate}, {Louvet}, {Motte}, {Pezzuto}, {Schisano},
  {Shimajiri}, {Spinoglio}, {Ward-Thompson}, \& {White}}]{arabindo}
{Roy}, A., {Andr{\'e}}, P., {Arzoumanian}, D., {et~al.} 2015, \aap, 584, A111

\bibitem[{{Rygl} {et~al.}(2013){Rygl}, {Benedettini}, {Schisano}, {Elia},
  {Molinari}, {Pezzuto}, {Andr{\'e}}, {Bernard}, {White}, {Polychroni},
  {Bontemps}, {Cox}, {Di Francesco}, {Facchini}, {Fallscheer}, {di Giorgio},
  {Hennemann}, {Hill}, {K{\"o}nyves}, {Minier}, {Motte}, {Nguyen-Luong},
  {Peretto}, {Pestalozzi}, {Sadavoy}, {Schneider}, {Spinoglio}, {Testi}, \&
  {Ward-Thompson}}]{kazy}
{Rygl}, K.~L.~J., {Benedettini}, M., {Schisano}, E., {et~al.} 2013, \aap, 549,
  L1

\bibitem[{{Sadavoy} {et~al.}(2012){Sadavoy}, {di Francesco}, {Andr{\'e}},
  {Pezzuto}, {Bernard}, {Bontemps}, {Bressert}, {Chitsazzadeh}, {Fallscheer},
  {Hennemann}, {Hill}, {Martin}, {Motte}, {Nguyen Luong}, {Peretto}, {Reid},
  {Schneider}, {Testi}, {White}, \& {Wilson}}]{sarahB1}
{Sadavoy}, S.~I., {di Francesco}, J., {Andr{\'e}}, P., {et~al.} 2012, \aap,
  540, A10

\bibitem[{{Sadavoy} {et~al.}(2014){Sadavoy}, {Di Francesco}, {Andr{\'e}},
  {Pezzuto}, {Bernard}, {Maury}, {Men'shchikov}, {Motte}, {Nguyen-Lu'o'ng},
  {Schneider}, {Arzoumanian}, {Benedettini}, {Bontemps}, {Elia}, {Hennemann},
  {Hill}, {K{\"o}nyves}, {Louvet}, {Peretto}, {Roy}, \& {White}}]{sarah2}
{Sadavoy}, S.~I., {Di Francesco}, J., {Andr{\'e}}, P., {et~al.} 2014, \apjl,
  787, L18

\bibitem[{{Sadavoy} {et~al.}(2010){Sadavoy}, {Di Francesco}, {Bontemps},
  {Megeath}, {Rebull}, {Allgaier}, {Carey}, {Gutermuth}, {Hora}, {Huard},
  {McCabe}, {Muzerolle}, {Noriega-Crespo}, {Padgett}, \& {Terebey}}]{sarah2010}
{Sadavoy}, S.~I., {Di Francesco}, J., {Bontemps}, S., {et~al.} 2010, \apj, 710,
  1247

\bibitem[{{Salpeter}(1955)}]{salpeter}
{Salpeter}, E.~E. 1955, \apj, 121, 161

\bibitem[{{Schisano} {et~al.}(2020){Schisano}, {Molinari}, {Elia},
  {Benedettini}, {Olmi}, {Pezzuto}, {Traficante}, {Brescia}, {Cavuoti}, {di
  Giorgio}, {Liu}, {Moore}, {Noriega-Crespo}, {Riccio}, {Baldeschi},
  {Becciani}, {Peretto}, {Merello}, {Vitello}, {Zavagno}, {Beltr{\'a}n},
  {Cambr{\'e}sy}, {Eden}, {Li Causi}, {Molinaro}, {Palmeirim}, {Sciacca},
  {Testi}, {Umana}, \& {Whitworth}}]{eugenio2020}
{Schisano}, E., {Molinari}, S., {Elia}, D., {et~al.} 2020, \mnras, 492, 5420

\bibitem[{{Schisano} {et~al.}(2014){Schisano}, {Rygl}, {Molinari}, {Busquet},
  {Elia}, {Pestalozzi}, {Polychroni}, {Billot}, {Carey}, {Paladini},
  {Noriega-Crespo}, {Moore}, {Plume}, {Glover}, \&
  {V{\'a}zquez-Semadeni}}]{eugenio}
{Schisano}, E., {Rygl}, K.~L.~J., {Molinari}, S., {et~al.} 2014, \apj, 791, 27

\bibitem[{{Schnee} {et~al.}(2008){Schnee}, {Li}, {Goodman}, \&
  {Sargent}}]{schnee}
{Schnee}, S., {Li}, J., {Goodman}, A.~A., \& {Sargent}, A.~I. 2008, \apj, 684,
  1228

\bibitem[{{Schneider} {et~al.}(2011){Schneider}, {Bontemps}, {Simon},
  {Ossenkopf}, {Federrath}, {Klessen}, {Motte}, {Andr{\'e}}, {Stutzki}, \&
  {Brunt}}]{nicolaAv}
{Schneider}, N., {Bontemps}, S., {Simon}, R., {et~al.} 2011, \aap, 529, A1

\bibitem[{{Schneider} {et~al.}(2015){Schneider}, {Ossenkopf}, {Csengeri},
  {Klessen}, {Federrath}, {Tremblin}, {Girichidis}, {Bontemps}, \&
  {Andr{\'e}}}]{nicola}
{Schneider}, N., {Ossenkopf}, V., {Csengeri}, T., {et~al.} 2015, \aap, 575, A79

\bibitem[{{Scholz} {et~al.}(2013){Scholz}, {Geers}, {Clark}, {Jayawardhana}, \&
  {Muzic}}]{ICNGCIMF}
{Scholz}, A., {Geers}, V., {Clark}, P., {Jayawardhana}, R., \& {Muzic}, K.
  2013, \apj, 775, 138

\bibitem[{{Stanke} {et~al.}(2006){Stanke}, {Smith}, {Gredel}, \&
  {Khanzadyan}}]{stanke}
{Stanke}, T., {Smith}, M.~D., {Gredel}, R., \& {Khanzadyan}, T. 2006, \aap,
  447, 609

\bibitem[{{Strom} {et~al.}(1974){Strom}, {Strom}, \& {Carrasco}}]{strom}
{Strom}, S.~E., {Strom}, K.~A., \& {Carrasco}, L. 1974, \pasp, 86, 798

\bibitem[{{Swift} \& {Beaumont}(2010)}]{swiftBeau}
{Swift}, J.~J. \& {Beaumont}, C.~N. 2010, \pasp, 122, 224

\bibitem[{{Testi} \& {Sargent}(1998)}]{1998ApJ...508L..91T}
{Testi}, L. \& {Sargent}, A.~I. 1998, \apjl, 508, L91

\bibitem[{{Ward-Thompson} {et~al.}(2007){Ward-Thompson}, {Andr{\'e}},
  {Crutcher}, {Johnstone}, {Onishi}, \& {Wilson}}]{2007prpl.conf...33W}
{Ward-Thompson}, D., {Andr{\'e}}, P., {Crutcher}, R., {et~al.} 2007, in
  Protostars and Planets V, ed. B.~{Reipurth}, D.~{Jewitt}, \& K.~{Keil}, 33

\bibitem[{{Wright} {et~al.}(2010){Wright}, {Eisenhardt}, {Mainzer}, {Ressler},
  {Cutri}, {Jarrett}, {Kirkpatrick}, {Padgett}, {McMillan}, {Skrutskie},
  {Stanford}, {Cohen}, {Walker}, {Mather}, {Leisawitz}, {Gautier}, {McLean},
  {Benford}, {Lonsdale}, {Blain}, {Mendez}, {Irace}, {Duval}, {Liu}, {Royer},
  {Heinrichsen}, {Howard}, {Shannon}, {Kendall}, {Walsh}, {Larsen}, {Cardon},
  {Schick}, {Schwalm}, {Abid}, {Fabinsky}, {Naes}, \& {Tsai}}]{wise}
{Wright}, E.~L., {Eisenhardt}, P.~R.~M., {Mainzer}, A.~K., {et~al.} 2010, \aj,
  140, 1868

\bibitem[{{Young} {et~al.}(2015){Young}, {Young}, {Lai}, {Dunham}, \&
  {Evans}}]{c2d2015}
{Young}, K.~E., {Young}, C.~H., {Lai}, S.-P., {Dunham}, M.~M., \& {Evans}, II,
  N.~J. 2015, \aj, 150, 40

\bibitem[{{Zari} {et~al.}(2016){Zari}, {Lombardi}, {Alves}, {Lada}, \&
  {Bouy}}]{zari}
{Zari}, E., {Lombardi}, M., {Alves}, J., {Lada}, C.~J., \& {Bouy}, H. 2016,
  \aap, 587, A106

\bibitem[{{Zucker} {et~al.}(2018){Zucker}, {Schlafly}, {Speagle}, {Green},
  {Portillo}, {Finkbeiner}, \& {Goodman}}]{zucker}
{Zucker}, C., {Schlafly}, E.~F., {Speagle}, J.~S., {et~al.} 2018, \apj, 869, 83

\bibitem[{{Zucker} {et~al.}(2019){Zucker}, {Speagle}, {Schlafly}, {Green},
  {Finkbeiner}, {Goodman}, \& {Alves}}]{zucker2}
{Zucker}, C., {Speagle}, J.~S., {Schlafly}, E.~F., {et~al.} 2019, \apj, 879,
  125

\end{thebibliography}
\bibpunct{(}{)}{;}{a}{}{,}

\appendix
\section{The catalogue\label{appCat}}

The online material gives data for all 816 candidates cores, in the form of an atlas, one page per object. For each core we provide:
\begin{itemize}
\item a figure showing a 2\arcmin$\times$2\arcmin\ cutout around each source at 70~$\mu$m (top left), 160~$\mu$m (top middle), 250~$\mu$m (top right), 350~$\mu$m (bottom left), 500~$\mu$m (bottom middle), and from the high-resolution column density map (bottom right). If the source is detected by \textsl{getsources}, a green ellipse shows the source size from the extraction. We note that some sources were not observed at all wavelengths, because PACS and SPIRE have slightly different fields of view;
\item the observed SED (black points) and its best-fit modified black body function (blue curve). We also give the best-fit values for dust temperature ($T$) and mass ($M$) from the SED fit. The source radius ($R$), geometrical mean of major and minor FWHMs, is measured from the high-resolution column density map and is given in arcsec, first two rows, and in parsec, third row. The size in arcsec are before and after the deconvolution with the instrumental FWHM (18\arcsec). If a source is unresolved, we use a source size of $<6\farcs1$ (or 1/3 of the beam). In this case the physical radius in pc is computed assuming $R=6\farcs1$.
\end{itemize}

\begin{figure}
\includegraphics[scale=.3]{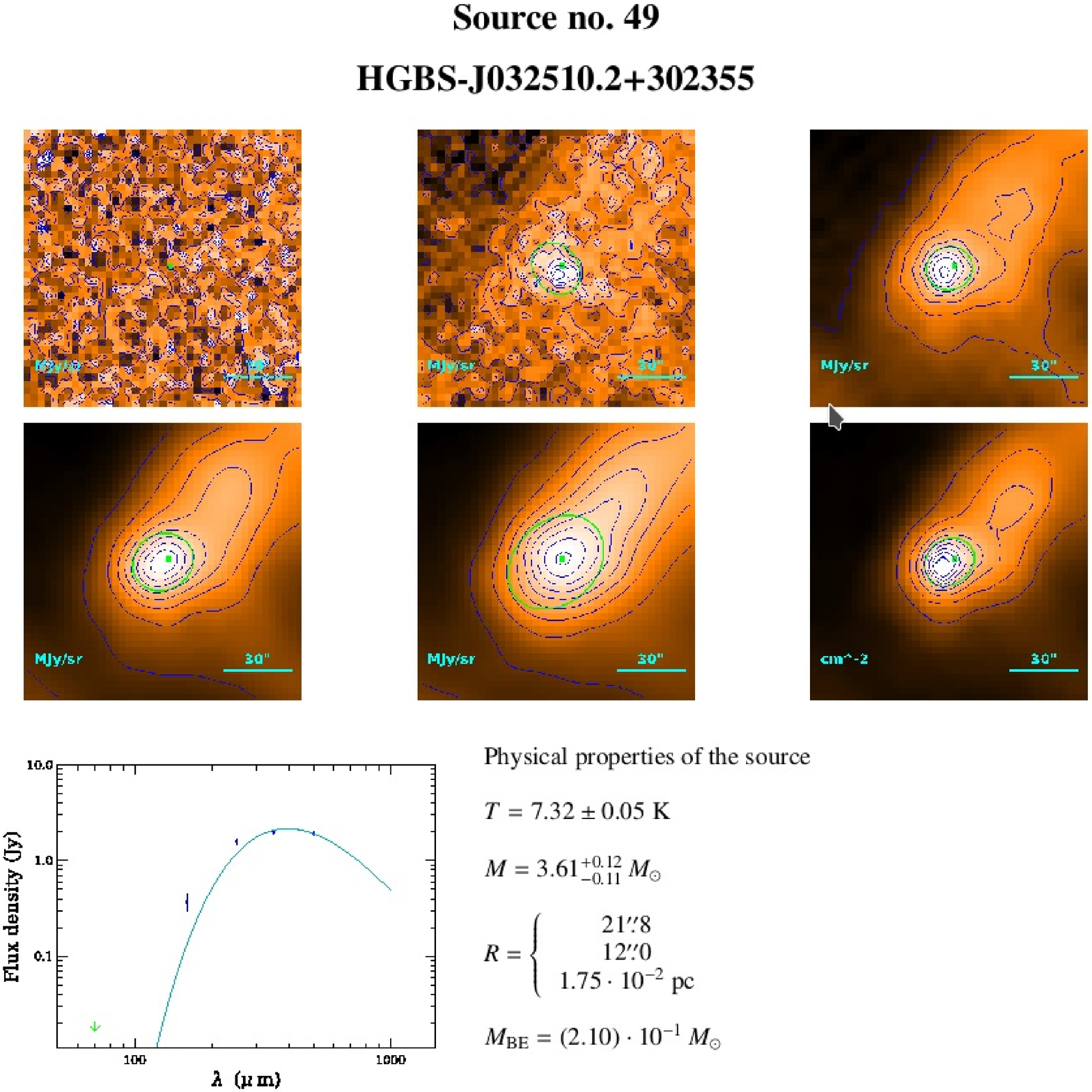}
\caption{An example of the atlas of the Perseus cores showing the prestellar core \#49. The original size of the page is A4, here the page has been shrunk to one third.\label{numero49}}
\end{figure}

\begin{figure}
\includegraphics[scale=.3]{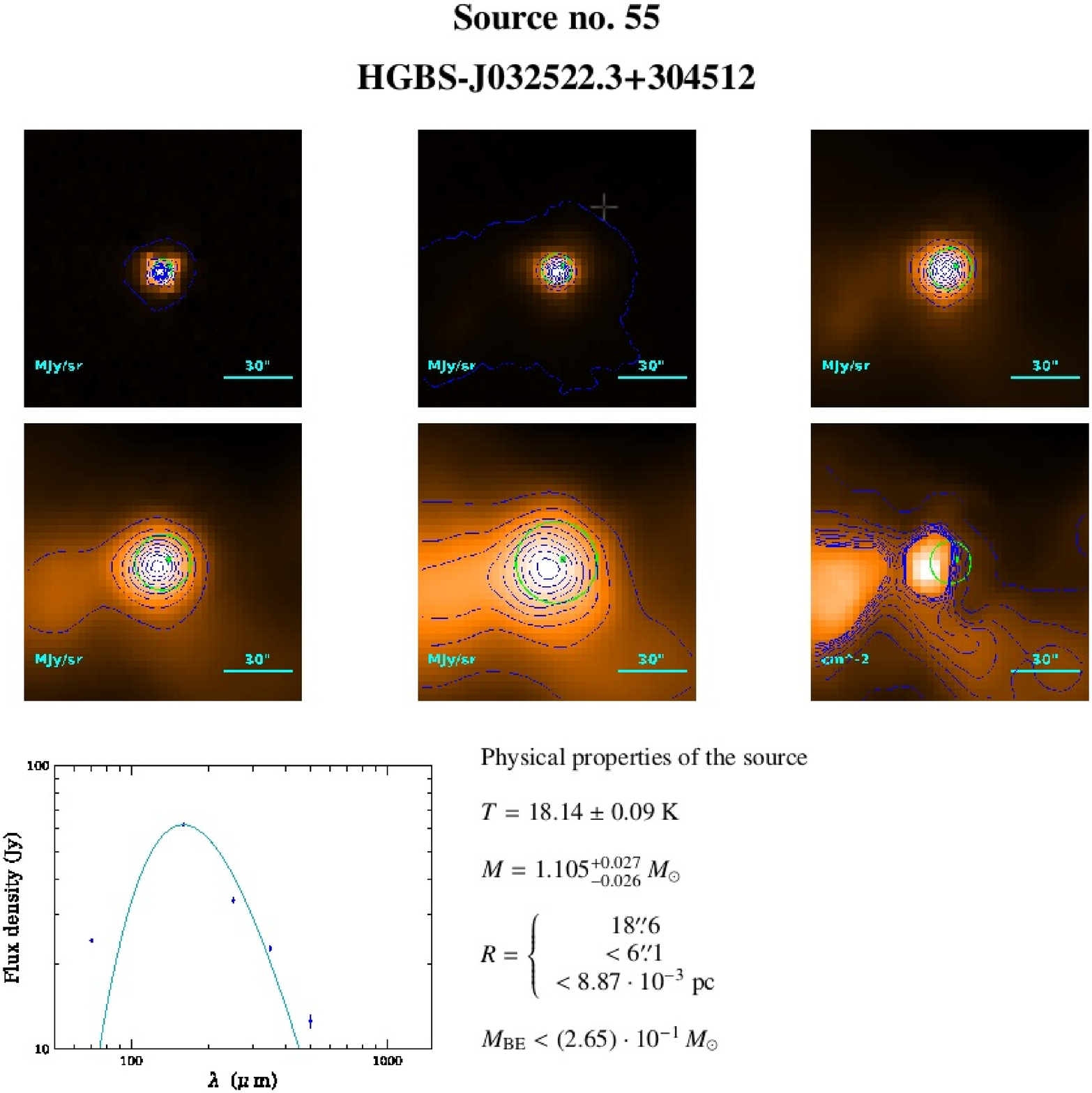}
\caption{An example of the atlas of the Perseus cores showing the protostar core \#55. Also in this case the page has been shrunk.\label{numero55}}
\end{figure}

Figure~\ref{numero49} shows an example of these products for a prestellar core (\#49) and Fig.~\ref{numero55} shows an example for a protostellar core (\#55).

\begin{figure}
\begin{tabular}{c}
\includegraphics[scale=.4]{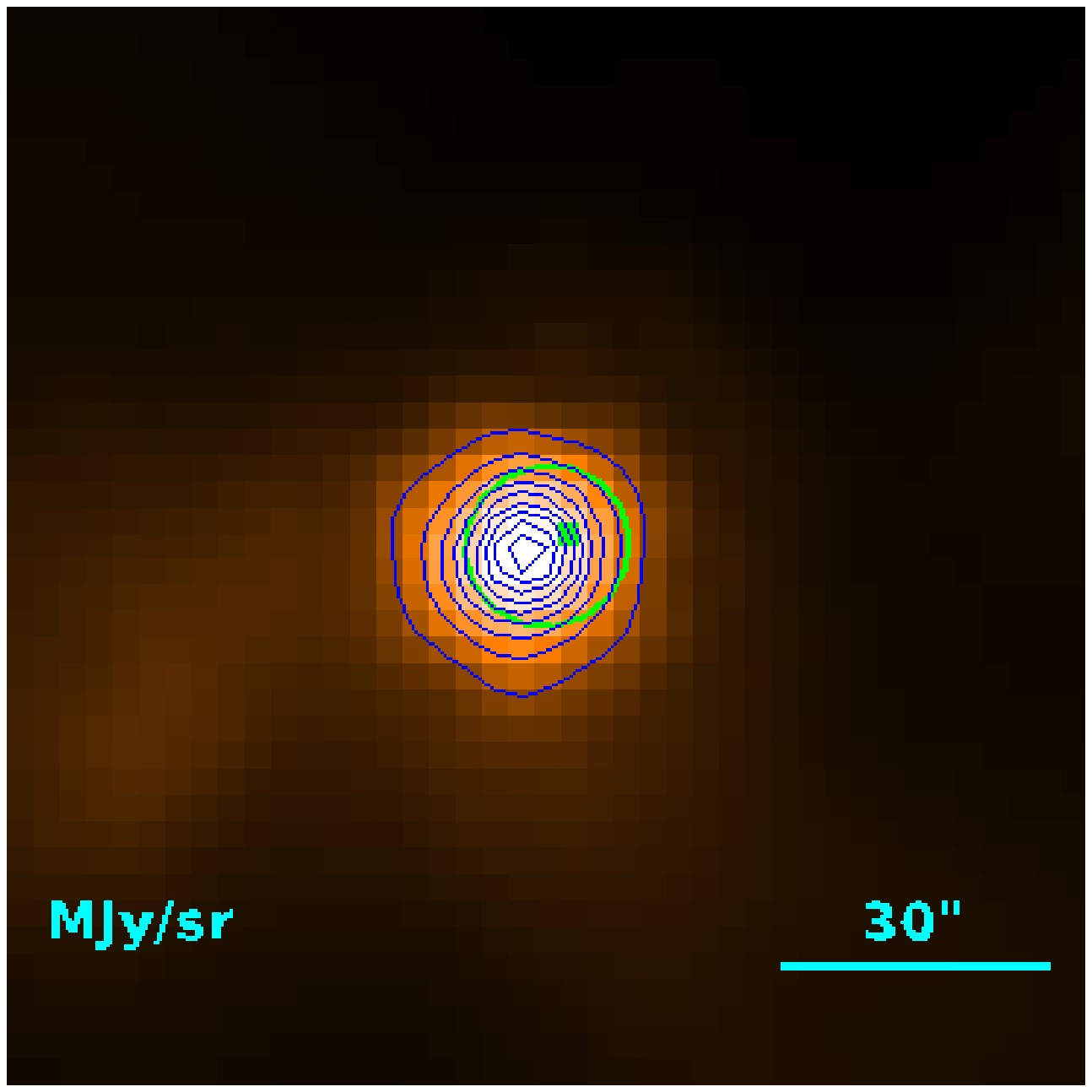}\\
\includegraphics[scale=.433]{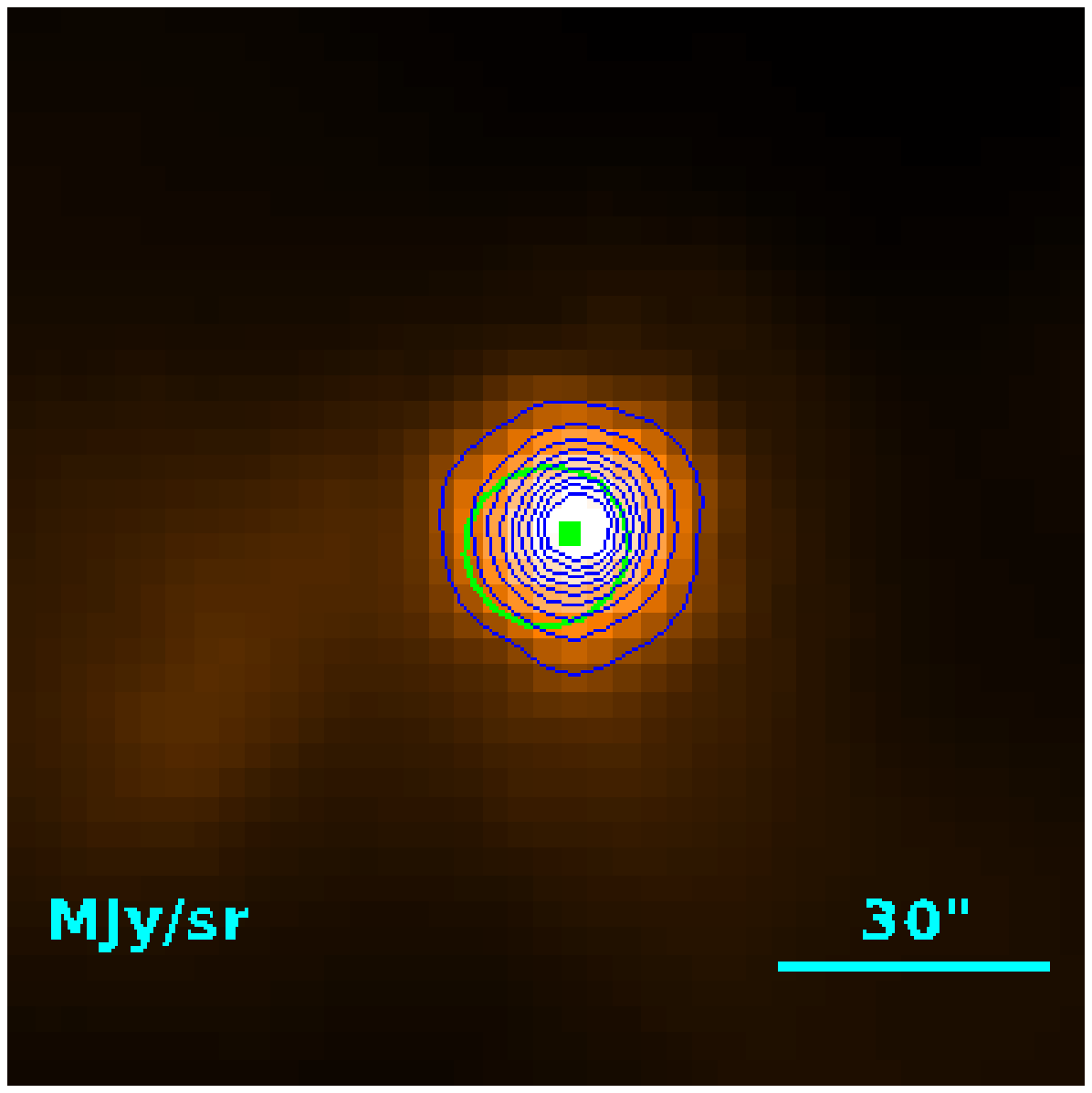}
\end{tabular}
\caption{Coordinates of source \#55 at 250~$\mu$m as derived in the protostars (top) and starless cores (bottom) catalogues, respectively.\label{confrCoord}}
\end{figure}

Protostars were detected using the 70~$\mu$m intensity map only. Since a residual misalignment may remain among the maps (see Sect.~2), the 70~$\mu$m coordinates of the centre may appear offset in the SPIRE intensity maps. The misalignment is smaller than the instrument spatial resolution, but before running \textsl{getsources} all the maps are reprojected onto a grid with pixel size 3\arcsec, so the displacement, if present, seems enhanced. Adopting the coordinates derived during the starless cores detection the protostar would appear well centred at the SPIRE wavelengths, but offset in the PACS bands. To show this effect we use Fig.~\ref{confrCoord} where the background image in both panels is the 250~$\mu$m cutout. In the upper panel, we show the ellipse and centre from the protostar catalogue (obtained from the PACS extraction), whereas in the bottom panel, we show the ellipse and centre from the starless core catalogue (obtained from the SPIRE extraction). Clearly, at 250~$\mu$m the source is better centred in the SPIRE extraction. So, for protostars, data in the PACS bands were taken from the prostostars catalogue, while data in the SPIRE bands were taken from the starless catalogue.

\begin{figure}
\begin{tabular}{cc}
\includegraphics[scale=.2]{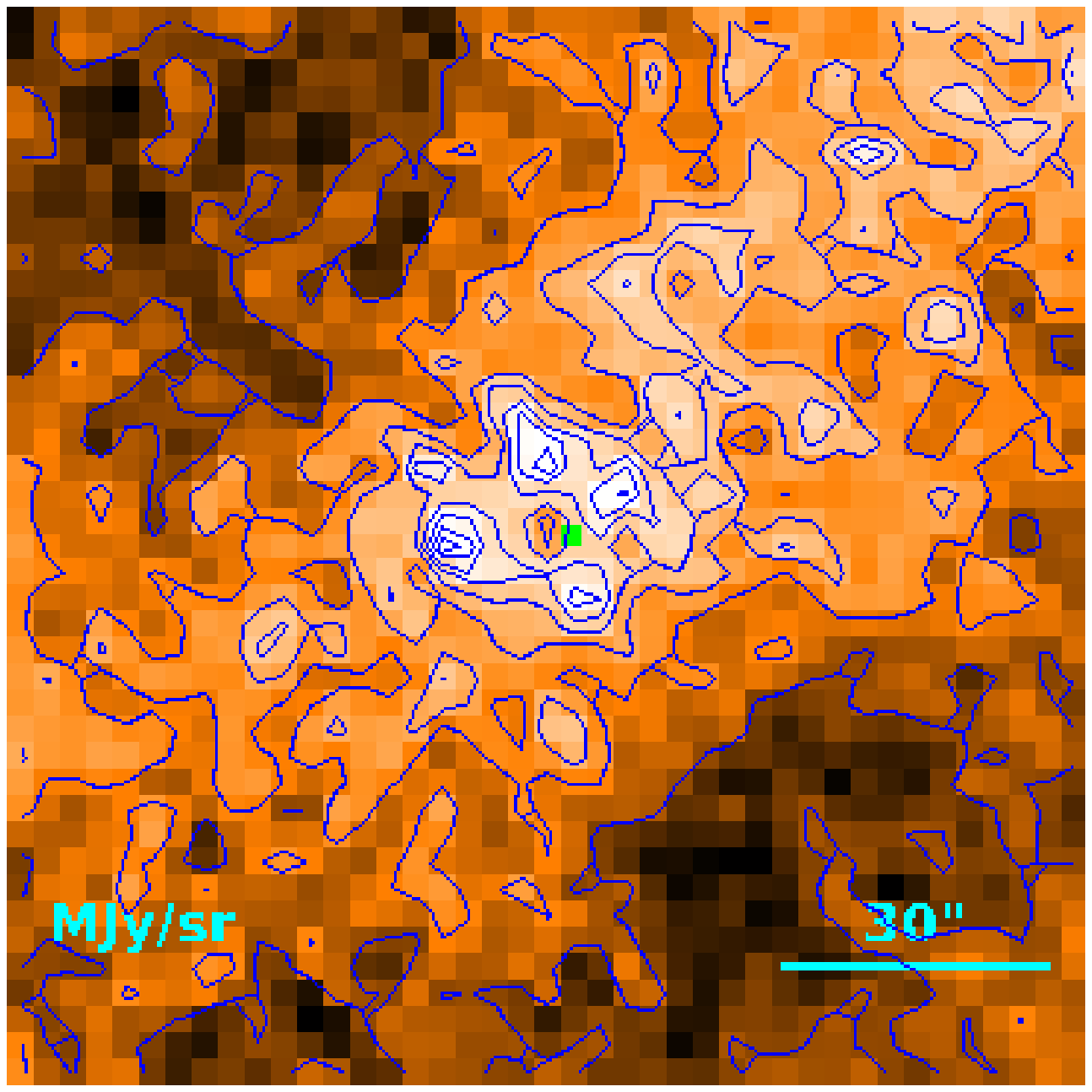}&\includegraphics[scale=.2]{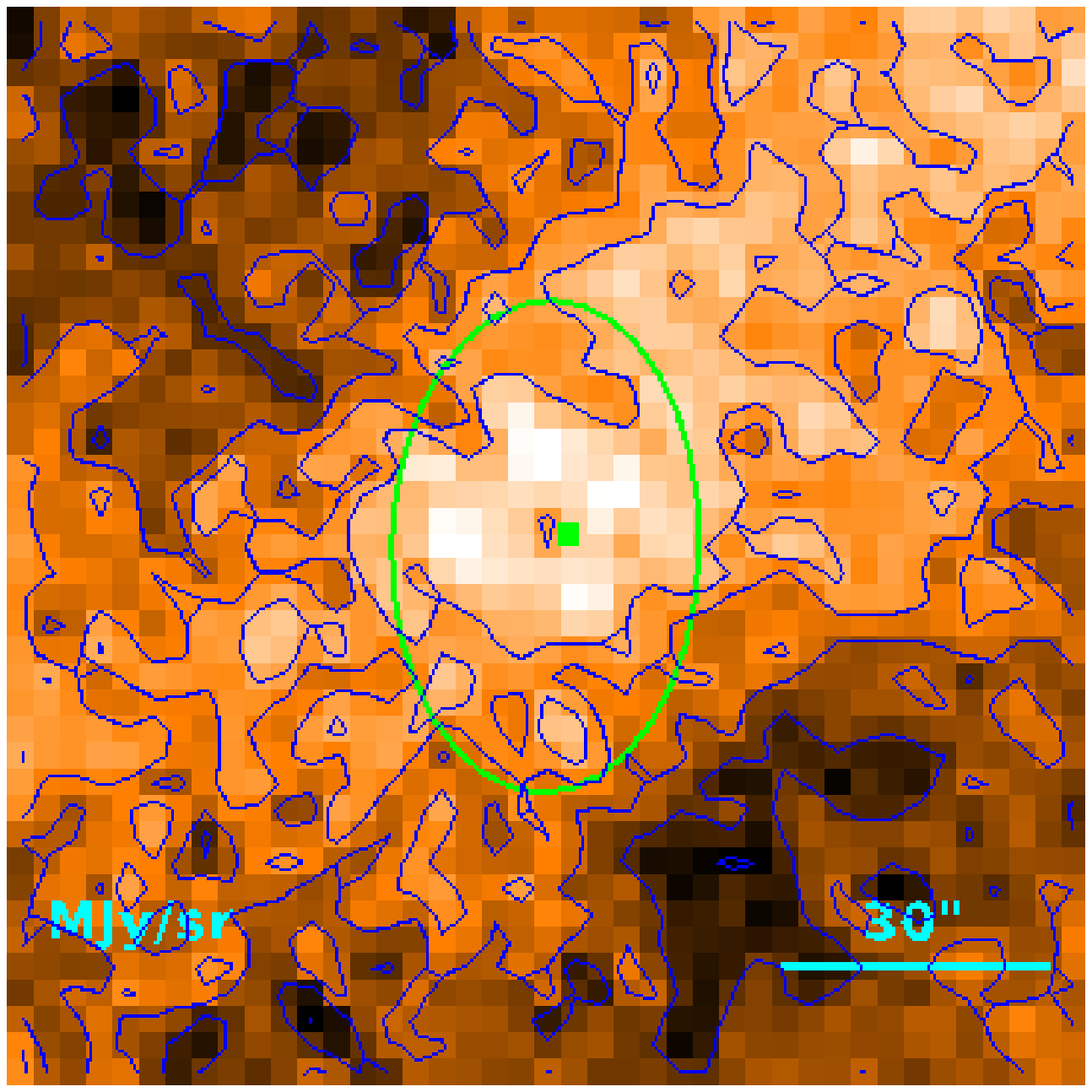}\\
\includegraphics[scale=.18]{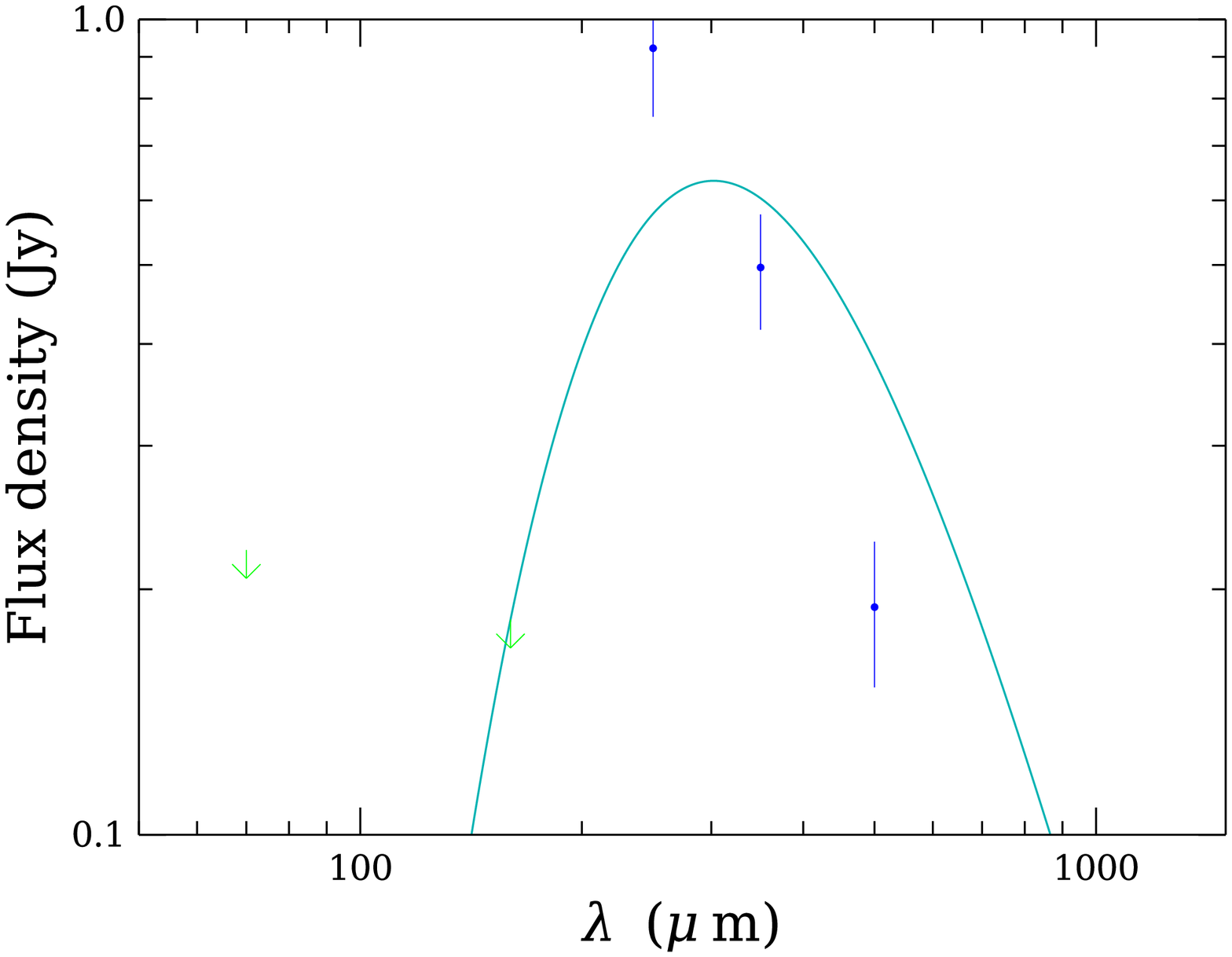}&\includegraphics[scale=.18]{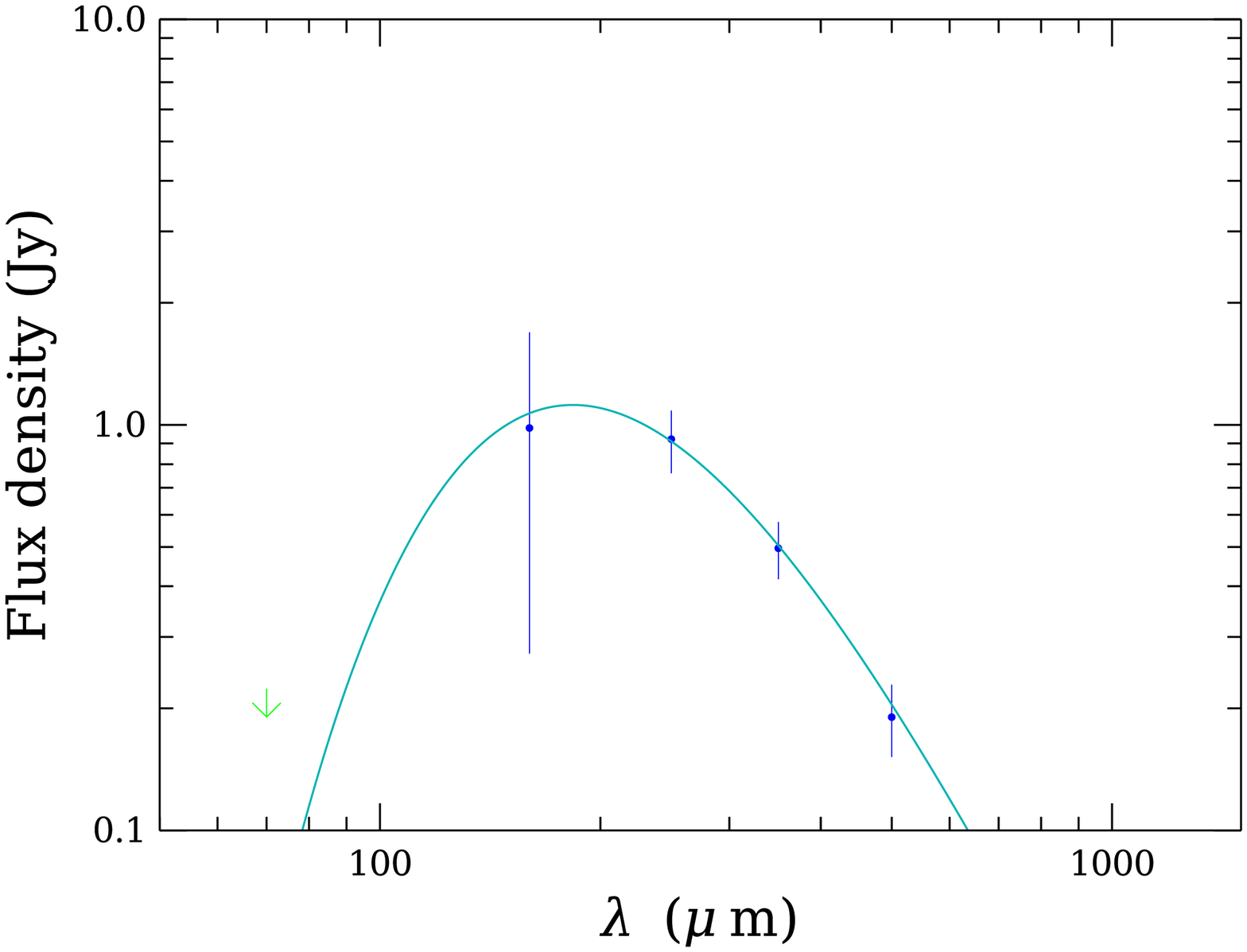}
\end{tabular}
\caption{Effects on the SED when using an alternative extraction strategy (see text). Top panels: undetected and detected source at 160~$\mu$m; bottom panels: SEDs with an upper limit and measure at 160~$\mu$m.\label{daLSaM}}
\end{figure}

A similar problem occurs at 160~$\mu$m for starless cores. In this case, the detection is done in the SPIRE maps, in the high-resolution column density map and in the 160~$\mu$m temperature-corrected map, but not in the 160~$\mu$m intensity map that is used for photometry extraction at the position found in the other maps. This strategy could be the reason why many sources have only an upper limit at 160~$\mu$m and in some cases the upper limit poses a problem when fitting the SED. In Fig.~\ref{daLSaM} we show such an example. The top-left panel shows the cutout in the PACS red band where \textsl{getsources} could not detect any source. However, already by eye it is possible to see that some emission is present at the position of the source that was not recognised as a real object. The bottom-left panel shows the corresponding SED with the 3$\sigma$ upper limit estimated with \textsl{getsources}.

We test our strategy by performing an alternative extraction using all 5 \textit{Herschel} intensity maps simultaneously. The benefit of this approach, is that the direct detection at 160~$\mu$m may help to extract photometry at this wavelength for weak sources. This alternative catalogue was used only to see if we could assign a measurement to sources having a stringent upper limit at 160~$\mu$m, like the one in Fig.~\ref{daLSaM}, with the conventional strategy. We found measurements for 12 sources using the alternative extraction. The top-right panel of Fig.~\ref{daLSaM} shows that now the starless core is detected also at 160~$\mu$m. In the bottom-right panel we show how the SED fitting benefits of this measurement.

The new 160~$\mu$m measurement has a large impact on the source physical parameters, with the temperature increasing from 10.42~K to 16.14~K and the mass decreasing from 0.627~$M_\sun$ to 0.073~$M_\sun$. Since there are 22 sources in the final catalogue with masses $0.6\le M/M_\sun<0.7$, two of which have no reliable SED fit, it is clear that removing even one single source may have some impact on the CMF.

\begin{figure}
\begin{tabular}{cc}
\includegraphics[scale=.18]{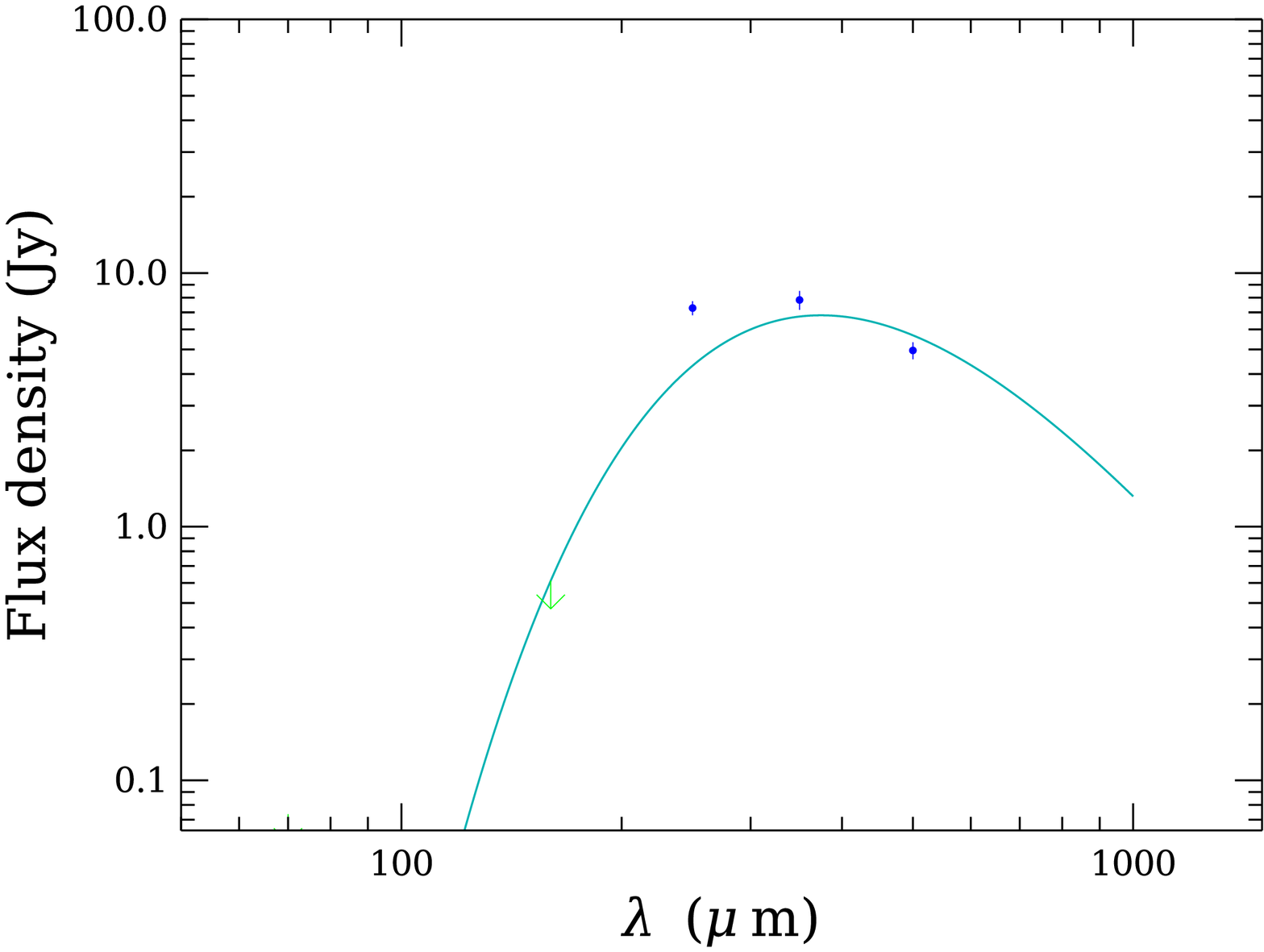}&\includegraphics[scale=.18]{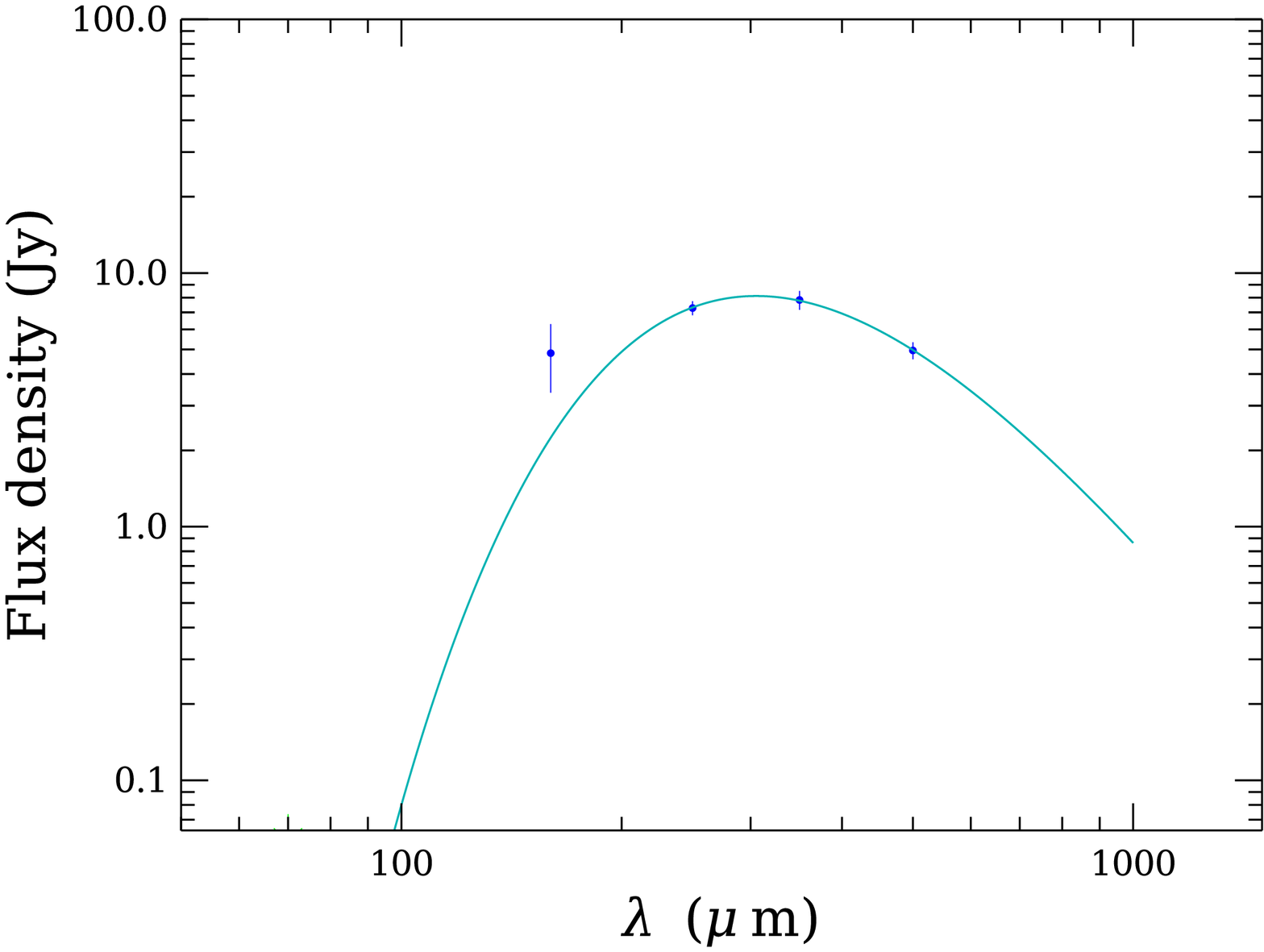}
\end{tabular}
\caption{Effects on the SED when the measurement at 160~$\mu$m is used instead of the upper limit for undetected sources (see text). Left panel: best-fit with upper limit; right panel: best-fit with a measure.\label{senzaLS}}
\end{figure}

Another approach we tried to solve the problem of the upper limits at 160~$\mu$m was to exclude this value during the fitting procedure. When \textsl{getsources} can not detect at a certain wavelength a source found in other images, it sets the so-called monochromatic significance to the special value \verb+9.999E-31+. When we found evident that the upper limit at 160~$\mu$m was influencing the SED fit much more than the SPIRE measures, we changed this value to \verb+9.999E-30+. In this case the fitting code uses the ``extracted'' 160 um intensity as if it is reliable, but assigns it a very low weight such that this band is not used to constrain the fit. This strategy was already used in the paper on Lupus (Benedettini et al., submitted). We show in Fig.~\ref{senzaLS} the result of this procedure applied to source \#58.

In the left panel we show the SED and best-fit modified blackbody function of this core as derived by \textsl{getsources}: it is clear that the upper limit in the red band forces the best-fit to have a low temperature. The best-fit has $T=7.74\pm0.02$~K and $M=8.59\pm0.45\,M_\sun$, a large mass indeed. On the other hand, using the total flux density at 160~$\mu$m that \textsl{getsources} measured at the position of the core in the other bands gives the SED shown in the right panel of the figure. The 160~$\mu$m value has a very low weight so the best-fit is now determined by the SPIRE intensities only and the parameters become $T=11.39^{+0.04}_{-0.05}$~K and $M=0.330\pm0.014\,M_\sun$, much reasonable values for Perseus.

We \textit{corrected} in this way the SED of 139 unbound cores with reliable SED fitting, 51 prestellar and 5 likely bound sources. A much more simple approach would be that of ignoring completely the upper limit, but we found that in some cases the best-fit model agrees well with the 160~$\mu$m measurement even if it was not relevant during the fitting procedure. This means that in this case \textsl{getsources} was able to make a reasonable measure even without detecting directly the source. The reader will recognize the \textit{corrected} sources because in this case the ellipse showing the size of the core (see Figs.~\ref{numero49} and \ref{numero55}) are drawn in red instead than in green.

In the next section on the completeness limit, we will make a precise assessment about the validity of these two approaches to the 160~$\mu$m upper limit problem, because in that case we deal with simulated sources for which we know the true values of the physical parameters.

The master catalogue listing the observed properties of all the cores is available in the online version of this paper. A template is provided in Table~\ref{esempioCat} to illustrate its form and content which follows \citet{Aquila}. There are cases in which two \textit{Spitzer} sources can be associated with one \textit{Herschel} protostar. In this case we chose the \textit{Spitzer} source detected at 70~$\mu$m, exploiting the fact that in no case both sources had detections at 70~$\mu$m. If none was revealed in this band, we associated the source with the higher 24~$\mu$m detection.

The derived properties (physical radius, mass, SED dust temperature, peak column density at the resolution of the 500~$\mu$m data, average column density, peak volume density, and average density) are given in another table also available online. Table~\ref{esempio2} gives an example. Note that for protostars the derived properties refer to the external envelope. A more detailed SED fitting will be object of a forthcoming paper.

We give in a third table, see the sample in Table~\ref{tab3}, additional data for each source. Columns~(2) and (3) give the source luminosity in two ways. First, for a modified blackbody the bolometric luminosity $L_\mathrm{bol}$, can be computed analytically as shown by \citet{grey}. For $\beta=2$
\begin{displaymath}
L_\mathrm{bol}=\frac{30}{63}\pi R^2\left(\frac{4\pi k}{h\nu_0}\right)^2\sigma T^6
\end{displaymath}
where $\sigma$ is the Stefan-Boltzmann constant, $k$ and $h$ are the Boltzmann and Planck constants, $T$ is the dust temperature, $\nu_0=c/\lambda_0$ where $\lambda_0$ is the wavelength above which the model becomes optically thin. For a given dust opacity parametrized with a power law of parameters $(\lambda_\mathrm{ref},\kappa_\mathrm{ref},\beta)$ $\lambda_0$ is \citep{FHSC}
\begin{equation}
\lambda_0 = \lambda_\mathrm{ref}\left(\kappa_\mathrm{ref}\frac{M}{\pi R^2}\right)^{1/\beta}=0.756\left(\frac{R}{1\mathrm{pc}}\right)^{-1}\sqrt{\frac{M}{1\mathrm{M}_\sun}}\,\,\mu m\label{lambda0}
\end{equation}
with $R$ and $M$ radius and mass of the (spherical) source.

The bolometric luminosity then becomes
\begin{equation}
L_\mathrm{bol}=\frac{30}{63}\kappa_\mathrm{ref}M\left(\frac{4\pi k\lambda_\mathrm{ref}}{ch}\right)^2\sigma T^6=9.45\times10^{-8}\left(\frac{M}{1\mathrm{M}_\sun}\right)T^6\,\,L_\sun\label{lbol}
\end{equation}

Equation~(\ref{lbol}) implies an integration of the whole SED, even at $\lambda<\lambda_0$ where the modified blackbody becomes optically thick. \citet{grey} have, however, shown that for Equation~(\ref{lbol}) to be valid, it is not necessary that the SED is optically thin at all wavelengths, it is enough that optical depth is less than 1 for $\lambda=\lambda_\star$ where $\lambda_\star$ is defined by the condition
\begin{displaymath}
\int_0^\infty\frac{x^{3+\beta}}{\mathrm{e}^x-1}\mathrm{d}x\approx\int_0^{x_\star}\frac{x^{3+\beta}}{\mathrm{e}^x-1}\mathrm{d}x
\end{displaymath}
with $x_\star=hc/k\lambda_\star T$. For $\beta=2$ the integral becomes negligible ($< 0.01$) at $x_\star=15$. This implies that the dusty envelope should be already optically thin at $\lambda_\star=960/T$~$\mu$m.

Columns~(4) and (5) in Table~\ref{tab3} report for each model $\lambda_0$ and $\lambda_\star$, while Col.~(3) gives the luminosity obtained by numerical integration of the SED at \textit{Herschel} wavelengths. For the first source in the table, the model becomes optically thin at 82~$\mu$m, while $\lambda_\star$ is 131~$\mu$m. In other words, the analitical formula for $L_\mathrm{bol}$ does a good job as long as the model is optically thin at 131~$\mu$m which is correct given that $\lambda_0=82$~$\mu$m.

The second source is a protostar. In this case $L_\mathrm{bol}$ cannot be computed with Equation~(\ref{lbol}) so that Col.~(2) is not correct and one can use Col.~(3) as a measure of the FIR luminosity. The third source does not have a reliable SED fit so that $M$ and $T$ were not derived from the SED. In this case Cols.~(2), (4) and (5) report a zero while Col.~(3) gives a more precise FIR luminosity.

Column (6) reports the kind of core: 0 is unbound, 1 prestellar, 2 protostellar and 3 candidate. Column (7) gives the reliability of the SED fitting: 1 means that the SED could be fit with a modified blackbody, 0 otherwise. Column (8) gives the region to which each source is associated, the numerical correspondence is given in the caption. These last three Cols. allow to make selections on the sample of cores easily: for instance, to select prestellar cores in IC348 one can look for sources with (1,1,9) in these Cols.

\begin{landscape}
\begin{table}
\tiny
\centering
\caption{Catalogue of dense cores identified in the HGBS maps of Perseus molecular complex (full catalogue is at CDS).\label{esempioCat}}
\begin{tabular}{cccccccccccccc}
\hline\hline
No.&Name&RA$_\mathrm{2000}$&Dec$_\mathrm{2000}$&Sig$_{070}$&\multicolumn{2}{c}{$S^\mathrm{peak}_{070}$}&$S^\mathrm{peak}_{070}/S_\mathrm{bg}$&$S^\mathrm{conv\,500}_{070}$&\multicolumn{2}{c}{$S^\mathrm{tot}_{070}$}&r$^\mathrm{a}_{070}$&r$^\mathrm{b}_{070}$&PA$_{070}$\\
&HGBS\_J&(h m s)&($^\circ$ \arcmin\ \arcsec)&&\multicolumn{2}{c}{(Jy/beam)}&&(Jy/beam$_{500}$)&\multicolumn{2}{c}{(Jy)}&(\arcsec)&(\arcsec)&($^\circ$)\\
(1)&(2)&(3)&(4)&(5)&\multicolumn{2}{c}{(6) $\pm$ (7)}&(8)&(9)&\multicolumn{2}{c}{(10) $\pm$ (11)}&(12)&(13)&(14)\\\hline
49&032510.2+302355&03:25:10.28&+30:23:55.4&0.000&2.93e--03&1.9e--03&0.88&3.19e--03&2.14e--01&1.4e--01&8&8&90\\
55&032522.3+304512&03:25:22.31&+30:45:12.0&735.700&1.48e+01&2.2e--02&227.14&1.61e+01&2.42e+01&3.7e--02&8&8&49\\
60&032529.2+303108&03:25:29.27&+30:31:08.4&0.000&5.70e--03&3.1e--03&1.04&2.13e--02&-7.21e-02&3.9e--02&25&12&38\\\hline
&&&&&&&&&&&&&\\
&&&&&&&&&&&&&\\
&&&&&&&&&&&&&
\end{tabular}
\begin{tabular}{cccccccccccccccccccc}
\hline\hline
Sig$_{160}$&\multicolumn{2}{c}{$S^\mathrm{peak}_{160}$}&$S^\mathrm{peak}_{160}/S_\mathrm{bg}$&$S^\mathrm{conv\,500}_{160}$&\multicolumn{2}{c}{$S^\mathrm{tot}_{160}$}&r$^\mathrm{a}_{160}$&r$^\mathrm{b}_{160}$&PA$_{160}$&Sig$_{250}$&\multicolumn{2}{c}{$S^\mathrm{peak}_{250}$}&$S^\mathrm{peak}_{250}/S_\mathrm{bg}$&$S^\mathrm{conv\,500}_{250}$&\multicolumn{2}{c}{$S^\mathrm{tot}_{250}$}&r$^\mathrm{a}_{250}$&r$^\mathrm{b}_{250}$&PA$_{250}$\\
&\multicolumn{2}{c}{(Jy/beam)}&&(Jy/beam$_{500}$)&\multicolumn{2}{c}{(Jy)}&(\arcsec)&(\arcsec)&($^\circ$)&&\multicolumn{2}{c}{(Jy/beam)}&&(Jy/beam$_{500}$)&\multicolumn{2}{c}{(Jy)}&(\arcsec)&(\arcsec)&($^\circ$)\\
(15)&\multicolumn{2}{c}{(16) $\pm$ (17)}&(18)&(19)&\multicolumn{2}{c}{(20) $\pm$ (21)}&(22)&(23)&(24)&(25)&\multicolumn{2}{c}{(26) $\pm$ (27)}&(28)&(29)&\multicolumn{2}{c}{(30) $\pm$ (31)}&(32)&(33)&(34)\\\hline
6.897&2.02e--01&4.2e--02&0.52&4.80e--01&3.70e--01&7.7e--02&25&20&53&23.880&8.09e--01&4.8e--02&1.01&8.93e--01&1.59e+00&9.4e--02&21&19&--18\\
1673&5.44e+01&1.5e--01&43.22&5.92e+01&6.20e+01&1.7e--01&14&14&0&1426&3.20e+01&6.9e--01&7.22&3.16e+01&3.36e+01&7.2e--01&19&18&--29\\
0.000&3.82e--02&3.2e--02&0.11&1.45e--01&8.45e--02&7.1e--02&35&30&--89&6.138&1.64e--01&3.5e--02&0.31&2.17e--01&2.55e--01&5.5e--02&30&18&83\\\hline
&&&&&&&&&&&&&\\
&&&&&&&&&&&&&\\
&&&&&&&&&&&&&
\end{tabular}
\begin{tabular}{ccccccccccccccccccc}
\hline\hline
Sig$_{350}$&\multicolumn{2}{c}{$S^\mathrm{peak}_{350}$}&$S^\mathrm{peak}_{350}/S_\mathrm{bg}$&$S^\mathrm{conv\,500}_{350}$&\multicolumn{2}{c}{$S^\mathrm{tot}_{350}$}&r$^\mathrm{a}_{350}$&r$^\mathrm{b}_{350}$&PA$_{350}$&Sig$_{500}$&\multicolumn{2}{c}{$S^\mathrm{peak}_{500}$}&$S^\mathrm{peak}_{500}/S_\mathrm{bg}$&\multicolumn{2}{c}{$S^\mathrm{tot}_{500}$}&r$^\mathrm{a}_{500}$&r$^\mathrm{b}_{500}$&PA$_{500}$\\
&\multicolumn{2}{c}{(Jy/beam)}&&(Jy/beam$_{500}$)&\multicolumn{2}{c}{(Jy)}&(\arcsec)&(\arcsec)&($^\circ$)&&\multicolumn{2}{c}{(Jy/beam)}&&\multicolumn{2}{c}{(Jy)}&(\arcsec)&(\arcsec)&($^\circ$)\\
(35)&\multicolumn{2}{c}{(36) $\pm$ (37)}&(38)&(39)&\multicolumn{2}{c}{(40) $\pm$ (41)}&(42)&(43)&(44)&(45)&\multicolumn{2}{c}{(46) $\pm$ (47)}&(48)&\multicolumn{2}{c}{(49) $\pm$ (50)}&(51)&(52)&(53)\\\hline
54.300&1.49e+00&7.1e--02&1.50&1.56e+00&2.02e+00&9.6e--02&28&25&--32&66.400&1.67e+00&7.9e--02&1.66&1.95e+00&9.1e--02&47&36&--43\\
849.200&2.02e+01&4.8e--01&4.87&2.00e+01&2.27e+01&5.4e--01&25&25&--10&407.900&1.17e+01&6.4e--01&4.82&1.26e+01&6.9e--01&36&36&--8\\
6.427&1.05e--01&3.4e--02&0.19&1.29e--01&1.46e--01&4.8e--02&43&25&78&0.000&6.68e--02&2.5e--02&0.11&4.95e--02&1.8e--02&40&36&74\\\hline
&&&&&&&&&&&&&\\
&&&&&&&&&&&&&\\
&&&&&&&&&&&&&
\end{tabular}
\begin{tabular}{cccccccccccccc}
\hline\hline
Sig$_{N_{H_2}}$&$N^\mathrm{peak}_{H_2}$&$N^\mathrm{peak}_{H_2}/N_\mathrm{bg}$&$N^\mathrm{conv\,500}_{H_2}$&$N^\mathrm{bg}_{H_2}$&$r^\mathrm{a}_{H_2}$&$r^\mathrm{b}_{H_2}$&PA$_{H_2}$&N$_\mathrm{SED}$&\textsl{CuTEx}&Core type&SIMBAD&\textit{Spitzer}&Comments\\
&($10^{21}$~cm$^{-2}$)&&($10^{21}$~cm$^{-2}$)&($10^{21}$~cm$^{-2}$)&(\arcsec)&(\arcsec)&($^\circ$)\\
(54)&(55)&(56)&(57)&(58)&(59)&(60)&(61)&(62)&(63)&(64)&(65)&(66)&(67)\\\hline
166.200&14.190&3.32&4.591&4.272&24&20&--41&4&1&prestellar&[KJT2007] SMM J032516+30238\\
883.600&64.580&2.80&16.770&23.080&19&18&--89&5&1&protostellar&&J032522.32+304513.9\\
8.689&0.683&0.53&0.256&1.284&35&18&76&2&1&starless&&&no SED fit\\\hline
\end{tabular}
\tablefoot{Cols.: (1) core running number; (2) name = HGBS\_J followed by J2000 source coordinates in sexagesimal format; (3) and (4) J2000 RA and Dec of source centre; (5), (15), (25), (35) and (45): detection significance  from monochromatic single scales in the \textit{Herschel} bands (0.0 when the core is not visible in clean single scales); (6)$\pm$(7), (16)$\pm$(17), (26)$\pm$(27), (36)$\pm$(37) and (46)$\pm$(47): peak flux density and its uncertainty estimated by \textsl{getsources}; (8), (18), (28), (38) and (48): contrast over the local background, defined as the ratio of the background-subtracted peak intensity to the local background intensity; (9), (19), (29) and (39): peak flux density smoothed to a 36\farcs3 beam; (10)$\pm$(11), (20)$\pm$(21), (30)$\pm$(31), (40)$\pm$(41) and (49)$\pm$(50): integrated flux density and its uncertainty estimated by \textsl{getsources}; (12)-(13), (22)-(23), (32)-(33), (42)-(43) and (51)-(52): major \& minor FWHM of the core, respectively, estimated by \textsl{getsources}; (14), (24), (34), (44) and (53): position angle, measured east of north, of the core major axis; (54): detection significance in the high-resolution column density map; (55) peak H$_2$ column density as estimated by \textsl{getsources} in the high-resolution column density map;(56) column density contrast over the local background estimated by \textsl{getsources} in the high-resolution column density map; (57) peak column density smoothed to a 36\farcs3 beam; (58) local background H$_2$ column density estimated by \textsl{getsources} in the high-resolution column density map; (59)-(60)-(61): major \& minor FWHM and position angle of the core, respectively, estimated by \textsl{getsources} in the high-resolution column density map; (62): number of \textit{Herschel} bands in which the core is significant (Sig$_\lambda>5$) and has a positive flux density, excluding the column density plane; (63): 2 means that \textsl{CuTEx} \citep{cutex} found a source whose position falls within the ellipse defined by the FWHMs estimated on the high-resolution column density map, if the distance between the two peaks is less than 6\arcsec\ the flag is 1, while 0 means that no counterpart was found by \textsl{CuTEx}; (64): core classification, starless means unbound core; (65) closest counterpart found in SIMBAD, if any, up to 6\arcsec\ from the Herschel peak position (the identifier is copied as is in the SIMBAD archive); (66) closest \textit{Spitzer}-identified YSO from the c2d survey \citep{evans} within 6\arcsec\ from the Herschel peak position, if any. When present, the leading part of the name (SSTc2d) has been removed; (67) comments.}
\end{table}
\end{landscape}
                        
\begin{landscape}
\begin{table}
\tiny
\centering
\caption{Catalogue of dense cores identified in the HGBS maps of Perseus molecular complex (full catalogue is at CDS).\label{esempio2}}
\begin{tabular}{ccccccccccccccccccc}
\hline\hline
No.&Name&RA$_\mathrm{2000}$&Dec$_\mathrm{2000}$&\multicolumn{2}{c}{$R_\mathrm{core}$}&\multicolumn{2}{c}{$M_\mathrm{core}$}&\multicolumn{2}{c}{$T_\mathrm{dust}$}&$N^\mathrm{peak}_{H_2}$&\multicolumn{2}{c}{$N^\mathrm{ave}_{H_2}$}&$n^\mathrm{peak}_{H_2}$&\multicolumn{2}{c}{$n^\mathrm{ave}_{H_2}$}&$\alpha_\mathrm{BE}$&Core type&Comments\\
&HGBS\_J&(h m s)&($^\circ$ \arcmin\ \arcsec)&\multicolumn{2}{c}{(pc)}&\multicolumn{2}{c}{($M_\sun$)}&\multicolumn{2}{c}{(K)}&($10^{21}$~cm$^{-2}$)&\multicolumn{2}{c}{($10^{21}$~cm$^{-2}$)}&($10^4$~cm$^{-3}$)&\multicolumn{2}{c}{($10^4$~cm$^{-3}$)}\\
(1)&(2)&(3)&(4)&(5)&(6)&\multicolumn{2}{c}{(7) $\pm$ (8)}&\multicolumn{2}{c}{(9) $\pm$ (10)}&(11)&(12)&(13)&(14)&(15)&(16)&(17)&(18)&(19)\\\hline
49&032510.2+302355&03:25:10.28&+30:23:55.4&1.75e--02&3.17e--02&3.611&0.120&7.3&0.1&36.333&168.27&51.12&20.949&234.31&39.24&0.058&prestellar\\
55&032522.3+304512&03:25:22.31&+30:45:12.0&8.87e--03&2.70e--02&1.105&0.028&18.1&0.1&33.650&199.34&21.45&19.402&546.05&19.27&0.239&protostellar\\
60&032529.2+303108&03:25:29.27&+30:31:08.4&2.56e--02&3.68e--02&0.047&0.025&10.4&1.0&0.262&1.01&0.49&0.151&0.96&0.32&9.421&starless&no SED fit\\\hline
\end{tabular}
\tablefoot{Cols.: (1) core running number; (2) name = HGBS\_J followed by J2000 source coordinates in sexagesimal format; (3) and (4) J2000 RA and Dec of source centre; (5) and (6) geometrical average of the two FWHMs measured in the high-resolution column density map, after and before deconvolution with HPBW of 18\farcs2, respectively (NB: both values provide estimates of the object's outer radius when the core can be approximately described by a Gaussian distribution, as is the case for the critical Bonnor-Ebert spheroid); (7)$\pm$(8) estimated core mass; (9)$\pm$(10) SED dust temperature; (11) peak H$_2$ column density, at the resolution of 500~$\mu$m intensity map, derived from a modified blackbody SED fit to the core peak flux densities measured in a common 36\farcs3 beam at all wavelengths; (12) average column density, calculated as $M_\mathrm{core}/(\mu m_\mathrm{H}\pi R_\mathrm{core}^2)$, where $M_\mathrm{core}$ is the estimated core mass (Col. 7), $R_\mathrm{core}$ the estimated core radius before deconvolution (Col. 6), $\mu$=2.8; (13) average column density, calculated same way as for Col.~(12), but using the deconvolved radius (Col.~5); (14) beam-averaged peak volume density at the resolution of 500~$\mu$m intensity map, derived from the peak column density (Col.~11) assuming a Gaussian spherical distribution: $n^\mathrm{peak}_{\mathrm{H}_2}=\sqrt{\frac{4\ln2}{\pi}}\frac{N^\mathrm{peak}_{\mathrm{H}_2}}{\mathrm{FWHM}_{500}}$; (15) average volume density, calculated as $n^\mathrm{ave}_{\mathrm{H}_2}=\frac{M_\mathrm{core}}{4/3\pi R_\mathrm{core}^3}\frac1{\mu m_\mathrm{H}}$ using the estimated core radius before deconvolution; (16) average volume density calculated as previous Col. but using the deconvolved core radius; (17) Bonnor-Ebert mass ratio $\alpha_\mathrm{BE}=M_\mathrm{BE,crit}/M_\mathrm{obs}$; (18) core type: starless, prestellar or protostellar; (19) comments.}
\end{table}

\end{landscape}

\begin{table}
\tiny
\centering
\caption{Catalogue of dense cores identified in the HGBS maps of Perseus molecular complex (full catalogue is at CDS).\label{esempio3}}
\begin{tabular}{cccccccc}
\hline\hline
No.&$L_\mathrm{bol,ana}$&$L_\mathrm{SED}$&$\lambda_0$&$\lambda_\ast$&Core&SED&Region\\
&$(L_\sun)$&$(L_\sun)$&($\mu$m)&($\mu$m)&type&fit&\\
(1)&(2)&(3)&(4)&(5)&(6)&(7)&(8)\\\hline
49&0.052&0.049&82.307&131&1&1&1\\
55&3.722&4.152&89.583&53&2&1&2\\
60&0.000&0.002&0.000&0&0&0&2\\\hline
\end{tabular}
\tablefoot{Cols.: (1) core running number; (2) bolometric luminosity found from Equation~(\ref{lbol}); (3) bolometric luminosity found by numerical integration of the SED; (4) $\lambda_0$ as given by Equation~(\ref{lambda0}); (5) $\lambda_\ast$ see text; (6) core type: 0=starless; 1=prestellar; 2=protostellar; 3=candidate bound; (7) SED fit: 0=no reliable SED fit; 1=reliable SED fit; (8) region: 1=L1451; 2=L1448; 3=L1455; 4=NGC1333; 5=Perseus6; 6=B1; 7=B1E; 8=L1468; 9=IC348; 10=B5; 11=HPZ1; 12=HPZ2; 13=HPZ3; 14=HPZ4; 15=HPZ5; 16=HPZ6; 0=outside all the regions.\label{tab3}}
\end{table}

\section{Simulation of prestellar cores}\label{complSec}
Simulated 432 prestellar cores with Bonnor-Ebert density profile and mass between 0.05~$M_\sun$ and 0.96~$M_\sun$ were injected in the source-free intensity maps to verify the correctness of our extraction procedure and to derive the mass completeness limit. The simulations were performed as in previous HGBS works \citep{Aquila,2016MNRAS.459..342M,milena2}. The extraction of the simulated sources was performed nominally and the post-selection checks for cores described in Sect.~\ref{estrazione} were executed. This catalogue was then cross-checked against the truth table of the synthetic cores and 315 sources out of the 432 were recovered.

\subsection{Restoring the 160$\mu$m intensity}
As discussed in the previous section, we found some problem when dealing with the 160~$\mu$m upper limit often found in the standard extraction. To avoid that the upper limit forces the SED fitting to too low temperatures, we adopted two different strategies previously described and here we discuss their validity.

First, we made an alternative extraction using all five of the \textit{Herschel} bands simultaneously. Compared to the nominal extraction, two sources with only upper limits at 160~$\mu$m ended up with intensity measurements in the alternative extraction. For one source, the mass derived from the SED fitting using the upper limit at 160~$\mu$m is 1.46 times the true mass. Using the measurement from the alternative catalogue the derived mass is 1.36 the true mass, slightly better. For the second source, using the upper limit causes the derived mass to be overestimated by more than a factor 3. Using the measured flux density in the alternative catalogue the ratio is 0.75, which means that the mass in only underestimated by 25\%, in any case a better result.

The second approach was to pretend that the upper limit in the nominal extraction is indeed a real measurement but with a negligible weight during the SED fitting so that it does not force the fitting. For 96 sources out of 147 simulated cores with an upper limit, nothing changed either because the best-fit model fullfilled the upper limit, or because the SED could not be fit anyway. For the remaining 51 sources, the mean ratio between (SED-fitting derived mass)/(true mass) is 1.5$\pm$1.1 when using the upper limit, and 0.92$\pm$0.78 when the 160~$\mu$m intensity is weighted down in the SED fit. The large standard deviation of the means shows that the results are valid for the whole sample and not applicable to the single sources. Nevertheless, the upper limit causes, on average, an overestimation of the mass by 50\% while weighing down the 160~$\mu$m intensity generally agrees with the input mass, on average, of about 8\%. We concluded then that the result of the SED fitting improves when using both approaches.

\subsection{Completeness limit in mass}\label{appComplMass}
\begin{figure}
\includegraphics[scale=1.5]{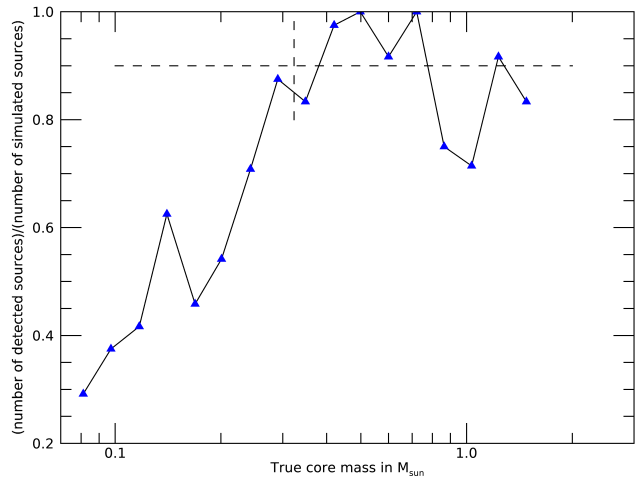}
\caption{Number of detected sources over number of simulated sources vs. true core mass. The completeness limit is defined as the mass at which completeness is 90\%: in our case this happens at 0.323~$M_\sun$, limit shown as intersection between the two dashed lines. The decreasing of detected sources at high mass is discussed in the text (see also next figure).\label{complM}}
\end{figure}

In Fig.~\ref{complM} we show the fraction of detected sources with respect to the number of simulated sources as a function of the true core mass. To improve the statistics we summed the number of detected sources in three adjacent bins having similar masses, while the mass of each bin was the average of the three masses. The last two points are the sum/average of four bins.

The horizontal dashed lines shows the 90\% level of completeness while the vertical line corresponds to a mass of 0.323~$M_\sun$, found by linear interpolation of the two bins at 0.290~$M_\sun$ and 0.420~$M_\sun$. %, and ignoring the bin at 0.348~$M_\sun$.

\begin{figure}
\includegraphics[scale=1.5]{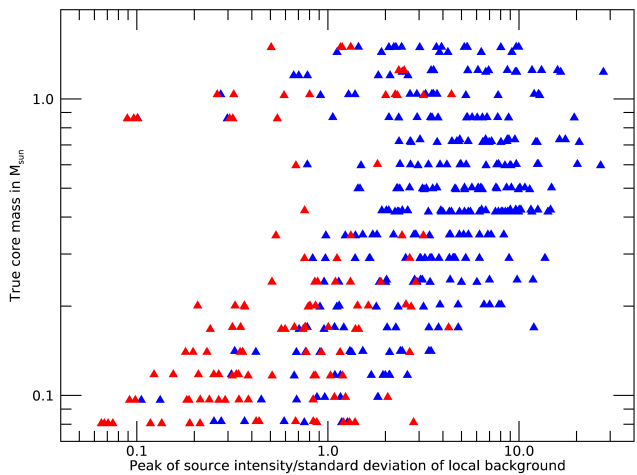}
\caption{Contrast level (see text) for detected sources (blue triangles) and undetected sources (red triangles).\label{SNR}}
\end{figure}

Very interesting is also the trend at high-mass in Fig.~\ref{complM} where the number of detected sources decreases. As shown by \citet{Aquila}, the likelihood of detecting a sources does not depend only on its intensity, but also on the contrast, i.e., how well the source stands out above the emission of the diffuse medium. In general, high-mass cores are found in high column density regions which are very bright. To estimate the contrast we used the signal-to-noise ratio (SNR) expressed as the peak intensity of the simulated sources over the standard deviation of the local background, both in MJy\,sr$^{-1}$. This ratio, for the fiducial wavelength of 250~$\mu$m, is shown in Fig.~\ref{SNR}: blue and red triangles refer to detected/undetected sources, respectively. We can see that for SNR$\ge3$ we detect $\sim$98\% of the sources independent on their mass. The percentage is as low as 31\% for SNR$\le1$. At high mass we explored low levels of SNR that resulted in a small fraction of detected sources. This figure shows that not necessarily high-mass cores are found easier than low-mass cores. Interested reader are invited to read the in-depth discussion on this topic in \citet{Aquila}.

\section{Definition of Perseus sub-regions borders through \element[][13]{C}\element[][]{O} observations\label{GaiaCom}}
The Perseus molecular complex is known to host different regions with distances varying between $\ga$200~pc and $\la$350~pc. From our intensity maps, because the dust emission is optically thin, it is not possible to estimate which regions are closer or further. Also, their borders are arbitrary defined.

\begin{figure*}
\sidecaption
\begin{tabular}{c}
\includegraphics[scale=1.1]{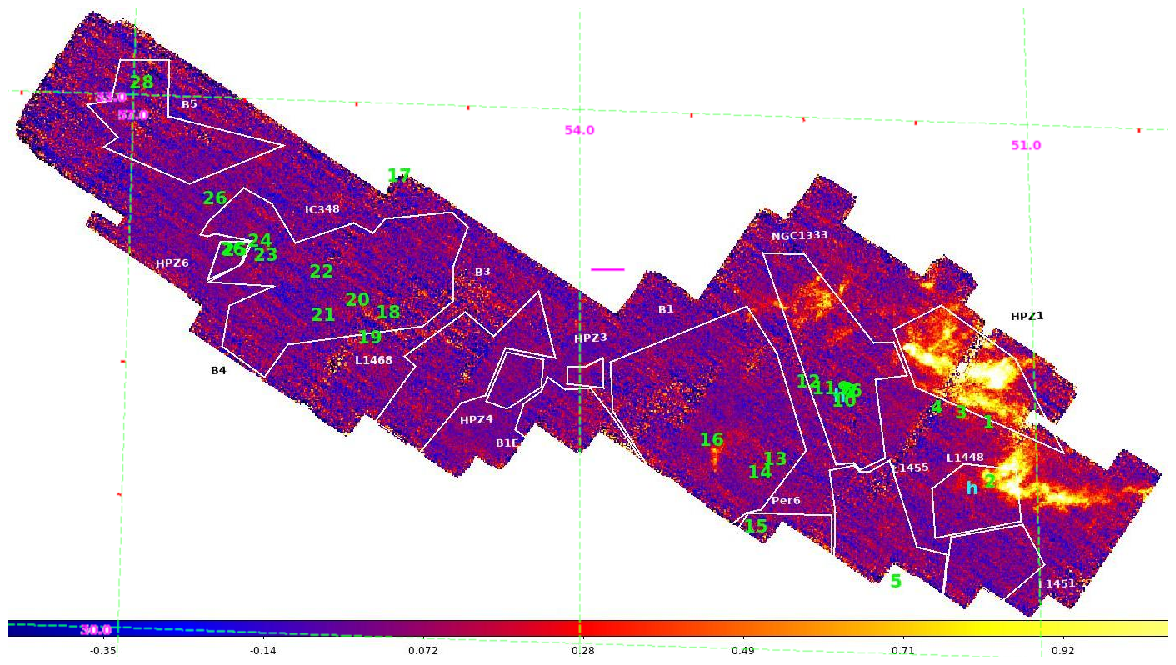}\\
\\
\includegraphics[scale=1.1]{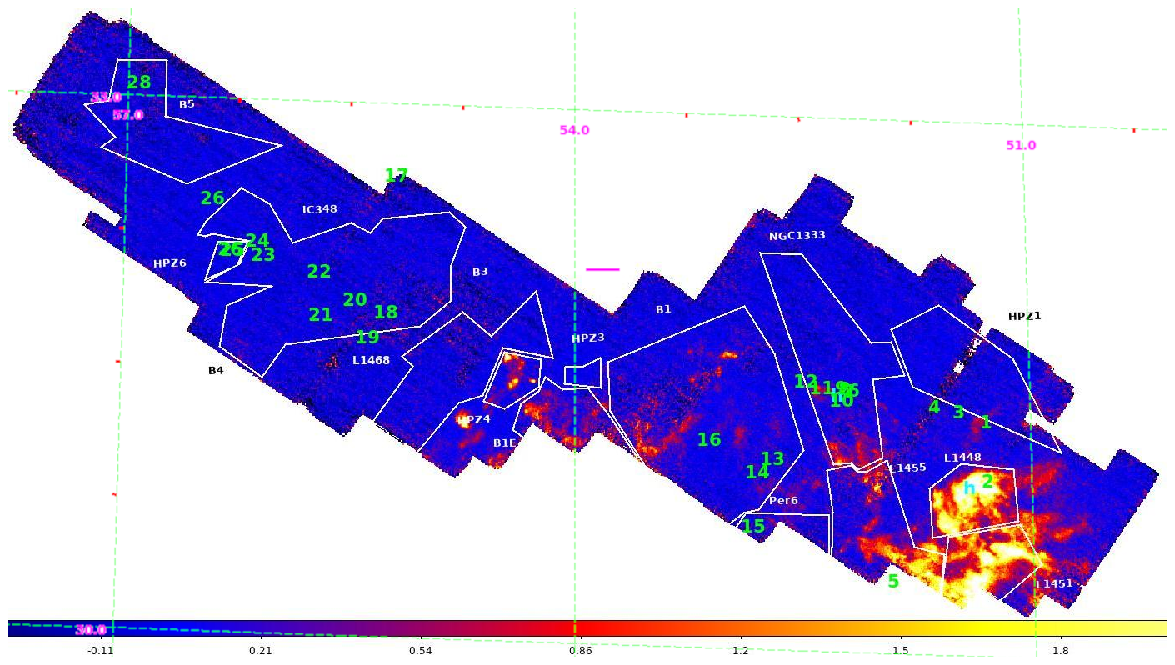}\\
\\
\includegraphics[scale=1.1]{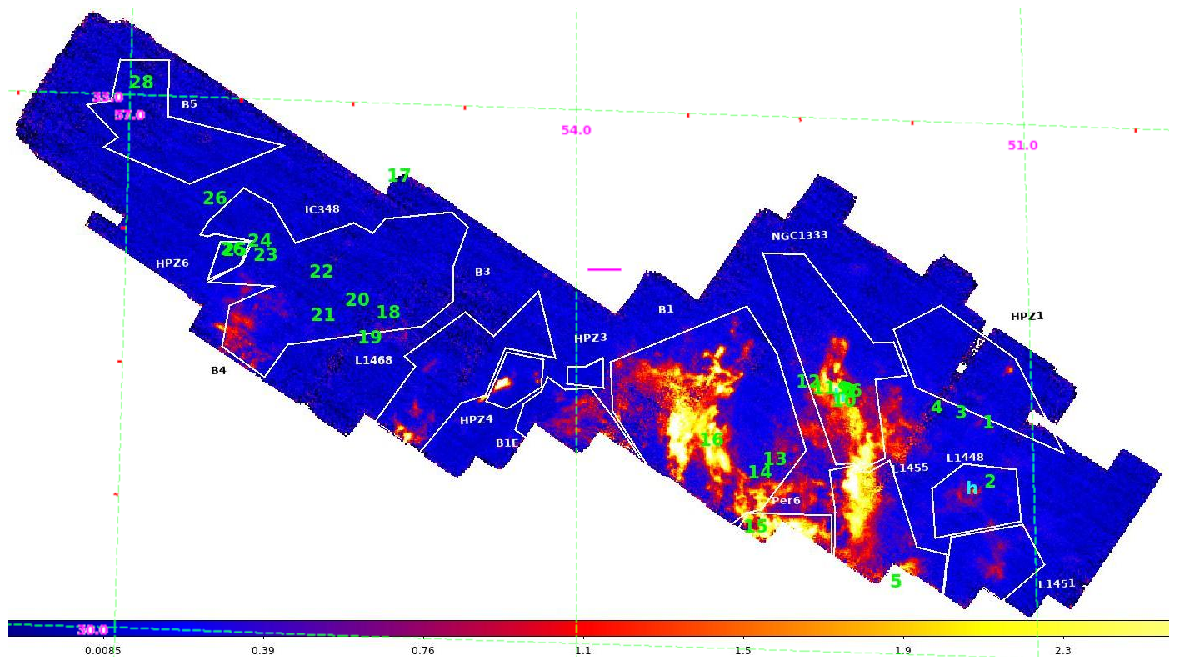}\\
\end{tabular}
\caption{Maps of \element[][13]{C}\element[][]{O} in three velocity channels: 1.1~Km~s$^{-1}$ (top), 4.35~Km~s$^{-1}$ (centre), 6.0~Km~s$^{-1}$ (bottom); colour bars are uncorrected antenna temperature in K. Green labels mark the position of GAIA sources (see Table~\ref{distanze}); \textit{h} labels show the position of \citet{hirota8} and \citet{hirota1} sources (SVS13 is in the cloud of sources 6-10). The magenta lines at the centre of each map shows the angular scale corresponding to 1~pc at 300~pc; J2000.0 coordinates grid is shown.\label{mappe13COa}}
\end{figure*}

\begin{figure*}
\sidecaption
\begin{tabular}{cc}
\includegraphics[scale=1.1]{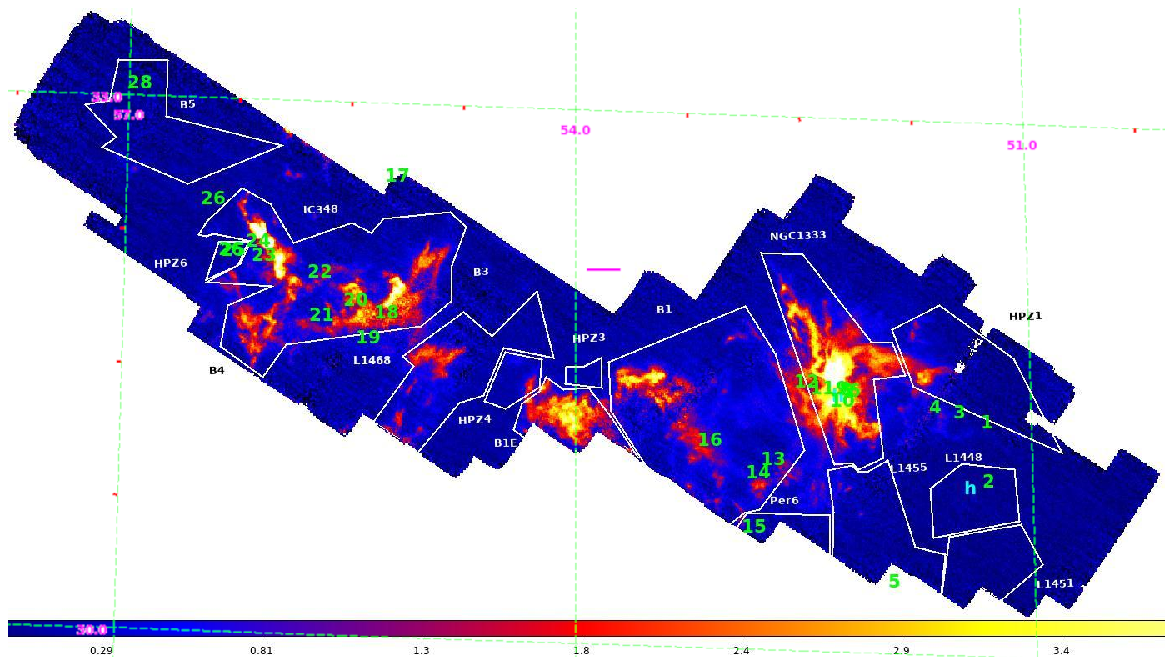}\\
\\
\includegraphics[scale=1.1]{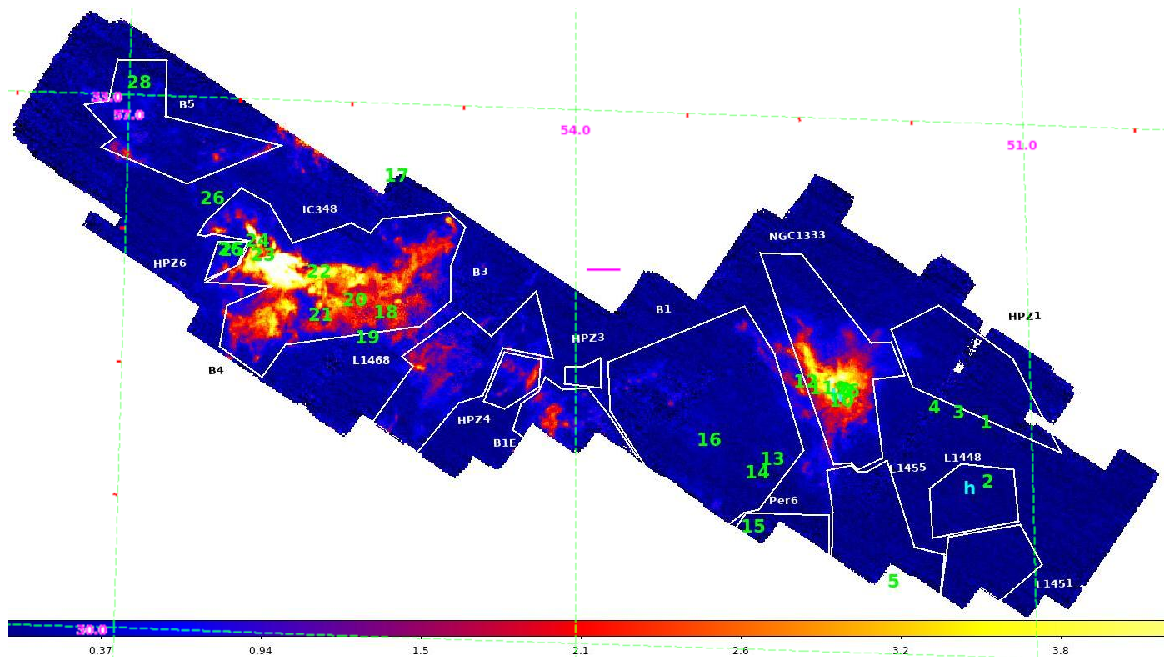}\\
\\
\includegraphics[scale=1.1]{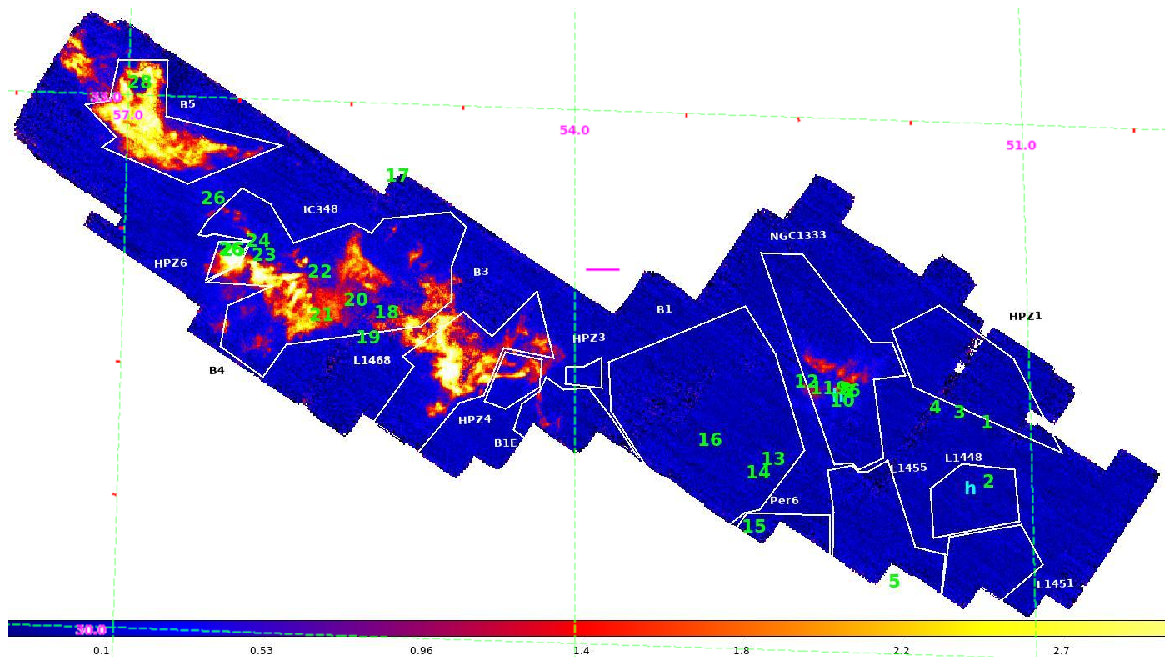}\\
\end{tabular}
\caption{Maps of \element[][13]{C}\element[][]{O} in three velocity channels: 7.9~Km~s$^{-1}$ (top), 8.8~Km~s$^{-1}$ (centre), 10.0~Km~s$^{-1}$ (bottom). In the latter panel, north of B5 and slightly covered by label 28, the bubble CPS12 \citep{bubbles} is evident. The magenta lines at the centre of each map shows the angular scale corresponding to 1~pc at 300~pc; J2000.0 coordinates grid is shown. Labels as in Fig.~\ref{mappe13COa}.\label{mappe13COb}}
\end{figure*}

\begin{figure*}
\centering
\begin{tabular}{ccc}
\includegraphics[scale=.25]{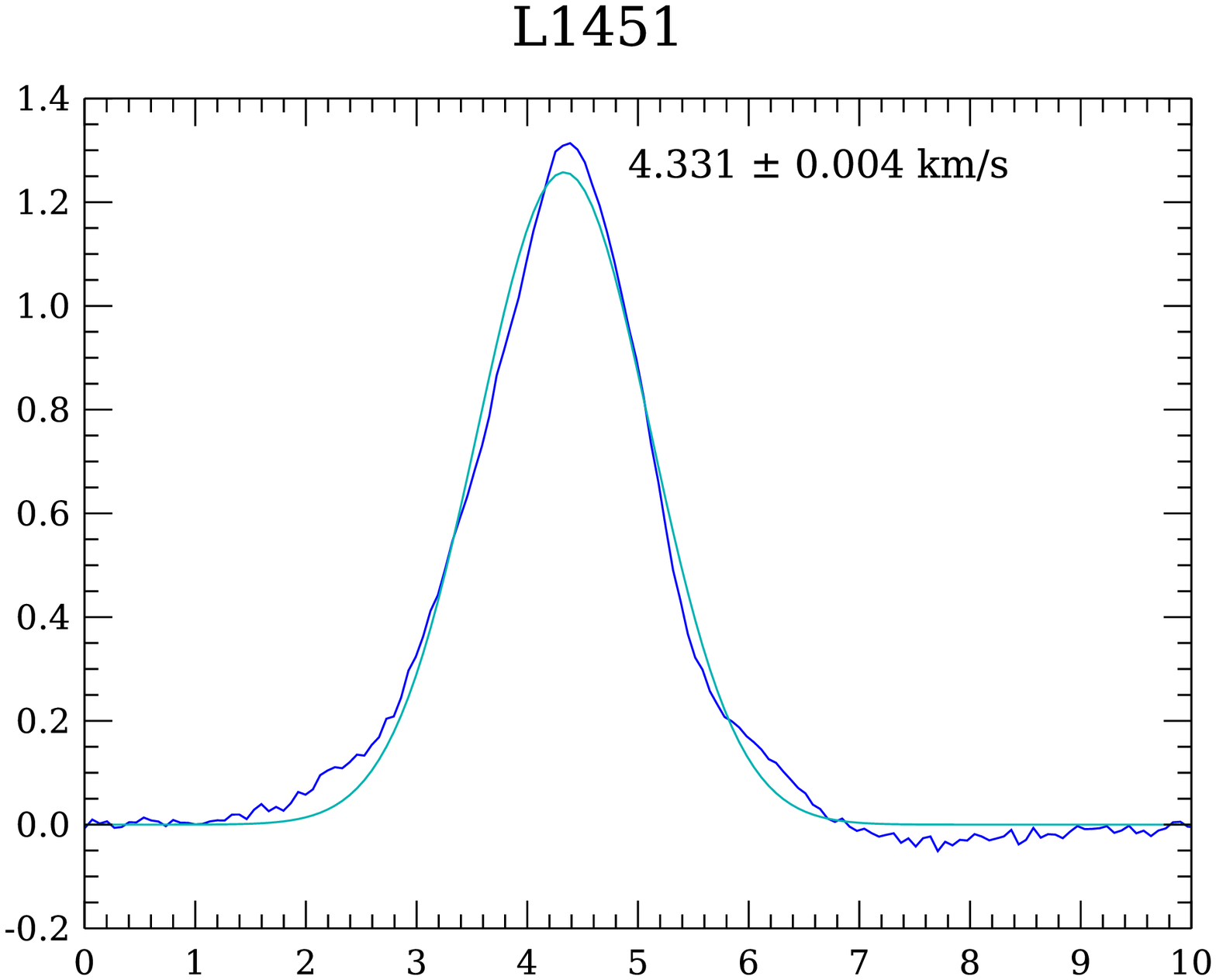}&
\includegraphics[scale=.25]{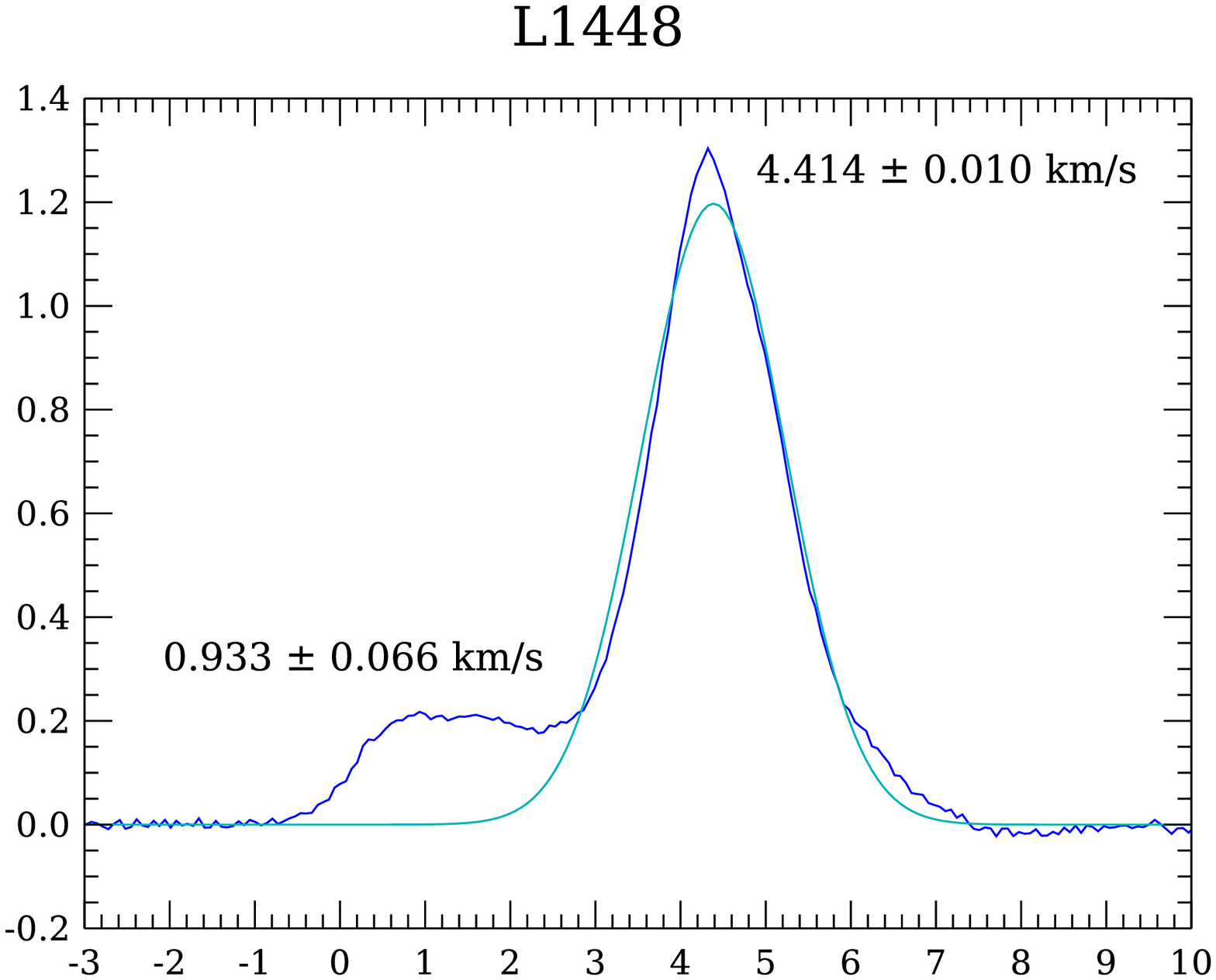}&
\includegraphics[scale=.25]{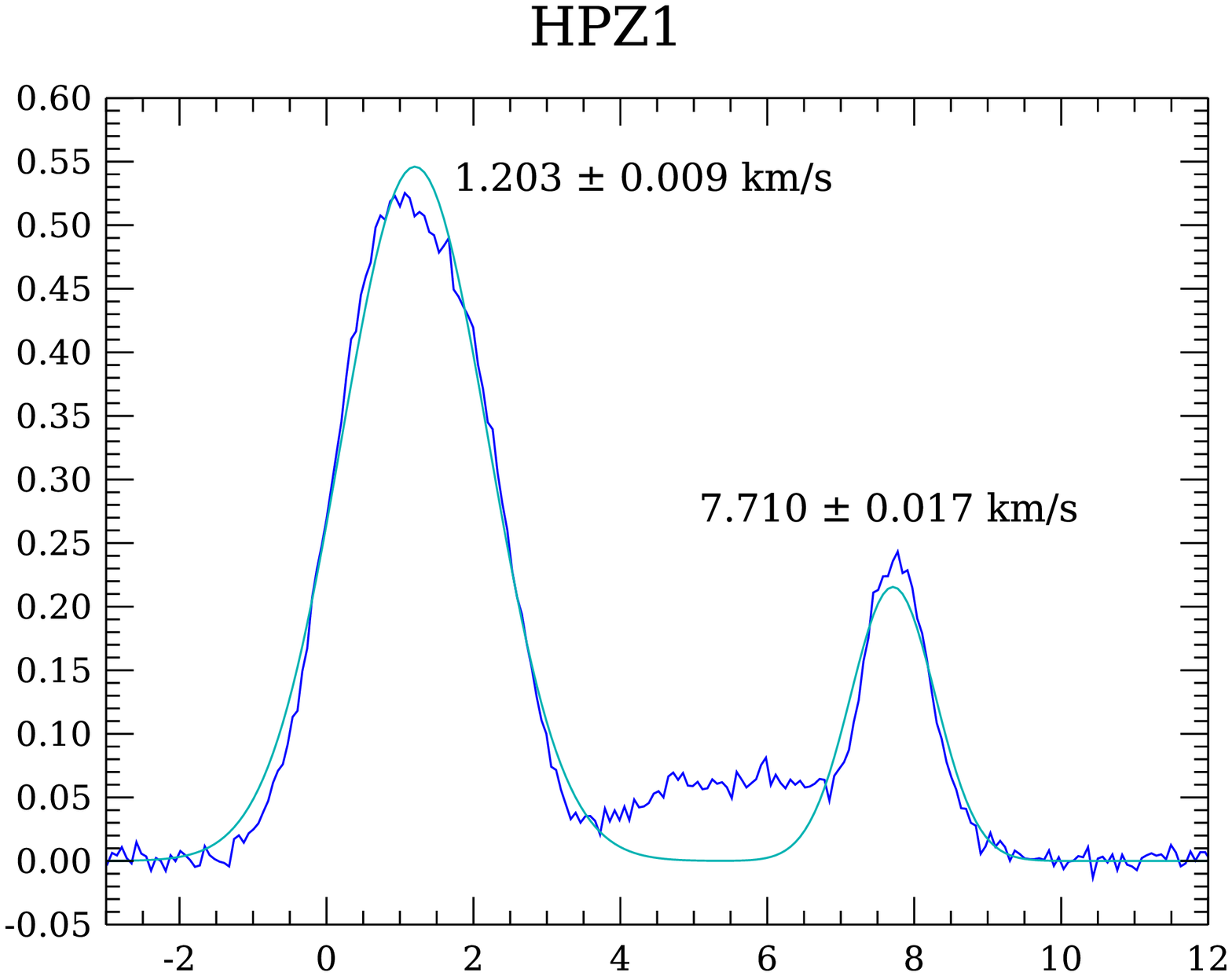}\\
\includegraphics[scale=.25]{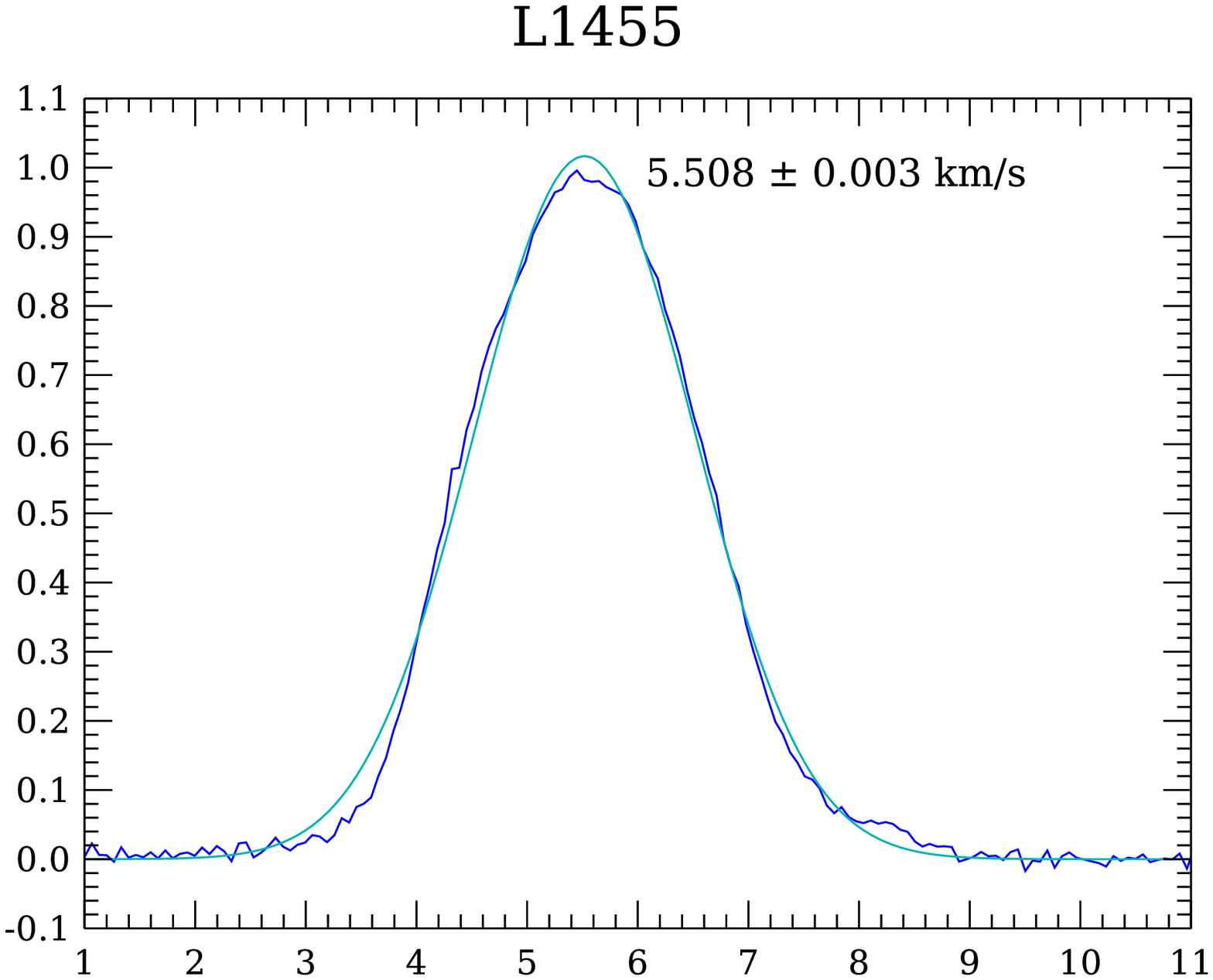}&
\includegraphics[scale=.25]{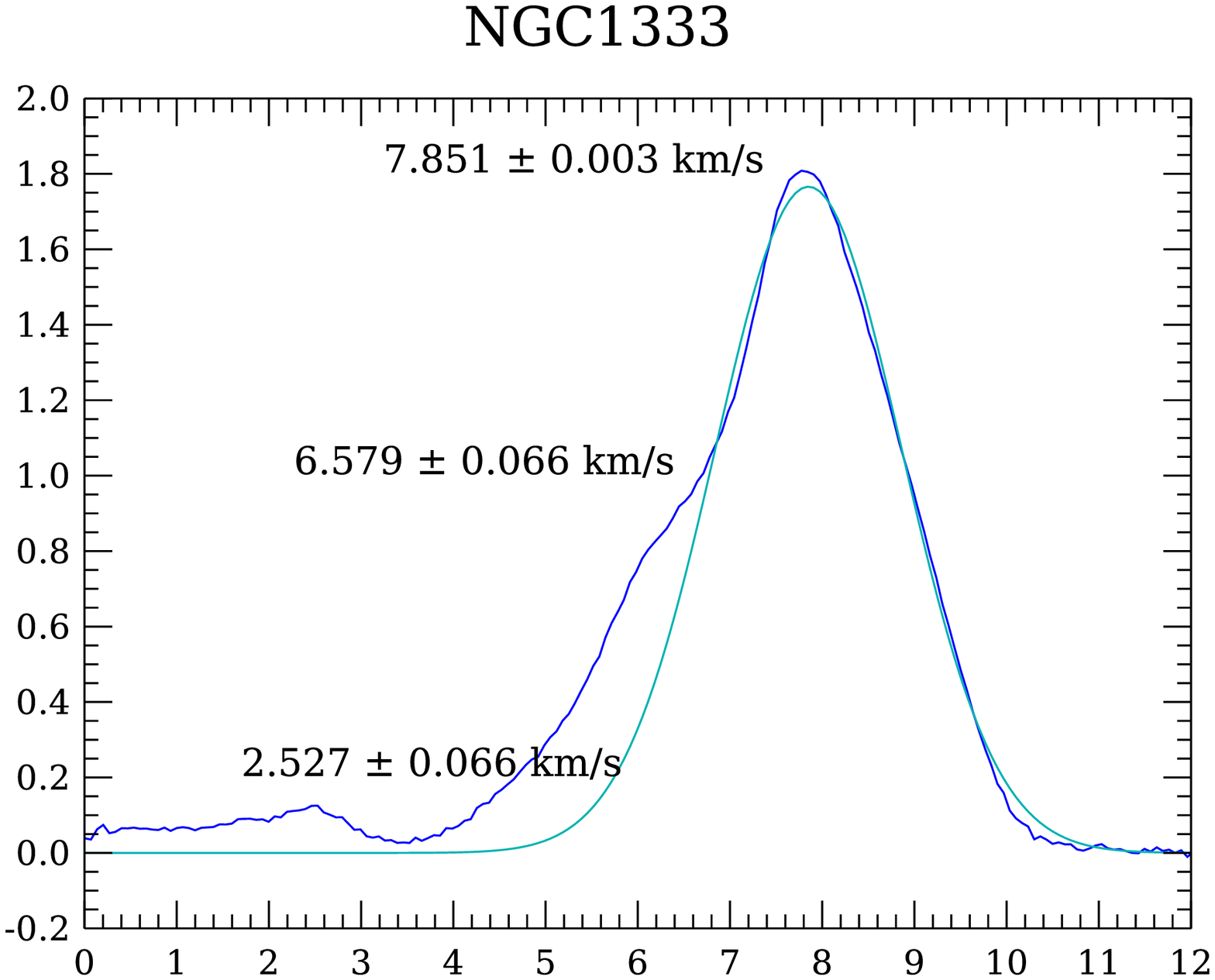}&
\includegraphics[scale=.25]{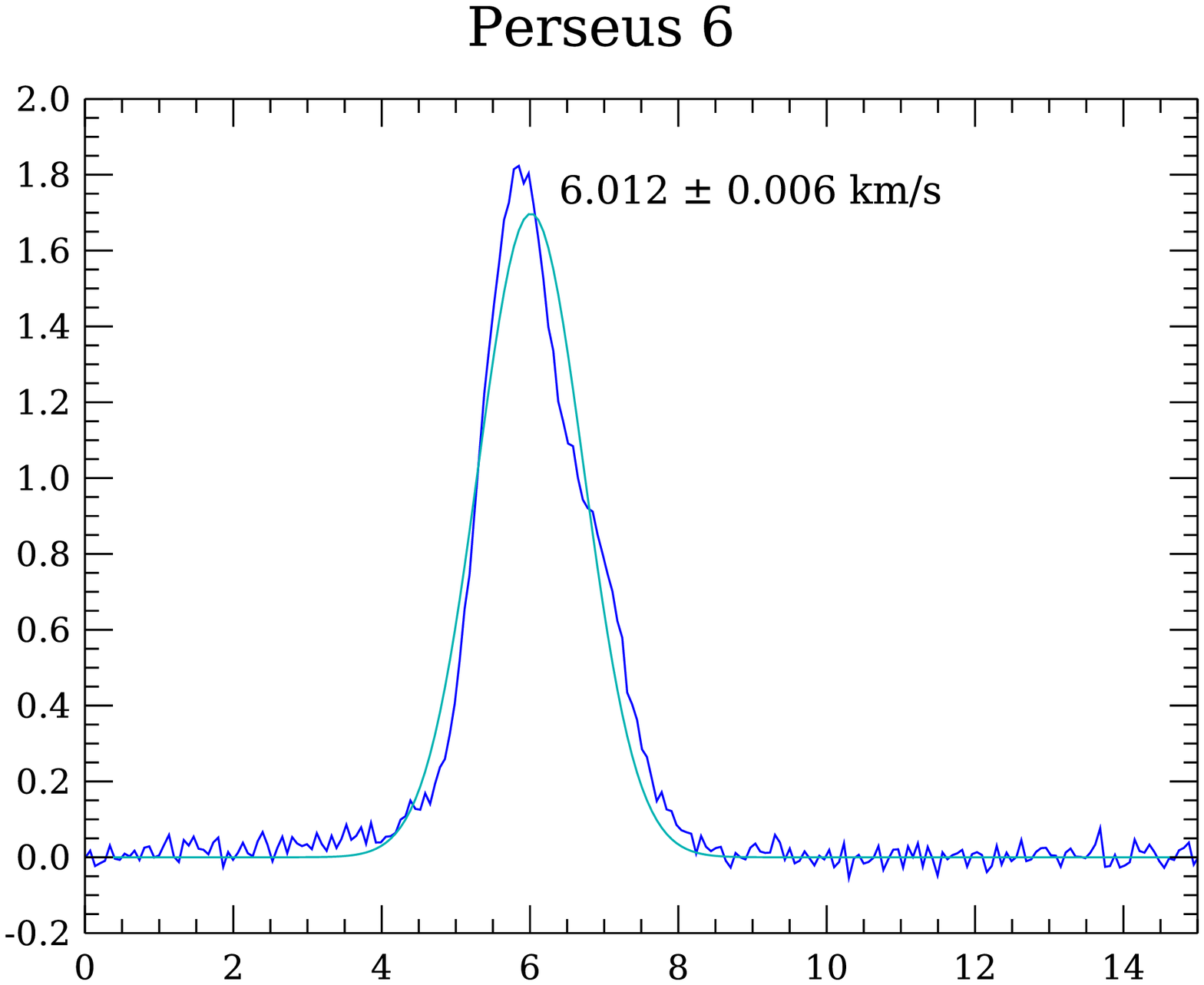}\\
\includegraphics[scale=.25]{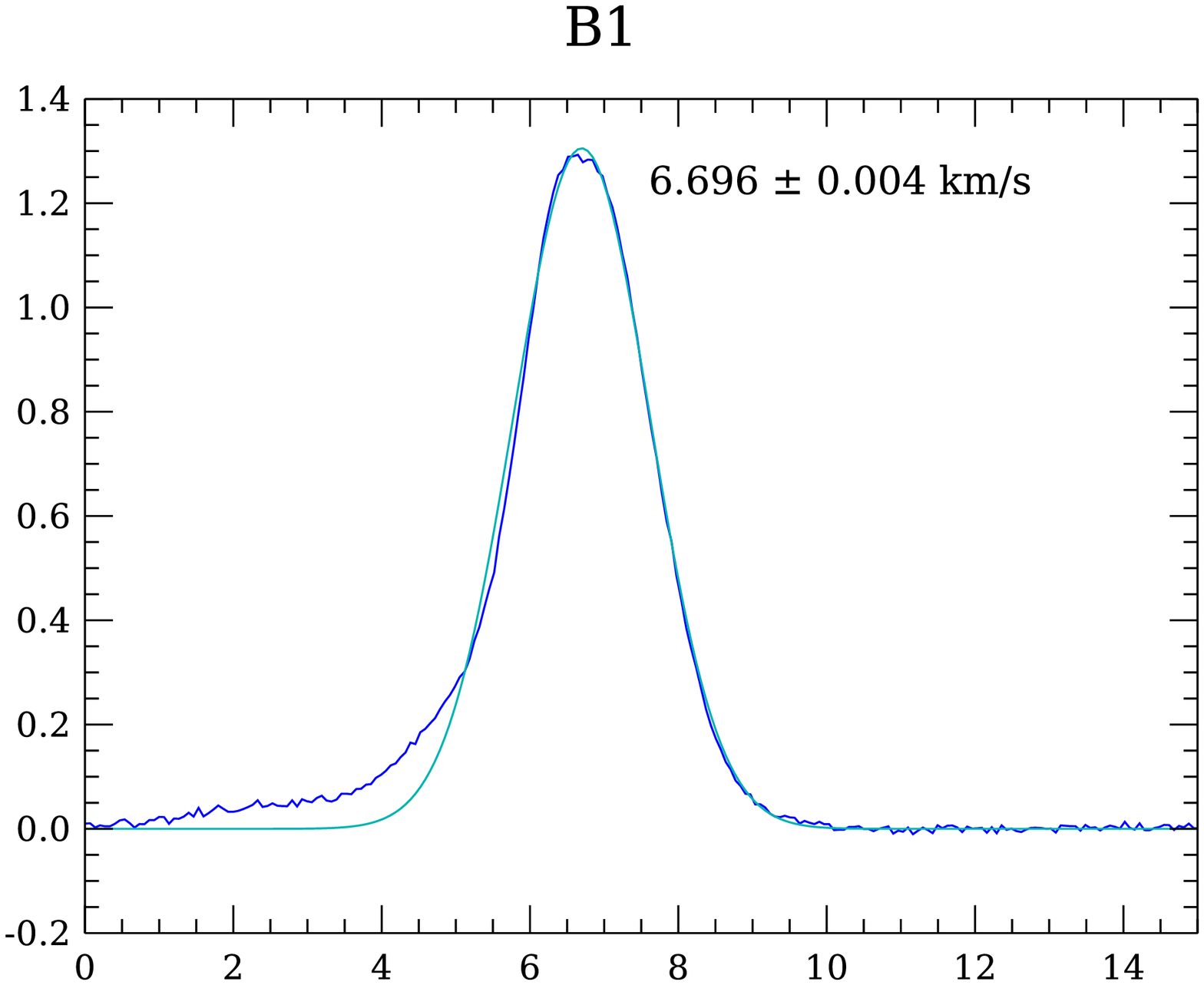}&
\includegraphics[scale=.25]{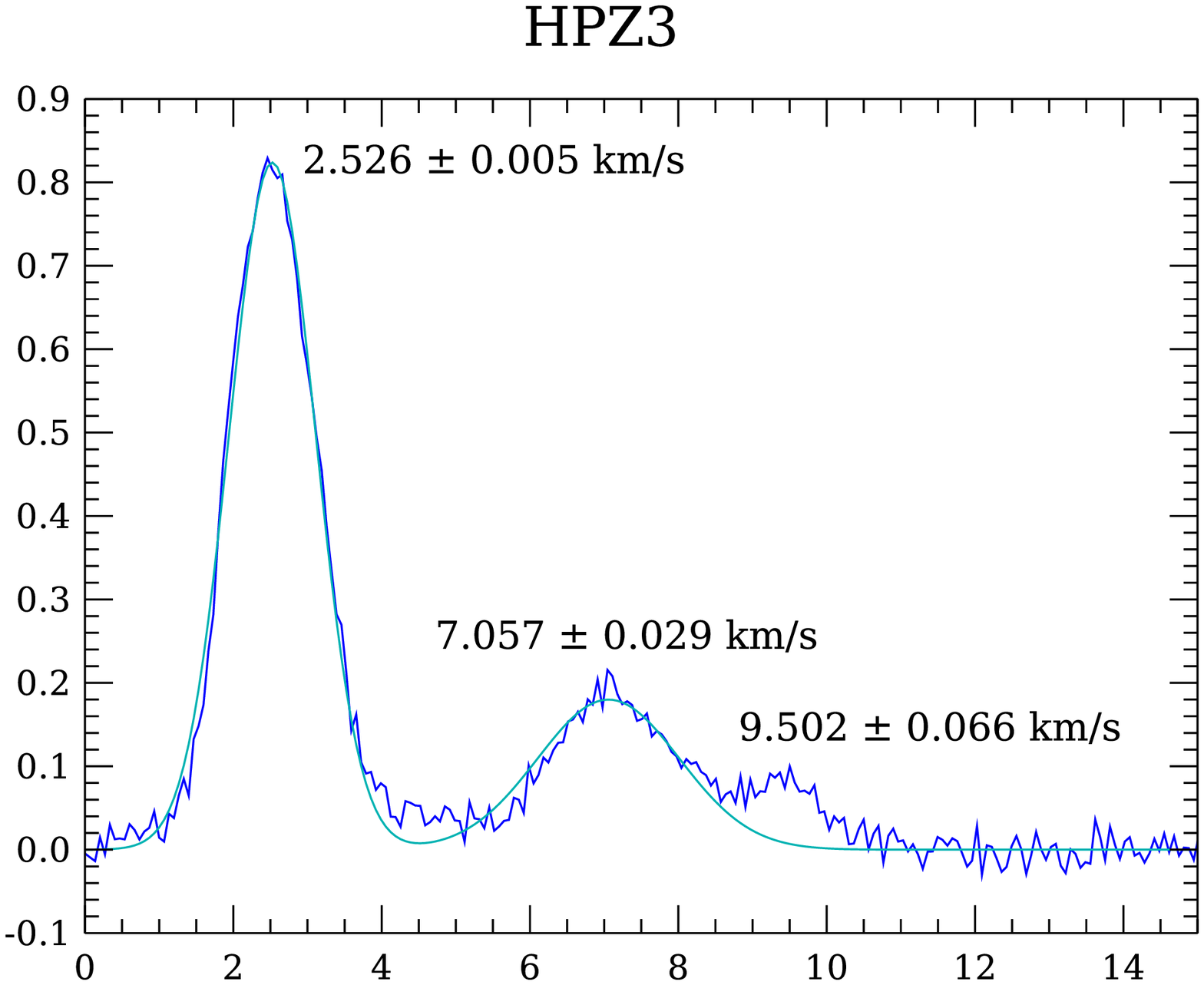}&
\includegraphics[scale=.25]{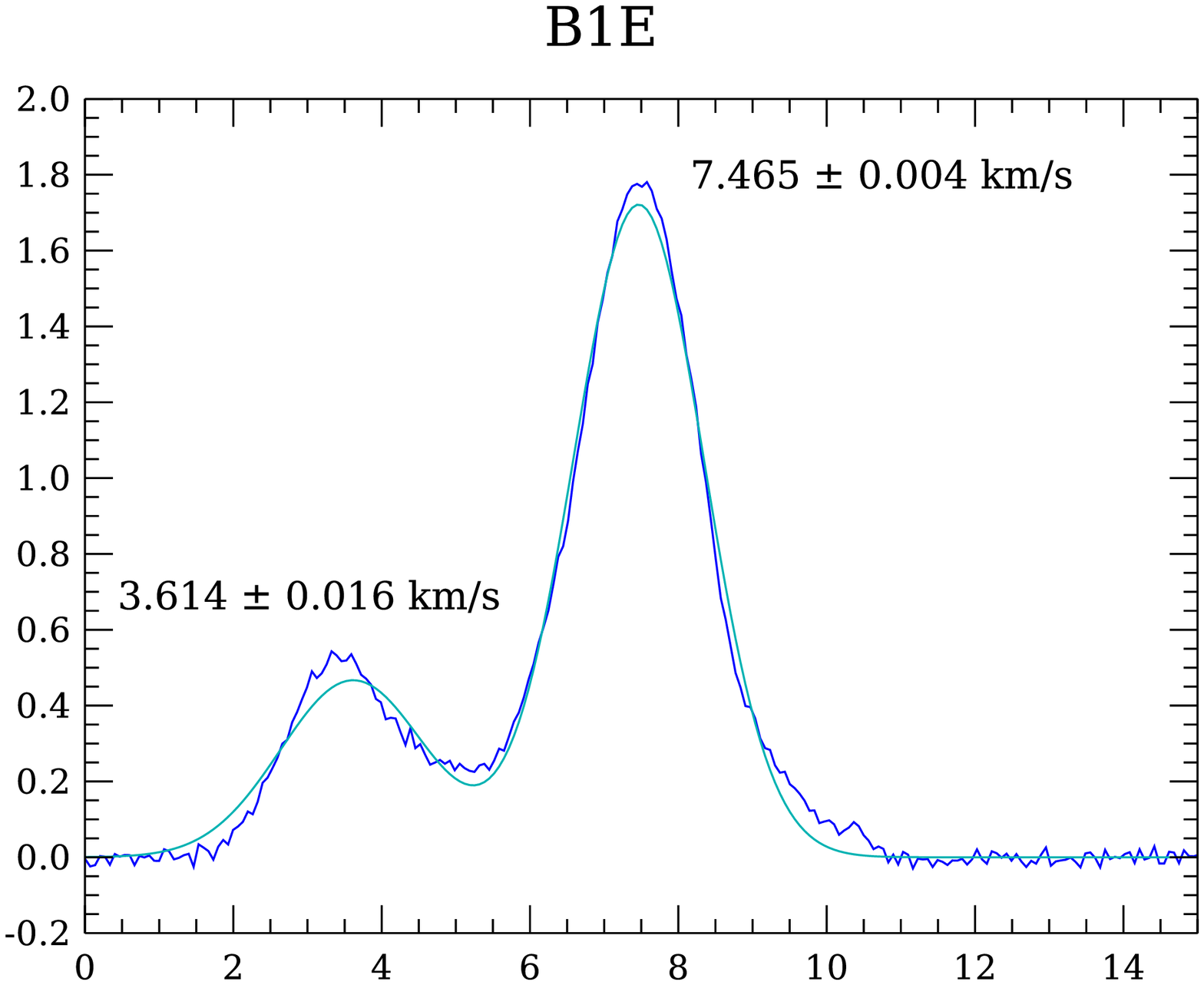}\\
\includegraphics[scale=.25]{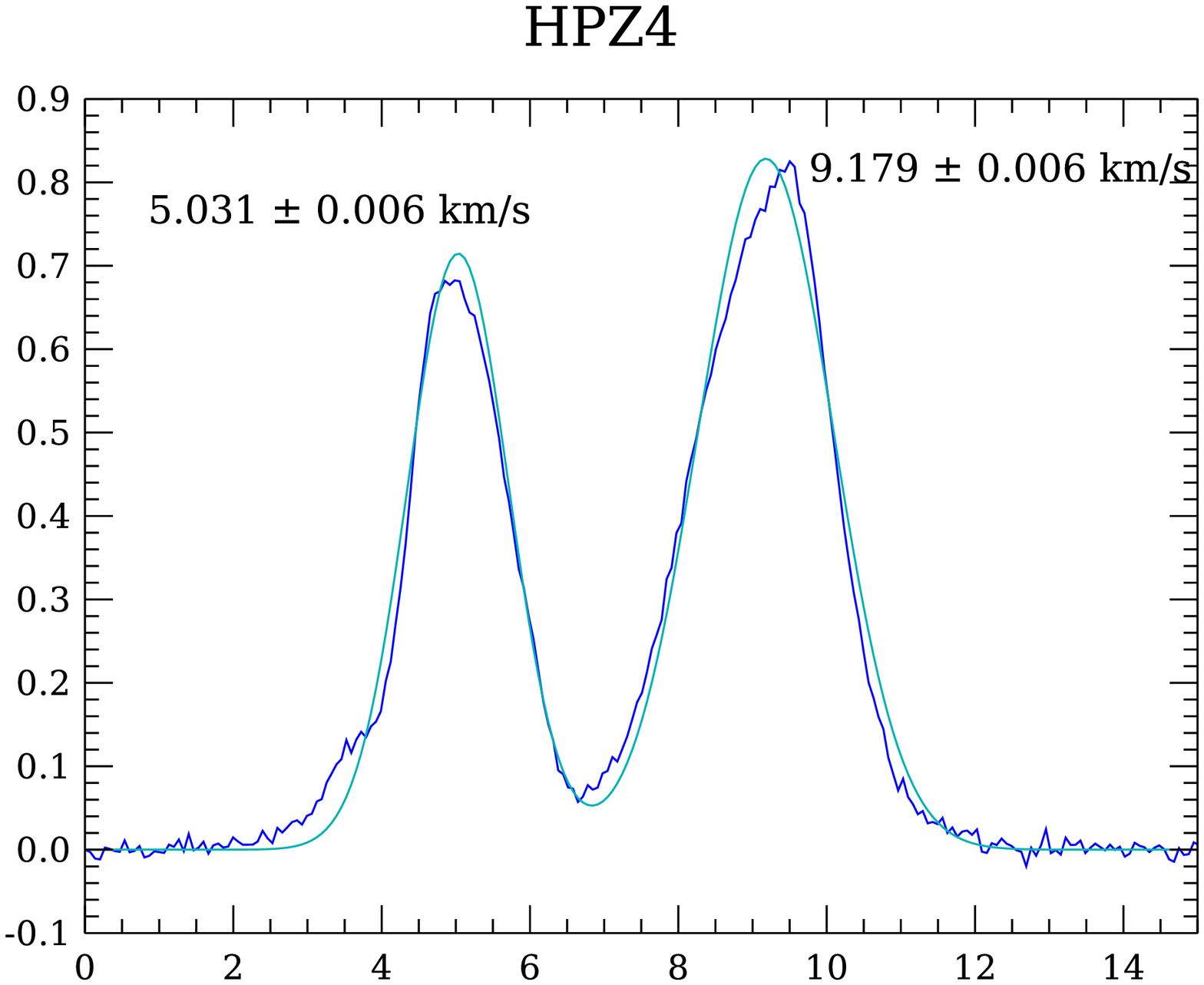}&
\includegraphics[scale=.25]{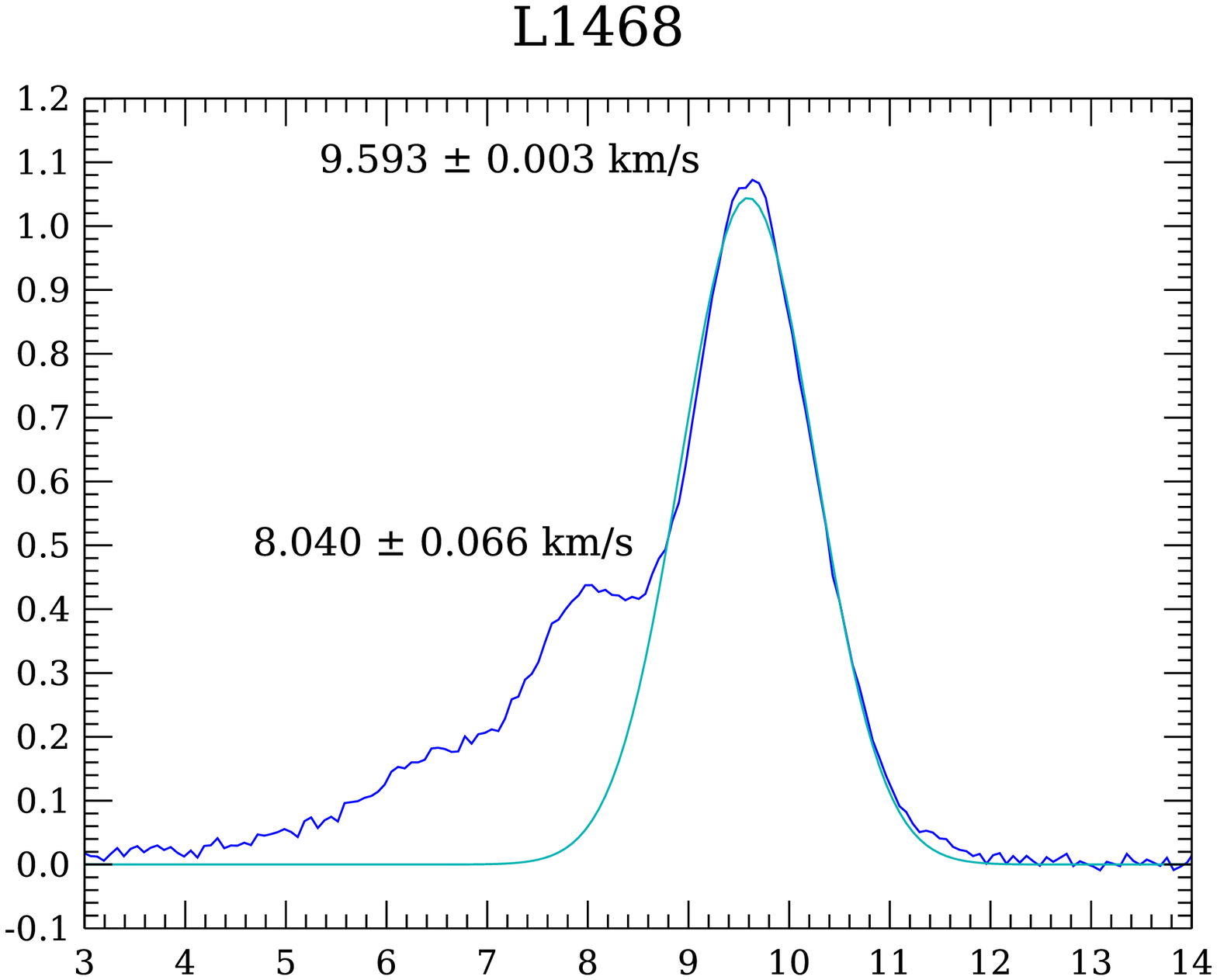}&
\includegraphics[scale=.25]{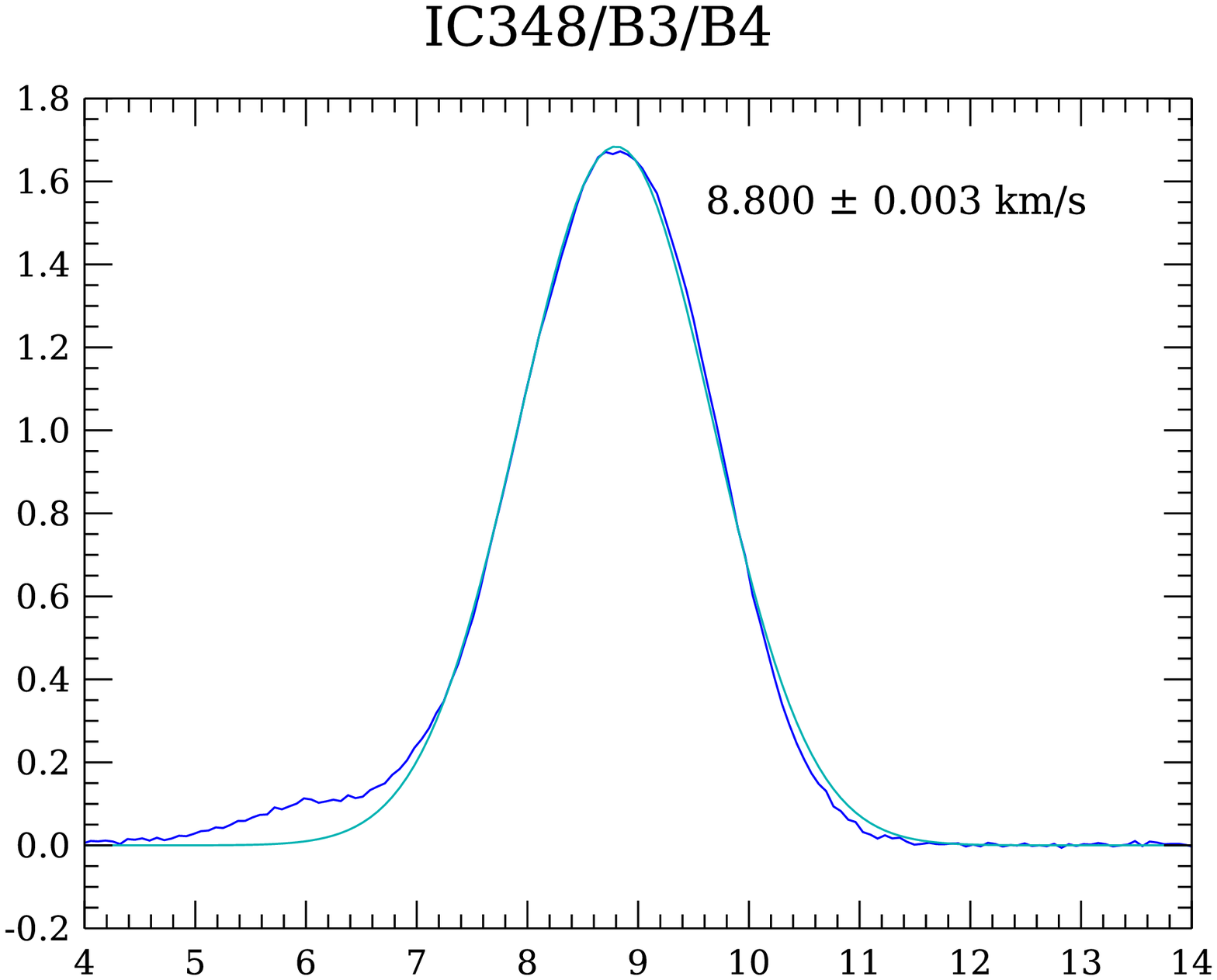}\\
\includegraphics[scale=.25]{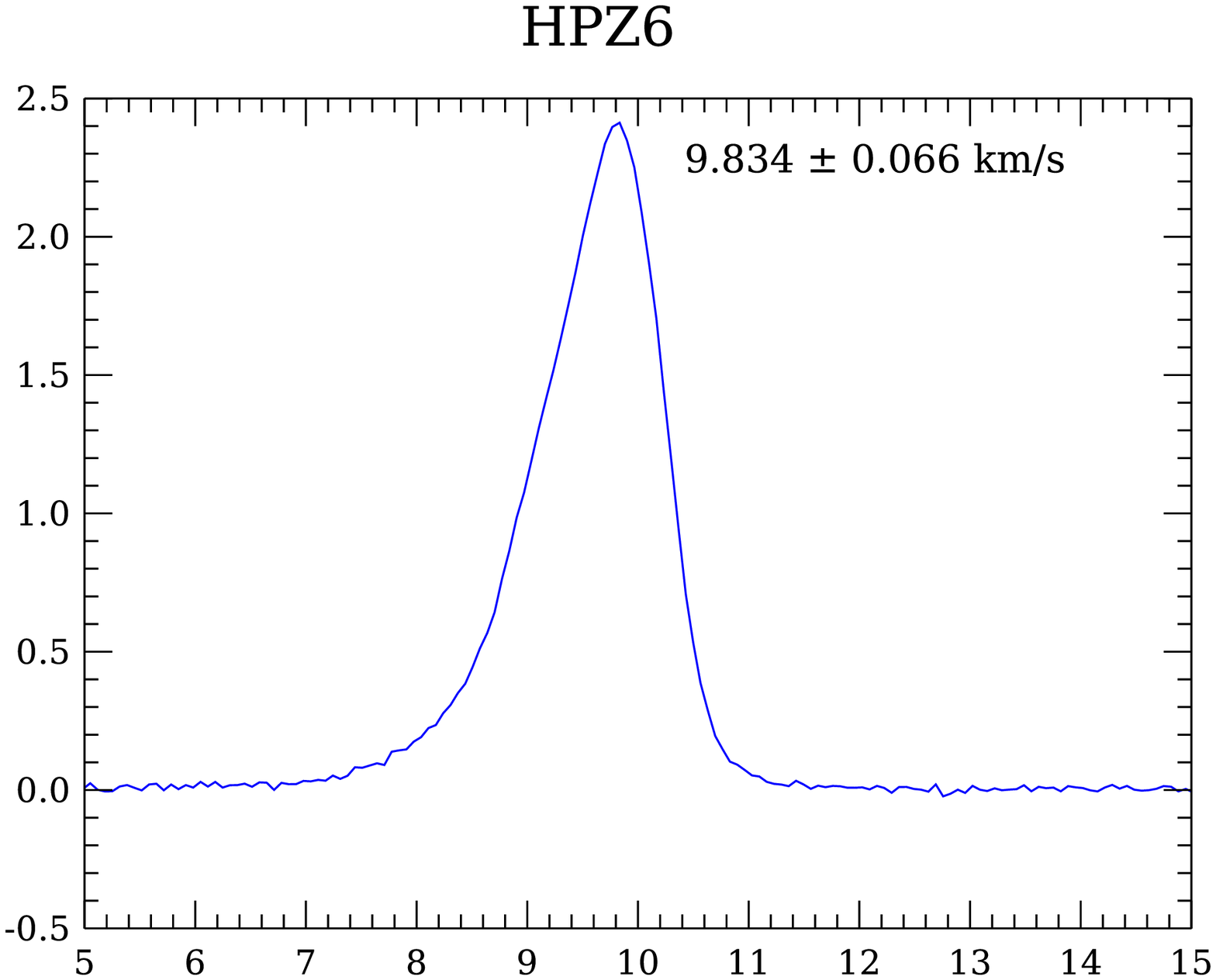}&
\includegraphics[scale=.25]{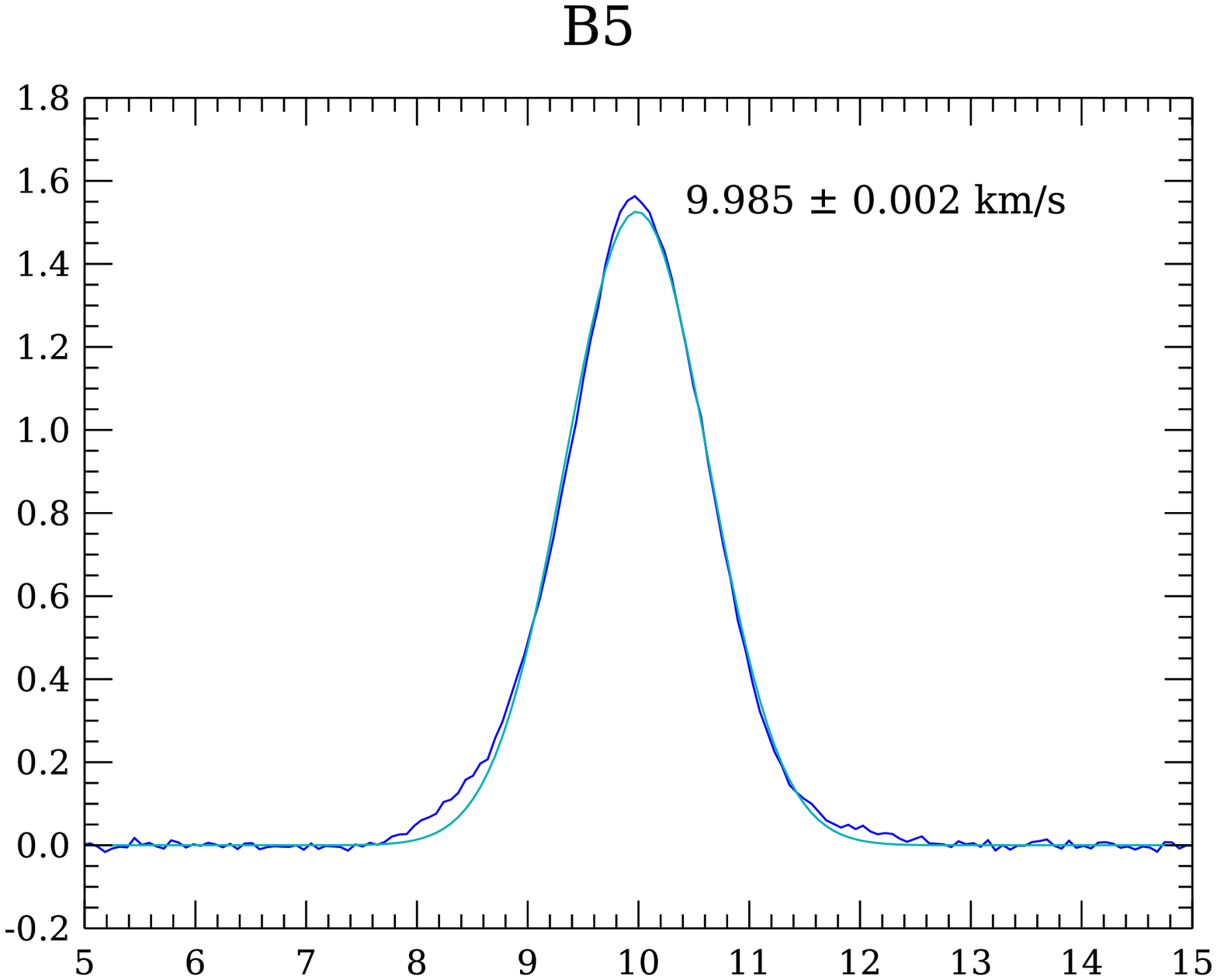}\\
\end{tabular}
\caption{\element[][13]{C}\element[][]{O} spectra for all the subregions we defined in Perseus. Whenever possible we made a Gaussian fit to one or two profiles; for the broad bumps we just derived the position of the maximum and assigned it an uncertainty of 0.066~km\,s$^{-1}$, the step in the velocity grid.\label{righe13CO}}
\end{figure*}

To get some insight in these problems, especially for tracing the borders, we combined our column density map (Sect.~\ref{cMap}) with the \element[][13]{C}\element[][]{O} map of Perseus \citep{10075_2012}. In particular, we started from the $N($H$_2)$ contour at $3\times10^{21}$~cm$^{-2}$ and modified it in order to maximise, wherever possible, one single velocity component in each region. The result of this exercise is visualised in Figs.~\ref{mappe13COa}--\ref{mappe13COb}, where we show how the borders have been defined to accomodate the observed velocity components, and in Fig.~\ref{righe13CO} that shows the \element[][13]{C}\element[][]{O} spectra within these borders. The regions are shown over the column density map in Fig.~\ref{zoneNH2}.

From these figures it is clear that only in a few cases a region is associated to one single \element[][13]{C}\element[][]{O} component. In general, two or even three components are present, often in the form of broad bump. The most complex case is NGC1333 which appears bright from 6 to 10~km\,s$^{-1}$. It is not possible to assess if these different velocity components are associated to different distances or simply reflect the presence of complex motions of gas at the same distance.

Border coordinates are here reported in DS9 format.

{\small
\begin{verbatim}
# Region file format: DS9 version 4.1
global color=green dashlist=8 3 width=1 font=
"helvetica 10 normal roman" select=1 highlite=1 dash=0
fixed=0 edit=1 move=1 delete=1 include=1 source=1
image
# text(1487.4314,333.61838) color=blue
font="helvetica 10 bold roman" text={L1451}
polygon(1333.8089,437.23983,1440.7546,458.68454,
1470.8901,400.07136,1379.8181,321.59092,1318.014,
336.61325) # color=blue width=2
# text(1502.1189,499.27846) color=blue
font="helvetica 10 bold roman" text={L1448}
polygon(1356.7824,546.66324,1434.3763,539.49322,
1439.1677,463.43743,1310.8333,436.43609,1308.4301,
509.10723) # color=blue width=2
# text(1190.5906,242.44105) color=blue
font="helvetica 10 bold roman" text={L1455}
polygon(1157.7149,533.49812,1193.5852,541.55673,
1201.6303,532.283,1215.238,531.67128,1246.7729,
549.00541,1283.1111,422.1613,1327.4841,411.09164,
1304.5877,251.04107,1145.3085,315.92795,1158.7509,
368.22467,1165.0377,472.01562) # color=blue width=2
# text(1126.212,879.76219) color=blue
font="helvetica 10 bold roman" text={NGC1333}
polygon(1071.1585,853.15932,1132.7315,852.35554,
1230.2714,722.39349,1259.8445,725.13296,1277.862,
676.20236,1230.1883,669.20757,1240.7268,553.33499,
1203.1152,535.40696,1193.8331,545.29856,1168.4754,
541.57477) # color=blue width=2
# text(984.80575,409.03949) color=blue
font="helvetica 10 bold roman" text={Per6}
polygon(1036.4774,468.77807,1159.281,466.4625,
1156.1584,374.55727,1035.721,398.26535,1015.8924,
441.03821) # color=blue width=2
# text(925.20483,769.29545) color=blue
font="helvetica 10 bold roman" text={B1}
polygon(1044.0405,774.40647,1086.8576,705.50003,
1124.9583,562.11768,1054.5893,474.75889,1032.5161,
469.56833,980.24576,429.13633,935.86235,461.59737,
908.00317,506.37669,840.6034,618.84583,839.77572,
690.94226) # color=blue width=2
# text(805.68456,485.18367) color=blue
font="helvetica 10 bold roman" text={B1E}
polygon(746.29823,669.50423,771.72844,651.05987,
808.98863,653.10202,840.60823,610.13084,877.69561,
550.77242,804.47781,506.58305,696.42301,591.8366,
709.0804,630.66426,736.00938,648.1077) # color=blue
width=2
# text(489.66258,695.47969) color=blue
font="helvetica 10 bold roman" text={L1468}
polygon(627.59045,765.79554,668.84339,729.81106,
737.21644,796.78156,758.23062,698.28841,684.74148,
712.86762,652.8725,643.16733,617.50019,632.44604,
531.17747,534.78311,482.46366,598.33846,550.86898,
686.697,536.24192,701.54936) # color=blue width=2
# text(422.30564,920.67911) color=blue
font="helvetica 10 bold roman" text={IC348}
polygon(239.96759,887.13841,252.9806,907.15254,
274.84348,925.7651,306.5413,955.20126,347.81727,
930.38764,380.28695,871.58782,467.83656,900.06909,
495.59982,884.78731,515.42993,905.80607,613.92733,
913.78255,636.89789,891.3492,613.76131,834.94829,
610.25222,781.98471,564.80593,745.37937,363.56661,
721.83811,327.29322,675.35843,266.47791,722.34789,
279.74514,781.64591,345.61729,808.68617,250.32306,
816.90053,300.46231,842.71522,317.67854,875.81041,
263.01753,887.93356,251.07107,885.34014) # color=blue
width=2
# text(231.55717,1080.3695) color=blue
font="helvetica 10 bold roman" text={B5}
polygon(131.84123,1149.1204,204.32139,1146.0994,
199.21689,1063.8127,368.68022,1016.0499,227.01718,
963.66495,101.94755,1020.7039,123.39263,1035.5521,
78.991784,1084.8691,115.25351,1089.8388) # color=blue
width=2
# text(1458.7042,766.89968) color=blue
font="helvetica 10 bold roman" text={HPZ1}
polygon(1259.8775,743.24297,1331.4864,780.605,1438.2509,
703.67412,1504.4839,566.14416,1288.6928,658.07238)
# color=blue width=2
# text(944.78788,261.33954) color=blue
font="helvetica 10 bold roman" text={HPZ2}
polygon(979.47848,380.01503,1009.5859,377.65325,
1091.2147,328.57301,1091.2463,264.39894,1030.2487,
250.1076,979.51784,301.58003) # color=blue width=2
polygon(269.65329,875.23531,314.0925,875.68734,
296.44116,841.19846,249.80269,820.81436) # color=blue
width=2
# text(811.58094,726.68431) color=blue
font="helvetica 10 bold roman" text={HPZ3}
polygon(801.04278,684.50226,828.78401,698.12952,
828.70223,652.90765,774.89423,659.87374,777.25972,
684.49736) # color=blue width=2
# text(638.77426,606.92082) color=blue
font="helvetica 10 bold roman" text={HPZ4}
polygon(688.10315,706.15275,740.73908,694.85417,
736.00504,655.23816,697.21552,630.51863,684.53517,
623.01812,653.71632,634.77201) # color=blue width=2
# text(685.39798,440.72586) color=blue
font="helvetica 10 bold roman" text={HPZ5}
polygon(587.24505,484.82277,660.47976,537.87714,
709.20818,522.27317,700.28996,493.47845,604.90108,
444.89766) # color=blue width=2
# text(197.62231,846.05505) color=blue
font="helvetica 10 bold roman" text={HPZ6}
polygon(269.65329,875.23531,314.0925,875.68734,
296.44116,841.19846,249.80269,820.81436) # color=blue
width=2
\end{verbatim}}

\section{\textit{Herschel} intensity maps\label{immagini}}
\begin{figure*}
\centering
\begin{tabular}{c}
\includegraphics[scale=1]{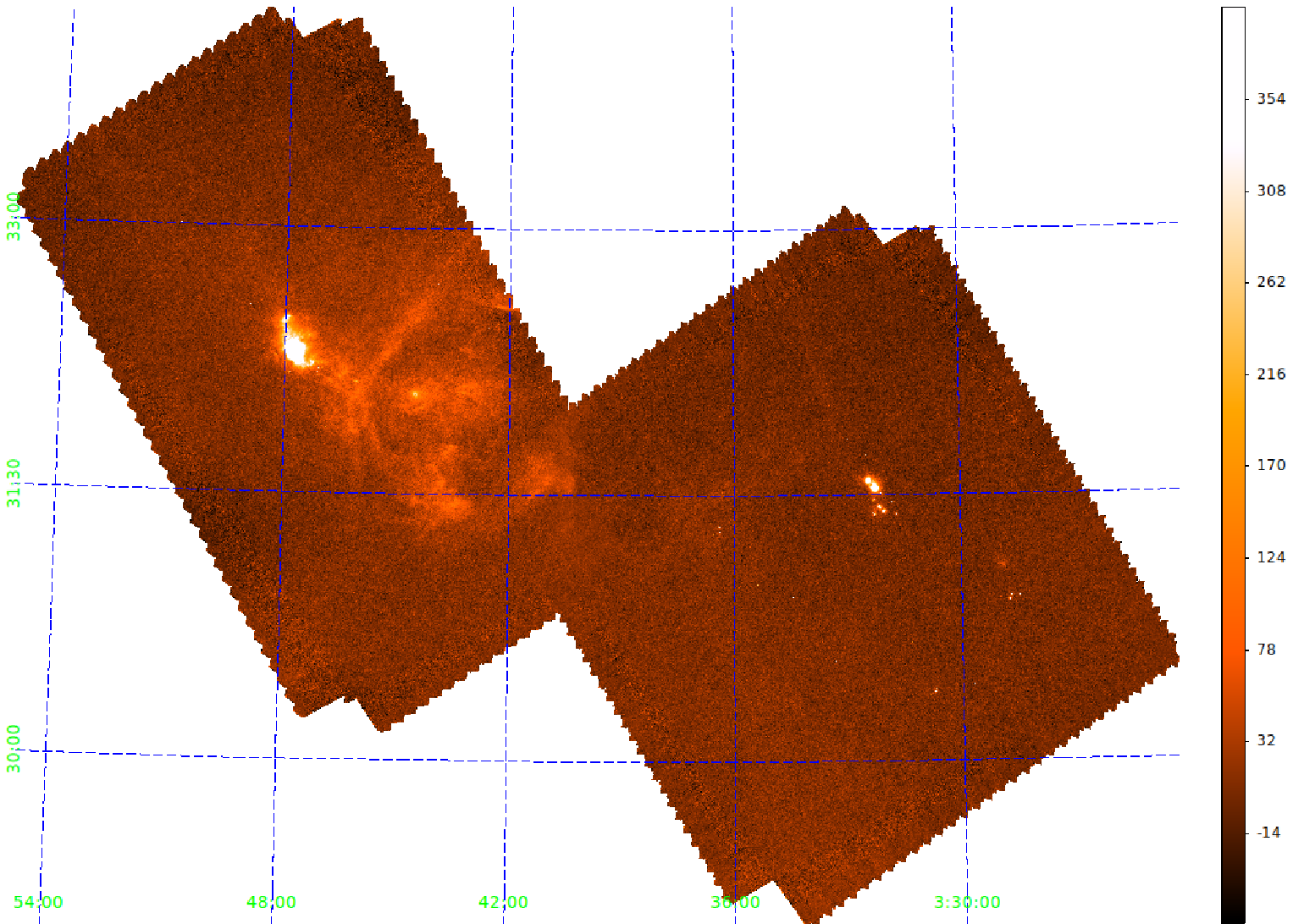}\\%&
\includegraphics[scale=2.5]{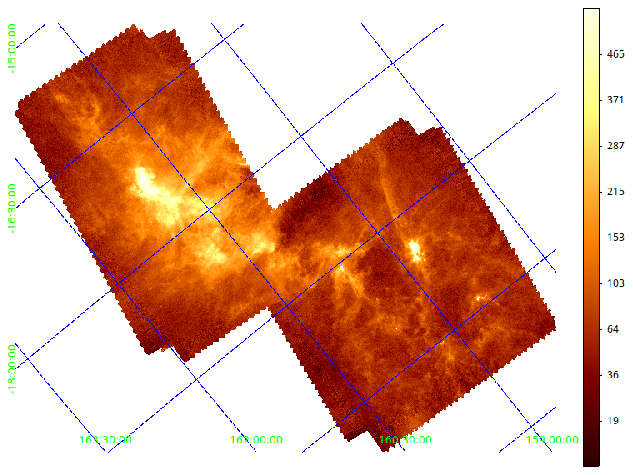}\\
\end{tabular}
\caption{The combined maps of Perseus at 70~$\mu$m (top, grid of equatorial J2000.0 coordinates, HPBW 8\farcs4) and 160~$\mu$m (bottom, grid of Galactic coordinates, HPBW 13\farcs5). Colourbars in MJy/sr.\label{70}}
\end{figure*}

\begin{figure*}
\centering
\begin{tabular}{c}
\end{tabular}
\includegraphics[scale=2.1]{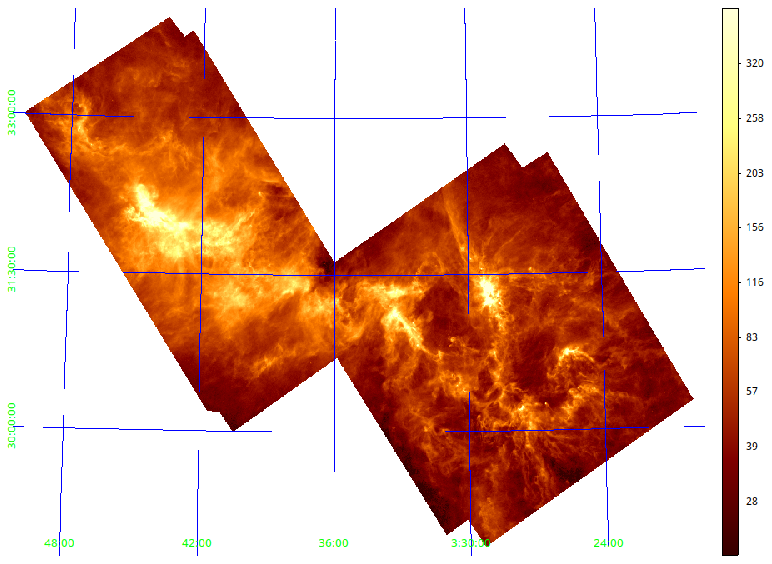}\\%&
\includegraphics[scale=2.15]{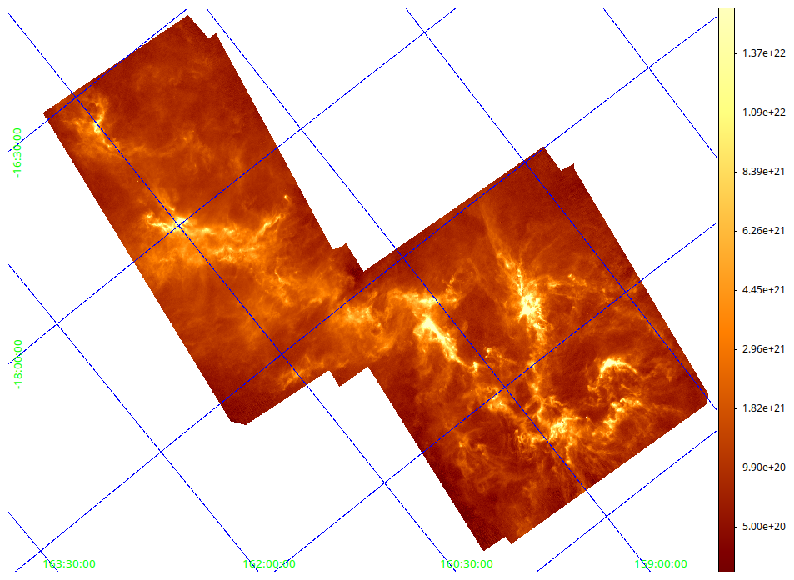}\\%&
\caption{SPIRE intensity map of Perseus at 250~$\mu$m (top, grid of equatorial J2000.0 coordinates) and the high resolution column density map (bottom, grid of Galactic coordinates). For both maps the HPBW is 18\farcs2. Colorbars in MJy/sr (top) and cm$^{-2}$ (bottom).\label{SPIRE1}}
\end{figure*}

\begin{figure*}
\centering
\begin{tabular}{c}
\includegraphics[scale=2.1]{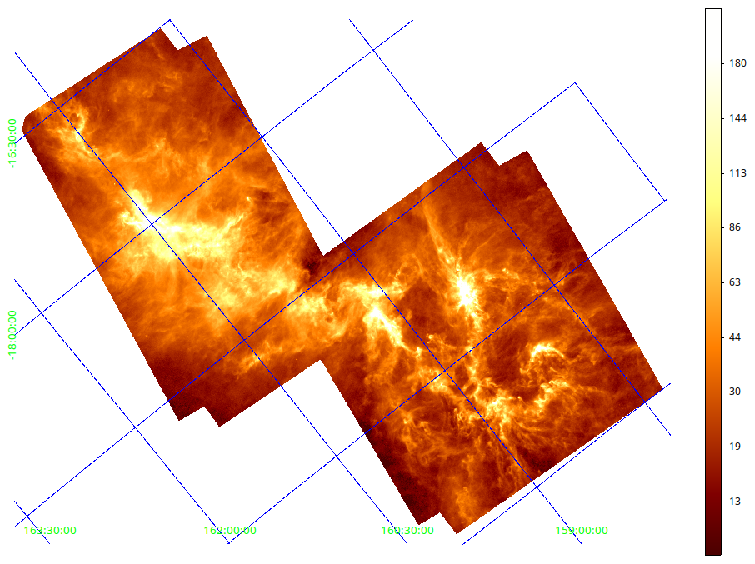}\\
\includegraphics[scale=2.1]{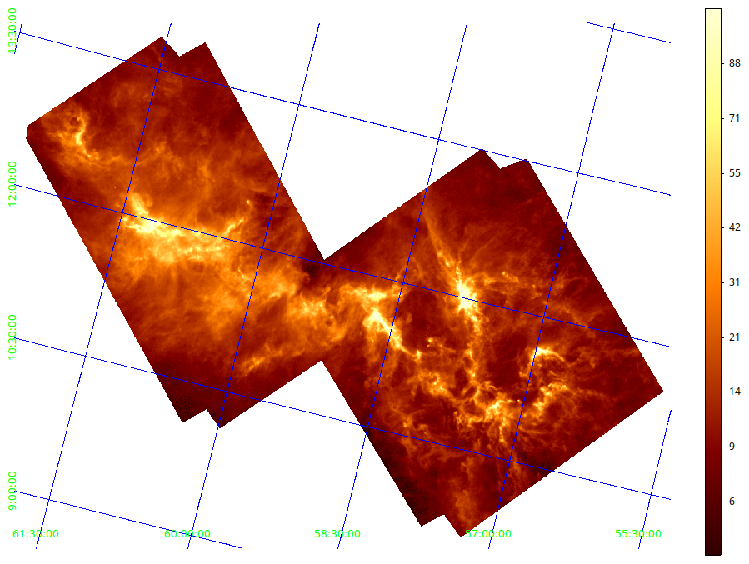}%\\
\end{tabular}
\caption{SPIRE intensity maps of Perseus at 350~$\mu$m (top, grid of Galactic coordinates, HPBW 24\farcs9) and 500~$\mu$m (bottom, grid of ecliptic coordinates, HPBW 36\farcs1). Colourbars in MJy/sr.\label{SPIRE2}}
\end{figure*}

In this appendix we show the PACS intensity maps at 70~$\mu$m and 160~$\mu$m in Figure~\ref{70}; SPIRE intensity map at 250~$\mu$m and the high resolution column density map with a resolution of 18\farcs2 in Figure~\ref{SPIRE1} (the column density map shown in Figure~\ref{NHT} is at 36\farcs1); and SPIRE intensity maps at 350~$\mu$m and 500~$\mu$m in Figure~\ref{SPIRE2}. To help to localise the region of Perseus in other surveys, we give a grid of coordinates in different systems: FK5 (equatorial) for 70~$\mu$m and 250~$\mu$m, galactic for 160~$\mu$m, 350~$\mu$m and for the high resolution column density map, ecliptic for 500~$\mu$m.

Maps centres, as given in FITS header and converted to the specific coordinate system are: (3$^\mathrm{h}$36$^\mathrm{m}$43$^\mathrm{s}$,+31$^{\degr}$35\arcmin27\arcsec), or (54\fdg1775,+31\fdg5908) for blue band; (159.5082,--19.3010) for red band; (3$^\mathrm{h}$35$^\mathrm{m}$06$^\mathrm{s}$,+31$^{\degr}$31\arcmin20\arcsec), or (53\fdg7732,+31\fdg5221) for PSW band; (159.3639,--19.4821) for the column density map; (159.4321,--19.5946) for PMW band; (59\fdg3449,+11\fdg7751) for PLW band.

From Figures~\ref{70}--\ref{SPIRE2} it is evident that the diffuse emission intensity increases at longer wavelengths, being marginally visible at $\lambda=70$~$\mu$m.

The zero-level intensities for the diffuse emission were found through comparison with the predicted Planck+IRAS emission \citep{JPB} for the same region. For each map and wavelength, the zero-level intensity was found separately for West and East Perseus and the offset added to the map. Finally, the two maps were combined with a simple pixel-by-pixel mean (the two maps were created onto the same spatial grid). The offsets are reported in Table~\ref{offsets}.

\begin{table}
\centering
\caption{The zero-levels of the diffuse emission for each map.\label{offsets}}
\begin{tabular}{rlr}
\hline
\multicolumn{1}{c}{$\lambda$}&\multicolumn{1}{c}{East}&\multicolumn{1}{c}{West}\\
\multicolumn{1}{c}{($\mu$m)}&\multicolumn{1}{c}{(MJy\,sr$^{-1}$)}&\multicolumn{1}{c}{(MJy\,sr$^{-1}$)}\\\hline
70&\phantom{0}3.320&$-$3.492\\
160&54.39&68.77\phantom{0}\\
250&66.32&45.32\phantom{0}\\
350&31.69&23.06\phantom{0}\\
500&13.35&10.14\phantom{0}\\\hline
\end{tabular}
\end{table}

\section{Additional catalogue of sources}\label{addCat}
Here we report basic data for sources eliminated from the main catalogue; since these data are not part of the official HGBS delivered products they are reported here. The Cols. in the following tables are a subset of the Cols. in Table~\ref{esempioCat}, of which they keep the numbering: so, for instance, meaning of Col.~(12) in Table~\ref{ancCat} is the same of Col.~(12) in Table~\ref{esempioCat}. The only difference is units in Col.~(55): $10^{21}$~cm$^{-2}$ in Table~\ref{esempioCat} and $10^{20}$~cm$^{-2}$ here.

The other difference is Col.~(2) in this table: here an identifier is given that explains why source was removed (acronyms as in Simbad database): G -- galaxy; W -- counterpart in WISE catalogue; 2 -- counterpart in 2MASS; HH -- Herbig-Haro object; M -- star from catalogue MBO; HL -- candidate YSO from [HL2013]; DS -- source in [DS95]; LRL -- source in ``Cl* IC 348 LRL''; PSZ -- source in [PSZ2003]; NTC -- source in ``Cl* IC 348 NTC''; U -- W UMa star; $\delta$ -- $\delta$ Scuti star.

In particular, W and 2 means that an infrared sources was found in WISE or 2MASS catalogues within 6\arcsec\ from an \textit{Herschel} source not detected at 70~$\mu$m nor in the millimetre band. No attempt was made to distinguish between physical and projected associations.

\begin{landscape}
\begin{table}
\tiny
\centering
\caption{Catalogue of additional sources excluded from the main catalogue: Cols. have the same meaning as in Table~\ref{esempioCat}.\label{ancCat}}
\begin{tabular}{rccccrcrcrrrcrcrcrrr}
\hline\hline
No.&Type&RA$_\mathrm{2000}$&Dec$_\mathrm{2000}$&Sig$_{070}$&\multicolumn{2}{c}{$S^\mathrm{peak}_{070}$}&\multicolumn{2}{c}{$S^\mathrm{tot}_{070}$}&\multicolumn{1}{c}{r$^\mathrm{a}_{070}$}&\multicolumn{1}{c}{r$^\mathrm{b}_{070}$}&\multicolumn{1}{c}{PA$_{070}$}&Sig$_{160}$&\multicolumn{2}{c}{$S^\mathrm{peak}_{160}$}&\multicolumn{2}{c}{$S^\mathrm{tot}_{160}$}&\multicolumn{1}{c}{r$^\mathrm{a}_{160}$}&\multicolumn{1}{c}{r$^\mathrm{b}_{160}$}&\multicolumn{1}{c}{PA$_{160}$}\\
&&(degree)&(degree)&&\multicolumn{2}{c}{(Jy/beam)}&\multicolumn{2}{c}{(Jy)}&\multicolumn{1}{c}{(\arcsec)}&\multicolumn{1}{c}{(\arcsec)}&\multicolumn{1}{c}{($^\circ$)}&&\multicolumn{2}{c}{(Jy/beam)}&\multicolumn{2}{c}{(Jy)}&\multicolumn{1}{c}{(\arcsec)}&\multicolumn{1}{c}{(\arcsec)}&\multicolumn{1}{c}{($^\circ$)}\\
(1)&(2)&(3)&(4)&(5)&\multicolumn{2}{c}{(6) $\pm$ (7)}&\multicolumn{2}{c}{(10) $\pm$ (11)}&\multicolumn{1}{c}{(12)}&\multicolumn{1}{c}{(13)}&\multicolumn{1}{c}{(14)}&(15)&\multicolumn{2}{c}{(16) $\pm$ (17)}&\multicolumn{2}{c}{(20) $\pm$ (21)}&\multicolumn{1}{c}{(22)}&\multicolumn{1}{c}{(23)}&\multicolumn{1}{c}{(24)}\\\hline
 1&W&51.03003&30.27615&0.000&  2.02e--2&1.3e--2&--1.58e--2&5.6e--2& 60.8& 47.2&133.5&0.000&  1.81e--1&4.1e--2&  1.30e+0 &1.1e--1& 47.9& 37.0&149.7\\
 2&W&51.18427&30.27827&0.000&  2.75e--2&1.5e--2&  4.30e--1&6.6e--2& 54.0& 47.5& 69.3&0.000&  6.13e--2&3.2e--2&  8.90e--1&8.6e--2& 57.0& 51.0& 52.4\\
 3&W&51.30787&30.76503&0.000&  1.53e--2&1.3e--2&--1.42e--1&4.7e--2& 46.1& 40.1& 58.2&6.779&  9.46e--2&8.3e--2&  5.66e--1&1.2e--1& 51.9& 18.6&108.4\\
 4&HH&51.30889&30.78195&0.000&  1.81e--2&1.4e--2&--7.24e--2&3.4e--2& 35.3& 23.6& 67.6&0.000&  5.55e--2&4.4e--2&  2.87e--1&7.4e--2& 32.3& 28.1& 57.1\\
 5&W&51.41323&31.32864&0.000&  6.48e--3&8.1e--3&--5.62e--2&6.0e--2& 91.8& 84.2& 80.8&7.148&  2.09e--1&4.7e--2&  3.29e+0 &2.0e--1& 65.8& 45.4& 50.7\\
\hline
\end{tabular}
\end{table}
\end{landscape}

\setcounter{table}{0}
\begin{landscape}
\begin{table}
\tiny
\centering
\caption{\textbf{continued}}
\begin{tabular}{rccccrcrcrrrcrcrcrrr}
\hline\hline
No.&Type&RA$_\mathrm{2000}$&Dec$_\mathrm{2000}$&Sig$_{070}$&\multicolumn{2}{c}{$S^\mathrm{peak}_{070}$}&\multicolumn{2}{c}{$S^\mathrm{tot}_{070}$}&\multicolumn{1}{c}{r$^\mathrm{a}_{070}$}&\multicolumn{1}{c}{r$^\mathrm{b}_{070}$}&\multicolumn{1}{c}{PA$_{070}$}&Sig$_{160}$&\multicolumn{2}{c}{$S^\mathrm{peak}_{160}$}&\multicolumn{2}{c}{$S^\mathrm{tot}_{160}$}&\multicolumn{1}{c}{r$^\mathrm{a}_{160}$}&\multicolumn{1}{c}{r$^\mathrm{b}_{160}$}&\multicolumn{1}{c}{PA$_{160}$}\\
&&(degree)&(degree)&&\multicolumn{2}{c}{(Jy/beam)}&\multicolumn{2}{c}{(Jy)}&\multicolumn{1}{c}{(\arcsec)}&\multicolumn{1}{c}{(\arcsec)}&\multicolumn{1}{c}{($^\circ$)}&&\multicolumn{2}{c}{(Jy/beam)}&\multicolumn{2}{c}{(Jy)}&\multicolumn{1}{c}{(\arcsec)}&\multicolumn{1}{c}{(\arcsec)}&\multicolumn{1}{c}{($^\circ$)}\\
(1)&(2)&(3)&(4)&(5)&\multicolumn{2}{c}{(6) $\pm$ (7)}&\multicolumn{2}{c}{(10) $\pm$ (11)}&\multicolumn{1}{c}{(12)}&\multicolumn{1}{c}{(13)}&\multicolumn{1}{c}{(14)}&(15)&\multicolumn{2}{c}{(16) $\pm$ (17)}&\multicolumn{2}{c}{(20) $\pm$ (21)}&\multicolumn{1}{c}{(22)}&\multicolumn{1}{c}{(23)}&\multicolumn{1}{c}{(24)}\\\hline
 54&W&52.29768&31.31080&195.7&  7.57e--3&1.1e--1&  4.33e--1&1.5e--1&19.2& 8.4&  6.9&378.4&  9.66e+0 &2.0e--1&1.79e+1 &2.0e--1& 19.8&14.0& 88.2\\
 55&W&52.29803&31.30691&674.2&  5.55e+0 &4.5e--2&  3.18e+1 &9.2e--2&29.0&12.2& 62.8&842.1&  2.14e+1 &1.9e--1&6.53e+1 &4.9e--1& 27.0&16.7& 81.0\\
 56&W&52.29877&31.43993&0.000&  2.23e--2&2.4e--2&  4.67e--1&6.4e--2&37.1&29.3&135.8&15.21&  4.79e--1&6.6e--2&1.12e+0 &9.7e--2& 23.9&15.3&174.2\\
 57&W&52.30389&31.61624&3.026&--4.58e--2&1.5e--2&--6.65e--3&4.0e--2&44.2&19.3& 41.0&0.000&  7.78e--2&3.3e--2&3.21e--1&8.4e--2& 48.7&43.1&  2.7\\
 58&2&52.30625&31.39899&23.45&  7.17e--2&9.2e--2&  2.64e--1&1.8e--1&42.6& 9.7& 98.8&96.97&  2.24e+0 &6.6e--1&8.01e+0 &1.0e+0 & 40.6&16.5& 78.2\\
\hline
\end{tabular}
\end{table}
\end{landscape}

\setcounter{table}{0}
\begin{landscape}
\begin{table}
\tiny
\centering
\caption{\textbf{continued}}
\begin{tabular}{cccccrcrcrrrcrcrcrrr}
\hline\hline
No.&Type&RA$_\mathrm{2000}$&Dec$_\mathrm{2000}$&Sig$_{070}$&\multicolumn{2}{c}{$S^\mathrm{peak}_{070}$}&\multicolumn{2}{c}{$S^\mathrm{tot}_{070}$}&\multicolumn{1}{c}{r$^\mathrm{a}_{070}$}&\multicolumn{1}{c}{r$^\mathrm{b}_{070}$}&\multicolumn{1}{c}{PA$_{070}$}&Sig$_{160}$&\multicolumn{2}{c}{$S^\mathrm{peak}_{160}$}&\multicolumn{2}{c}{$S^\mathrm{tot}_{160}$}&\multicolumn{1}{c}{r$^\mathrm{a}_{160}$}&\multicolumn{1}{c}{r$^\mathrm{b}_{160}$}&\multicolumn{1}{c}{PA$_{160}$}\\
&&(degree)&(degree)&&\multicolumn{2}{c}{(Jy/beam)}&\multicolumn{2}{c}{(Jy)}&\multicolumn{1}{c}{(\arcsec)}&\multicolumn{1}{c}{(\arcsec)}&\multicolumn{1}{c}{($^\circ$)}&&\multicolumn{2}{c}{(Jy/beam)}&\multicolumn{2}{c}{(Jy)}&\multicolumn{1}{c}{(\arcsec)}&\multicolumn{1}{c}{(\arcsec)}&\multicolumn{1}{c}{($^\circ$)}\\
(1)&(2)&(3)&(4)&(5)&\multicolumn{2}{c}{(6) $\pm$ (7)}&\multicolumn{2}{c}{(10) $\pm$ (11)}&\multicolumn{1}{c}{(12)}&\multicolumn{1}{c}{(13)}&\multicolumn{1}{c}{(14)}&(15)&\multicolumn{2}{c}{(16) $\pm$ (17)}&\multicolumn{2}{c}{(20) $\pm$ (21)}&\multicolumn{1}{c}{(22)}&\multicolumn{1}{c}{(23)}&\multicolumn{1}{c}{(24)}\\\hline
107&W&53.36533&31.37536&0.000&  5.93e--2&1.8e--2&  6.52e--1&8.1e--2& 50.9&47.9& 44.6&7.039&1.73e--1&4.1e--2&7.67e--1&1.0e--1& 38.9& 31.3& 75.2\\ 
108&2&53.42230&31.22838&9.578&  1.37e--1&1.9e--2&  1.94e--1&3.2e--2& 29.6& 8.4& 68.1&6.864&2.02e--1&3.5e--2&2.36e--1&5.5e--2& 21.6& 13.5& 76.7\\ 
109&W&53.44435&30.46080&0.000&  2.54e--3&7.6e--3&  1.71e-+0&5.4e--2&100.7&72.4&163.3&7.073&1.57e--1&6.5e--2&3.09e+0 &2.6e--1& 83.9& 48.7&126.0\\ 
110&2&53.46041&31.43651&0.000&  1.30e--2&1.0e--2&  3.76e--1&4.4e--2& 58.3&48.2& 35.8&0.000&1.01e--1&3.6e--2&1.32e+0 &9.9e--2& 55.2& 50.0&117.7\\ 
111&W&53.46134&31.20830&6.387&--5.84e--2&9.1e--3&  6.48e--3&2.8e--2& 70.4&21.0&158.1&0.000&1.42e--1&3.3e--2&1.90e+0 &1.0e--1& 62.1& 46.1& 63.9\\
\hline
\end{tabular}
\end{table}
\end{landscape}

\setcounter{table}{0}
\begin{landscape}
\begin{table}
\tiny
\centering
\caption{\textbf{continued}}
\begin{tabular}{cccccrcrcrrrcrcrcrrr}
\hline\hline
No.&Type&RA$_\mathrm{2000}$&Dec$_\mathrm{2000}$&Sig$_{070}$&\multicolumn{2}{c}{$S^\mathrm{peak}_{070}$}&\multicolumn{2}{c}{$S^\mathrm{tot}_{070}$}&\multicolumn{1}{c}{r$^\mathrm{a}_{070}$}&\multicolumn{1}{c}{r$^\mathrm{b}_{070}$}&\multicolumn{1}{c}{PA$_{070}$}&Sig$_{160}$&\multicolumn{2}{c}{$S^\mathrm{peak}_{160}$}&\multicolumn{2}{c}{$S^\mathrm{tot}_{160}$}&\multicolumn{1}{c}{r$^\mathrm{a}_{160}$}&\multicolumn{1}{c}{r$^\mathrm{b}_{160}$}&\multicolumn{1}{c}{PA$_{160}$}\\
&&(degree)&(degree)&&\multicolumn{2}{c}{(Jy/beam)}&\multicolumn{2}{c}{(Jy)}&\multicolumn{1}{c}{(\arcsec)}&\multicolumn{1}{c}{(\arcsec)}&\multicolumn{1}{c}{($^\circ$)}&&\multicolumn{2}{c}{(Jy/beam)}&\multicolumn{2}{c}{(Jy)}&\multicolumn{1}{c}{(\arcsec)}&\multicolumn{1}{c}{(\arcsec)}&\multicolumn{1}{c}{($^\circ$)}\\
(1)&(2)&(3)&(4)&(5)&\multicolumn{2}{c}{(6) $\pm$ (7)}&\multicolumn{2}{c}{(10) $\pm$ (11)}&\multicolumn{1}{c}{(12)}&\multicolumn{1}{c}{(13)}&\multicolumn{1}{c}{(14)}&(15)&\multicolumn{2}{c}{(16) $\pm$ (17)}&\multicolumn{2}{c}{(20) $\pm$ (21)}&\multicolumn{1}{c}{(22)}&\multicolumn{1}{c}{(23)}&\multicolumn{1}{c}{(24)}\\\hline
160&LRL&56.28257&32.00969&13.66&--5.20e--2&3.8e--3&--8.86e--2&7.8e--3&57.4&10.8& 64.8&13.42&2.63e--1&4.2e--2&1.53e+0 &9.8e--2&36.2&27.3& 95.3\\ 
161&G&56.28887&33.13569&10.76&  2.04e--1&2.1e--2&  3.34e--1&4.8e--2&27.9& 8.4&129.9&14.11&3.78e--1&3.4e--2&4.82e--1&9.3e--2&13.5&13.5& 56.7\\
162&PSZ&56.33264&32.04403&0.000&  5.29e--2&2.3e--2&  1.55e--1&7.5e--2&41.7&36.5&172.3&9.917&1.65e--1&5.3e--2&8.38e--1&8.7e--2&38.9&20.6&165.6\\ 
163&W&56.33779&31.72232&0.000&  4.83e--2&1.9e--2&--4.78e--2&7.1e--2&46.2&40.0& 92.5&0.000&6.34e--2&3.3e--2&3.37e--1&7.4e--2&45.8&37.6& 88.7\\ 
164&W&56.48066&32.68635&0.000&  2.50e--6&1.2e--2&  7.17e--5&5.9e--2&62.5&54.1&106.3&0.000&2.93e--2&2.7e--2&3.55e--1&7.0e--2&54.8&37.0&112.2\\
\hline 
\end{tabular}
\end{table}

\begin{table}
\tiny
\centering
\caption{Same as Table E.1 for 250~$\mu$m and 350~$\mu$m.}
\begin{tabular}{rcrcrcrrrcrcrcrrr}
\hline\hline
No.&Sig$_{250}$&\multicolumn{2}{c}{$S^\mathrm{peak}_{250}$}&\multicolumn{2}{c}{$S^\mathrm{tot}_{250}$}&\multicolumn{1}{c}{r$^\mathrm{a}_{250}$}&\multicolumn{1}{c}{r$^\mathrm{b}_{250}$}&\multicolumn{1}{c}{PA$_{250}$}&Sig$_{350}$&\multicolumn{2}{c}{$S^\mathrm{peak}_{350}$}&\multicolumn{2}{c}{$S^\mathrm{tot}_{350}$}&\multicolumn{1}{c}{r$^\mathrm{a}_{350}$}&\multicolumn{1}{c}{r$^\mathrm{b}_{350}$}&\multicolumn{1}{c}{PA$_{350}$}\\
&&\multicolumn{2}{c}{(Jy/beam)}&\multicolumn{2}{c}{(Jy)}&\multicolumn{1}{c}{(\arcsec)}&\multicolumn{1}{c}{(\arcsec)}&\multicolumn{1}{c}{($^\circ$)}&&\multicolumn{2}{c}{(Jy/beam)}&\multicolumn{2}{c}{(Jy)}&\multicolumn{1}{c}{(\arcsec)}&\multicolumn{1}{c}{(\arcsec)}&\multicolumn{1}{c}{($^\circ$)}\\
(1)&(25)&\multicolumn{2}{c}{(26) $\pm$ (27)}&\multicolumn{2}{c}{(30) $\pm$ (31)}&\multicolumn{1}{c}{(32)}&\multicolumn{1}{c}{(33)}&\multicolumn{1}{c}{(34)}&(35)&\multicolumn{2}{c}{(36) $\pm$ (37)}&\multicolumn{2}{c}{(40) $\pm$ (41)}&\multicolumn{1}{c}{(42)}&\multicolumn{1}{c}{(43)}&\multicolumn{1}{c}{(44)}\\\hline
 1&9.220&2.64e--1&3.9e--2&  1.05e+0 &7.1e--2&35.8&30.5& 67.6&10.40&2.58e--1&3.9e--2&  6.09e--1&5.5e--2&36.2&32.1& 91.2\\
 2&7.151&2.02e--1&2.8e--2&  6.30e--1&4.3e--2&37.4&24.2& 75.1&8.910&2.21e--1&1.8e--2&  4.18e--1&2.2e--2&35.2&26.9& 70.2\\
 3&18.46&5.30e--1&1.9e--1&  1.41e+0 &2.3e--1&42.2&18.2&102.8&29.90&9.23e--1&1.7e--1&  1.76e+0 &1.8e--1&42.8&24.9&106.7\\
 4&12.45&2.62e--1&1.0e--1&  4.85e--1&1.0e--1&28.9&19.2& 32.4&15.93&3.08e--1&1.2e--1&  3.96e--1&1.0e--1&31.4&24.9& 42.0\\
 5&12.08&4.41e--1&6.4e--2&  4.32e+0 &1.6e--1&64.6&43.5& 53.1&16.96&5.37e--1&6.6e--2&  2.70e+0 &1.2e--1&64.1&40.9& 58.5\\
\hline
\end{tabular}
\end{table}
\end{landscape}

\setcounter{table}{1}
\begin{landscape}
\begin{table}
\tiny
\centering
\caption{\textbf{continued}}
\begin{tabular}{ccrcrcrrrcrcrcrrr}
\hline\hline
No.&Sig$_{250}$&\multicolumn{2}{c}{$S^\mathrm{peak}_{250}$}&\multicolumn{2}{c}{$S^\mathrm{tot}_{250}$}&\multicolumn{1}{c}{r$^\mathrm{a}_{250}$}&\multicolumn{1}{c}{r$^\mathrm{b}_{250}$}&\multicolumn{1}{c}{PA$_{250}$}&Sig$_{350}$&\multicolumn{2}{c}{$S^\mathrm{peak}_{350}$}&\multicolumn{2}{c}{$S^\mathrm{tot}_{350}$}&\multicolumn{1}{c}{r$^\mathrm{a}_{350}$}&\multicolumn{1}{c}{r$^\mathrm{b}_{350}$}&\multicolumn{1}{c}{PA$_{350}$}\\
&&\multicolumn{2}{c}{(Jy/beam)}&\multicolumn{2}{c}{(Jy)}&\multicolumn{1}{c}{(\arcsec)}&\multicolumn{1}{c}{(\arcsec)}&\multicolumn{1}{c}{($^\circ$)}&&\multicolumn{2}{c}{(Jy/beam)}&\multicolumn{2}{c}{(Jy)}&\multicolumn{1}{c}{(\arcsec)}&\multicolumn{1}{c}{(\arcsec)}&\multicolumn{1}{c}{($^\circ$)}\\
(1)&(25)&\multicolumn{2}{c}{(26) $\pm$ (27)}&\multicolumn{2}{c}{(30) $\pm$ (31)}&\multicolumn{1}{c}{(32)}&\multicolumn{1}{c}{(33)}&\multicolumn{1}{c}{(34)}&(35)&\multicolumn{2}{c}{(36) $\pm$ (37)}&\multicolumn{2}{c}{(40) $\pm$ (41)}&\multicolumn{1}{c}{(42)}&\multicolumn{1}{c}{(43)}&\multicolumn{1}{c}{(44)}\\\hline
35&80.78&2.13e+0 &1.1e--1&  3.57e+0 &1.3e--1&22.7&19.2& 81.7&80.77&2.19e+0 &2.1e--1&  2.82e+0 &2.0e--1&27.9&24.9& 66.7\\
36&27.49&5.97e--1&1.0e--1&  7.22e--1&9.8e--2&21.9&18.2& 29.9&26.90&7.51e--1&1.6e--1&  8.28e--1&1.4e--1&24.9&24.9& 67.5\\
37&91.96&2.00e+0 &1.1e--1&1.99e+0 &1.1e--1&18.2&18.2&166.3&82.59&1.71e+0 &1.0e--1&1.53e+0 &9.8e--2&24.9&24.9&171.2\\
38&39.96&1.50e+0 &2.0e--1&1.98e+0 &1.8e--1&21.9&18.2& 79.2&30.46&2.09e+0 &1.7e--1&3.41e+0 &1.6e--1&32.3&27.8& 95.2\\
39&10.36&2.72e--1&6.3e--2&1.36e+0 &1.0e--1&58.9&30.6& 17.3&15.61&4.08e--1&6.9e--2&1.01e+0 &8.6e--2&61.6&28.2& 16.1\\
\hline
\end{tabular}
\end{table}
\end{landscape}

\setcounter{table}{1}
\begin{landscape}
\begin{table}
\tiny
\centering
\caption{\textbf{continued}}
\begin{tabular}{rcrcrcrrrcrcrcrrr}
\hline\hline
No.&Sig$_{250}$&\multicolumn{2}{c}{$S^\mathrm{peak}_{250}$}&\multicolumn{2}{c}{$S^\mathrm{tot}_{250}$}&\multicolumn{1}{c}{r$^\mathrm{a}_{250}$}&\multicolumn{1}{c}{r$^\mathrm{b}_{250}$}&\multicolumn{1}{c}{PA$_{250}$}&Sig$_{350}$&\multicolumn{2}{c}{$S^\mathrm{peak}_{350}$}&\multicolumn{2}{c}{$S^\mathrm{tot}_{350}$}&\multicolumn{1}{c}{r$^\mathrm{a}_{350}$}&\multicolumn{1}{c}{r$^\mathrm{b}_{350}$}&\multicolumn{1}{c}{PA$_{350}$}\\
&&\multicolumn{2}{c}{(Jy/beam)}&\multicolumn{2}{c}{(Jy)}&\multicolumn{1}{c}{(\arcsec)}&\multicolumn{1}{c}{(\arcsec)}&\multicolumn{1}{c}{($^\circ$)}&&\multicolumn{2}{c}{(Jy/beam)}&\multicolumn{2}{c}{(Jy)}&\multicolumn{1}{c}{(\arcsec)}&\multicolumn{1}{c}{(\arcsec)}&\multicolumn{1}{c}{($^\circ$)}\\
(1)&(25)&\multicolumn{2}{c}{(26) $\pm$ (27)}&\multicolumn{2}{c}{(30) $\pm$ (31)}&\multicolumn{1}{c}{(32)}&\multicolumn{1}{c}{(33)}&\multicolumn{1}{c}{(34)}&(35)&\multicolumn{2}{c}{(36) $\pm$ (37)}&\multicolumn{2}{c}{(40) $\pm$ (41)}&\multicolumn{1}{c}{(42)}&\multicolumn{1}{c}{(43)}&\multicolumn{1}{c}{(44)}\\\hline
 87&6.218&3.12e--1&7.9e--2&3.15e+0 &2.5e--1&56.7&44.5&110.1&11.94&3.26e--1&1.0e--1&1.40e+0 &1.9e--1&73.7&33.1&179.3\\
 88&12.48&2.50e--1&1.6e--2&2.11e--1&1.5e--2&18.2&18.2&155.7&13.23&2.72e--1&1.6e--2&2.09e--1&1.4e--2&24.9&24.9&130.7\\
 89&7.045&1.89e--1&4.3e--2&4.28e--1&4.9e--2&41.2&18.2&174.5&7.230&1.62e--1&3.3e--2&2.46e--1&3.4e--2&39.8&24.9&175.2\\
 90&28.73&5.93e--1&1.9e--2&4.26e--1&1.7e--2&18.2&18.2&170.3&21.46&4.43e--1&3.5e--2&3.44e--1&3.2e--2&24.9&24.9&171.2\\
 91&6.648&1.67e--1&7.1e--2&2.34e--1&7.0e--2&31.4&18.7& 63.1&10.41&2.06e--1&8.9e--2&2.47e--1&8.1e--2&33.4&24.9& 71.2\\
\hline
\end{tabular}
\end{table}
\end{landscape}

\setcounter{table}{1}
\begin{landscape}
\begin{table}
\tiny
\centering
\caption{\textbf{continued}}
\begin{tabular}{ccrcrcrrrcrcrcrrr}
\hline\hline
No.&Sig$_{250}$&\multicolumn{2}{c}{$S^\mathrm{peak}_{250}$}&\multicolumn{2}{c}{$S^\mathrm{tot}_{250}$}&\multicolumn{1}{c}{r$^\mathrm{a}_{250}$}&\multicolumn{1}{c}{r$^\mathrm{b}_{250}$}&\multicolumn{1}{c}{PA$_{250}$}&Sig$_{350}$&\multicolumn{2}{c}{$S^\mathrm{peak}_{350}$}&\multicolumn{2}{c}{$S^\mathrm{tot}_{350}$}&\multicolumn{1}{c}{r$^\mathrm{a}_{350}$}&\multicolumn{1}{c}{r$^\mathrm{b}_{350}$}&\multicolumn{1}{c}{PA$_{350}$}\\
&&\multicolumn{2}{c}{(Jy/beam)}&\multicolumn{2}{c}{(Jy)}&\multicolumn{1}{c}{(\arcsec)}&\multicolumn{1}{c}{(\arcsec)}&\multicolumn{1}{c}{($^\circ$)}&&\multicolumn{2}{c}{(Jy/beam)}&\multicolumn{2}{c}{(Jy)}&\multicolumn{1}{c}{(\arcsec)}&\multicolumn{1}{c}{(\arcsec)}&\multicolumn{1}{c}{($^\circ$)}\\
(1)&(25)&\multicolumn{2}{c}{(26) $\pm$ (27)}&\multicolumn{2}{c}{(30) $\pm$ (31)}&\multicolumn{1}{c}{(32)}&\multicolumn{1}{c}{(33)}&\multicolumn{1}{c}{(34)}&(35)&\multicolumn{2}{c}{(36) $\pm$ (37)}&\multicolumn{2}{c}{(40) $\pm$ (41)}&\multicolumn{1}{c}{(42)}&\multicolumn{1}{c}{(43)}&\multicolumn{1}{c}{(44)}\\\hline
140&8.379&2.22e--1&4.4e--2&5.87e--1&5.1e--2&43.2&19.2&159.4&11.49&2.94e--1&5.9e--2&4.81e--1&6.1e--2&42.8&24.9&153.7\\
141&10.23&3.00e--1&6.8e--2&7.67e--1&8.8e--2&40.0&18.4& 32.2&7.819&2.59e--1&6.4e--2&5.17e--1&7.3e--2&42.3&24.9& 33.0\\
142&12.96&3.10e--1&6.4e--2&2.84e--1&5.6e--2&18.2&18.2&135.0&6.493&1.44e--1&6.0e--2&1.62e--1&5.4e--2&27.1&24.9&129.3\\
143&7.395&2.08e--1&6.8e--2&1.18e+0 &1.2e--1&50.9&33.8& 74.5&8.008&2.10e--1&5.0e--2&5.95e--1&6.8e--2&47.1&32.3& 85.8\\
144&6.079&2.55e--1&7.8e--2&1.82e+0 &1.6e--1&57.6&34.0& 10.9&7.791&2.18e--1&8.5e--2&9.83e--1&1.4e--1&64.2&37.9&  4.2\\
\hline
\end{tabular}
\end{table}
\end{landscape}

\begin{table*}
\tiny
\centering
\caption{Same as Table E.1 for 500~$\mu$m and high-resolution column density map.}
\begin{tabular}{rcccccccccccccccccc}
\hline\hline No.&Sig$_{500}$&\multicolumn{2}{c}{$S^\mathrm{peak}_{500}$}&\multicolumn{2}{c}{$S^\mathrm{tot}_{500}$}&\multicolumn{1}{c}{r$^\mathrm{a}_{500}$}&\multicolumn{1}{c}{r$^\mathrm{b}_{500}$}&\multicolumn{1}{c}{PA$_{500}$}&Sig$_{N_{H_2}}$&\multicolumn{1}{c}{$N^\mathrm{peak}_{H_2}$}&\multicolumn{1}{c}{r$^\mathrm{a}_{H_2}$}&\multicolumn{1}{c}{r$^\mathrm{b}_{H_2}$}&\multicolumn{1}{c}{PA$_{H_2}$}\\
&&\multicolumn{2}{c}{(Jy/beam)}&\multicolumn{2}{c}{(Jy)}&\multicolumn{1}{c}{(\arcsec)}&\multicolumn{1}{c}{(\arcsec)}&\multicolumn{1}{c}{($^\circ$)}&&\multicolumn{1}{c}{($10^{20}$~cm$^{-2}$)}&\multicolumn{1}{c}{(\arcsec)}&\multicolumn{1}{c}{(\arcsec)}&\multicolumn{1}{c}{($^\circ$)}\\(1)&(45)&\multicolumn{2}{c}{(46) $\pm$ (47)}&\multicolumn{2}{c}{(49) $\pm$ (50)}&\multicolumn{1}{c}{(51)}&\multicolumn{1}{c}{(52)}&\multicolumn{1}{c}{(53)}&(54)&\multicolumn{1}{c}{(55)}&\multicolumn{1}{c}{(59)}&\multicolumn{1}{c}{(60)}&\multicolumn{1}{c}{(61)}\\\hline
 1&8.059&  1.77e--1&4.8e--2&  2.49e--1&4.7e--2&44.3&37.6&131.8&12.36&8.45e--2&37.2&31.0& 88.2\\
 2&7.044&  1.38e--1&2.2e--2&  1.63e--1&2.0e--2&49.4&36.3& 78.7&10.68&6.56e--2&41.0&32.8& 87.7\\
 3&24.31&  8.16e--1&2.0e--1&  8.29e--1&1.8e--1&38.6&36.3&106.4&61.64&4.17e--1&34.5&18.2&113.3\\
 4&13.34&  4.69e--1&1.2e--1&  5.12e--1&1.1e--1&38.2&36.3&  3.0&24.97&1.34e--1&22.6&18.2& 49.7\\
 5&14.81&  4.36e--1&7.0e--2&  1.10e+0 &9.5e--2&63.0&42.3& 54.9&22.97&1.69e--1&61.1&42.3& 52.9\\
\hline
\end{tabular}
\end{table*}

\setcounter{table}{2}
\begin{table*}
\tiny
\centering
\caption{\textbf{continued}}
\begin{tabular}{rcccccccccccccccccc}
\hline\hline No.&Sig$_{500}$&\multicolumn{2}{c}{$S^\mathrm{peak}_{500}$}&\multicolumn{2}{c}{$S^\mathrm{tot}_{500}$}&\multicolumn{1}{c}{r$^\mathrm{a}_{500}$}&\multicolumn{1}{c}{r$^\mathrm{b}_{500}$}&\multicolumn{1}{c}{PA$_{500}$}&Sig$_{N_{H_2}}$&\multicolumn{2}{c}{$N^\mathrm{peak}_{H_2}$}&\multicolumn{1}{c}{r$^\mathrm{a}_{H_2}$}&\multicolumn{1}{c}{r$^\mathrm{b}_{H_2}$}&\multicolumn{1}{c}{PA$_{H_2}$}\\
&&\multicolumn{2}{c}{(Jy/beam)}&\multicolumn{2}{c}{(Jy)}&\multicolumn{1}{c}{(\arcsec)}&\multicolumn{1}{c}{(\arcsec)}&\multicolumn{1}{c}{($^\circ$)}&&\multicolumn{2}{c}{(Jy/beam)}&\multicolumn{1}{c}{(\arcsec)}&\multicolumn{1}{c}{(\arcsec)}&\multicolumn{1}{c}{($^\circ$)}\\
(38)&(39)&\multicolumn{2}{c}{(40) $\pm$ (41)}&\multicolumn{2}{c}{(42) $\pm$ (43)}&\multicolumn{1}{c}{(44)}&\multicolumn{1}{c}{(45)}&\multicolumn{1}{c}{(46)}&(47)&\multicolumn{2}{c}{(48) $\pm$ (49)}&\multicolumn{1}{c}{(50)}&\multicolumn{1}{c}{(51)}&\multicolumn{1}{c}{(52)}\\\hline
 73&19.81&5.04e--1&5.4e--2&  1.05e+0 &7.0e--2&60.1&36.3&100.0&35.28&  2.43e--1& 46.3&25.2& 90.9\\
 74&6.735&2.61e--1&5.6e--2&  3.64e--1&5.1e--2&43.7&38.1&118.7&7.408&  1.08e--1& 23.9&20.7& 43.1\\
 75&0.000&5.35e--3&2.3e--2&  1.41e--3&2.1e--2&36.3&36.3&106.8&3.413&--9.23e--3& 33.1&18.7& 70.4\\
 76&13.44&3.74e--1&1.3e--1&  5.99e--1&1.3e--1&54.1&37.4&137.3&28.05&  1.41e--1&3.4e--2& 49.6&26.8&146.9\\
 77&9.046&1.66e--1&2.5e--2&  1.51e--1&2.3e--2&36.3&36.3& 71.3&15.30&  9.70e--2&1.1e--2& 25.7&24.6& 64.6\\
\hline
\end{tabular}
\end{table*}

\setcounter{table}{2}
\begin{table*}
\tiny
\centering
\caption{\textbf{continued}}
\begin{tabular}{ccccccccccccccccccc}
\hline\hline No.&Sig$_{500}$&\multicolumn{2}{c}{$S^\mathrm{peak}_{500}$}&\multicolumn{2}{c}{$S^\mathrm{tot}_{500}$}&\multicolumn{1}{c}{r$^\mathrm{a}_{500}$}&\multicolumn{1}{c}{r$^\mathrm{b}_{500}$}&\multicolumn{1}{c}{PA$_{500}$}&Sig$_{N_{H_2}}$&\multicolumn{2}{c}{$N^\mathrm{peak}_{H_2}$}&\multicolumn{1}{c}{r$^\mathrm{a}_{H_2}$}&\multicolumn{1}{c}{r$^\mathrm{b}_{H_2}$}&\multicolumn{1}{c}{PA$_{H_2}$}\\
&&\multicolumn{2}{c}{(Jy/beam)}&\multicolumn{2}{c}{(Jy)}&\multicolumn{1}{c}{(\arcsec)}&\multicolumn{1}{c}{(\arcsec)}&\multicolumn{1}{c}{($^\circ$)}&&\multicolumn{2}{c}{(Jy/beam)}&\multicolumn{1}{c}{(\arcsec)}&\multicolumn{1}{c}{(\arcsec)}&\multicolumn{1}{c}{($^\circ$)}\\
(38)&(39)&\multicolumn{2}{c}{(40) $\pm$ (41)}&\multicolumn{2}{c}{(42) $\pm$ (43)}&\multicolumn{1}{c}{(44)}&\multicolumn{1}{c}{(45)}&\multicolumn{1}{c}{(46)}&(47)&\multicolumn{2}{c}{(48) $\pm$ (49)}&\multicolumn{1}{c}{(50)}&\multicolumn{1}{c}{(51)}&\multicolumn{1}{c}{(52)}\\\hline
145&0.000&5.39e--2&4.8e--2&  3.18e--2&4.4e--2&36.3&36.3& 69.3&6.051&  3.48e--2&1.4e--2&37.0&18.2&157.4\\
146&0.000&1.45e--1&6.8e--2&  1.05e--1&6.2e--2&36.3&36.3& 36.0&6.769&  4.29e--2&1.7e--2&25.7&18.2& 71.5\\
147&7.163&1.83e--1&2.3e--2&  3.08e--1&2.5e--2&51.9&36.3&149.0&6.210&  4.19e--2&5.0e--3&49.2&21.3&130.7\\
148&0.000&8.19e--2&5.3e--2&  5.81e--2&4.8e--2&36.9&36.3&113.4&6.691&  4.03e--2&6.9e--3&27.0&18.2&120.1\\
149&30.61&1.05e+0 &4.4e--1&  2.01e+0 &5.7e--1&46.8&43.2& 59.4&6.124&  1.11e--1&3.7e--2&45.7&36.1& 64.3\\
\hline
\end{tabular}
\end{table*}

\end{document}